\documentclass[a4paper,12pt]{article}
\pdfoutput=1
\usepackage{jheppub}
\usepackage{ifpdf}
\usepackage[bb=boondox]{mathalfa}
\graphicspath{ {figures/} }
\usepackage{amsfonts}
\usepackage{amsmath}
\usepackage{amssymb}
\usepackage{epsfig}
\usepackage{tensor}
\usepackage{pdfpages}
\usepackage{bbm}
\usepackage{graphicx,epstopdf}
\usepackage{subcaption}
\usepackage[numbers]{natbib} 
\usepackage[makeroom]{cancel}
\usepackage{hyperref}
\usepackage{array}
\usepackage[export]{adjustbox}

\usepackage[normalem]{ulem}
\usepackage{romannum}

\usepackage{ulem}

\usepackage{longtable}

\usepackage{enumitem}

\newcommand{\eg}{{\it e.g.,}\ }
\newcommand{\ie}{{\it i.e.,}\ }
\newcommand{\viz}{{\it viz,}\ }
\newcommand{\mt}[1]{\textrm{\tiny #1}}
\newcommand{\reef}[1]{(\ref{#1})}

\renewcommand{\(}{\left(}
\renewcommand{\)}{\right)}
\renewcommand{\[}{\left[}
\renewcommand{\]}{\right]}

\newcommand{\mS}{\mathcal{S}}
\newcommand{\GN}{G_\mt{N}}

\newcommand{\mB}{\mathcal{B}}

\newcommand{\mH}{\mathcal{H}}

\newcommand{\tg}{\tilde{g}}

\newcommand{\be}{\begin{equation}}
	\newcommand{\ee}{\end{equation}}
\newcommand{\f}{\frac}
\newcommand{\s}{\sqrt}
\newcommand{\p}{\partial}

\newcommand{\bea}{\begin{eqnarray}}
	\newcommand{\eea}{\end{eqnarray}}
\newcommand{\ba}{\begin{align}}
	\newcommand{\ea}{\end{align}}

\newcommand{\la}{\langle}
\newcommand{\ra}{\rangle}
\newcommand{\DLD}{\Delta L_{\mt{D}}}
\newcommand*{\rom}[1]{\expandafter\@slowromancap\romannumeral #1@}
\newcommand{\TDCFT}{T_{\mt{DCFT}}}

\newcommand{\beq}{\begin{equation}}
	\newcommand{\eeq}{\end{equation}}
\newcommand{\beqa}{\begin{eqnarray}}
	\newcommand{\eeqa}{\end{eqnarray}}

\newcommand{\ellB}{\ell_\mt{B}}
\newcommand{\brane}{\mathcal{B}}

\newcommand{\arctanh}{\text{arctanh}}
\newcommand{\arccosh}{\text{arccosh}}
\newcommand{\arccoth}{\text{arccoth}}
\newcommand{\arcsinh}{\text{arcsinh}}

\newcommand{\RN}[1]{%
	\textup{\uppercase\expandafter{\romannumeral#1}}%
}

\definecolor{browna}{rgb}{0.76,0.72,0.65}

\subheader{\today}
\title{\boldmath Double Holography of Entangled Universes}

\author[a]{Robert C. Myers,}
\author[b]{Shan-Ming Ruan}
\author[b,c]{and Tomonori Ugajin}

\affiliation[a]{Perimeter Institute for Theoretical Physics, \\
	Waterloo, ON N2L 2Y5, Canada}
\affiliation[b]{Center for Gravitational Physics and Quantum Information, \\
	Yukawa Institute for Theoretical Physics, Kyoto University,\\
	Kitashirakawa Oiwakecho, Sakyo-ku, Kyoto 606-8502, Japan}
\affiliation[c]{Department of Physics, Rikkyo University, Toshima, Tokyo 171-8501, Japan  }

\emailAdd{rmyers@perimeterinstitute.ca}
\emailAdd{ruan.shanming@yukawa.kyoto-u.ac.jp}
\emailAdd{ugajin@rikkyo.ac.jp }

\abstract{We employ double holography to examine a system of two entangled gravitating universes that live on two codimension-one branes in an asymptotically AdS$_3$ spacetime with two disjoint conformal boundaries. There are distinct brane configurations depending on the temperature of the thermoﬁeld double (TFD) state between the left and right systems. The topology transition between two branes is naturally identified with the emergence of an Einstein-Rosen bridge connecting the two entangled universes. This doubly holographic construction offers a holographic perspective on gravitational collapse and black hole formation in brane universes. Through this holographic framework, we analyze the quantum information structure of the two gravitating universes. Specifically, we calculate the mutual information between defects present in the boundary theories on the left and right sides. Furthermore, we investigate the decoupling process in the Hayden-Preskill protocol applied to the two copies of the defect field theory and discuss the interpretation of the Yoshida-Kitaev decoding protocol.}

\arxivnumber{2403.XXXX}

\begin{document} 
	
	\begin{flushright}
		\hfill{ YITP-24-33}
	\end{flushright}

	\maketitle
	\flushbottom

\section {Introduction}\label{sec:introduction}
%

Recent studies have shown that many quantum information-theoretic ideas naturally appear in quantum gravity and its holographic realizations. The basic example of this phenomenon is the Ryu-Takayanagi (RT) formula \cite{Ryu:2006bv,Ryu:2006ef,Hubeny:2007xt} in the AdS/CFT, which relates the entanglement entropy of the boundary CFT to the area of an extremal surface in the bulk geometry of the dual gravity theory. The island formula \cite{Almheiri:2019hni} for the entropy of Hawking radiation is obtained by suitably generalizing the RT formula to include quantum corrections \cite{Engelhardt:2014gca,Faulkner:2013ana}. This enables a controlled calculation of the Page curve from the semi-classical description of spacetime (plus quantum field theory on top) \cite{Almheiri:2019psf,Penington:2019npb}. Therefore, we have gained surprising new insights into why semi-classical gravity is consistent with the unitarity of the underlying quantum theory.  One of the surprising outcomes of the island formula implies that the area of the black hole interior called "island" is encoded in the degrees of freedom of Hawking radiation as a quantum error-correcting code, and thus is reconstructable from it in principle. The island formula is proven by a replica calculation of the radiation entropy \cite{Penington:2019kki, Almheiri:2019qdq}. The gravitational path integral for its R\'enyi entropies contains saddles called replica wormholes, and they dominate the path integral after the Page time, resulting in the formation of the island region.

In the original setup of the island formula, the bath collecting Hawking quanta is  non-gravitational.  Upon activating gravity within the bath system, a geometric connection between the black hole and the gravitating bath system shows up and modifies the formula for the entropy when the entanglement between two systems is large, as shown in \cite{Balasubramanian:2021wgd,Miyata:2021qsm} (see also \cite{Anderson:2020vwi,Geng:2020fxl, Geng:2021iyq,Balasubramanian:2021xcm,Liu:2022pan,Balasubramanian:2023xyd} for related studies). In these studies, the authors considered the following setup: First, we start with two disjoint gravitating universes denoted by L and R, where both are asymptotically AdS and each contains an eternal  black hole. On these gravitating universes, we define the bulk effective quantum field theory as QFT$_{\mt L}$, QFT$_{\mt{R}}$, and take these to be the same theory\footnote{For simplicity, the gravity sector was treated as classical.}. The Hilbert space of this sector is given by $\mathcal{H}_{\rm QFT_{\mt{L}}} \otimes \mathcal{H}_{\rm QFT_{\mt R}}$. On this Hilbert space, we consider the following thermofield double (TFD) state, 
\begin{equation}\label{eq:TFDstate}
	|\Psi \ra = \f{1}{\s{Z(\beta)}}\,\sum_{i=0}^{\infty} e^{-\beta E_{i}/2} \,| \psi_{i} \ra_{\mt{\mt{\mt{L}}}}\,  |\psi_{i} \ra_{\mt{R}}  \,, 
\end{equation}
where the inverse temperature $\beta$ is now a parameter controlling the entanglement between these two universes.  This generalizes the setup of \cite{Balasubramanian:2020coy,Balasubramanian:2020xqf}  for the island formula where  one of two universes is non-gravitating.
The main statement in \cite{Balasubramanian:2021wgd} was that in the high temperature limit $\beta \rightarrow 0$ the entanglement entropy of the universe L
\begin{equation}
	S(\rho_{\mt{\mt{L}}}) = -{\rm tr} \rho_{\mt{L}} \log \rho_{\mt{L}}, \quad \rho_{\mt{L}} = {\rm tr}_{\mt{R}} |\Psi \ra \la \Psi | \,,
\end{equation}
is given by the generalized entropy $S_{\rm gen} (\mathrm{L/R})$ on a geometry denoted $\mathrm{L/R}$, which is constructed by suitably connecting the two universes together through a wormhole,  
\begin{equation}
	S(\rho_{\mt{L}}) =S_{\rm gen} (\mathrm{L/R}) =  \min_{C} \left[ \underset{C}{\text{ext}}  \( \f{{\rm Area}_{\mt{L/R}}[\p C]}{4G_{N}} +S_{\rm{eff}}[C] \)\right]  \,,
\end{equation}
where  $C$ is a codimension-two surface in the wormhole on $\mathrm{L/R}$,  ${\rm Area}_{\mt{L/R}}[\p C]$ denotes the area of the boundary of $C$ on  the connected geometry $\mathrm{L/R}$, and   $S_{\rm{eff}}[C]$ is the entanglement entropy of the bulk effective QFT on $\mathrm{L/R}$.  If each of these universes is an asymptotic AdS with two boundaries, then $\mathrm{L/R}$ again has two  boundaries, and its metric is obtained by solving Einstein equations with the boundary condition for $\rm{L}$ on one of these boundaries and for $\rm{R}$ on the other. Notably, this calculation is consistent with the ER=EPR hypothesis \cite{Maldacena:2013xja}, which suggests that strong quantum entanglement can lead to the emergence of an Einstein-Rosen bridge in quantum gravity.  

Doubly holographic framework \cite{Almheiri:2019hni,Chen:2020uac,Chen:2020hmv,Hernandez:2020nem,Grimaldi:2022suv} is an efficient tool to investigate quantum extremal surfaces and the black hole information paradox for AdS black holes, \eg see \cite{Chen:2019uhq,Rozali:2019day,Balasubramanian:2020hfs,Chen:2020jvn,Caceres:2020jcn,Krishnan:2020fer,Sully:2020pza,Omiya:2021olc,Neuenfeld:2021bsb,Neuenfeld:2021wbl,Chu:2021gdb,Li:2021dmf,Wang:2021xih,Geng:2021wcq,Ling:2021vxe,Geng:2022slq,Suzuki:2022xwv,Karch:2022rvr,Deng:2022yll,Anous:2022wqh,Neuenfeld:2023svs,Chang:2023gkt,Liu:2023ggg,Basak:2023bnc,Jeong:2023lkc,Kawamoto:2023wzj} for various studies about double holography. The configuration involves a $d+1$ dimensional asymptotically anti-de Sitter (AdS) space, acting as the dual to a holographic conformal field theory governing Hawking radiation. Additionally, there is a codimension-one brane housing a black hole and extending to the conformal boundary. The asymptotic AdS space's conformal boundary acts as a heat bath, collecting radiation emitted by the black hole. This perspective is termed the bulk perspective, as it realizes the original system of the evaporating black hole on the codimension-one surface of the $d+1$ dimensional semi-classical spacetime. The alternative description, known as the CFT perspective, is non-gravitational. It involves a boundary CFT on the $d$-dimensional conformal boundary coupled to defects that are dual to the black hole on the brane. In this framework, the radiation's entropy corresponds to the entanglement entropy of the boundary CFT in the presence of these defects (dual to the brane). The holographic calculation utilizes the standard RT formula, revealing the dominance of the extremal surface ending on the brane at sufficiently late times, leading to the formation of the island region in the black hole interior.

This paper delves into the dynamics of two entangled gravitating universes in two dimensions, from the three-dimensional bulk perspective. In contrast to the discussion in \cite{Balasubramanian:2021wgd}, which focused on two-dimensional JT gravity with limited calculations due to multiple saddles in the gravitational path integral, our approach, based on the three-dimensional bulk perspective, allows for a comprehensive exploration of the entanglement entropy.\footnote{The effective gravitational action on the brane takes a nonlocal Polyakov form \cite{Chen:2020uac,Chen:2020hmv,Grimaldi:2022suv} in the present paper (see the discussion in Appendix \ref{sec:braneaction}), but the analysis is easily extended to Jackiw-Teitelboim (JT) gravity gravity, as discussed in \cite{Grimaldi:2022suv}.}  The power of holographic duality enables us to follow the entire temperature dependence. The analysis uncovers a diverse phase structure dependent on $\beta$ and the distance between the brane endpoints. We model the matter's entangled state \eqref{eq:TFDstate} using a BTZ black hole in three-dimensional bulk spacetime, which is dual to the thermofield double state. The two-dimensional universe $L$ corresponds to two defects, $D^{(\mt{L})}_{a}$ and $D^{(\mt{L})}_{b}$, situated on the left boundary. Similarly, the universe R is represented by the defects $D^{(\mt{R})}_{a}$ and $D^{(\mt{R})}_{b}$ on the opposite boundary. 

\subsection{Double Holography}
Motivated by the framework explored in \cite{Balasubramanian:2021wgd}, we investigate a doubly holographic setup involving two gravitating branes entangled by matter degrees of freedom. This scenario offers three equivalent perspectives for understanding the system \cite{Chen:2020uac}.

{\bf The brane perspective:} From this point of view, the system comprises two disjoint asymptotically AdS$_2$ spacetimes, denoted as the left universe $U_{\mt{L}}$ and the right universe $U_{\mt{R}}$. Each possesses two one-dimensional boundaries, represented by $D^{(\mt{L})}_{a}$, $D^{(\mt{L})}_{b}$ for $U_{\mt{L}}$, and $D^{(\mt{R})}_{a} $, $D^{(\mt{R})}_{b} $ for $U_{\mt{R}}$. Additionally, a two-dimensional non-gravitational bath (${\bf Bath}_{\mt{L}}$ and ${\bf Bath}_{\mt{R}}$) is prepared for each spacetime. ${\bf Bath}_{\mt{L}}$ is attached to $U_{\mt{L}}$ at its asymptotic boundaries $D^{(\mt{L})}_{a}$, $D^{(\mt{L})}_{b}$, and similarly for the right universe. The degrees of freedom of CFT matter can propagate in both the gravitational and non-gravitational regions. We assume these two non-gravitational baths are disjoint, ensuring no direct interactions between the total regions $U_{\mt{L}}\cup {\bf Bath}_{\mt{L}}$ and $U_{\mt{R}}\cup {\bf Bath}_{\mt{R}}$. The quantum state on the total system is thus the TFD state \eqref{eq:TFDstate}.

{\bf The boundary perspective:} Given that $U_{\mt{L}}$ and $U_{\mt{R}}$ are asymptotically AdS$_{2}$, it is natural to anticipate the dual holographic description. In this perspective, $U_{\mt{L}}$ is dual to ``defect degrees of freedom" on its asymptotic boundaries $D^{(\mt{L})}_{a} , D^{(\mt{L})}_{b}$. The total system involves ambient matter modes on ${\rm Bath}_{\mt{L}}$ interacting with modes on two defects $ D^{(\mt{L})}_{a} , D^{(\mt{L})}_{b}$. A similar system is constructed for ${\rm Bath}_{\mt{R}} \cup D^{(\mt{R})}_{a} \cup D^{(\mt{R})}_{b} $, entangled with the left. To distinguish from the standard CFT, we will refer to the boundary field theory as the defect conformal field theory, denoted by DCFT. The boundary setup as illustrated in figure \ref{fig:DFT01} is characterized by the inverse temperature $\beta$ as well as $L_{1}$ and $L_{2}$ measuring the distance between two defects.

{\bf The bulk perspective}: In this paper, we assume that the matter QFT is a conformal field theory with a large central charge. Then the boundary entangled state \eqref{eq:TFDstate} has a dual gravity description in AdS$_3$ bulk spacetime. The dual geometry for the bulk spacetime is either the thermal AdS$_3$ or BTZ black hole, depending on the temperature. Two asymptotic boundaries of the AdS$_3$ bulk spacetime correspond to the bath regions ${\bf Bath}_{\mt{L}}$, ${\bf Bath}_{\mt{R}}$. Furthermore, in this setup the 2D gravitating universe in the brane perspective is realized as a codimension-one brane in AdS$_3$ bulk spacetime, which anchors on the asymptotic boundary at the locations of the defects. Because we require the brane to be sufficiently massive, it will backreact to the 3D bulk geometry. The backreaction of the codimension-one brane is treated by first preparing two AdS$_{3}$ (the masses $m_{1}, m_{2}$ of two spacetimes can be different in general) and then gluing two spacetime regions (say $\mathcal{S}_{1}$ and $\mathcal{S}_{2}$) along the brane, as depicted in figure \ref{fig:simp1}. The brane profile is determined by solving the junction conditions between the two sides, which are given by the so-called Israel junction condition.

\subsection{Boundary perspective: entangled DCFTs}
\begin{figure}[t]
	\centering
	\includegraphics[width=3.5in]{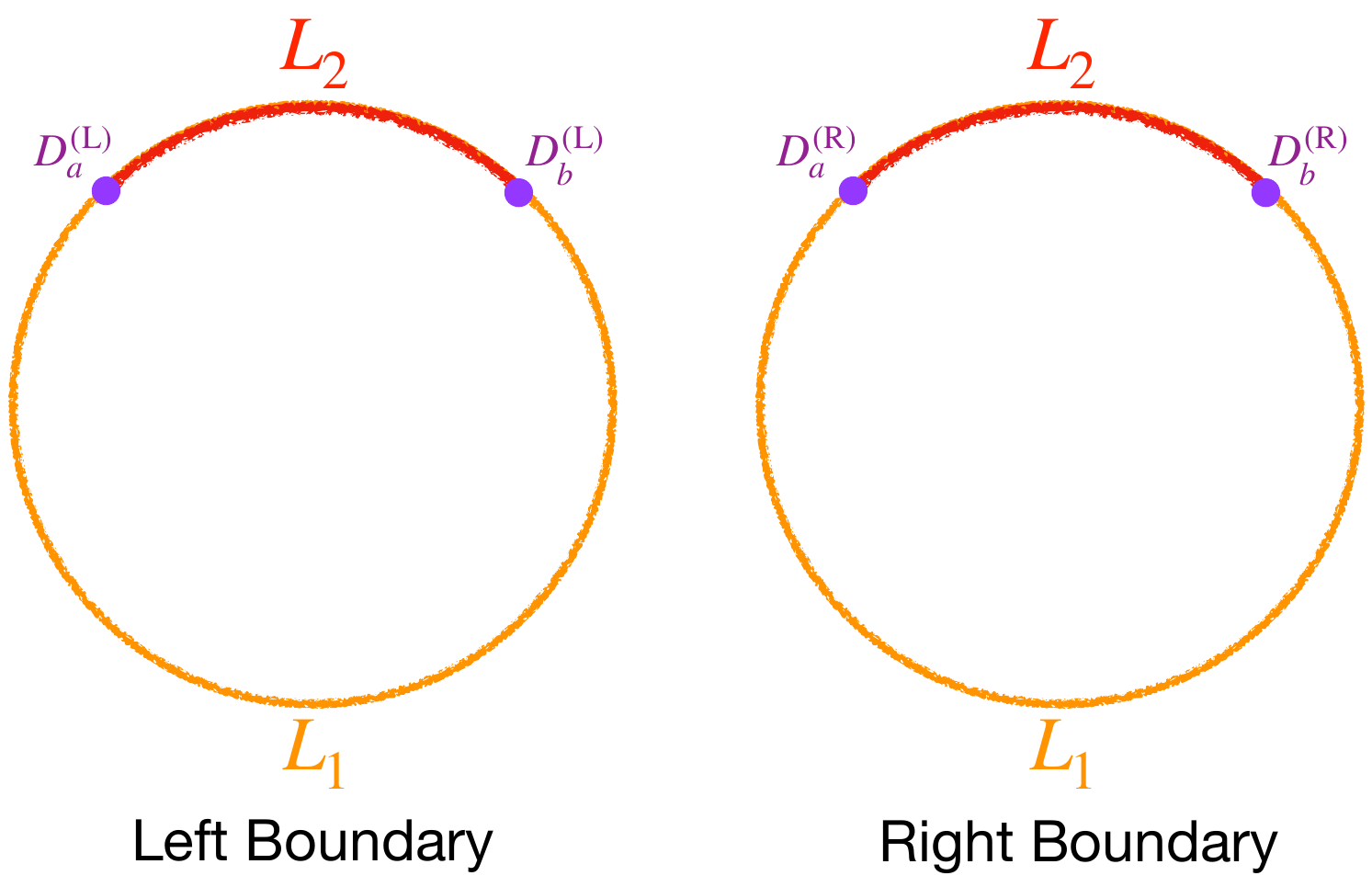}
	\caption{Boundary perspective: The two defect field theories at the temperature $T_{\mt{DCFT}}$ are on living on the left and right boundaries, respectively and are entangled together.}\label{fig:DFT01}
\end{figure}

Here, we briefly discuss the information we need to fix the boundary setup. In the main text, the dual bulk spacetime would be determined by using the boundary data. For the left/right boundary, we consider a boundary defect field theory living on a circle as shown in figure \ref{fig:DFT01}. The metric characterizing the conformal boundary takes the form:
\begin{equation}
ds^2 \big|_{\rm{bdy}}  =  -dt^2 +   R_{\rm bdy}^2  d \phi^2  \,.
	\label{bdymetric}
\end{equation}
The presence of two defects on the boundary circle partitions the boundary theories into distinct regions, each with a fixed length $L_1$ and $L_2$. In other words, the two-dimensional boundary theory is living on an infinitely long cylinder whose circumference is fixed as $L_{\rm bdy} \equiv L_1 +L_2= 2\pi R_{\rm bdy}$. To maintain simplicity, we posit that the defect field theories residing within the two regions share an identical central charge, namely $c_1 = c = c_2$. Within this context, the set of boundary data encompasses five independent  parameters: $(c,T_{\mathrm{DCFT}}, L_1, L_2)$ along with the defect entropy $2\log g$, which is determined by the boundary condition at the defects and measures the ground state degeneracy of the defect -- see eq.~\reef{eq:defg}. However, because the boundary theory is conformal, the physics will only depend on dimensionless combinations of these parameters. Hence we can parametrize the boundary physics by the four following combinations:
\begin{equation}\label{eq:parameters}
\left\{ c\,, \quad \log g\,, \quad 	T_{\mt{DCFT}} \(   L_1 +   L_2 \)\,, \quad  \frac{  L_1}{  L_2} \right\}
\end{equation}
with each of these parameters retaining its scaling-invariant nature.
	
\subsection{Bulk perspective: summary of three possible phases}
\begin{figure}[t]
	\centering
	\includegraphics[width=5.5in]{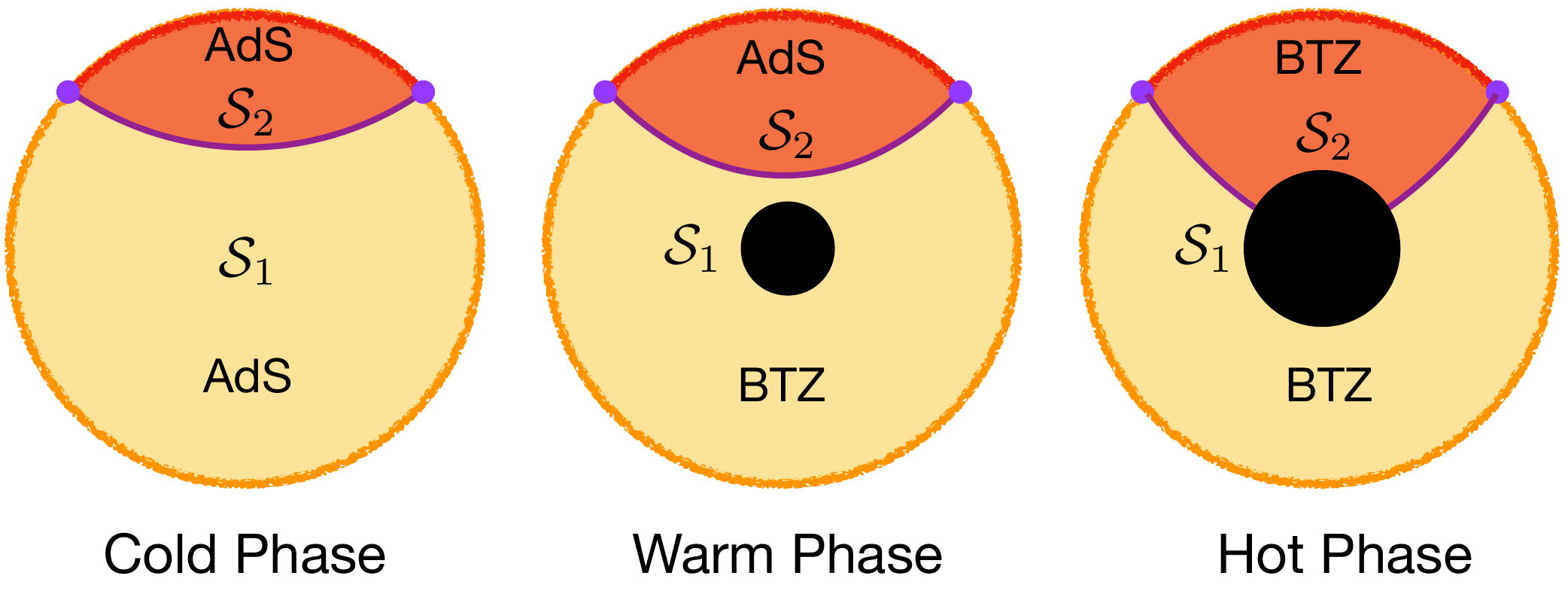}
	\caption{Sketch illustrating a time slice in the bulk for the three possible phases. We are only showing a single asymptotic boundary. }\label{fig:simp1}
\end{figure}

In the following section, we will discuss potential phases from the 3D bulk perspective, focusing on plausible bulk brane profiles that connect four defects on the boundary, given a specific set of boundary parameters \eqref{eq:parameters}. Generally, three distinct phases characterize the bulk spacetime \cite{Bachas:2021fqo}, as illustrated in figure \ref{fig:simp1}.

{\bf Cold phase}: In this scenario, when the temperature is sufficiently low, the 3D bulk geometry consists of two disconnected pieces, denoted as $\mathcal{M}_{\mt{L}}$ and $\mathcal{M}_{\mt{R}}$. In this phase, each brane connects two defects on the same boundary. This implies that the bath CFT is in a confined phase, and the correlation primarily exists between $D^{(\mt{L})}_{a}$ and $D^{(\mt{L})}_{b}$. Consequently, the mutual information between the left and right defects (denoted as $I_{\mt{L:R}}$) is of the order of $O(c^0)$, indicating a low correlation between the left and right.

{\bf Warm phase}: As the temperature increases, the bath CFT becomes deconfined. In the dual gravity description, this corresponds to the Hawking-Page transition, and the bulk geometry transforms into a black hole. At this middle stage, the bulk BTZ black hole remains small and the branes connecting the two defects on the left/right boundary stay outside the black hole horizon. Consequently, the entanglement entropy between the two sides increases, but the correlation between the left and right defects remains negligible.

{\bf Hot phase}: As the system heats up further, the black hole expands, eventually swallowing parts of the branes. In this case, each brane connects two defects - one on the left boundary and the other on the right. Due to the topology change of the brane profiles, the correlation between the left and right defects becomes significant in this hot phase.

\subsection{Entanglement structure of the holographic gravitating universes}
We use this holographic construction to perform a detailed study of the quantum information structure of the two gravitating universes. Our primary focus is on assessing the entanglement between the two branes and how it evolves through the phase transition. In the CFT perspective, this translates into the entanglement between two groups of defects $D^{(\mt{L})}_{a, b}$, with the corresponding quantity being the mutual information between them. We holographically compute this mutual information in each phase by meticulously determining the dominant RT surface for the entanglement entropy of each subregion. The summarized results are shown in figure \ref{fig:Table}. The key observation is that the mutual information is non-trivial only in the hot phase. This is consistent with expectations since each brane in the bulk can be interpreted as a ``bit thread'' \cite{Freedman:2016zud}, providing a visualization of the entanglement structure of the DCFT state.

\subsection{The Hayden-Preskill setup and the Yoshida-Kitaev protocol}
The Hayden-Preskill Gedanken experiment provides a valuable framework for investigating how information thrown into an evaporating black hole is recovered from Hawking radiation, particularly after the Page time. The crucial mechanism for achieving recovery is the phenomenon of decoupling, where the remaining black hole $C$ and the reference system $R$, initially maximally entangled with the diary, rapidly lose correlation as the size of the late Hawking radiation becomes comparable to that of the diary. We demonstrate that the setup of Hayden and Preskill finds a natural realization in our two entangled DCFTs setup. This is accomplished by establishing an explicit correspondence between subsystems in these two setups (see figure \ref{fig:HP}). We then use holography to study this decoupling in the DCFT setup by computing the mutual information $I_{R:C}$ between the radiation and the black hole. Our analysis shows that in the hot phase, the contribution of defects to this mutual information vanishes, indicating a successful decoupling.

Yoshida and Kitaev \cite{Yoshida:2017non} proposed an actual recovery protocol for the diary from early Hawking radiation in the Hayden-Preskill setup. In this protocol, the postselection to the maximally entangled state of the late Hawking radiation and its copy plays a key role, inducing the quantum teleportation of the diary to early Hawking radiation. We show that this postselection procedure has a natural realization in the DCFT setup, achieved by changing the distance between two defects.

\subsection{Organization of this paper}
The rest of the paper is organized as follows: Section \ref{sec:bulk} reviews the holographic construction of the system of interest, \ie the thermofield double state entangling two copies of a two-dimensional defect CFT. Further, we describe the various phases that the system goes through as the temperature increases. This review is primarily based on \cite{Bachas:2021fqo}. While the underlying structure is not simple enough that it can be described analytically, the final expressions for, \eg the phase diagram can be determined numerically. However, there are two domains where the system can be understood fully analytically: $L_1=L_2$ and $L_1\gg L_2$, which are described in sections \ref{sec:Phasediagram} and \ref{sec:sym}  respectively. In section \ref{sec:mutual}, we then turn to examine the quantum entanglements between various subsystems, \eg the quantum correlations between the various defects. We use mutual information as our measure of the strength of these correlations and examine how the mutual information varies in various phases. In the cold and warm phases, the mutual information between the various subsystems is independent of time. In the hot phase, we examine the time dependence of the entanglement and the appearance of quantum extremal islands. We again provide analytical results for $L_1=L_2$ and $L_1\gg L_2$ in sections \ref{sec:mutsym} and \ref{sec:mutualinformation}, respectively. In section \ref{sec:HP}, we demonstrate that this doubly holographic setup can be used to realize the Hayden-Preskill protocol. We conclude in section \ref{sec:discuss} with a discussion of our results and possible future directions. Appendix \ref{sec:brane} provides studies on various aspects of the brane perspective, such as phase transitions on the brane, renormalized action and so on. In appendix \ref{sec:corner}, we describe some technical details of the UV regulator for the bulk branes. Appendix \ref{sec:notes} provides an alternative analysis of the thermodynamics of the defect CFT in the large tension limit.

\section{Bulk Perspective: Three Phases} \label{sec:bulk}
%

In this section, we apply the bulk perspective to examine the holographic dual of entangled defect CFTs on the boundary. As discussed above, the bulk gravitational description involves two codimension-one branes extending through the AdS$_3$ geometry and anchored to the asymptotic boundaries at the location of the boundary defects. In the following discussion, it will be useful to introduce $\mS_{1,2}$ to denote the bulk regions bounded by the branes and the asymptotic boundary regions of length $L_{1,2}$ (as measured by the boundary metric \reef{bdymetric}) -- see figures \ref{fig:DFT01} and \ref{fig:simp1}. In principle, we could consider `interfaces' where the CFTs on either side of the defects are distinct. However, for simplicity, we will assume that the two CFTs are the same, which implies that in the bulk, the cosmological curvature scale $\ell$ is constant throughout the AdS$_3$ geometry.\footnote{The first half of this section is largely a review based on \cite{Bachas:2021fqo}. However, we note this reference considers a more general set-up where the AdS scale differs on either side of the brane.}  Hence the central charges of the two CFTs between the defects are identical, \ie $c_1=c_2 = c$, with
\begin{equation}
	c = \frac{3\ell}{2 \GN} \,. 
	\label{eq:central}
\end{equation} 
While the bulk spacetime is locally equivalent to AdS$_3$ everywhere (and in all of the different phases), the backreaction of the two-dimensional branes enlarges the geometry in the vicinity of the branes, \eg as illustrated in figure \ref{fig:ColdPhase}. 

The boundary theory is also characterized by Affleck-Ludwig entropy \cite{Affleck:1991tk} associated with the defects. This constant can be interpreted as the central charge of the conformal degrees of freedom living on the defects. As we demonstrate below (see eq.~\reef{eq:Sbdy}), it is determined by the brane tension \cite{Grimaldi:2022suv}\footnote{Following \cite{Grimaldi:2022suv}, we include a factor of 2 here because we are considering a defect (with two sides) rather than a boundary. With this normalization, our result matches the holographic calculation of $\log g$ in the usual AdS/BCFT correspondence \cite{Takayanagi:2011zk,Fujita:2011fp}.}
\begin{equation}
	S_{\rm def} \equiv 2\log g = \frac{c}{6}\, \log \( \frac{1+  T_o \ell}{1 - T_o \ell}  \) = \frac{c}{3}\, \arctanh \(  T_o \ell \) \,.  \label{eq:defg}
\end{equation}

The bulk gravitational action has three components 
\begin{equation}\label{eq:3Daction}
		I_{\rm tot} =I_{\rm{bulk}}+I_{\rm{brane}} +I_{\rm{boundary}}\,.
\end{equation}
We describe these terms in Euclidean signature since this choice is relevant for evaluating the action and determining the phase diagram in the subsequent sections.
Then the bulk action is, of course, simply given by the Einstein action with a negative cosmological constant,
\begin{equation}\label{eq:Ibulk}
	I_{\rm{bulk}}	=-\frac{1}{16\pi \GN} \int
	d^3x\,\sqrt{g}\(\mathcal{R}(g) + \frac{2}{\ell^2} \)\,.
\end{equation}
Note that the overall minus sign is associated with the Euclidean signature. We write the brane contribution as
\beq
I_{\rm{brane}} =-\frac{1}{8 \pi \GN} \int
d^2\sigma\,\sqrt{h}\( \Delta K -2 T_o\) \,.
\label{Ibrane}
\eeq
where $\Delta K_{ab}   \equiv K_{ab}^{(1)} + K_{ab}^{(2)}$ denotes the discontinuity of the extrinsic curvature across the brane. One can think of this Gibbons-Hawking-York contribution as representing the curvature contribution that would be created in the bulk Einstein term by the $\delta$-function source of the brane \cite{Israel:1966rt}.  Of course, $h_{ab}$ denotes the induced metric on the brane surface, and the extrinsic curvatures are given by $K_{ab}^{(i)}= h^\mu_a h^\nu_b \nabla_\mu n_\nu^{(i)}$ with the normal vector $n_{\mu}^{(i)}$ pointing outward from the region $\mS_{i}$. Eq.~\reef{Ibrane} also contains the usual worldvolume action of the brane proportional to the tension $T_o$.\footnote{Note the present normalization differs from that in \cite{Chen:2020uac,Chen:2020hmv,Hernandez:2020nem,Grimaldi:2022suv}. That is, $T_o^\mt{(present)}=4\pi\GN\,T_o^\mt{(previous)}$.} 
Finally, we have the boundary terms on the asymptotic boundary,
\beq
I_{\rm{boundary}}=-\frac{1}{8\pi \GN} \int_{\partial(\mS_1 \cup \mS_2)} \sqrt{h}\(K -\frac1\ell\) \,.
\label{Iboundary}
\eeq
with the usual Gibbons-Hawking-York term and the boundary counterterm for $d=2$ \cite{Emparan:1999pm}. As discussed in appendix \ref{sec:corner}, the boundary action may also contain two additional terms associated with the branes, depending on the choice of the cut-off surface near the defects. The first is a Hayward term \cite{Hayward:1993my,Brill:1994mb} for the joint where the asymptotic boundaries of $\mS_{1,2}$ meet at the brane. The second is a boundary counterterm for the gravitational theory on the brane. However, these two contributions precisely cancel (after taking $\epsilon \to 0$) and so for simplicity, we have not included them above in eq.~\reef{Iboundary}. Appendix \ref{sec:corner} provides an explicit discussion of the joint terms and their cancellation.


The dual bulk spacetime is ultimately obtained by gluing the bulk regions $\mS_{i}$ along the two branes $\brane_i$. Since we are in three dimensions, the bulk regions are simply  AdS$_3$ geometries locally because the bulk action \eqref{eq:3Daction} is nothing but the Einstein-Hilbert term involving a negative cosmological constant. Hence we must now determine to profile of the branes. (Note that we are employing Lorentzian signature for this discussion.)

Referring to the bulk coordinates in bulk region $\mS_i$ as $(t_{i}, r_{i}, \phi_{i})$, we can express the metric of each region as:
\begin{equation}\label{eq:AdS3}
	ds^{2} = - \( \frac{r^2_i}{\ell^2} - m_i  \) dt_i^2 + \frac{dr^2_i}{\frac{r^2}{\ell^2} - m_i } + r_i^2 d\phi_i^2 \,,
\end{equation}
where the dimensionless parameter $m_i$ is related to the ADM mass of the gravitational region by $M_{i} = \f{m_{i} }{8G}$. In the following discussion, we consistently consider the cut-off surface positioned at 
\begin{equation}
	r_i \approx \frac{\ell^2}{\epsilon}  + A_i + B_i \epsilon + \mathcal{O}(\epsilon^2) \to \infty \,.
\end{equation}
where we introduce some higher-order terms in order to ensure that the cut-off surface is smooth when it crosses the brane -- see Appendix \ref{sec:corner} for more discussion about this point. For physical quantities that are independent of the choice of the regulator, any extra terms involving the coefficients $A_i, B_i$ will automatically vanish after taking $\epsilon \to 0$. The boundary metric for each interval is fixed as
\begin{equation}
	ds^2\big|_{\rm bdy} = - dt^2_i+ \ell^2 d\phi^2_i \,.
\end{equation}
Given the thermal state on the conformal boundary, we note that the Euclidean time will be periodic.

It is important to highlight that the two mass parameters, $m_1$ and $m_2$, are typically distinct since the boundary sizes, $L_1$ and $L_2$, will generally be different. Moreover, the sign of $m_i$ is not predetermined either. In cases where $m_i$ is negative, the bulk region $\mS_i$ essentially corresponds to a portion of global AdS$_3$. In such instances, it is always possible to reformulate the bulk metric using global coordinates $(T, R, \Omega)$ with 
\begin{equation}
		ds^{2} = - \( \frac{R^2}{\ell^2} +1 \) dT^2 + \( \frac{R^2}{\ell^2} +1 \)^{-1} dR^2 + R^2 d\Phi^2 \,, 
		\label{eq:ads3}
\end{equation}
which is accomplished by rescaling the coordinates as follows
\begin{equation}\label{eq:rescale}
	T= \sqrt{-m_i} \, t_i \,, \quad  R = \frac{r_i}{\sqrt{-m_i}}\,, \quad \Phi = \sqrt{-m_i} \, \phi_i\,.
\end{equation}
Note that in this case, we always adopt the `periodicity' of $\Phi$ to be $2\pi$ and hence of $\phi_i$ to be $2\pi/\sqrt{-m_i}$. Hence there is no conical singularity at the `center' of the geometry, \ie $r_i=0$. 

With positive $m_i$, the bulk metric corresponds to the BTZ black hole, \ie 
\begin{equation}\label{eq:BTZ}
		ds^{2} = - \frac{r^2 - r_h^2}{\ell^2}  \, dt^2 + \frac{\ell^2}{r^2 - r_h^2  } \, dr^2+ r^2 d\phi^2 \,, \quad \text{with} \quad m_i >0\,,
\end{equation}
where the black hole horizon is located at $r=r_h = \sqrt{m_i}\ell$. The temperature of the black hole reads 
\begin{equation}\label{eq:temprature}
	T_{\mt{BH}} \equiv \frac{r_h}{2\pi \ell^2} = \frac{\sqrt{m_i}}{2\pi \ell}= T_{\mt{DCFT}}\,,
\end{equation}
where the latter equality stands as a consequence of the holographic duality.

Now, the position of the branes in the bulk $\mS_i$ is derived by solving the the so-called Israel junction conditions \cite{Israel:1966rt}: 
\begin{equation}\label{eq:Israel}
	 \Delta K_{ab} -\Delta K \, h_{ab} -2 T_o\, h_{ab}=0\,.
\end{equation}
Recall that $\Delta K_{ab}$ was introduced below eq.~\reef{Ibrane} and denotes the discontinuity of the extrinsic curvature across the brane. Implicitly above, we have been using that the induced metric on both sides of the brane must match, \ie $h_{ab}^{(1)}= h_{ab}^{(2)} =h_{ab}$, to successfully glue the two regions $\mS_{1,2}$ together. 

The static solutions for the brane have been extensively discussed in \cite{Bachas:2021fqo}, and we will offer a concise overview of the pertinent results in the following discussion. First, let us denote the brane coordinates as $(t,\sigma)$. Then given our focus on static brane profiles, we can make the gauge choice, that the timelike brane coordinate coincides with that in the bulk, \ie setting $t=t_1 =t_2$. In this context, we can thus write the brane metric as 
\begin{equation}
	ds^2\big|_{\mB} = -f(\sigma) dt^2 + g(\sigma) d\sigma^2\,. 
\end{equation}
The profile of the brane within the bulk spacetime is then determined by four functions, namely $\{ \phi_{i} (\sigma),r_{i}(\sigma)\}$ with $i=1,2$. 

Considering the pullback of the metric to the brane from both sides then yields the following equations:
\begin{equation}
	 f(\sigma) =   \frac{r^2_1}{\ell^2} - m_1   =  \frac{r^2_2}{\ell^2} - m_2 
\end{equation}
and 
\begin{equation}
 g(\sigma) =   \frac{r_1'^2}{f} + r_1^2 \phi'^2_1  = \frac{r_2'^2}{f} + r_2^2 \phi'^2_2 \,.
\end{equation}
Moreover, an additional independent constraint arises from the Israel junction conditions. 
For instance, considering the $(tt)$-component of eq.~\eqref{eq:Israel} yields 
\begin{equation}
 \sqrt{\frac{f(\sigma)}{g(\sigma)}} \frac{r_1^2 \phi'_1+r_2^2 \phi'_2}{\ell^2} = - 2T_o f(\sigma) \,.
\end{equation}
These are three independent equations for four variables. The undetermined degree of freedom corresponds to reparametrization of the spacelike worldvolume coordinate $\sigma$. It is convenient to fix it by imposing a gauge condition $f(\sigma) =\sigma$. In this gauge, we simply obtain
\begin{equation}\label{eq:gauge}
\frac{r_{i} (\sigma)}{\ell} =  \sqrt{\sigma+ m_i} \,, 
\end{equation}
where the {\it dimensionless} coordinate $\sigma$ plays the role of the `radius' on the brane. 
The asymptotic boundary is evidently defined as $\sigma \to \infty$. For the BTZ black hole with positive $m_i$, its black hole horizon is situated at $\sigma=0$ with $r=\sqrt{m_i} \ell \equiv r_h$. 

The induced metric of the brane can now be reformulated as
\begin{equation}
		ds^2\big|_{\mB} = - \sigma dt^2 + \frac{\ell^2}{4(1- T_o^2\ell^2) (\sigma - \sigma_+)(\sigma- \sigma_-)}  d\sigma^2\,, 
\end{equation}
with the two poles in the ($\sigma,\sigma$)-component are given by\footnote{As we will see only the pole at $\sigma_+$ will be physical and it represents a mere coordinate singularity since the $\sigma$ coordinate cannot fully parameterize a complete brane. In appendix \ref{sec:brane}, we introduce an alternative global coordinate system on the brane to avoid this issue.}
	\begin{equation}\label{eq:sigmapm}
		\sigma_\pm = \frac{1}{4(1- T_o^2\ell^2)}  \(  -(m_1+m_2) \pm  \sqrt{ \(\frac{m_1-m_2}{T_o \ell} \)^2+ 4m_1 m_2   }\)\,.
	\end{equation}
The physical solutions can only be obtained under the condition
\begin{equation}
	  |T_o| \le \frac{1}{\ell}\,, 
\end{equation}
which ensures that the brane geometry is asymptotically AdS$_2$, as we will demonstrate later. It is also straightforward to verify that $\sigma_+ \geq 0$ and $\sigma_- < 0$.
Finally, the profiles of the branes are parameterized by $\phi_i (\sigma)$ and can be obtained by integrating
\begin{equation}\label{eq:phiprime}
	\begin{split}
		\phi'_i &= - \frac{4 (T_o\ell)^2 \sigma +m_i -m_j}{ 8T_o\ell (\sigma+ m_i) \sqrt{(1-T_o^2\ell^2)\sigma (\sigma-\sigma_+)(\sigma-\sigma_-)}}\,.\\ 
	\end{split}
\end{equation}
An explicit expression can be found for $\phi_{i} (\sigma)$ involving elliptic functions. We note that
\begin{equation}
	\lim\limits_{\sigma \to \sigma_+} \frac{d\sigma}{d\phi_i} = 0 \,,
\end{equation}
due to the pole in eq.~\reef{eq:phiprime} at $\sigma=\sigma_+$. This implies that the point at $\sigma=\sigma_+$ can be interpreted as a turning point positioned at the middle of the brane $\mB$. Consequently, as the worldvolume radius ranges over $\sigma \in [\sigma_+, +\infty)$, one-half of the brane is covered by the brane extending from this middle point to the boundary defect on the asymptotic boundary. 

\subsection{Three phases of bulk spacetime} \label{sec:threeP}

In the preceding analysis, the bulk regions $\mS_1$ and $\mS_2$ are not fully determined yet, because we have not specified the dimensionless parameters $m_1$ and $m_2$. Depending on the sign of these mass parameters, we can define three distinct phases for the system: the cold phase with $m_1<0$ and $m_2 <0$, the warm phase with $m_1>0$ and $m_2<0$, and the hot phase with $m_1>0$ and $m_2>0$. The significance of these phase names will become evident in the phase diagram shown in figure \ref{fig:PhaseDiagram}. Each of these bulk geometries is characterized by two dimensionless parameters, namely the masses $m_{1}$ and $m_{2}$ of the two sides of the brane. In cases where the brane crosses the horizon, as occurs for $m_i > 0$, we have an additional dimensionless parameter $\Delta \phi_{{\rm Hor}}$, which signifies the area of the black hole horizon in the bulk region $\mS_i$. 

From the perspective of the boundary, the boundary DCFT setup is described by four independent and dimensionless parameters, as described at eq.~\reef{eq:parameters}.
Serving as a holographic duality, the geometric characteristics of the dual bulk spacetime can be deduced from these boundary parameters. For example, the central charge $c$ is determined by the dimensionless ratio $\ell/\GN$, as shown in eq.~\reef{eq:central}, while the boundary entropy $\log g$ is determined by the dimensionless bulk parameter $T_o\ell$. In the following, we will delve into the explicit relations between bulk information and the CFT data ${T_{\mt{DCFT}} (L_1+L_2), L_1/L_2 }$ for each bulk phase. The bulk phase diagram will be determined by evaluating the gravitational path integral using the total action \eqref{eq:3Daction} through the saddle point approximation in the semi-classical limit as $\GN \rightarrow 0$, with the three bulk phases represented by competing saddle points.

\subsubsection{Cold phase ($m_i<0$)}
\begin{figure}[t]
	\centering
	\includegraphics[width=4in]{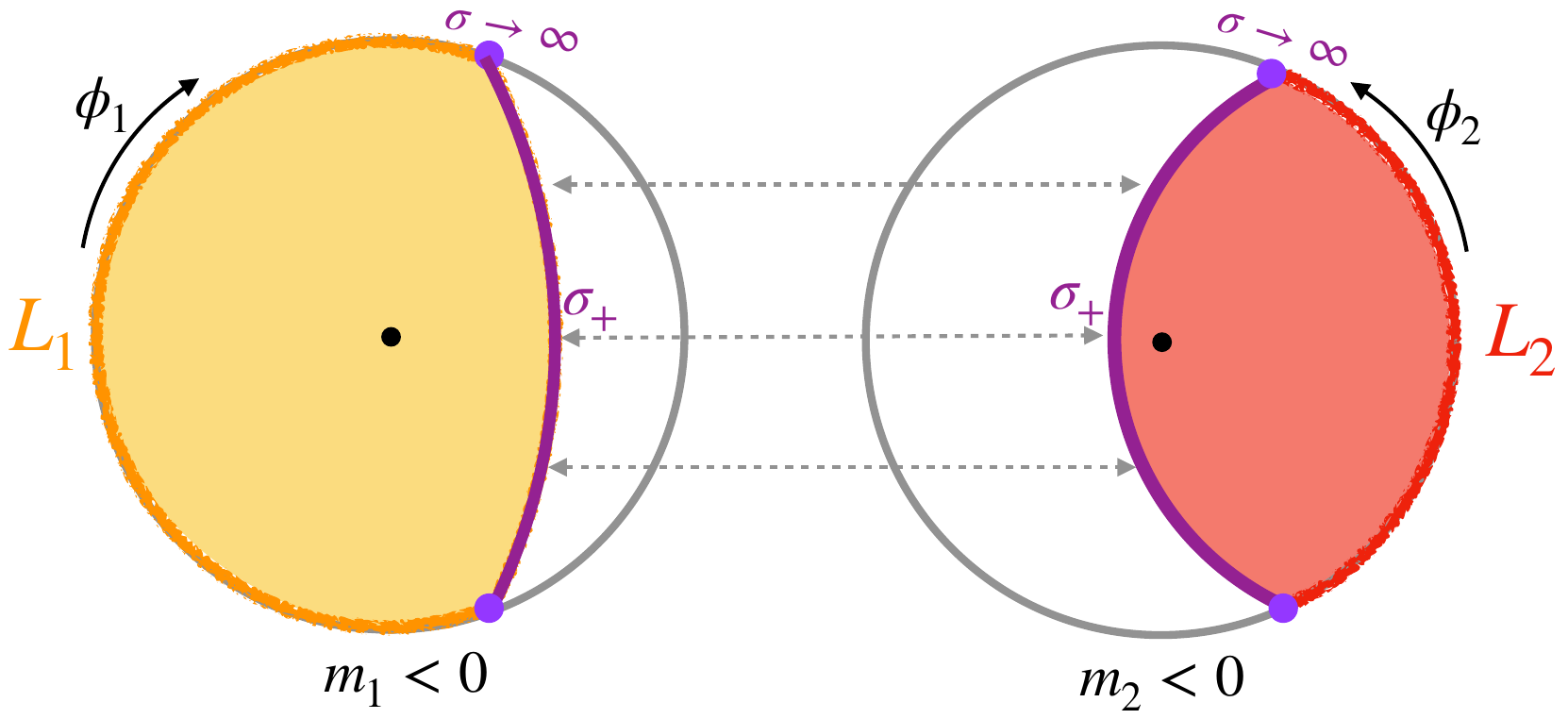}
	\caption{Bulk perspective in the cold phase: the bulk dual spacetime for both the left and right sides is created by joining two portions of AdS$_3$ vacuum geometry with $m_i <0$, along the two-dimensional brane. The codimension-one branes, labeled as $\mB_i$, possess a constant tension $T_o$ and are represented by the purple curve. The shaded regions, colored orange and red, correspond to the bulk regions $\mS_1$ and $\mS_2$, respectively. This configuration characterizes the bulk spacetime in the cold phase. The middle gray dot (not a conical defect) indicates the center of the geometry at $r_i=0$. }\label{fig:ColdPhase}
\end{figure}

We begin with the cold phase by taking $m_1<0, m_2<0$. The construction of the bulk geometry then involves gluing two AdS$_3$ vacua, described by the metric \reef{eq:AdS3} with $m_i < 0$, along the constant-tension branes, as illustrated in figure \ref{fig:ColdPhase}.  In accordance with the analysis in eq.~\eqref{eq:rescale}, the periodicity of the angular coordinate is set as
\begin{equation}
	\phi_{i} \sim \phi_{i} + \frac{2\pi}{\sqrt{-m_i}} \,, 
\end{equation}
to avoid a conical singularity in the bulk $\mS_i$. The profile of the brane, determining the interior boundary of $\mS_i$, is derived from eq.~\eqref{eq:phiprime}. The crucial feature among various bulk geometries for $\mS_i$ is whether the center of the original AdS$_3$, situated at $r=0$, is encompassed by the brane $\mB_i$. If the center of the bulk geometry $\mS_i$ is included by the brane $\mB_i$, the spatial cycle contracts, leading to $\phi'_i (\sigma_+) <0$. Utilizing eq.~\eqref{eq:phiprime}, it is straightforward to show that the sign of $\phi_1$ is proportional to 
\begin{equation*}
 	\phi'_1 (\sigma_+) \propto m_2+ m_1 \( 2T_o^2\ell^2 -1 \)  + T_o \ell \sqrt{ (m_1-m_2)^2 + 4 m_1 m_2 T_o^2\ell^2} \le m_1 \( 2T_o-1\)\(T_o\ell+1 \)\,, 
\end{equation*}
where the middle expression is a monotonically increasing function of $m_2$. Thus, it can be concluded that\footnote{For the regime with $T_o\ell < \frac{1}{2}$, we have $\phi'_1 (\sigma_+)<0$ for $m_2< m_1\( 1- 4T_o^2\ell^2 \)$ and  $\phi'_2 (\sigma_+)<0$ for $m_1< m_2\( 1- 4T_o^2\ell^2 \)$}
\begin{equation}
	\phi'_1 (\sigma_+) \le 0 \,, \qquad \text{when} \qquad  T_o\ell \ge  \frac{1}{2} \,.
\end{equation}
In a similar fashion, for the $\mS_{2}$ region to comply, it must also include the center $r_{2}=0$ if $T_o\ell \ge \frac{1}{2}$. This type of configuration, illustrated in figure \ref{fig:ColdPhase}, is denoted as $[E_1,E_1]$ in \cite{Bachas:2021fqo}.

Given that both bulk regions $\mS_1$ and $\mS_2$ encompass the center, the dictionary between CFT data $L_{1},L_{2}$ and the bulk parameters $m_{1},m_{2}$ is given by
\begin{equation}
	\begin{split}
			\frac{2\pi}{\sqrt{-m_1}} &= \frac{L_1}{\ell} + 2\int_{\infty}^{\sigma_{+}} \phi_{1}'(\sigma) d\sigma  \,,\\
		\frac{2\pi}{\sqrt{-m_2}} &= \frac{L_2}{\ell} + 2\int_{\infty}^{\sigma_{+}} \phi_{2}'(\sigma) d\sigma  \,,\\		
	\end{split}
\end{equation}
where both $\sigma_{+}$ and $\phi'_{i}(\sigma)$ depend on mass parameters $m_{i}$, as defined in eqs.~\eqref{eq:sigmapm} and \eqref{eq:phiprime}, respectively.  For example, 
the explicit expression of the boundary size $L_i$ in terms of bulk information $m_i$ and $m_j$ (with $i=1$, $j=2$ or $i=2$, $j=1$) can be written as follows:
\begin{equation}\label{eq:boundarybulk}	
	 \begin{split}
	 L_i  &= \frac{2\pi \ell }{\sqrt{-m_i}} - 2\Delta \phi_i \ell  \,, 
	 \end{split}
\end{equation}
with 
\begin{equation}\label{eq:deltaphi}
	\Delta \phi_i  \equiv \int_{\infty}^{\sigma_{+}} \phi_{i}'(\sigma) d\sigma  =  \frac{\left[ 4 T_o^2\ell^2 \, \mathbf{\Pi} \(- \frac{m_i}{\sigma_+}, \frac{\sigma_-}{\sigma_+}\) +\frac{m_{i}-m_{j}}{m_{i}} \left( \mathbf{K}\( \frac{\sigma_-}{\sigma_+}\) - \mathbf{\Pi} \(-\frac{m_i}{\sigma_+}, \frac{\sigma_-}{\sigma_+}\)  \right)\right]}{4T_o\ell \sqrt{(1-T_o^2 \ell^2)\ \sigma_+}  }  \,, \\
\end{equation}
in which $\Delta \phi_i$ represents the angular integral from the conformal boundary to the turning point on the brane, $\mathbf{\Pi}(z|m)$ refers to the incomplete elliptic integral of the third kind, and $\mathbf{K}(m)$ represents the complete elliptic integral of the first kind.

As suggested by the term ``cold phase," this configuration prevails at low temperatures with respect to fixed boundaries. To demonstrate this, a comparison of the Euclidean action $I_{\mt{E}}$ for different phases is required. Notably, the periodicity of the Euclidean time $t_{\mt E}$ is given by the inverse temperature $\frac{1}{T_{\mt{DCFT}}}$, just as in the boundary theory. Upon performing the Euclidean integral in eq.~\eqref{eq:3Daction}, which includes the appropriate boundary counterterms, it is straightforward to show that the renormalized Euclidean action in the cold phase is given by 
\begin{equation}\label{eq:ColdAction}
	\begin{split}
		I^{\rm cold}_{\mt{E}} &= \frac{1}{16 \pi \GN T_{\mt{DCFT}} \ell }  \(  L_1 m_1 + L_2 m_2  \) \\
		&= - \frac{c}{6\pi} \(  \frac{(\pi -\sqrt{-m_1} \, \Delta\phi_1)^2}{T_{\mt{DCFT}}L_1} + \frac{(\pi - \sqrt{-m_2}\, \Delta\phi_2)^2}{T_{\mt{DCFT}}L_2}  \)  \,.
	\end{split}
\end{equation} 
where the mass parameters $m_i$, as well as $\Delta \phi_i$, are understood to be functions of boundary sizes $L_i$ by solving eq.~\eqref{eq:boundarybulk}. The expression is then recast in terms of the boundary data in the second line. It should be noted that obtaining the $m_i $ in terms of the boundary data is generally challenging. Nevertheless, we will delve into two analytical limits in the later subsections -- see also appendix \ref{sec:notes}.

\subsubsection{Warm phase ($m_1>0, m_2<0$)} \label{lukewarm}

\begin{figure}[t]
	\centering
	\includegraphics[width=4in]{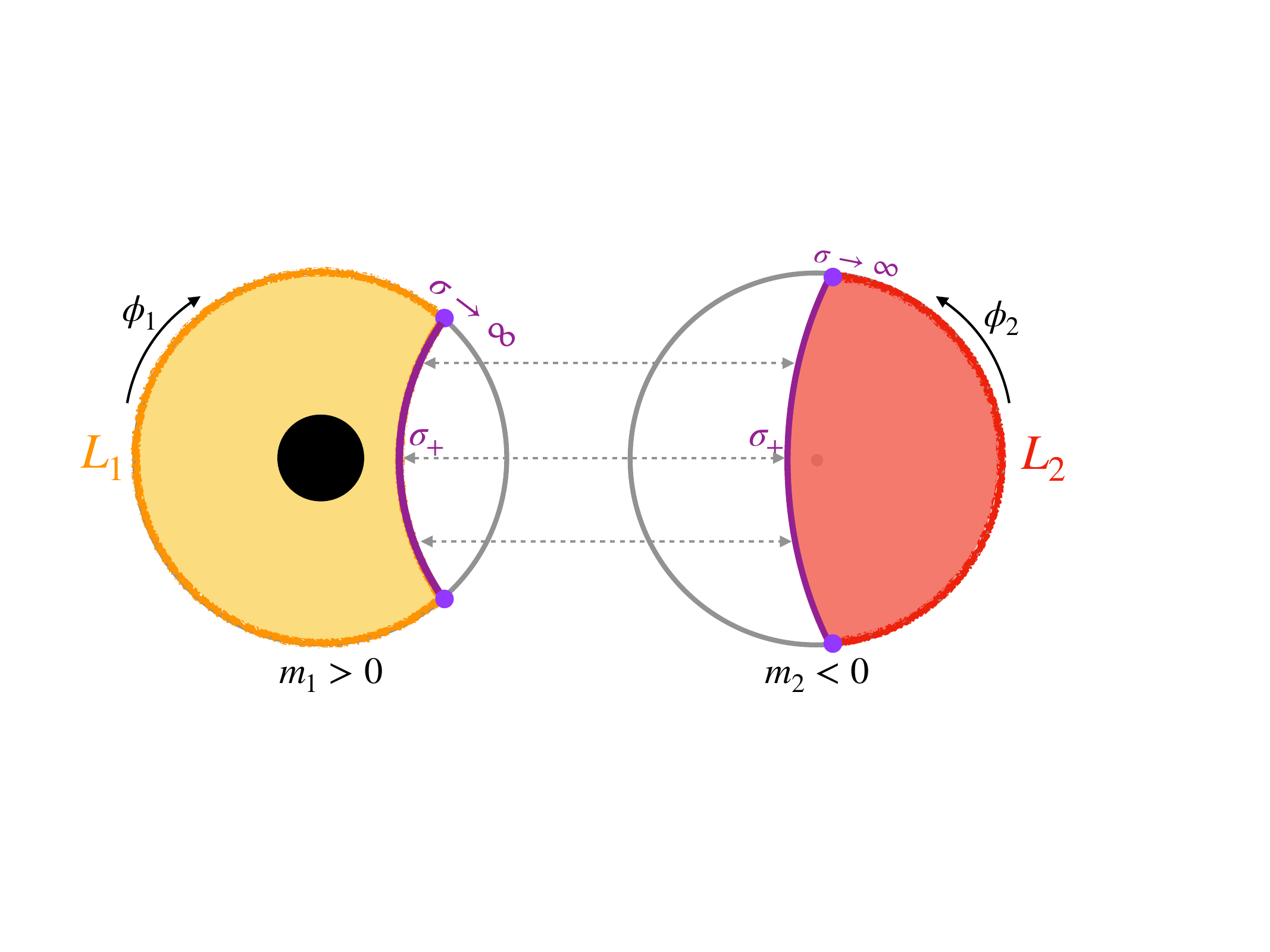}
	\caption{Bulk perspective within the warm phase. The bulk dual spacetime for left/right side is constructed by gluing a portions of BTZ black hole with $m_1>0$ and AdS$_3$ vacuum with $m_2 <0$ along a two dimensional branes.}\label{fig:WarmPhase}
\end{figure}

As described by the standard Hawking-Page transition, the AdS$_3$ bulk geometry is dominated by a BTZ black hole solution rather than a vacuum AdS$_3$ when the temperature surpasses a critical value $T_{\mt{HP}}$. Applying this concept to the construction of bulk spacetime corresponding to the TFD state of DCFT at temperature $T_{\mt{DCFT}}$, it is reasonable to anticipate the potential inclusion of a black hole geometry when the thermal temperature is sufficiently high. Consider the scenario where only one of the bulk regions, say $\mS_1$, contains a black hole. This configuration corresponds to selecting $m_1 > 0$ and $m_2 < 0$ and is termed the warm phase. The dual bulk spacetime is thus given by gluing part of the BTZ spacetime $\mS_1$ containing a black hole and a portion of AdS$_3$ vacuum $\mS_2$, as sketched in figure \ref{fig:WarmPhase}.

Given the configuration $m_1 > 0$ and $m_2 < 0$, we need to establish a relationship between these bulk parameters and the corresponding boundary data, similar to what was done for the cold phase. In the context of the BTZ black hole as defined in \eqref{eq:BTZ}, it is apparent that the temperature of the black hole $T_{\mt{BH}}$ matches that of the boundary DCFT, as indicated by eq.~\eqref{eq:temprature}. This implies that we can set the first mass parameter $m_1$ as:
\begin{equation}\label{eq:m1T}
	m_1 = \(  2\pi  \ell\, T_{\mt{BH}} \)^2   = \(  2\pi  \ell \,T_{\mt{DCFT}} \)^2 \,. 
\end{equation}
Furthermore, it is also worth noting that, in the warm phase, the turning point $\sigma = \sigma_+$ satisfies the condition:
\begin{equation}
	\sigma_+  >  \frac{|m_1 + m_2| - (m_1 +m_2)}{4\( 1 - T_o^2 \ell^2 \)} \ge 0\,.
\end{equation}
This indicates that the brane $\mB_1$ remains a certain distance away from the horizon, as illustrated in figure \ref{fig:WarmPhase}.

On the other hand, the size of the initial planar BTZ black hole, as defined by the metric \eqref{eq:BTZ}, is arbitrary. We can specify the area of the black hole horizon included in the bulk region $\mS_1$ as $\Delta \phi_1^{\rm Hor} r_h$. Since the topology of the black hole horizon in $\mS_1$ forms a closed circle, it is possible to obtain a relation between the boundary size and the horizon area, \ie 
\begin{equation}\label{eq:warmL1}
	 \frac{L_1}{\ell} = \Delta \phi_1^{\rm Hor} - 2\int_{\infty}^{\sigma_{+}} \phi_{1}'(\sigma) d\sigma  \equiv \Delta \phi_1^{\rm Hor}  - 2 \Delta \phi_1 \,,\\
\end{equation}
where the explicit form of the integral $\Delta \phi_1 $ along $\phi_1$ direction has been provided in \eqref{eq:deltaphi}. This equation establishes the connection between the bulk parameter $\Delta \phi_1^{\rm Hor}$ and the boundary data $L_1$.

Considering another bulk region $\mS_2$ with $m_2 \le 0$, it can be verified that the turning point on the brane satisfies:
\begin{equation}
	\phi'_2 (\sigma_+) <  0 \,, \qquad \text{when}  \qquad m_2<  \frac{m_1}{1-4 T_o^2 \ell^2} < 0\,,  
\end{equation}
by employing eq.~\eqref{eq:phiprime}. Similar to the situation in the cold phase, this indicates that the corresponding bulk region $\mS_2$ also includes the center at $r_{2} = 0$. As a result, the boundary size $L_2$ can be expressed as 
\begin{equation}\label{eq:warmL2}
	 L_2  = \frac{2\pi \ell }{\sqrt{-m_2}} -  2\Delta \phi_2 \ell  \,, 
\end{equation}
where $\Delta \phi_2$  is shown explicitly in eq.~\eqref{eq:boundarybulk}. On the contrary, the condition $\phi'_2 (\sigma+) > 0$ arises for the following parameter ranges\footnote{These conditions are derived for arbitrary configurations of $m_1, m_2$. And of course the condition $m_2 < m_1$ is trivial in the warm phase considered here.},
\begin{equation}
	\begin{split}
	 \frac{m_1}{1-4 T_o^2 \ell^2} < &m_2 < m_1 \,, \qquad \text{for}    \qquad   T_o  > \frac{1}{2\ell}  \,, \\
	 &m_2 <m_1\,, \qquad  \text{for}  \qquad   T_o  <  \frac{1}{2\ell} \,. \\
\end{split}
\end{equation}
It implies that $\mS_2$ does not contain the center point and the boundary-bulk dictionary is instead established as 
\begin{equation}\label{eq:warmL22}
L_2^{(+)} = + 2 \ell \int^{\infty}_{\sigma_{+}} \phi_{2}'(\sigma) d\sigma  \equiv  - 2\Delta \phi_2 \ell  \,. 
\end{equation}
These equations outline how the boundary sizes $L_2$ and $L_2^{(+)}$ are connected with the bulk parameter $\Delta \phi_2$ for the warm phase.

Since the mass parameter $m_1$ in the warm phase is fixed by the temperature, as in eq.~\eqref{eq:m1T} above, the boundary size $L_2$ will be determined by $m_2$ according to eq.~\eqref{eq:warmL2} or \eqref{eq:warmL22}. It is important to note that this relationship between $L_2$ and $m_2$ may not be a one-to-one map in general. Finally, let us note that we can focus on the case where the center is included in $\mS_2$ in the large tension limit $T_o \sim \frac{1}{\ell}$. This is because the boundary-bulk map \eqref{eq:warmL22} is ruled out by the constraint that $L_2^{(+)} > 0$ in the large tension limit. The critical tension could be derived by solving 
\begin{equation}
	0 =  \lim\limits_{m_2 \to \frac{m_1}{1-4 T_o^2 \ell^2}} L_2^{(+)}  = \frac{\ell}{\sqrt{-m_2}} \(  \pi - \frac{ 2 T_o \ell \,\mathbf{K} \(  \frac{T_o\ell }{T_o\ell -1}   \)  }{\sqrt{1- T_o^2 \ell^2}}  \) \,, 
\end{equation}  
which gives an approximate solution of $T_o \ell \approx 0.7927$. For brane tensions larger than this critical value, the bulk integral in eq.~\eqref{eq:warmL22} would always lead to $L_2^{(+)} < 0$, which is unphysical.  Because we are interested in the large tension limit for doubly holographic construction, this paper will focus on the scenario where $\mS_2$ in warm phase also contains the center and uses the relation $L_2(m_2)$ defined in eq.~ \eqref{eq:warmL2} or its inverse function $m_2(L_2)$.

 To determine the saddle point among the competing phases, the gravitational action of the warm phase needs to be evaluated. The final result can be expressed as follows \cite{Bachas:2021fqo}:
 \begin{equation}\label{eq:WarmAction}
 	\begin{split}
 		I^{\rm warm}_{\mt{E}} &= \frac{1}{16 \pi \GN T_{\mt{BH}} \ell }  \(  (L_1 - 2  \ell \Delta \phi_1^{\rm Hor} ) m_1 + L_2 m_2  \) \\
 		&= -  \frac{c }{6}  \(    \pi T_{\mt{DCFT}}   \(  L_1 + 4 \ell \Delta \phi_1  \)   +\frac{(\pi - \sqrt{-m_2}\, \Delta\phi_2)^2}{\pi T_{\mt{DCFT}}L_2}   \)\,. 
 	\end{split}
 \end{equation} 
In this equation, $\Delta \phi_1$ and $\Delta \phi_2$ can be interpreted as functions of $T_{\mt{DCFT}}L_2$ by solving the inverse function $m_2(L_2)$ from equation \eqref{eq:warmL2}. This expression provides a way to compare the warm phase with other phases, which is essential for determining the dominant saddle point in the semi-classical limit.

\subsubsection{Hot phase ($m_i>0$)} 

\begin{figure}[t]
	\centering
	\includegraphics[width=4in]{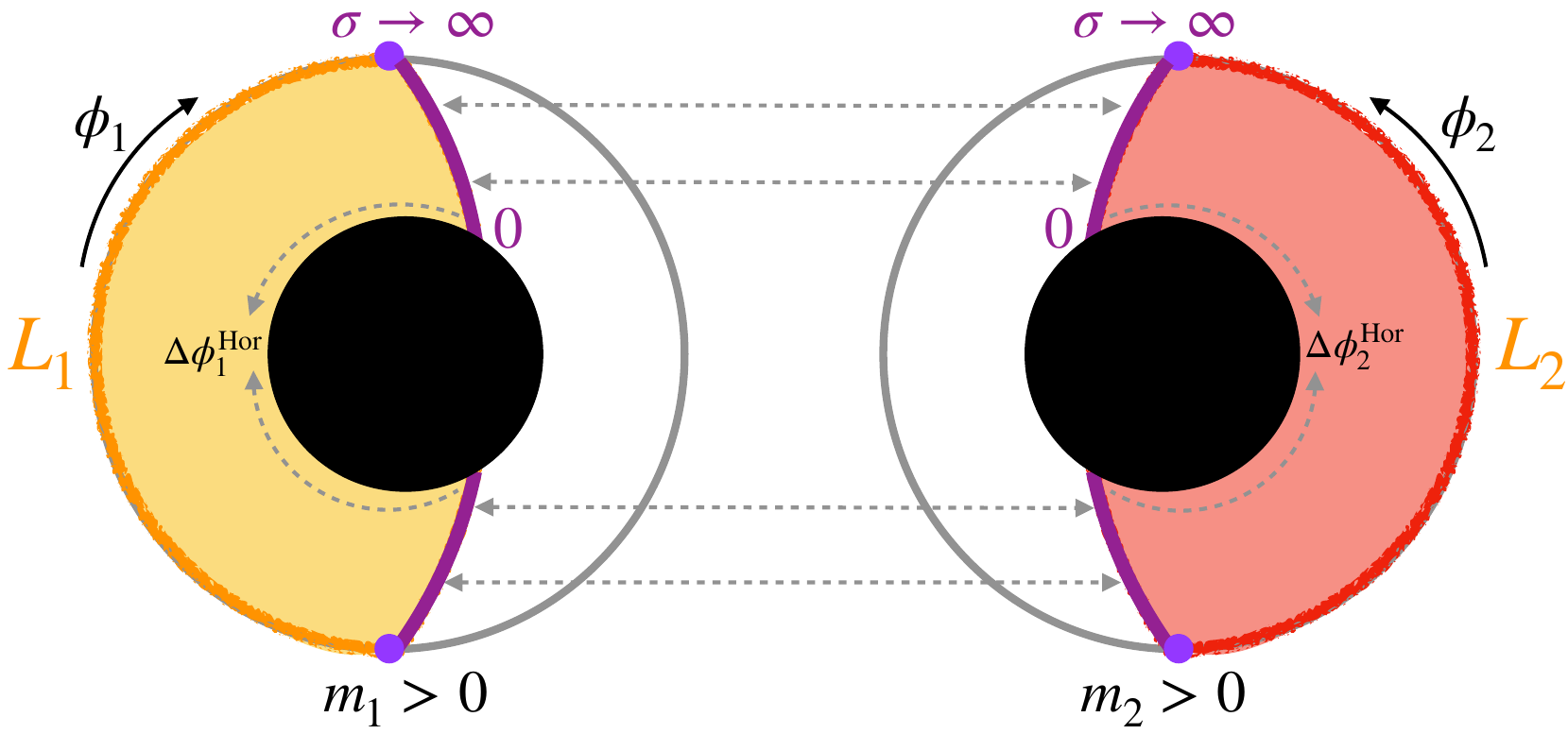}
	\caption{Bulk perspective within hot phase. The bulk dual spacetime for left/right side is constructed by gluing a portion of BTZ black hole and a portion of another BTZ black hole with the same mass. The horizon area encompassed within each part is denoted as  $\Delta \phi_{i}^{\rm Hor}$, respectively.}\label{fig:HotPhase}
\end{figure}

We proceed to the third phase, referred to as the hot phase, characterized by $m_1>0$ and $m_2>0$. This phase emerges as the dominant classical geometry in the regime of elevated temperatures. Within this phase, our construction involves the initiation of two BTZ black hole geometries. By fusing the two designated portions, denoted as $\mathcal{S}_1$ and $\mathcal{S}_2$, along the brane, we establish the corresponding holographic dual spacetime. This configuration is visually represented in figure \ref{fig:HotPhase}.

Recalling our earlier analyses for the $\mathcal{S}_1$ region for the warm phase, it becomes evident that the temperature of the black hole should match that of the thermal DCFT on the boundary due to the identical Euclidean time periodicity. Consequently, the two mass parameters are constrained to be equal, \viz 
\begin{equation}
	m_1  = \(  2\pi  \ell T_{\mt{BH}} \)^2   = \(  2\pi  \ell T_{\mt{DCFT}} \)^2 = m_2 \,. 
	\label{parrot77}
\end{equation}
This matching is inherently expected, given our consideration of static profiles that correspond to a state of thermal equilibrium. Importantly, this equivalence leads to a straightforward outcome: the turning point on the brane is positioned at
\begin{equation}\label{eq:zerosigma}
	\sigma_+ \big|_{m_1=m_2} = 0 \,, 
\end{equation}
which coincides with the location of the black hole horizon at $r=r_h$ -- see comments below eq.~\reef{eq:gauge}. This implies that, at higher temperatures, the black hole horizon expands, exerting a stronger gravitational pull that attracts the branes towards this enlarged horizon. Drawing a parallel with the warm phase scenario, we label the area of the black hole horizon encompassed within the $\mathcal{S}_i$ region as $\Delta \phi_i^{\rm Hor} r_h$. These two parameters capture the crucial bulk information that characterizes the holographic spacetime.

Furthermore, upon closer examination, it becomes evident that the angular integral in the hot phase (with $m_i>0$) can be simplified into a more explicit expression. Specifically, we have
\begin{equation}\label{eq:deltaphihot}
	\Delta \phi_i  \equiv 	 \int_{\infty}^{0} \phi_{i}'(\sigma) d\sigma =  \int^{\infty}_{0} \frac{T_o \ell}{2(m_i +\sigma) \sqrt{ m_i + (1- T_o^2 \ell^2)\sigma }} d\sigma = \frac{\arctanh \( T_o \ell \) }{\sqrt{m_i}} \,. 
\end{equation}
Analogous to the relation in eq.~\eqref{eq:warmL1}, the connection between the bulk parameters $\Delta \phi_i^{\rm Hor}$ and the boundary data can be rendered in a simpler form:
\begin{equation}\label{eq:hotmap}
	L_i= \Delta \phi_i^{\rm Hor}   \ell -  \frac{1}{\pi T_{\mt{DCFT}}} \arctanh \( T_o\ell \)\,.
\end{equation}
where $\Delta \phi_i^{\rm Hor}$ represents the size of the horizon within the bulk region $\mathcal{S}_i$, as illustrated in figure \ref{fig:HotPhase}.

Evaluating the renormalized Euclidean action in the hot phase then yields
\begin{equation}\label{eq:HotAction}
	\begin{split}
		I_{\mt{E}}^{\rm hot} &=\frac{1}{16 \pi \GN T_{\mt{BH}} \ell }  \(  \(L_1 - 2   \Delta \phi_1^{\rm Hor}  \ell \) m_1 +  \(  \(L_2- 2  \Delta \phi_2^{\rm Hor}   \ell \) m_2   \) \)   \\ 
		&=- \frac{c }{6} \pi T_{\mt{BH}}  \(    L_1 +   L_2   \) - \frac{2c }{3}  \log \(   \sqrt{   \frac{ 1+T_o\ell}{1-T_o\ell}  }\)   \,,\\
		&=- \frac{c \pi }{6} T_{\mt{DCFT}}  \(   L_1 +   L_2  \) -   2\, S_{\rm def}   \,,\\
	\end{split}
\end{equation}
where we have identified the boundary entropy associated with one defect as 
\begin{equation}\label{eq:Sbdy}
S_{\rm def}\equiv 2\log g = \frac{c }{6}  \log \(   {   \frac{ 1+T_o\ell}{1-T_o\ell}  }\)   \,. \\
\end{equation}
as given in  eq.~\reef{eq:defg}. This relationship reproduces the same result in the AdS/BCFT correspondence \cite{Takayanagi:2011zk,Fujita:2011fp}. The factor of $4$ that precedes the boundary entropy in eq.~\eqref{eq:HotAction} accounts for the total number of defects existing on both the left and right boundaries. In subsequent subsections, we will further confirm that the aforementioned boundary entropy \eqref{eq:Sbdy} corresponds to the holographic entanglement entropy of a single defect.

\subsection{Phase diagram}\label{sec:Phasediagram}
\begin{figure}[t]
	\centering
	\includegraphics[width=5in]{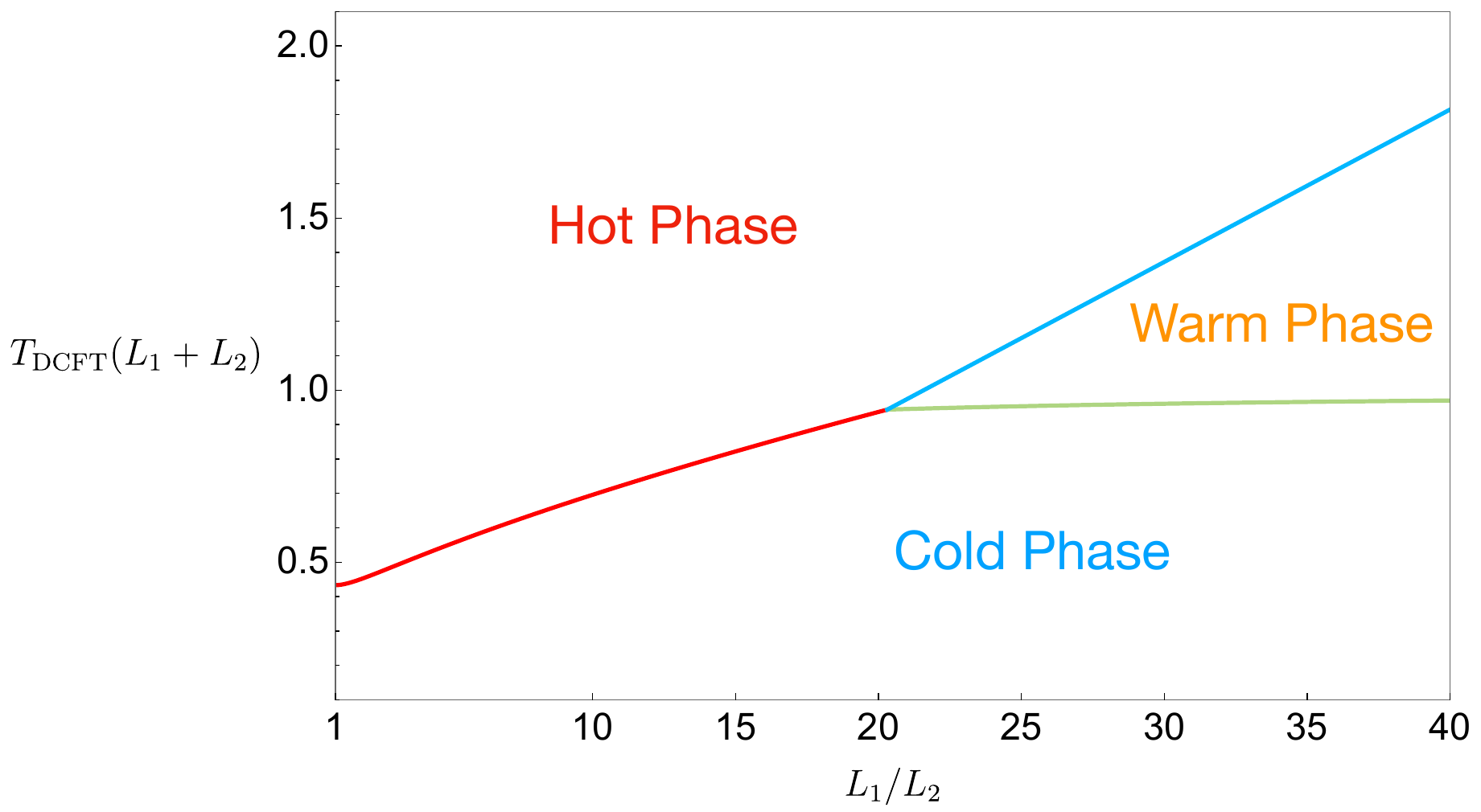}
	\caption{Phase diagram with respect to the dimensionless boundary data $\{T_{\mt{DCFT}} \( L_1 +L_2 \), \frac{L_1}{L_2} \}$ with choosing $T_o \ell = 0.9$.}\label{fig:PhaseDiagram}
\end{figure}

In the previous subsection, we reviewed the possible solutions for the bulk spacetime and evaluated the gravitational action \eqref{eq:3Daction} for the corresponding phases. These solutions encompass three distinct phases, namely the cold, warm and hot phases. Upon specifying the boundary data, we now determine the dominant saddle point by comparing the Euclidean actions of these different phases. As discussed at eq.~\reef{eq:parameters}, the boundary configurations are specified by four dimensionless parameters. The first two $c$ and $\log g$ characterize the boundary theory and are fixed here. The last two $T_{\mt{DCFT}} \( L_1 +L_2 \)$ and ${L_1}/{L_2}$ describe the boundary geometry and the temperature and will be varied to determine the phase diagram. By using numerical solutions of the inverse functions inherent in the boundary-bulk dictionary, it is straightforward to evaluate the gravitational action for a particular choice of these parameters. Subsequently, we can draw the phase diagram as shown in figure \ref{fig:PhaseDiagram} with respect to these two dimensionless parameters.

To provide insight into the shape of the phase diagram depicted in figure \ref{fig:PhaseDiagram}, we will derive the approximate expressions for the boundaries between the various phases. Beginning with the transition between the warm and hot phases, we compare the renormalized actions in eqs.~\reef{eq:WarmAction} and \reef{eq:HotAction} and find that the corresponding boundary in the phase diagram is determined by solving
\begin{equation}
	4 \pi T_{\mt{DCFT}} \ell \Delta \phi_1  +\frac{(\pi - \sqrt{-m_2}\, \Delta\phi_2)^2}{\pi T_{\mt{DCFT}}L_2}   =  \pi T_{\mt{DCFT}} L_2 + 4\arctanh \( T_o \ell \)\,,
\end{equation}
where $\Delta \phi_i$ are defined with taking $m_1=(2\pi T_{\mt{DCFT}} \ell)^2$, and $m_2$ is obtained by the inverse function presented in eq.~\eqref{eq:warmL2}.
Since $L_1$ has dropped out of this equation, the corresponding phase boundary is precisely determined by 
\begin{equation}
	T_{\mt{DCFT}} L_2 = C_{\mt{WH}} (T_o\ell) \,, 
\end{equation}
which represents a constant governed by the boundary entropy or the combination $T_o \ell$. Consequently, the boundary between the warm and hot phases becomes a straight line in figure \ref{fig:PhaseDiagram}, \ie  $T_{\mt{DCFT}}(L_1+ L_2) = C_{\mt{WH}} \(1 + {L_1}/{L_2}\)$.

\begin{figure}[!]
	\centering
	\includegraphics[width=3.5in]{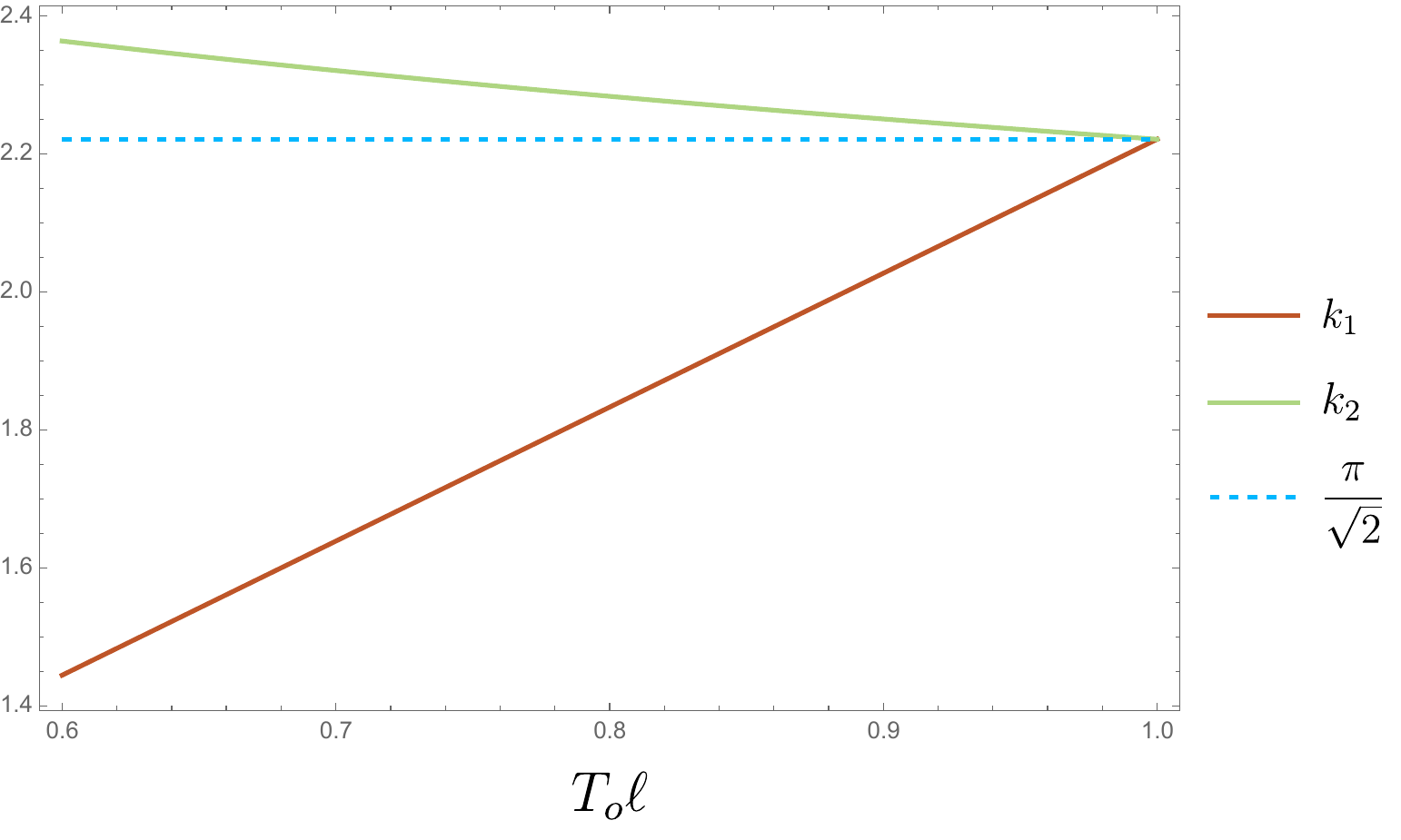}
	\caption{The coefficients $k_1, k_2$ as a function of the dimensionless tension $T_o\ell$.}\label{fig:K1K2}
\end{figure}

Although obtaining the analytical solution for the phase transitions is challenging, a notable observation is that the phase transitions associated with the warm phase simplify when $L_1 \gg L_2$. This regime can be approached by considering the limit $m_2 \to -\infty$, corresponding to the condition $T_{\text{DCFT}} L_2 \ll 1$. In this context, the angular integrals defined in eq.~\eqref{eq:deltaphi} can be approximated by 
\begin{equation}
\lim\limits_{m_2 \to -\infty }	\Delta \phi_1 \approx \frac{k_1}{\sqrt{-m_2}}  + \cdots  \,, \qquad \lim\limits_{m_2 \to -\infty }	\Delta \phi_2 \approx  \frac{k_2}{\sqrt{-m_2}}  + \frac{k_1 m_1}{2(-m_2)^{3/2}} + \cdots \,,  \label{eq:defk2}
\end{equation}
with the constants defined by
\begin{equation}\label{rooster4}
	\begin{split}
			k_1 &= 2 \sqrt{ \frac{T_o \ell}{1+ T_o\ell} }  \(    (1+T_o\ell ) \mathbf{E}\( \frac{T_o\ell -1 }{T_o\ell +1} \)  -  \mathbf{K}\( \frac{T_o\ell -1 }{T_o\ell +1} \)   \) \,, 
	\\		
		k_2 &= \frac{1}{2\sqrt{T_o\ell(1+ T_o\ell)}}  \(  \mathbf{K}\( \frac{T_o\ell -1 }{T_o\ell +1} \) + (4 T_o^2\ell^2 -1) \mathbf{\Pi}\(  4T_o\ell (1- T_o\ell) ; \frac{T_o\ell -1 }{T_o\ell +1} \)        \)\,. \\
	\end{split}
\end{equation}
We show the numerical plot for $k_1, k_2$ in figure \ref{fig:K1K2}. Further, we note that in the large tension limit, \ie $T_o\ell \to 1$ limit, one finds
\begin{equation}\label{rocket3}
	k_1 \approx  \frac{\pi}{\sqrt{2}} \(  1 - 7\,\frac{1-T_o\ell}{8} \) + \mathcal{O}((1-T_o\ell)^2)\,, \quad  k_2 \approx \frac{\pi}{\sqrt{2}} \(  1 + \frac{1-T_o\ell}{8} \) + \mathcal{O}((1-T_o\ell)^2)\,.
\end{equation}
Both of these approximations help to simplify the bulk-boundary dictionary. As an example, let us focus on $m_2\to-\infty$ for which the bulk-boundary dictionary in the warm phase simplifies
\begin{equation}
	L_1 \simeq \(  \Delta \phi^{\rm Hor}_1  - \frac{2k_1}{\sqrt{-m_2}}  \)\ell \,, \quad L_2 \simeq  \frac{(2\pi - 2k_2)\ell}{\sqrt{-m_2}} -  \frac{(2\pi T_{\mt{DCFT}}\ell)^2 k_1 \ell}{(-m_2)^{3/2}}  \,.
	\label{eq:regionsize}
\end{equation}
Hence disregarding higher-order terms in powers of $T_{\text{DCFT}} L_2$, we can approximate the phase transition between the hot and warm phases as:
\beqa\label{eq:WHtrans}
	 T_{\mt{DCFT}} L_2  \big|_{\mt{WH}}&=& C_{\mt{WH}}  \simeq  \frac{(\pi -k_2 )}{ (\pi -k_1 - k_2)\pi} \\
	 &&\times\ \( \sqrt{(\pi-k_2)(\pi -k_1-k_2) + 4 \arctanh^2 (T_o\ell)}  - 2\arctanh(T_o\ell)  \) \,. \nonumber
\eeqa
It is useful for later to noting that $C_{\mt{WH}} $ is a small parameter in the large tension limit  $T_o\ell \to 1$ because of  
\begin{equation}
 C_{\mt{WH}}  \simeq  \frac{(\pi - k_2)^2}{2 \pi} \  { \log^{-1}\! \( {  \frac{1+T_o \ell}{1-T_o\ell} }\)} \,. 
 \label{rocket7}
\end{equation}

The approximations with $|m_2| \gg 1$ can also be applied to explore the phase transition between the cold and warm phases. By comparing the renormalized actions in the cold and warm phases, as given by eqs.~\eqref{eq:ColdAction} and \eqref{eq:WarmAction}, one can describe the phase transition as follows 
 \begin{equation}\label{eq:Ccw}
   T_{\mt{DCFT}} L_1\big|_{\mt{CW}}  = C_{\mt{CW}} \simeq  1 -   \frac{k_1}{\pi - k_2} \frac{L_2}{L_1} + \(  \frac{k_1}{\pi - k_2} \frac{L_2}{L_1}\)^2 + \mathcal{O}\(  \(\frac{L_2}{L_1}\)^3 \)  \,. 
 \end{equation}
With $L_2/L_1 \to 0$, the phase transition is located at $T_{\mt{DCFT}}L_1=T_{\mt{DCFT}}L_{\rm bdy}=1$. 

Combining the above results, we can immediately find that the triple point of the phase diagram is situated at
\begin{equation}
	\text{triple point:} \qquad    T_{\mt{DCFT}} L_1  \simeq 1 -  \frac{k_1}{\pi -k_2} C_{\mt{WH}}  \,, \quad T_{\mt{DCFT}} L_2  = C_{\mt{WH}}   \,.
\end{equation}
Then the ratio of the boundary lengths at the triple point is approximated by $\frac{L_1}{L_2}\big|_{\rm triple} \simeq \frac{C_{\mt{CW}}}{C_{\mt{WH}}} $ with
\begin{equation}\label{eq:L1L2triple}
\frac{C_{\mt{CW}}}{C_{\mt{WH}}}   \simeq \frac{k_1( k_2 - \pi )+2 \pi  \arctanh (T_o\ell)  + \pi   \sqrt{(\pi-k_2)(\pi -k_1-k_2) + 4 \arctanh^2 (T_o\ell)}    }{(\pi - k_2)^2}  \,. 
 \end{equation}
Near the triple point, the phase transition between the cold and hot phases can be approximated by $ T_{\text{DCFT}} (L_1 + L_2)=C_{\mt{CH}} $ where the latter constant is given by 
\begin{equation}\label{eq:CWtrans}
 \sqrt{  \( \frac{L_2}{L_1}\frac{k_1 }{(k_2 - \pi)} + \(\frac{L_2}{L_1}+1\)  \(1+ \frac{L_1}{L_2}\(1 -\frac{k_2}\pi\)^2 \) \)  + \frac4{\pi^2}\arctanh^2 (T_o\ell)  }  - \frac2\pi \arctanh (T_o\ell) \,,
\end{equation}
in terms of the dimensionless parameter $L_1/L_2$. However, it is important to note that these approximations, derived from a series expansion with respect to $1/\sqrt{-m_2}$, are not valid in the vicinity of the regime $L_1 \approx L_2$. Instead, we will demonstrate the full analytical understanding of the symmetric setup where $L_1 = L_2$ below in subsection \ref{sec:sym}.
 
\subsection{Fusion limit}\label{rocket2}
\begin{figure}[t]
	\centering
	\includegraphics[width=3in]{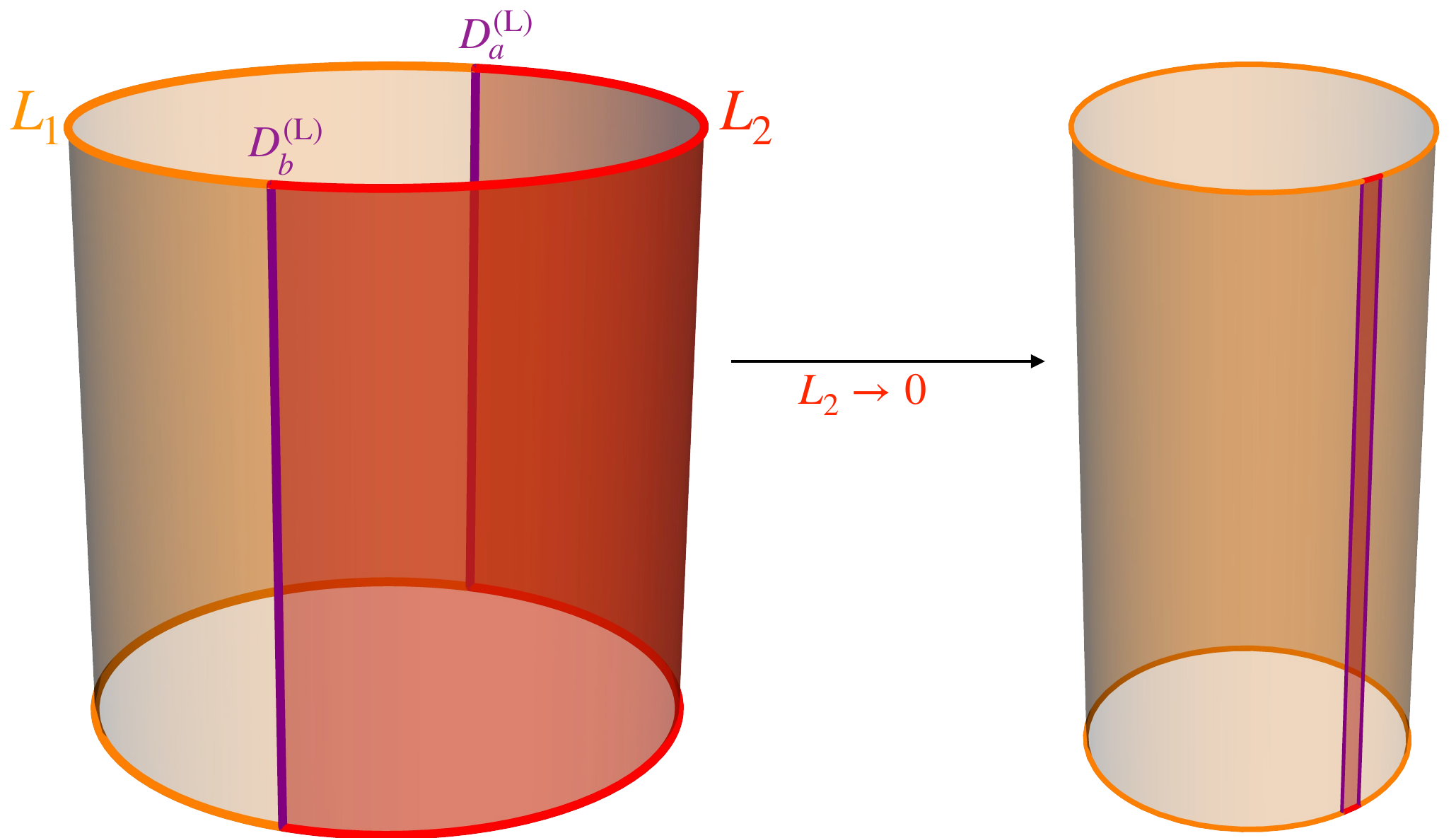}
	\caption{The fusion limit of boundary defect filed theory is defined by taking $L_2 \to 0$.}\label{fig:Fusion}
\end{figure}

Despite the challenge of finding analytical results for generic values of the parameters, there are several interesting limits where we can derive simple results. Considering the fusion limit \cite{Bachas:2007td} with $L_2/L_1 \to 0$, the boundary system behaviors like a thermal CFT$_2$ on a circle whose circumference is fixed as $L_1$. As shown in the phase diagram \ref{fig:PhaseDiagram}, the fusion limit leads us to the right boundary of the phase diagram. Moreover, there are only two phases in this limit, \ie the warm and cold phases.\footnote{We note that the three phases could all exist at the limit $L_1/L_2 \to \infty$. Whether we can probe the hot phase depends on the limit at which we approach $L_1/L_2 \to \infty$. In the fusion limit with $\TDCFT L_2 \to 0$, which is below the black hole threshold given by eq.~\eqref{eq:WHtrans}, the hot phase is absent. }
The thermal partition function of holographic CFT$_2$ is universal. If the CFT$_2$ resides on a circle of circumference $L_{\rm bdy}$ and the temperature is $T_{\mt{CFT}}$, we have, \eg \cite{Hartman:2014oaa}
\begin{equation}
	\ln Z_{\rm thermal}^{\mt{CFT}}  = 
	\begin{cases}
		  \frac{\pi\,c}6\, \frac{1}{T_{\mt{CFT}} L_{\rm bdy}}    \,, \qquad  T_{\mt{CFT}} L_{\rm bdy}<1\,,  \\ 
		\,\\
		   \frac{\pi\,c}6\, T_{\mt{CFT}} L_{\rm bdy} \,, \qquad     T_{\mt{CFT}} L_{\rm bdy}>1 \,,\\
	\end{cases}
\end{equation}
From the bulk perspective, the renormalized action at the fusion limit is given by 
\begin{equation}
	\begin{split}
		\lim\limits_{L_2/L_1 \to 0}	I_{\mt{E}}^{\rm cold} &= -\frac{c}{6}  \(    \frac{\pi}{T_{\mt{DCFT}} L_1} +  \frac{(\pi-k_2)^2}{\pi T_{\mt{DCFT}}L_2}\)  + \mathcal{O}(L_2)  \,,\\
		\lim\limits_{L_2/L_1 \to 0}	I_{\mt{E}}^{\rm warm} &=- \frac{c }{6}  \(    \pi T_{\mt{DCFT}} L_1 +    \frac{(\pi-k_2)^2}{\pi T_{\mt{DCFT}}L_2}\)  + \mathcal{O}(L_2)   \,,\\
	\end{split}
\end{equation}
where the second term, which is common to both phases, originates from the Casimir energy of the $\mS_2$ region and is divergent in this limit as the defects bounding the  region approach one another. The phase transition definitely happens at 
\begin{equation}
	T_{\mt{DCFT}}= \frac{1}{L_1} =  \lim\limits_{L_2 \to 0}	 \frac{1}{L_{\rm bdy}}\,, 
\end{equation}
which agrees with the Hawking-Page temperature of AdS$_3$. The phase transition is universal in the sense that it is not affected by the properties of the defects as it obviously does not depend on the boundary entropy $\log g$ or the brane tension $T_o$. 

Ignoring the corrections from the higher powers of $L_2$, one can also derive the energy and entropy of the system (see Appendix \ref{sec:notes} for more details) in the fusion limit as follows: 
\begin{equation} \label{rocket5}
	\begin{split}
		E^{\rm{cold}}&=-T_\mt{DCFT}^2\frac{\partial I^{\rm{cold}}_\mt{E}}{\partial T_\mt{DCFT}}\approx -\frac{\pi \, c}{6\,L_1}-\frac{c}{6}\frac{(\pi- k_2)^2}{\pi L_2}  + \mathcal{O}(L_2/L_1^2)\,, \\
		S^{\rm{cold}}&=-\frac{\partial (T_\mt{DCFT}\, I^{\rm{cold}}_\mt{E})}{\partial T_\mt{DCFT}}=0\,,
	\end{split}
\end{equation}
and 
\begin{equation}
	\begin{split}
		E^{\rm{warm}}&=-T_\mt{DCFT}^2\frac{\partial I^{\rm{warm}}_\mt{E}}{\partial T_\mt{DCFT}}\approx \frac{c}{6} \pi T^2_\mt{DCFT} L_1  -\frac{c}{6}   \frac{(\pi- k_2)^2}{\pi L_2}  + \mathcal{O}(T_\mt{DCFT}^2L_2)\,, \\
		S^{\rm{warm}}&=-\frac{\partial (T_\mt{DCFT}\, I^{\rm{warm}}_\mt{E})}{\partial T_\mt{DCFT}}\approx \frac{\pi c}{3} T_\mt{DCFT} L_1  +  \mathcal{O}(T_\mt{DCFT}L_2)\,.
	\end{split}
\end{equation}
Obviously, apart from the Casimir contributions appearing in both and whose coefficient is determined by the brane tension, the results are the same as that of a thermal CFT$_2$ living on a circle with a circumference $L_1$.

\subsection{Symmetric boundary} \label{sec:sym}
In the preceding analyses, two distinct parameters, $L_1$ and $L_2$, were introduced to define the sizes of the two boundary intervals. However, the analysis also simplifies for the specific scenario where $L_1=L_2$.\footnote{An extensive examination of this configuration can also be found in appendix C of \cite{Grimaldi:2022suv}.} From the perspective of the boundary theory, this choice produces a $Z_2$ symmetry between the two regions across the defect. By the symmetry argument, we can expect that the dual bulk spacetime should also exhibit a $Z_2$ symmetry around the branes $\mathcal{B}_i$. From the viewpoint of bulk spacetime, it is then apparent that only the cold and hot phases are compatible with this symmetry requirement, while the warm phase is precluded due to the absence of symmetry. The absence of the warm phase is also evident upon examining the left boundary of the phase diagram \ref{fig:PhaseDiagram}. Further, the symmetry demands that the mass parameters $m_i$ for the two bulk regions $\mathcal{S}_i$ are identical, \ie $m_1 = m = m_2$. 

This configuration not only simplifies various expressions but also enhances the clarity of the boundary-bulk dictionary. For instance, the angular integral $\Delta \phi_i$, defined in eq.~\eqref{eq:deltaphi}, reduces to
\begin{equation}\label{eq:deltaphisym}
	\Delta \phi_i \big|_{m_1=m_2} =   \begin{cases}
		 \frac{\pi}{2 \sqrt{-m}} \,, \qquad m<0\,,\\
		 \frac{\arctanh (T_o\ell)}{\sqrt{m}} \,, \qquad m> 0\,.\\
	\end{cases}
\end{equation}
Furthermore, the symmetry requires the brane profiles in $\mS_{1,2}$ to be the same and so the corresponding extrinsic curvatures are equal, \ie $K^{(1)}_{ab}=K^{(2)}_{ab}$. Hence, eq.~\eqref{eq:Israel} reduces to
\begin{equation}\label{eq:Neumann}
	K^{(1)}_{ab} - K^{(1)} \, h_{ab}  - T_o h_{ab}=0\,.
\end{equation}
The most general solutions for the brane profile satisfying eq.~\eqref{eq:Neumann} can be found in, \eg ref.~\cite{Kawamoto:2023wzj}.  In the subsequent discussion, we will concentrate on the corresponding brane solutions within the cold and hot phases, respectively.

In the cold phase, the bulk geometry is described by the AdS$_3$ vacuum geometry, parametrized as follows
\begin{equation}\label{eq:defectlAdS3}
	\begin{split}
		ds^2  &= - \( \frac{r^2 }{\ell^2} - m \) dt ^2  + \(  \frac{r^2 }{\ell^2}  - m \)^{-1} d r^2  + r^2 d \phi^2 \,. \\
	\end{split}
\end{equation}
The static brane profile in this geometry is derived as  
\begin{equation}\label{eq:globalbrane}
\frac{r}{ \ell \sqrt{-m}} \sin \(  \sqrt{-m} |  \phi  \pm \phi_{\rm bdy} | \)  = \frac{ T_o \ell}{ \sqrt{1- (T_o \ell)^2} } \,, 
\end{equation}
where the brane intersects the conformal boundary at $\phi = \pm \phi_{\text{bdy}}$.
In our conventions, the boundary region with a fixed length of $L_i$ is given by $\phi \in \[ - \phi_{\rm bdy}, \phi_{\rm bdy} \]$  with the assumption that the midpoint of the boundary lies at $\phi=0$. It is evident that the boundary size is $L_1 =L_2 = 2 \phi_{\text{bdy}} \ell $. For a regular bulk spacetime without defects, the periodicity condition, $\phi \sim \phi + \frac{2\pi}{\sqrt{-m}}$, fixes the free parameter $\phi_{\text{bdy}}$ as $\phi_{\text{bdy}} = \frac{\pi}{2\sqrt{-m}}$. Consequently, the boundary-bulk map in the cold phase can be summarized as:
\begin{equation}\label{royal3}
	 L_1 =L_2 = \frac{\pi \ell}{ \sqrt{-m}} \,. 
\end{equation}
This result can also be derived from eq.~\eqref{eq:boundarybulk} by employing $\Delta \phi_i$ from eq.~\eqref{eq:deltaphi}. It can be understood as a consequence of the fact that the brane profile in the global AdS$_3$ configuration always connects two antipodal points on the boundary circle. Finally, the total Euclidean action for the cold phase reduces to 
\begin{equation}\label{eq:coldaction}
	\begin{split}
		I_{\mt{E}}^{\rm cold} &=  \frac{c}{24 \pi \ell^2 T_{\mt{DCFT}}}  \(   L_1 m_1+  L_2 m_2   \) = -\frac{\pi\,c}{6} \frac{1}{T_{\mt{DCFT}} (L_1 +L_2)}  \,.
	\end{split}
\end{equation}

In the hot phase, characterized by $m_1=m_2=m>0$, the bulk spacetime is represented by a (static) BTZ black hole geometry whose line element is given by 
\begin{equation}\label{eq:BTZ02}
	ds^2 = -\frac{r^2 -r_h^2}{\ell^2} dt^2  + \frac{\ell^2}{r^2 -r_h^2} dr^2 + r^2 d \phi^2 \,,
\end{equation}
where $r_h$ denotes the radius of the horizon and is related to the positive mass parameter by $m = (r_h/\ell)^2$. Similar to eq.~\eqref{eq:globalbrane}, the static brane profile living in BTZ black hole spacetime \eqref{eq:BTZ02} is given by
\begin{equation}\label{eq:BTZbrane}
	\frac{r}{ r_h } \sinh \( \frac{r_h}{\ell}  | \phi  \pm \phi_{\rm bdy} | \)  = \frac{ T_o \ell}{ \sqrt{1- (T_o \ell)^2} } \,. 
\end{equation}
It is obvious that the brane always intersects the BTZ horizon at:
\begin{equation}
 \phi= \phi_{\rm hor} =    \pm \(  	\phi_{\rm bdy}    +   \frac{\ell}{r_h} \arctanh \( T_o \ell \)  \)  \,,
\end{equation}
which results in the boundary-bulk map shown in \eqref{eq:hotmap} by utilizing $L_i = 2\phi_{\rm bdy} \ell$ $\Delta \phi_i^{\rm Hor} = 2 \phi_{\rm hor}$, and $T_{\mt{DCFT}}= T_{\mt{BH}}= \frac{r_h}{2\pi \ell^2}$. By explicitly employing the BTZ metric and the brane profiles from eq.~\eqref{eq:BTZbrane}, it is straightforward to derive the renormalized action for the two-sided system:
\begin{equation}\label{eq:hotaction}
	\begin{split}
		I_{\mt{E}}^{\rm hot} &=- \frac{\pi\,c }{6} \,T_{\mt{DCFT}}  \(    L_1 +   L_2   \) - \frac{c }{3}  \log \(   {   \frac{ 1+T_o\ell}{1-T_o\ell}  }\)   \,. \\
	\end{split}
\end{equation}

The transition from the cold phase to the hot phase bears similarities to the Hawking-Page transition. Upon comparing the effective actions for this symmetric configuration (\ie $ L_1 = L_2$), as given in eqs.~\eqref{eq:coldaction} and \eqref{eq:hotaction}, one finds that the transition occurs at  
\begin{equation}\label{eq:ColdHot}
 C_{\mt{CH}} \equiv  - \frac{2\,\arctanh \( T_o\ell \)}{ \pi} + \sqrt{ 1 + \frac{4\,\arctanh^2 \( T_o\ell \)}{ \pi^2}   } \in [0,1] \,.
\end{equation}
The same result has been derived before in \eg \cite{Fujita:2011fp,Cooper:2018cmb,Grimaldi:2022suv}. In the limit of a large tension with $T_o\ell \to 1$, the phase transition point is shifted to a very low-temperature regime with $C_{\mt{CH}}    \simeq \frac{\pi}{4 \arctanh (T_o\ell)} \ll 1$. Following \cite{Grimaldi:2022suv}, we can also express this transition point in terms of the boundary data
\begin{equation}\label{eq:ColdHot2}
 C_{\mt{CH}} =  \sqrt{ 1 + \(\frac6\pi\,\frac{S_\mt{def}}{c}\)^2} - \frac6\pi\,\frac{S_\mt{def}}{c} \,.
\end{equation}
using the defect entropy in eq.~\reef{eq:defg}. As discussed above, the limit $T_o\ell \to 1$ 
corresponds to the regime where $S_\mt{def}/c\gg1$ and we find $C_{\mt{CH}}    \simeq
\frac{\pi\,c}{12\,S_\mt{def}}$.


\section{Holographic Entanglement of the Defects}\label{sec:mutual}
%

In this section, we holographically compute entanglement entropies of various subsystems in our setup. From the  CFT perspective, we are particularly interested in the mutual information $I_{\mt{L:R}}$
between the defects $D^{(\mt{L})}_{a,b}$ on the left boundary  and  $D^{(\mt{R})}_{a,b}$
on the right, \ie 
\begin{equation}
	I_{\mt{L:R}}:=  S_{\mt{EE}}  ( D^{(\mt{R})}_{a,b} )+S_{\mt{EE}}  ( D^{(\mt{L})}_{a,b} )   -	S_{\mt{EE}}  ( D^{(\mt{R,L})}_{a,b} ) \,.
\end{equation}
This mutual information is interesting because, in the brane perspective, this corresponds to measuring the correlation between two branes representing ``two gravitating universes". The entanglement entropy between two gravitating disjoint spacetimes in two dimensions was considered in \cite{Balasubramanian:2021wgd}. It was found that even if these two are originally disjoint,  as we increase the entanglement of the matter degrees of freedom on them, the two geometries become connected by an Einstein-Rosen bridge, which is reminiscent of the ER=EPR conjecture \cite{Maldacena:2013xja}. One of the purposes of this paper is to see the dynamics of this spacetime connection from the bulk perspective, and to provide independent support for the arguments given in \cite{Balasubramanian:2021wgd}.

To begin with, it is useful to recall that the entanglement entropy between the gravitating region and the non-gravitating region from the brane perspective which is computed with the conventional island formula and is equal to the entanglement entropy between the degrees of freedom of the boundary CFTs and those on the defects, from the boundary perspective.

\begin{figure}[!]
	\centering
	\includegraphics[width=4in]{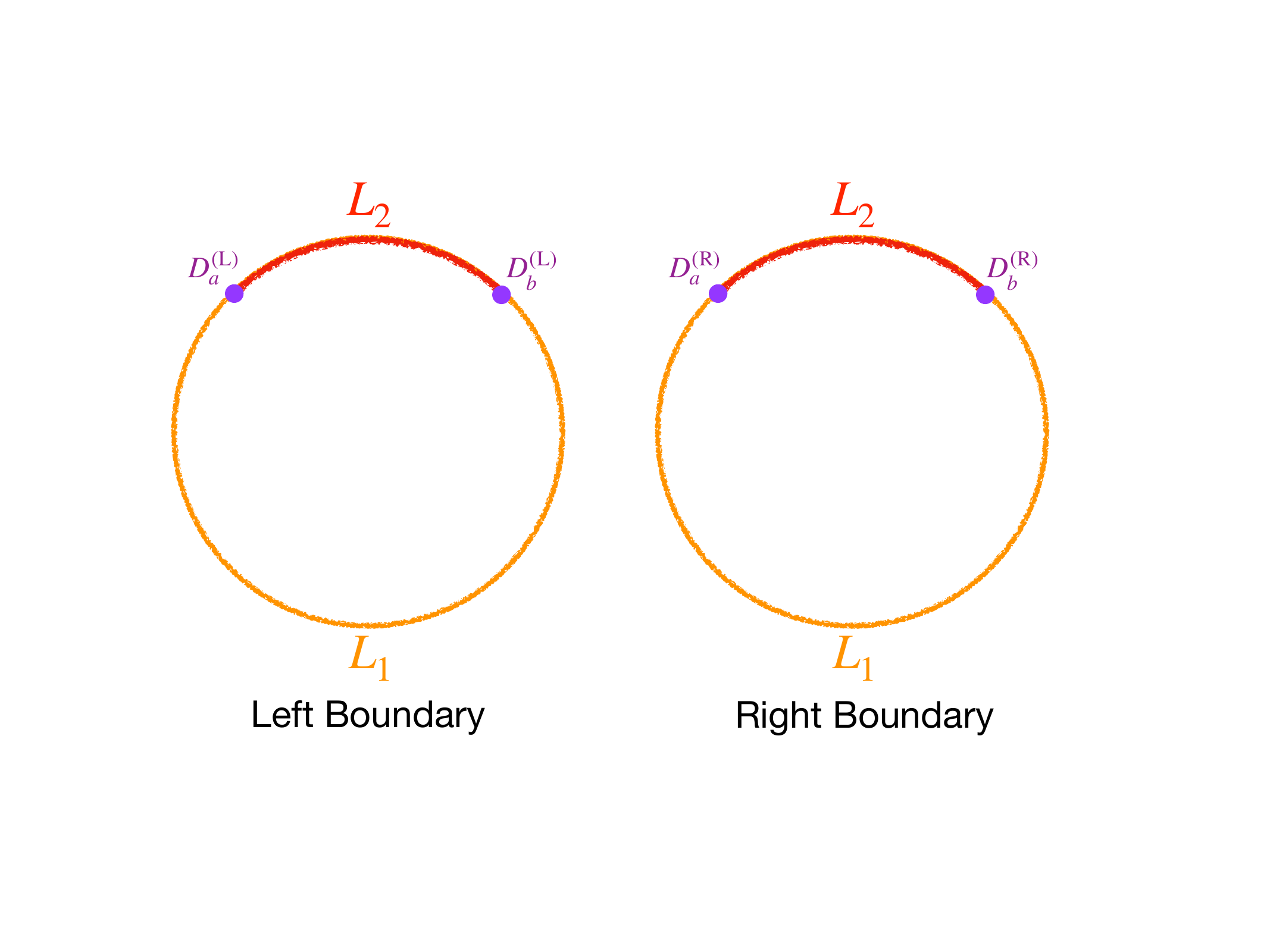}
	\caption{The defect denoted as $\{D^{(\mt{R})}_{a},D^{(\mt{R})}_{b},D^{(\mt{L})}_{a},D^{(\mt{L})}_{b} \}$ is regulated by including a boundary interval with a length $\Delta L_{\mt{D}} \ll \ell$.}\label{fig:DFTLeft}
\end{figure}

In the current case, we are interested in the entanglement between the two gravitating universes L and R communicating through exchanges of QFT degrees of freedom. From the boundary CFT perspective, this corresponds to the entanglement between two groups of  defects $D^{(\mt{L})}_{a,b}= D^{(\mt L)}_{a}  \cup D^{(\mt L)}_{b}  $ and  $D^{(\mt{R})}_{a,b}= D^{(\mt R)}_{a}  \cup D^{(\mt R)}_{b} $. More importantly, we need to regulate our holographic entanglement entropy calculations for the defects by considering a small interval around each defect, whose size is fixed as $\Delta L_{\mt D}$, as was done in \cite{Grimaldi:2022suv}. This setup is depicted in figure \ref{fig:DFTLeft}. From the boundary perspective, these small defect intervals are considered as the boundary dual of the two-dimensional braneworld. We would like to further note that the regulator length $\Delta L_{\mt D}$ should be considered as much smaller than the AdS radius or the boundary size, but it is still distinct from (and larger than) the UV cut-off scale. As a summary, the defect length satisfies 
\begin{equation}
	\epsilon \ll \DLD \ll \ell \,. \label{rooster2}
\end{equation}

Furthermore, these defects are coupled through the CFT degrees of freedom which correspond to degrees of freedom of  Hawking quanta from the brane perspective -- see figure \ref{fig:global}. Therefore, one way to quantify the entanglement between them is to compute the mutual information $I_{\mt{L:R}}$. This holographic mutual information obviously depends on the bulk phases which are determined by the dimensionless CFT parameters as shown in the phase diagram \ref{fig:PhaseDiagram}. In order to allow the existence of all three phases, we will focus on the regime with $L_1/L_2 \gg 1$, \ie the fusion limit discussed in subsection \ref{rocket2}, which also simplifies the analysis. But first, we will consider the symmetric case with $L_1=L_2$ in the next subsection, where we can analytically derive all expressions. This probes the regime where $L_1 \sim L_2$ and there are only cold and hot phases.

In the following, to compute the holographic entanglement entropy in different phases, we need to evaluate the length of various geodesics or extremal curves. This simplifies because the bulk spacetime is always locally an AdS$_3$ geometry. It is useful to regard AdS$_{3}$ as a hyperbola,
\begin{equation}
	X_{0}^{2} +X_{1}^{2}-X_{2}^{2} -X_{3}^{2} =\ell^{2},
\end{equation}
living in the embedding space with the metric 
\begin{equation}
	ds^{2} =  - (dX_{0})^{2}-(dX_{1})^{2} +  (dX_{2})^{2}  + (dX_{3})^{2}  \,. 
\end{equation}
It is then easy to find that the geodesic length $d_{\rm geodesic}$ between two points in AdS$_3$ is given by
\begin{equation}
	d_{\rm geodesic}  \equiv  \ell  \, \arccosh \( D_{\rm geodesic}  \)= \ell  \, \arccosh \(  \frac{X_0 X_0' + X_1 X_1' - X_2 X_2' - X_3 X_3'}{\ell^2}   \)   \,. 
\end{equation}
For example, in global coordinates \eqref{eq:ads3}, this geodesic distance   becomes
\begin{equation}
	D_{\rm global} = \sqrt{1 + \frac{R^2}{\ell^2}}\sqrt{1 + \frac{R'^2}{\ell^2}}\cos \( \frac{T-T'}{\ell}  \) - \frac{RR'}{\ell^2} \cos \(  \Phi- \Phi'  \)\,.
\end{equation}
Considering the BTZ black hole metric defined in \eqref{eq:BTZ}, the geodesic distance becomes
\begin{equation}\label{eq:BTZd1}
	D_{\mathrm{BTZ}}=\frac{r r' \cdot \cosh \left(\frac{r_h \Delta \phi}{\ell}\right)-\sqrt{\left(r^2-r_h^2\right)\left(r'^2-r_h^2\right)} \cosh \left(\frac{r_h \Delta t}{\ell^2}\right)}{r_h^2} \,,
\end{equation}
with $\Delta \phi \equiv \phi- \phi'$ and $\Delta t = t- t'$. One can also define the two-sided BTZ geometry by using the Kruskal coordinates. Correspondingly, we can obtain the length of a geodesic connecting the left and right regions as
\begin{equation}\label{eq:BTZd2}
	D_{\mathrm{BTZ}} =\frac{r_{\mt{R}} r_{\mt{L}} \cdot \cosh \left(\frac{r_h ( \phi_{\mt L }-\phi_{\mt{R}})}{\ell}\right)+\sqrt{\left(r_{\mt{R}}^2-r_h^2\right)\left(r_{\mt{L}}^2-r_h^2\right)} \cosh \left(\frac{r_h}{\ell^2}\left(t_{\mt{R}}-t_{\mt{L}}\right)\right)}{r_h^2} \,.
\end{equation}
Obviously, the latter is related to the one-sided geodesic \eqref{eq:BTZd1} by taking the analytic continuation $(r,r',t,t')\to (r_\mt{L},r_\mt{R},t_{\mt{L}},t_{\mt{R}} + i \frac{\pi \ell^2}{r_h})$. Note here that our convention for the time arrows on the two asymptotic boundaries is that $t_{\mt{L}}$ runs down on the left boundary of the Penrose diagram and $t_{\mt{R}}$ runs up on the right boundary.

\subsection{Entanglement with the symmetric boundary} \label{sec:mutsym}
In this subsection, our focus lies on the investigation of holographic entanglement entropy and mutual information in the symmetric setup with $L_1 = L_2$. We carry out these calculations analytically utilizing the explicit geodesic distances given above for any two points in AdS$_3$ spacetime. Notably, we will also need to find the extremal curves when the geodesics intersect the branes. This entails a process of extremization over the intersection points along the brane. For the sake of conciseness, we shall omit the intricate details of these calculations and provide the salient results for the holographic entanglement entropy. Given multiple extremal surfaces with respect to fixed boundary subregions, the final result of the holographic entanglement entropy is governed by the smallest distance among the choices of these extremal curves. Since there are only cold or hot phases for the symmetric setup, we need to evaluate the entanglement entropy and mutual information in both phases. To distinguish the two phases, we denote the transition temperature as $T_{\mt{CH}}$, \ie 
\begin{equation}\label{eq:TCH}
	T_{\mt{CH}} \equiv  \frac{C_\mt{CH}}{L_1+L_2}=  \frac{1}{ L_1 + L_2}\( \sqrt{ 1 +\frac4{\pi^2} \, \arctanh^2 \( T_o\ell \)   }  - \frac2\pi\, \arctanh \( T_o\ell \)  \) \,,
\end{equation}
which was derived in eq.~\eqref{eq:ColdHot}. In the following, we consider the left and right boundaries on the time slices $t_{\mt{L}}$ and  $t_{\mt{R}}$, respectively. We start with the holographic entanglement entropy of different defect regions and then proceed to the evaluation of the mutual information.

\subsubsection{Holographic entanglement entropy of defects}
We begin by examining the holographic entanglement entropy associated with different defects, which will subsequently be used to evaluate the mutual information. As explained above eq.~\reef{rooster2}, we begin by regularizing each defect with a small interval of fixed length $\Delta L_{\mt{D}}$, followed by taking the limit as $\Delta L_{\mt{D}} \to 0$ to derive the final results.

For the case of a single defect, such as $D^{(\mt{R})}_{a}$ or $D^{(\mt{R})}_{b}$, the corresponding holographic entanglement entropy is obtained from the extremal geodesic connecting the two endpoints of the defect interval. We will refer to this type of geodesic as ``connected" geodesics (as shown in the left column of figure \ref{fig:Table}). For both phases, the results are summarized as 
\begin{equation}
	S_{\mt{EE}}  (   D^{(\mt{R})}_{a}   )= 			\begin{cases}
		\frac{c}{3}  \[  \log \(  \frac{ L_{\rm bdy }}{\pi \epsilon }  \sin \( \frac{ \pi \Delta L_{\mt{D}}}{ L_{\rm bdy} }\)  \) +\log \(  \sqrt{\frac{1+ T_o\ell}{1- T_o\ell}}\)   \]  \,, \quad   T_{\mt{DCFT}}  \le  T_{\mt{CH}} 
		\,\\	
		\,\\
		\frac{c}{3}  \[  \log \(  \frac{\beta}{ \pi \epsilon}   \sinh \(  \frac{\pi}{\beta}  \Delta L_{\mt{D}} \) \)  +   \log \(   \sqrt{   \frac{ 1+T_o\ell}{1-T_o\ell}  }\) \]  \,, \quad  T_{\mt{DCFT}}  \ge  T_{\mt{CH}}   \,. \\ 
	\end{cases}
\end{equation}
Here, the notation for the inverse temperature $\beta = 1/T_{\mt{DCFT}} $ is used to maintain consistency with the established convention in other literature. Importantly, it is worth noting that the two expressions, pertaining to the cold and hot phases, can be interchanged using the following replacement 
\begin{equation}\label{eq:replacement}
	L_{\rm bdy}\equiv L_1 + L_2  \quad  \longleftrightarrow   \quad i \beta= \frac{i}{T_{\mt{DCFT}}}   \,. 
\end{equation}

However, it is obvious that the length of connected extremal geodesic becomes ill-defined if we take the limit as $\Delta L_{\mt{D}}$ approaches 0. However, as in eq.~\reef{rooster2}, while we wish $\Delta L_{\mt{D}}$ to be smaller than any macroscopic scales, we keep $\Delta L_{\mt{D}}\gg\epsilon$. Hence in evaluating the above expressions, we have
\begin{equation}\label{eq:canonical}
	 \frac{\Delta L_{\mt{D}}}{ \epsilon}  =	\lim\limits_{ \Delta L_{\mt{D}}/L_{\rm bdy} \to 0}  \(    \frac{L_{\rm bdy} }{ \pi \epsilon}   \sin \( \frac{ \pi \Delta L_{\mt{D}}}{ L_{\rm{bdy}} }\)   \)   =  \lim\limits_{ \Delta L_{\mt{D}}/\beta \to 0}  \(  \frac{\beta}{\pi \epsilon} \sinh \(  \frac{\pi}{\beta}  \Delta L_{\mt{D}} \)   \)   \,.
\end{equation}
It is worth noting that keeping ${\Delta L_{\mt{D}}}/{ \epsilon} $ as fixed in both phases ensures the entanglement entropy $S_{\mt{EE}} (D^{(\mt{R})}_{a} )$ is continuous at the transition point. This choice effectively regulates the entanglement entropy of each defect, yielding
\begin{equation}\label{eq:defectentropy}
	S_{\mt{EE}} ( D^{(\mt{R})}_{a})= 		
	\frac{c}{3}  \(   \log \(\frac{\Delta L_{\mt{D}}}{ \epsilon} \)+ \frac12\, \log \(     \frac{ 1+T_o\ell}{1-T_o\ell} \)  \)   \equiv S^\mt{(reg)}_{\rm{def}} \,, 
\end{equation}
which remains the same for both the cold and hot phases. Note that the second term is just the defect entropy $S_{\rm{def}}$ defined in eq.~\reef{eq:defg}, while the first term is simply the entanglement entropy that would be found for the small interval $\Delta L_{\mt{D}}$ in the CFT vacuum. Notably, this regulated defect entropy is independent of the bulk phases and the temperature, as it primarily characterizes the intrinsic information related to the boundary conditions on the defects.

Furthermore, let us consider the holographic entanglement entropy associated with two defects on the same side, \eg $D^{(\mt{R})}_{a,b} = D^{(\mt{R})}_{a} \cup D^{(\mt{R})}_{b}$. Apart from the connected geodesics introduced above, there are disconnected geodesics that can connect $D^{(\mt{R})}_{a}$ and $D^{(\mt{R})}_{b}$.  For the cold phase, the holographic entanglement entropy is given by the minimal area of the connected and disconnected extremal surfaces, \ie  
\begin{equation}
	S_{\mt{EE}}^{(\mt{C})}  ( D^{(\mt{R})}_{a,b})= \min 
	\begin{cases}
		\frac{c}{3} \log \(  \frac{ L_{\rm bdy} }{ \pi \epsilon }  \sin \(   \frac{ \pi (L_1-\Delta L_{\mt{D}})}{ L_{\rm bdy }}\)   \)  + \frac{c}{3} \log \(  \frac{ L_{\rm bdy} }{ \pi \epsilon }  \sin \(   \frac{ \pi (L_2-\Delta L_{\mt{D}})}{ L_{\rm bdy} }\)   \) \,, \\ 
		\,\\
		\frac{2c}{3}  \[  \log \(  \frac{ L_{\rm bdy} }{\pi \epsilon }  \sin \( \frac{ \pi \Delta L_{\mt{D}}}{ L_{\rm bdy} }\)  \) +\log \(  \sqrt{\frac{1+ T_o\ell}{1- T_o\ell}}\)   \]   \,, \\
	\end{cases} 
\end{equation}
which does not depend on the temperature $T_{\mt{DCFT}}$. Taking the defect limit with $\DLD \to 0$ yields 
\begin{equation}
	\lim\limits_{ \Delta L_{\mt{D}} \to 0}  		S_{\mt{EE}}^{(\mt{C})}  ( D^{(\mt{R})}_{a,b})= \min  \, \left\{  \frac{2c}{3} \log \(  \frac{ L_1 +L_2 }{ \pi \epsilon }    \)  \,, \, 2 \,S^\mt{(reg)}_{\rm{def}}   \right\} \,.
\end{equation}
In this limit, the first expression (coming from the connected geodesics) is minimal when 
\begin{equation}\label{eq:conditionalpha}
	\sqrt{\frac{1+ T_o\ell}{1- T_o\ell}}  \ge   \frac{L_1 +L_2}{\pi\DLD} \,, \qquad \text{condition-}\alpha
\end{equation}
is satisfied.

At high temperatures, the bulk spacetime for each side is given by a BTZ black hole at a temperature $T_{\mt{DCFT}}= T_{\mt{BH}}= \frac{r_{h}}{2\pi \ell^2}  $. The corresponding holographic entanglement entropy for two defects on one side, \eg $D^{(\mt{R})}_{a,b}$, is derived as 
\begin{equation}\label{eq:SEEsymDab}
	S_{\mt{EE}}^{(\mt{H})}   ( D^{(\mt{R})}_{a,b} )= \min 
	\begin{cases}
		\frac{c}{3} \log \(  \frac{\beta}{ \pi \epsilon}   \sinh \(  \frac{\pi \( L_1 - \Delta L_{\mt{D}}  \)}{\beta}  \)  \) +	\frac{c}{3} \log \(  \frac{\beta}{ \pi \epsilon}   \sinh \(  \frac{\pi\( L_2 - \Delta L_{\mt{D}}  \)}{\beta}  \)  \) +S_{\rm{horizon}}  \,,  \\ 
		\,\\
		\frac{2c}{3}  \[  \log \(  \frac{\beta}{ \pi \epsilon}   \sinh \(  \frac{\pi}{\beta}  \Delta L_{\mt{D}} \) \)  +   \log \(   \sqrt{   \frac{ 1+T_o\ell}{1-T_o\ell}  }\) \]  \,.\\
	\end{cases}
\end{equation}
Here, we also include the area of black hole horizon $S_{\rm horizon}$ in the contribution to the entanglement entropy because the disconnected-disconnected phase for the RT surface also encompasses the black hole sitting at the middle of the glued spacetime. The corresponding RT surfaces for $S_{\mt{EE}}^{(\mt{H})}   ( D^{(\mt{R})}_{a,b} )$ are represented by the  two plots to the far left in figure \ref{fig:RTHot01} after taking $L_1=L_2$.\footnote{Obviously, the RT surfaces shown in the right two plots in figure \ref{fig:RTHot01} can not dominate here since they are not symmetric.} The Bekenstein-Hawking entropy $S_{\rm horizon}$ is thus defined by the area of horizon, \ie 
\begin{equation}\label{eq:BHhot}
	\begin{split}
		S_{\rm{horizon}}  &\equiv  \frac{  r_h\Delta \phi^{\rm{Hor}}_1 +r_h \Delta \phi^{\rm{Hor}}_2}{4\GN}  
		= \frac{c }{3} \pi T_{\mt{BH}}  \(    L_1 +   L_2   \) + \frac{2c }{3}  \log \(   \sqrt{   \frac{ 1+T_o\ell}{1-T_o\ell}  }\)   \,.\\
	\end{split}
\end{equation}
where $\Delta \phi^{\rm{Hor}}_i$ is given in eq.~\eqref{eq:hotmap}. The latter expression corresponds to the entropy of the thermal CFT on the boundary of length $L_1+L_2$ plus the entropy $S_{\rm def}$ of the two defects.  Hence we note that $S_{\rm{horizon}}$ is not the black hole entropy $S_{\mt{BH}}$ of the original BTZ black hole when the tension of the brane is nonzero. Similar to the observation for  $S_{\mt{EE}}( D^{(\mt{R})}_{a} )$, it is obvious that the holographic entanglement entropy $S_{\mt{EE}}( D^{(\mt{R})}_{a,b} )$ in the cold phase (without $S_{\rm{horizon}}$ part) and hot phase are related to each other by the replacement in eq.~\eqref{eq:replacement}. Although there are two types of extremal surface in the hot phase, we can find that the minimal one is always given by the connected geodesics in the limit $\DLD \to 0$. As a result, the holographic entanglement entropy of the left/right defects in hot phases are nothing but twice the defect entropy $S_{\rm Defect}$, namely 
\begin{equation}
	\lim\limits_{ \Delta L_{\mt{D}} \to 0}  	S_{\mt{EE}}^{(\mt{H})}   ( D^{(\mt{R})}_{a,b} )
	=	
	\frac{2c}{3}  \(   \log \(\frac{\Delta L_{\mt{D}}}{ \epsilon} \)+ \frac12\, \log \(     \frac{ 1+T_o\ell}{1-T_o\ell} \)  \)   =2\, S^\mt{(reg)}_{\rm{def}}   \,.
	\label{hanger7}
\end{equation}
We also note that there are some other types of disconnected extremal surfaces which can cross the brane twice. However, we can confirm that their areas cannot yield the minimal value in the limit $\DLD \to 0$. 

Finally, we investigate the most symmetric case by including four defects on both the left and right boundary. Since the left and right bulk spacetime are disconnected in the cold phase, the holographic entanglement entropy of four defects is simply given by 
\begin{equation}
	S_{\mt{EE}}^{(\mt{C})}  \(    D^{(\mt{R,L})}_{a,b}  \) = 	S^{(\mt{C})} _{\mt{EE}}  (  D^{(\mt{R})}_{a,b}  )+S^{(\mt{C})} _{\mt{EE}}  (  D^{(\mt{L})}_{a,b}  ) \,. 
\end{equation}
However, we have several types of geodesics which can cross the horizon and connect the defects on the left and right sides. The candidates of RT surfaces are shown by the first two plots in figure \ref{fig:RTHot02} and the left plot in figure \ref{fig:RTHot03} (after taking $L_1=L_2$). The corresponding entanglement entropy is provided by the minimum among the three candidates, \ie 
\begin{equation}\label{eq:SEELRhot}
	S_{\mt{EE}}^{(\mt{H})}   ( D^{(\mt{R,L})}_{a,b} )= \min 
	\begin{cases}
		\frac{4c}{3} \log \(  \frac{\beta}{ \pi \epsilon}   \sinh \(  \frac{\pi  \( L_1 - \Delta L_{\mt{D}}  \)}{\beta}\)  \)  \,,  \\ 
		\,\\
		\frac{4c}{3}  \[  \log \(  \frac{\beta}{ \pi \epsilon}   \sinh \(  \frac{\pi}{\beta}  \Delta L_{\mt{D}} \) \)  +   \log \(   \sqrt{   \frac{ 1+T_o\ell}{1-T_o\ell}  }  \) \]  \,, \\
		\,\\
		\frac{4c}{3}\log \(  \frac{\beta}{ \pi \epsilon}    \cosh \left(  \frac{\pi}{\beta}  (t_{\mt{R}} -t_{\mt{L}} ) \right)   \)  \,. \\ 
	\end{cases}
\end{equation}
where the factor $  \cosh \left(  \frac{2\pi}{\beta}  (t_{\mt{R}} -t_{\mt{L}} ) \right)$ originates from the (almost) linear growth of extremal surface in the wormhole region. Similar to the simplifications shown before, our regularization for the defect limit $\Delta L_{\mt{D}} \to 0$ leads to  
\begin{equation}
	S_{\mt{EE}}^{(\mt{H})}   ( D^{(\mt{R,L})}_{a,b} )=
	\min \left\{	\frac{4c}{3} \log \(  \frac{\beta}{ \pi \epsilon}   \sinh \(  \frac{\pi L_{\rm bdy}}{2\beta}\)  \), 4\,S^\mt{(reg)}_{\rm{def}}  \,, \frac{4c}{3}\log \(  \frac{\beta}{ \pi \epsilon}    \cosh \left(  \frac{\pi}{\beta}  (t_{\mt{R}} -t_{\mt{L}} ) \right)   \)     \right \} \,, 
\end{equation}
which is bounded from above by $4S^\mt{(reg)}_{\rm{def}} $. In the regime with a large defect entropy, \ie in the large tension limit, one find that the entanglement entropy jumps to a lower value after the phase transition to the hot phase, signifying the formation of a wormhole connecting the left and right regions.

\subsubsection{Holographic mutual information}
With the results of entanglement entropy of different defect regions, it is straightforward to evaluate the mutual information between different subregions.

\begin{figure}[t!]
	\centering
	\includegraphics[width=2.9in]{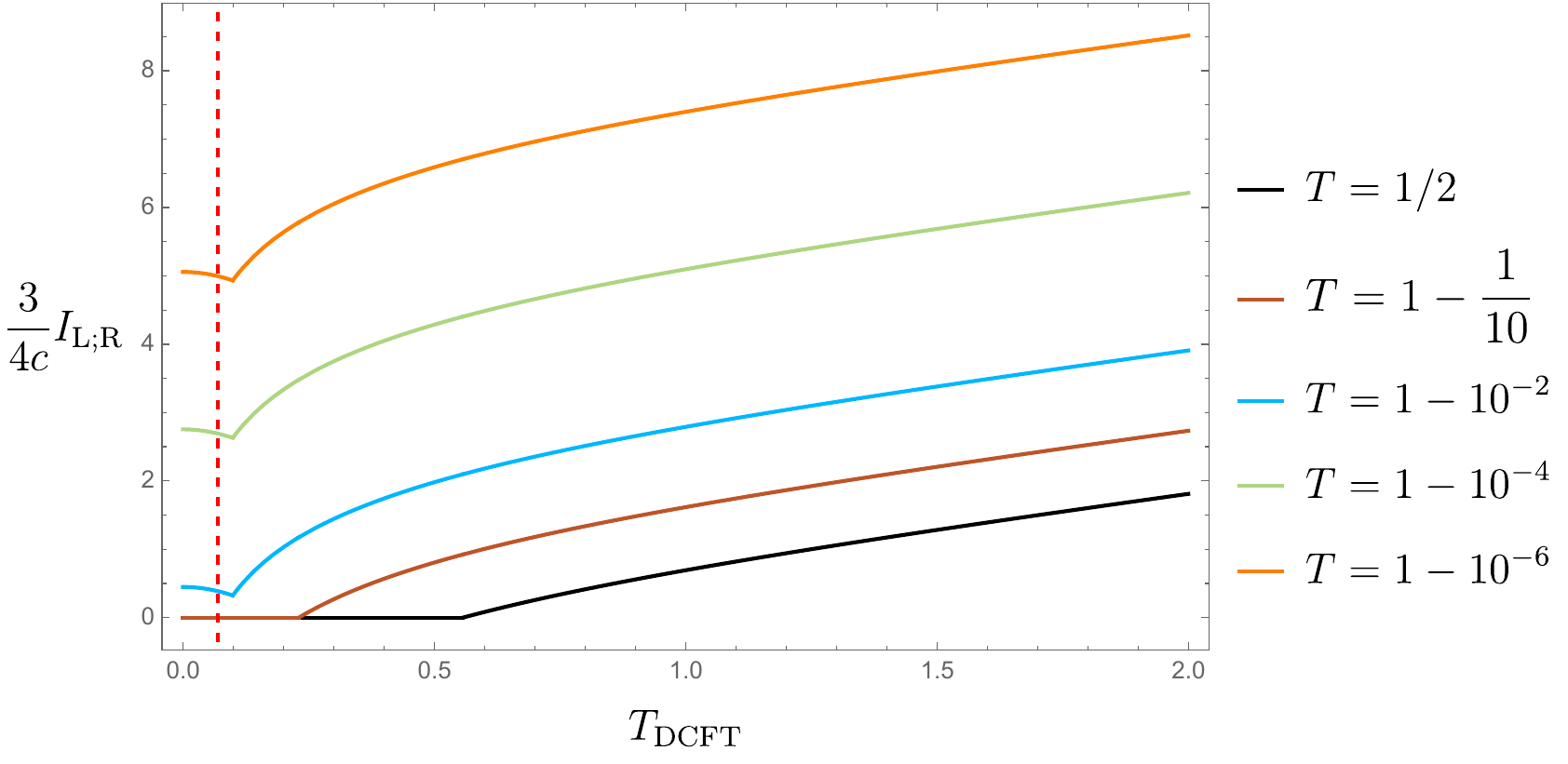}
	\includegraphics[width=2.9in]{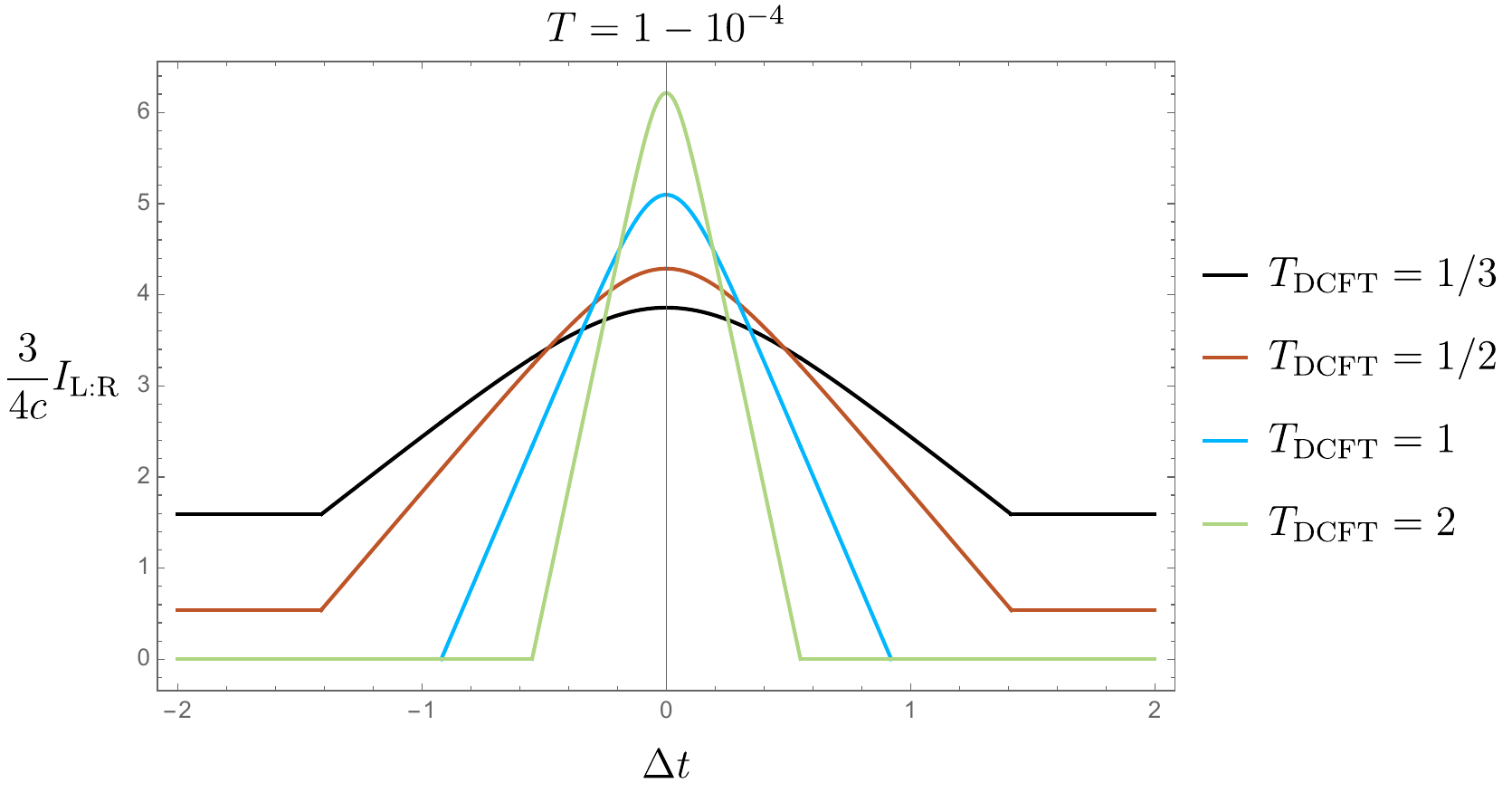}
	\caption{The mutual information between the left and right defects in the symmetric set-up. Left: the temperature dependence with fixing $t_{\mt{L}} =t_{\mt{R}}$  but varying the tension of the brane. Right: the time dependence with $\Delta t \equiv |t_{\mt{L}} - t_{\mt{R}}|$.}\label{fig:MutualLR}
\end{figure}

{ \bf Mutual Information $I_{\mt{R}_a:\mt{R}_b}$}: 
The mutual information between the two defects on one side is defined by 
\begin{equation}
	I_{\mt{R}_a:\mt{R}_b}=  S_{\mt{EE}}  (  D^{(\mt{R})}_{a}  )+S_{\mt{EE}}  (  D^{(\mt{R})}_{b}  )   -	S_{\mt{EE}}  ( D^{(\mt{R})}_{a,b} ) \,. 
\end{equation}
In the cold phase, this reduces to 
\begin{equation}\label{eq:IRaRb}
	I_{\mt{R}_a:\mt{R}_b}^{(\mt{C})}= 	\max \, \left \{ 0 \,,  
	\frac{2c}{3} \log \(   \frac{\pi  \Delta L_{\mt{D}}  }{L_1 +L_2}   \sqrt{ \frac{ 1+T_o\ell}{1-T_o\ell} }   \)   \right \}\,, 
\end{equation}
which is non-zero if the condition-$\alpha$  in eq.~\eqref{eq:conditionalpha} is satisfied. Recall to be in the cold phase, the temperature must also be lower than the critical one, \ie $T_{\mt{DCFT}} \le T_{\mt{CH}}$. 
On the other hand, in the hot phase (with $T_{\mt{DCFT}} \ge T_{\mt{CH}}$), the mutual information $I_{\mt{R}_a:\mt{R}_b}$ simply vanishes, \viz 
\begin{equation}
	I_{\mt{R}_a:\mt{R}_b}^{(\mt{H})}= 0 \,.
\end{equation}
This is expected because, in the hot phase, the branes $\mB_i$ extend from a defect on the right to another on the left, rather than between the two defects on the right side, \ie from  $D^{(\mt{R})}_{a}$ to $D^{(\mt{L})}_{a}$ rather than $D^{(\mt{R})}_{b}$.

{\bf Mutual Information $I_{\mt{R}_a:\mt{L}_a}$:} The mutual information of two defects on the left and right side is defined by 
\begin{equation}
	I_{\mt{R}_a: \mt{L}_a}:=  S_{\mt{EE}}  (  D^{(\mt{R})}_{a}  )+S_{\mt{EE}}  (  D^{(\mt{L})}_{a}  )   -	S_{\mt{EE}}  ( D^{(\mt{R})}_{a}   \cup  D^{(\mt{L})}_{a}  ) \,.
\end{equation}
After taking account of the regularization with $\DLD \to 0$, it reduces to 
\begin{equation}
	\begin{cases}
		I_{\mt{R}_a: \mt{L}_a}^{(\mt{C})} =	0 \,,  \\ 
		\,\\
		I_{\mt{R}_a: \mt{L}_a}^{(\mt{H})} =		\max \(  0\,,	\frac{2c}{3}\log \(  \frac{   \sinh \(  \frac{\pi}{\beta}  \Delta L_{\mt{D}}  \)  }{ \cosh \left(  \frac{\pi}{\beta}  (t_{\mt{R}} -t_{\mt{L}} ) \right) } \sqrt{ \frac{ 1+T_o\ell}{1-T_o\ell} }    \)    \)  \,.
	\end{cases}
\end{equation}

{\bf Mutual Information $I_{\mt{L:R}}$:} Finally, we are interested in the mutual information between the left and right defects, \ie 
\begin{equation}
	I_{\mt{L:R}}:=  S_{\mt{EE}}  ( D^{(\mt{R})}_{a,b} )+S_{\mt{EE}}  ( D^{(\mt{L})}_{a,b} )   -	S_{\mt{EE}}  ( D^{(\mt{R,L})}_{a,b} ) \,.
\end{equation}
It is straightforward to obtain 
\begin{equation}
	\begin{cases}
		I_{\mt{L:R}}^{(\mt{C})}= 	0 \,,   \\ 
		\,\\
		I_{\mt{L:R}}^{(\mt{H})}= 	\max \left\{  \frac{4c}{3} \log \(  \frac{   \sinh \(  \frac{\pi}{\beta}  \Delta L_{\mt{D}}  \)  }{ \sinh  \( \(  \frac{\pi}{\beta}  \( L_1 - \Delta L_{\mt{D}}  \)\)\) } \sqrt{ \frac{ 1+T_o\ell}{1-T_o\ell} }  \)\,,  0\,,	\frac{4c}{3}\log \(  \frac{   \sinh \(  \frac{\pi}{\beta}  \Delta L_{\mt{D}}  \)  }{ \cosh \left(  \frac{\pi}{\beta}  (t_{\mt{R}} -t_{\mt{L}} ) \right) } \sqrt{ \frac{ 1+T_o\ell}{1-T_o\ell} }    \)    \right\} 	 \,.
	\end{cases}
\end{equation}
Taking the symmetric time slice with $t_{\mt{L}} =t_{\mt{R}} $, we find that the mutual information $I_{\mt{L:R}}$ in the hot phase is dominated by the defect contributions
\begin{equation}
	I_{\mt{L:R}}^{(\mt{H})} \simeq 4\(\frac{\pi\,c}{3}\,   T_{\mt{DCFT}} \,\DLD +  S_{\rm def}\) +\cdots \,. 
\end{equation}
That is, we have $S_{\rm def}$ in eq.~\reef{eq:defg} for each defect plus the thermal entropy of the CFT in the interval around these. This comes from the extremal surfaces connecting the left and right boundaries in the hot phase. At the late time with $|t_{\mt{L}} - t_{\mt{R}}| \gg L_1$, the mutual information $ I_{\mt{L:R}}^{(\mt{H})}$ between two sides always reduces to a constant. These behaviours are also illustrated in the figure \ref{fig:MutualLR}.

%
\subsection{Holographic mutual information}\label{sec:mutualinformation}
\begin{figure}[t!]
	\centering
	\includegraphics[width=2in]{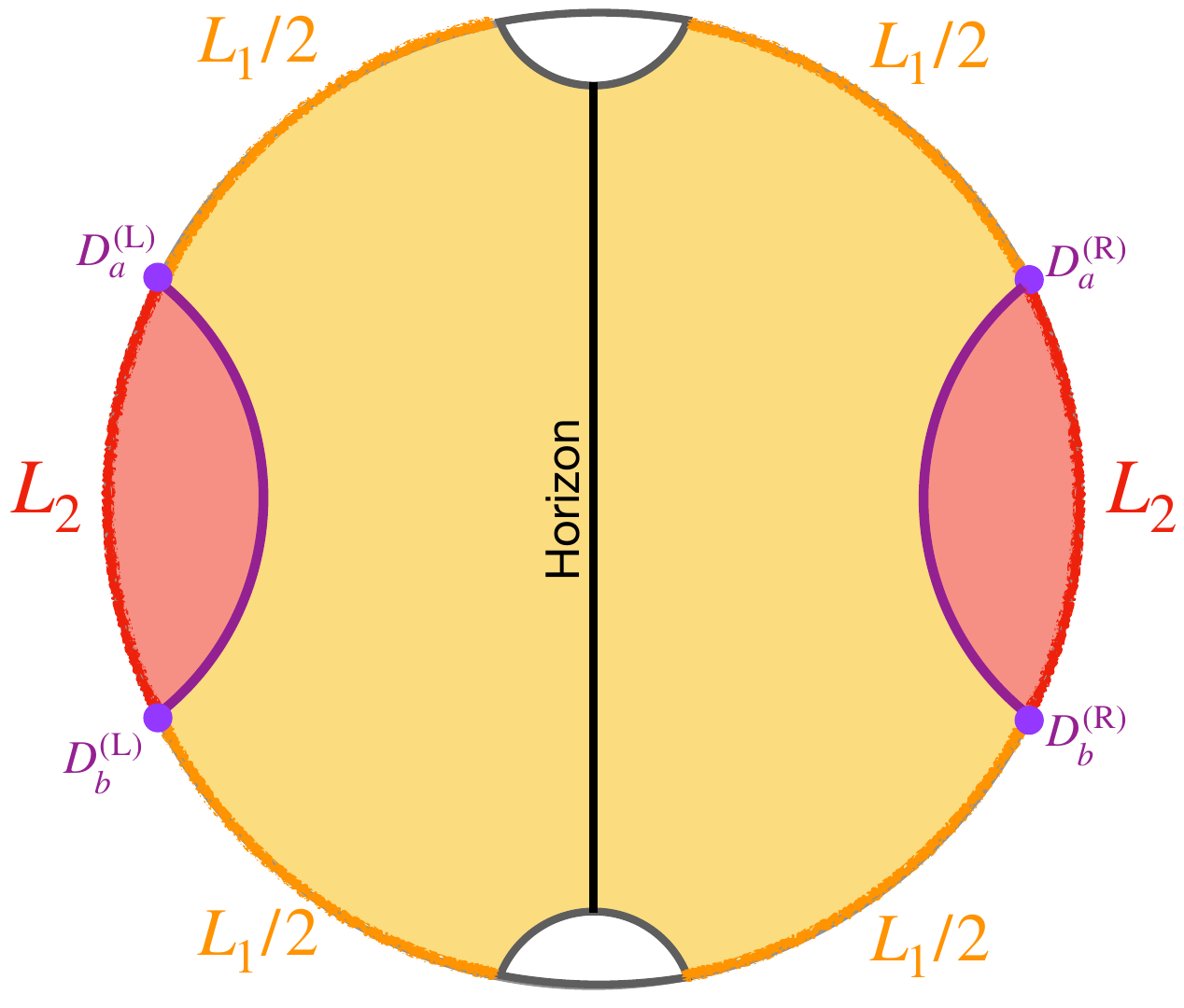}
	\quad 
	\includegraphics[width=2in]{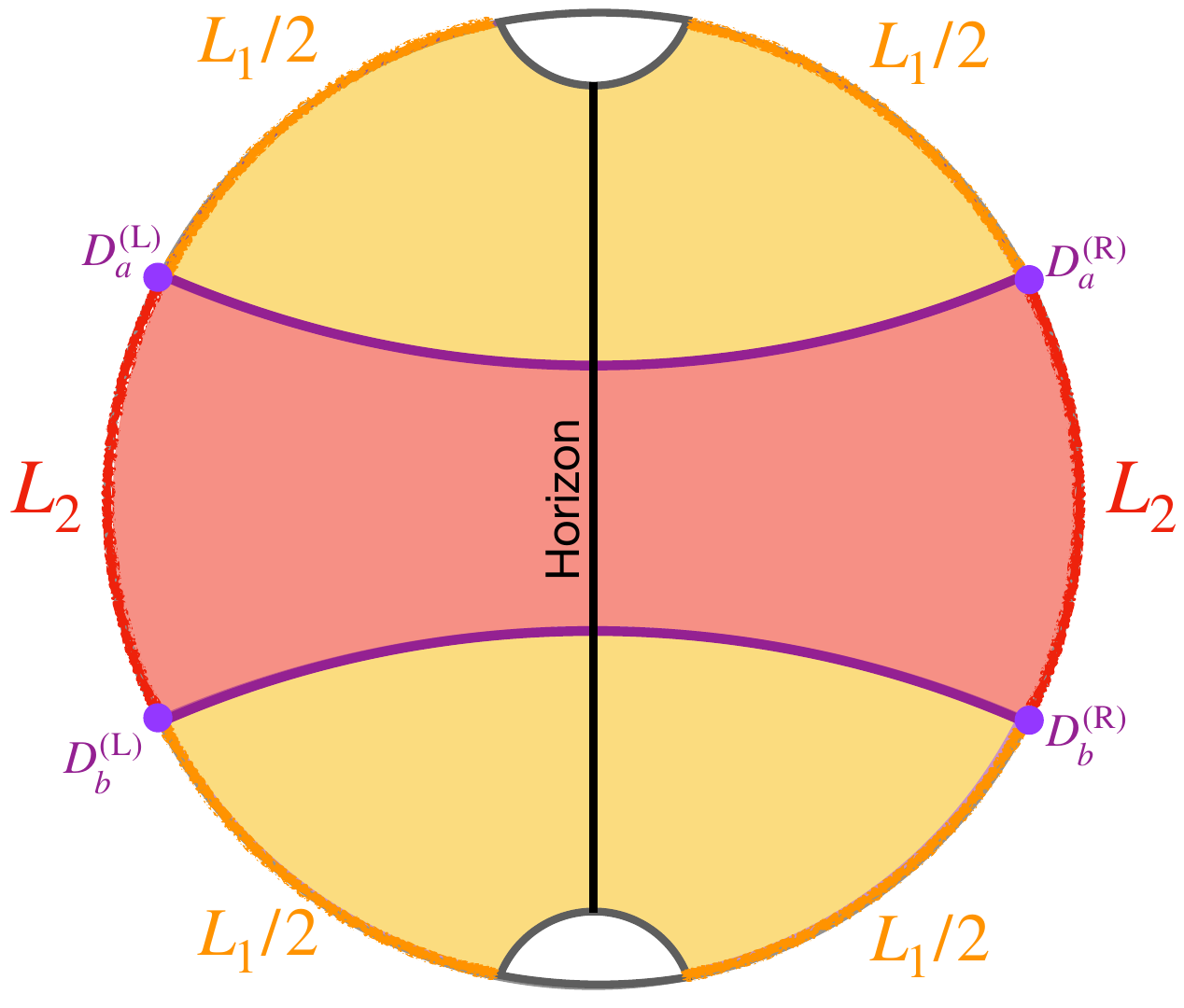}
	\caption{A time slice ($t_{\mt{L}}=0=t_{\mt{R}}$) of the bulk spacetime in the warm (left figure) and hot (right figure) phases. The gray curves denote the geodesics identified to make a quotient of the global AdS$_3$ with left and right boundaries. The black line in the middle indicates the position of the horizon of BTZ black hole, \ie $r=r_h$. }\label{fig:global}
\end{figure}

In contrast to the regime with $L_1 \sim L_2$ like the symmetric case above, the bulk spacetime allows three distinct phases when $L_1 \gg L_2$. To illustrate the corresponding entanglement, we consider the holographic calculation of the mutual information while assuming $L_1 \gg L_2$, which helps to simplify the calculations because the relevant extremal surfaces are limited. 

In the holographic calculations for the symmetric setup, we regulated the defects by including a small boundary interval around each defect, with a length $\DLD$ which is a small length measure compared with the AdS radius $\ell$ -- see eq.~\reef{rooster2}. This is motivated by our goal that the defect region is considered as the boundary dual of the braneworld in our doubly holographic construction. However, in order to make it true, one can expect that the defect region cannot be too small otherwise all information is localized inside the defects. Taking the symmetric case as an example, one can find that all mutual information between these defects vanishes if one simply takes $\DLD$ to be too small. For example, examing eq.~\eqref{eq:IRaRb}), we find that a critical size  given by $\frac{\DLD}{L_{\rm bdy}} =\frac1\pi\, \sqrt{\frac{1- T_o\ell}{1+ T_o\ell}} $ where $I_{\mt{R}_a:\mt{R}_b}^{(\mt{C})}=0$. Inspired by this result, we extend eq.~\reef{rooster2} to restrict the following holographic analysis to the following regime
\begin{equation}\label{eq:regime}
	\epsilon	\ll \ell \sqrt{\frac{1- T_o\ell}{1+ T_o\ell}}  \ll {\DLD} \ll \ell\,. 
\end{equation}
The first condition guarantees that the brane still resides below the UV cutoff surface in the bulk AdS$_3$ spacetime and the last two conditions will ensure that the defect is well approximated by the holographic gravity theory on the brane.\footnote{The AdS radius $\ellB$ of braneworld is fixed by the brane tension $T_o$ , as we discuss in appendix \ref{sec:brane}. From eq.~\eqref{eq:Lbrane}, we have $\ellB = \ell/\sqrt{1- (T_o\ell)^2}  \gg \ell$ in the large tension limit with $T_o \ell \sim 1$. From the brane perspective, this regime suppresses the contributions from the higher curvature terms in the effective gravitational action \reef{sock99} on the branes, \eg see \cite{Grimaldi:2022suv}.}

\subsubsection{Mutual information in the cold phase}
\begin{figure}[t]
	\centering
	\includegraphics[width=5in]{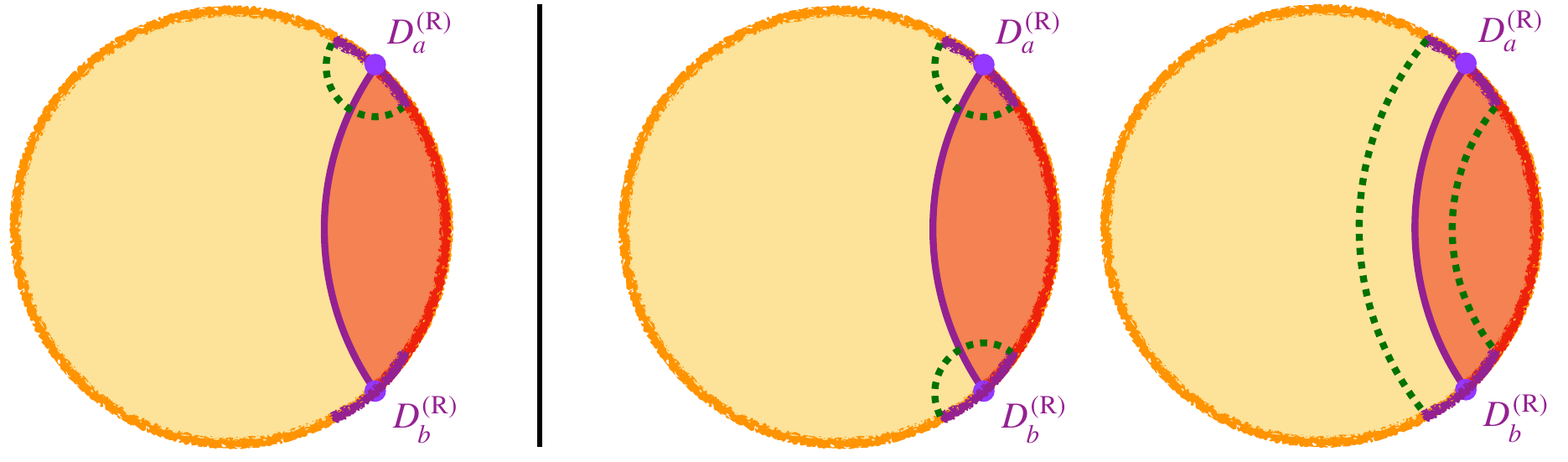}
	\caption{The extremal surfaces in the cold phase. Left: The connected extremal surface (green curves) for a single defect $D^{(\mt{R})}_{a}$. Right: Two types of extremal surfaces (green curves) with respect to one-sided defects, \ie $D^{(\mt{R})}_{a,b}$. The left and right one are referred to as the disconnected and connected phases, respectively.}\label{fig:RTCold}
\end{figure}

We begin with the cold phase which is the dominant solution at low temperatures. More precisely, it is controlled by the dimensionless boundary size, \ie $\TDCFT L_1  \le C_{\mt{CW}}$ as discussed around eq.~\eqref{eq:Ccw}. Noting that we are interested in the regime with $L_1 \gg L_2$, one  concludes that the cold phase arises with
\begin{equation}
	 \TDCFT L_2  \le  C_{\mt{CW}} \ \frac{L_2}{L_1} \approx  \frac{L_2}{L_1}  \ll 1 \,. 
\end{equation}

We first need to evaluate the holographic entanglement entropy for the individual defects, as we did in the symmetric setup. However, this calculation requires more work here since the two boundary intervals have different sizes. The extremal surface associated with a single defect, \eg $D^{(\mt{R})}_{a}$ is depicted in the left plot of figure \ref{fig:RTCold}. As before, since this surface crosses the brane and connects the two endpoints of the defect interval, we need to minimize over the intersection point where it meets the brane to find the minimal surface. However, since the defect region is small with $\DLD \ll \ell$, we expect to find that the leading contribution in the limits $\DLD \to 0$ and $T_o \ell \to 1$ is 
\begin{equation}\label{eq:lengths}
		S_{\mt{EE}} ( D^{(\mt{R})}_{a}) =
	\frac{c}{3}  \(   \log \(  \frac{\DLD}{\epsilon} \)+  \log \(   \sqrt{   \frac{ 1+T_o\ell}{1-T_o\ell}  }\)    + \mathcal{O} (1)   \)  \simeq S^\mt{(reg)}_{\rm{def}} \,, 
\end{equation}
where we have ignored subleading corrections and in the last expression, recast the formula as the regulated defect entropy defined in eq.~\eqref{eq:defectentropy}. This leading contribution is universal because near the conformal boundary at $r\to \infty$, the brane geometry is asymptotically given by AdS$_2$ and the tension fixes the angle between the AdS$_2$ brane and conformal boundary. We will also use the same formula for $	S_{\mt{EE}} ( D^{(\mt{R})}_{a}) $ in both warm and hot phases in the later subsections.

Considering the two defects on a single asymptotic boundary, the extremal surfaces associated with, \eg $S_{\mt{EE}} ( D^{(\mt{R})}_{a,b})$, come in two distinct classes, as shown in the right plot of figure \ref{fig:RTCold}. The corresponding holographic entanglement entropy is determined by the minimal  length among the two pairs of candidate surfaces, \ie 
\begin{equation}\label{eq:SEEDabCold}
	S_{\mt{EE}} ( D^{(\mt{R})}_{a,b})=
	 \min 
	\begin{cases}
	2S_{\mt{EE}} ( D^{(\mt{R})}_{a}) \simeq 2S^\mt{(reg)}_{\rm{def}}  \,, \\ 
		\,\\
	\frac{c}{3} \log \left( \f{L_{\mathcal{S}_{1}}}{\pi \epsilon}  \sin \f{\pi L_{1}}{L_{\mathcal{S}_{1}}}\right)+ \f{c}{3} \log \left( \f{L_{\mathcal{S}_{2}}}{\pi \epsilon}  \sin \f{\pi L_{2}}{L_{\mathcal{S}_{2}}}\right) \,, \\
	\end{cases} 
\end{equation}
where $L_{\mathcal{S}_{1}} =\frac{2\pi \ell}{\sqrt{-m_{1}}} $ and $L_{\mathcal{S}_{1}} =\frac{2\pi \ell}{\sqrt{-m_{2}}}$ are the boundary perimeter which the full bulk spacetimes associated with $\mathcal{S}_{1}$ and $\mathcal{S}_{2}$ would have had if they had not been cutoff by the brane in constructing the holographic dual of the defect CFT.
Generally, the minimization defined in eq.~\eqref{eq:SEEDabCold} is nontrivial and depends on all of the boundary parameters. However, it is easy to realize that the extremal surface crossing the brane (as shown in figure \ref{fig:RTCold}) is not preferred in the regime of present interest \eqref{eq:regime} because of the large contribution coming from the defect entropy. As a result, we always have non-vanishing mutual information between the two defects regions $D^{(\mt{R})}_{a}, D^{(\mt{R})}_{b}$. More explicitly, this mutual information can be expressed as   
\begin{equation}
	I_{\mt{R}_a:\mt{R}_b}^{(\mt{C})} \simeq    4\log g-	\f{c}{3} \log \left( \f{L_{\mathcal{S}_{1}}}{\pi \DLD}  \sin \f{\pi L_{1}}{L_{\mathcal{S}_{1}}}\right) - \f{c}{3} \log \left( \f{L_{\mathcal{S}_{2}}}{\pi \DLD}  \sin \f{\pi L_{2}}{L_{\mathcal{S}_{2}}}\right)  \,. \\
\end{equation}
The two defects are obviously connected by the brane living in the bulk spacetime. The non-vanishing mutual information here is then a manifestation of the ``ER=EPR" proposal \cite{Maldacena:2013xja}. Although it is hard to obtain a generic analytic answer, we can consider the limit $L_2/L_1 \to 0$ which is realized by taking $m_2\to -\infty$. From the results derived in eqs.~\eqref{eq:defk2} and \reef{rooster4}, it is straightforward to find the dominant contributions given the boundary perimeters, \ie 
\begin{equation}\label{eq:coldL1L2}
	\begin{split}
			L_{\mathcal{S}_{1}} &\equiv \frac{2\pi \ell}{ \sqrt{-m_1}} \simeq  L_1  + \frac{k_1 L_2}{\pi- k_2} + \mathcal{O}(L_2^{\,2}/L_1)\,,  \\
			L_{\mathcal{S}_{2}} &\equiv \frac{2\pi \ell}{ \sqrt{-m_2}} \simeq  \frac{\pi L_2}{\pi- k_2} + \mathcal{O}(L_2^{\,2}/L_1)\,.   \\
	\end{split}
\end{equation} 
Correspondingly, the mutual information $I_{\mt{R}_a:\mt{R}_b}^{(\mt{C})}$ reduces to 
\begin{equation}\label{eq:coldIab}
I_{\mt{R}_a:\mt{R}_b}^{(\mt{C})} \simeq  \frac{2c}{3}  \log \( \frac{\pi-k_2}{\sqrt{k_1 \sin (\pi -k_2)}}  \frac{\Delta L_{\mt{D}}  }{L_2}   \sqrt{ \frac{ 1+T_o\ell}{1-T_o\ell} }   \)  + \mathcal{O}\( \( \frac{L_2}{L_1} \)^2\) \,, 
\end{equation}
where we have substituted the simplified entanglement entropy
\begin{equation}\label{eq:ColdSDab}
	S_{\mt{EE}} ( D^{(\mt{R})}_{a,b}) \simeq \frac{c}{3} \log \(  \frac{\sin (\pi - k_2)}{\pi -k_2}\, \frac{L_2}{\epsilon} \) + \frac{c}{3} \log \(  \frac{k_1}{\pi- k_2} \frac{L_2}{\epsilon} \) + \mathcal{O}\( \( \frac{L_2}{L_1} \)^2\) \,. \\
	\end{equation}
We would like to comment on some features of the above approximations. First, the leading contributions do not depend on the boundary size $L_1$ because the extremal surfaces are far from the center of bulk spacetime when $L_2/L_1 \to 0$. Second, the square of the ratio $\frac{L_2}{L_1}$ suppresses the subleading corrections. Furthermore, note that the three phases only coexist in the regime where $\frac{L_2}{L_1}\ll1$, which implies that the leading contributions remain good approximations. 
It is interesting to note that even in the fusion limit $L_2 \ll L_1$, the entanglement entropy like $S_{\mt{EE}} ( D^{(\mt{R})}_{a,b}) $ does not reduce to the vacuum value. This nontrivial value of entropy exists because in this limit, the Casimir energy between two defects becomes large and negative. Hence, this can even affect the physics of the ultraviolet, resulting in the detectable difference between this state and the vacuum even in this short distance limit. 

Finally, let us mention that the mutual information $I_{\mt{L:R}}$ between the left and right defects simply vanishes in the cold phase because the corresponding bulk spacetimes are disconnected.

\subsubsection{Mutual information in the warm phase}\label{subsec:MIW}

Next, we turn our attention to the warm phase. As a reminder, the warm phase arises in the regime 
\begin{equation}\label{eq:WarmCondition}
	\frac{L_2}{L_1}	\approx  \frac{L_2}{L_1} C_{\mt{CW}} \le   T_{\mt{DCFT}} L_2 \le C_{\mt{WH}} \,.
\end{equation} 
Here the second inequality signifies the transition between the warm and hot phases, which is parametrized by
\begin{equation}
	C_{\mt{WH}} \simeq  \frac{(\pi -k_2 )\( \sqrt{(\pi-k_2)(\pi -k_1-k_2) + 4 \arctanh^2 (T_o\ell)}  - 2\arctanh(T_o\ell)  \) }{ (\pi -k_1 - k_2)\pi} \,, 
\end{equation}
which was given above in eq.~\eqref{eq:WHtrans}. The first inequality is equivalent to 
\begin{equation}\label{eq:WarmCondition2}
T_{\mt{DCFT}} L_1 \ge C_{\mt{CW}}  \approx 1 -   \frac{k_1}{\pi - k_2} \frac{L_2}{L_1} + \mathcal{O} \( (\frac{L_1}{L_2})^2 \) \,,
\end{equation}
which marks the transition between the warm and cold phases.
We have assumed that the boundary interval with a length $L_1$ is the bigger side. The boundary temperature is associated with the mass of the bulk black hole in the bulk region $\mS_1$ by $m_{1} =(2\pi\ell T_{\mt{DCFT}})^{2}$, as in eq.~\reef{parrot77}. The other bulk region $\mS_2$ is a portion of the AdS$_3$ spacetime with a mass denoted by $m_2<0$. 

\subsubsection*{Holographic entanglement entropy $S_{\mt{EE}}  (  D^{(\mt{R})}_{a,b} )$}
\begin{figure}[t]
	\centering
	\includegraphics[width=4in]{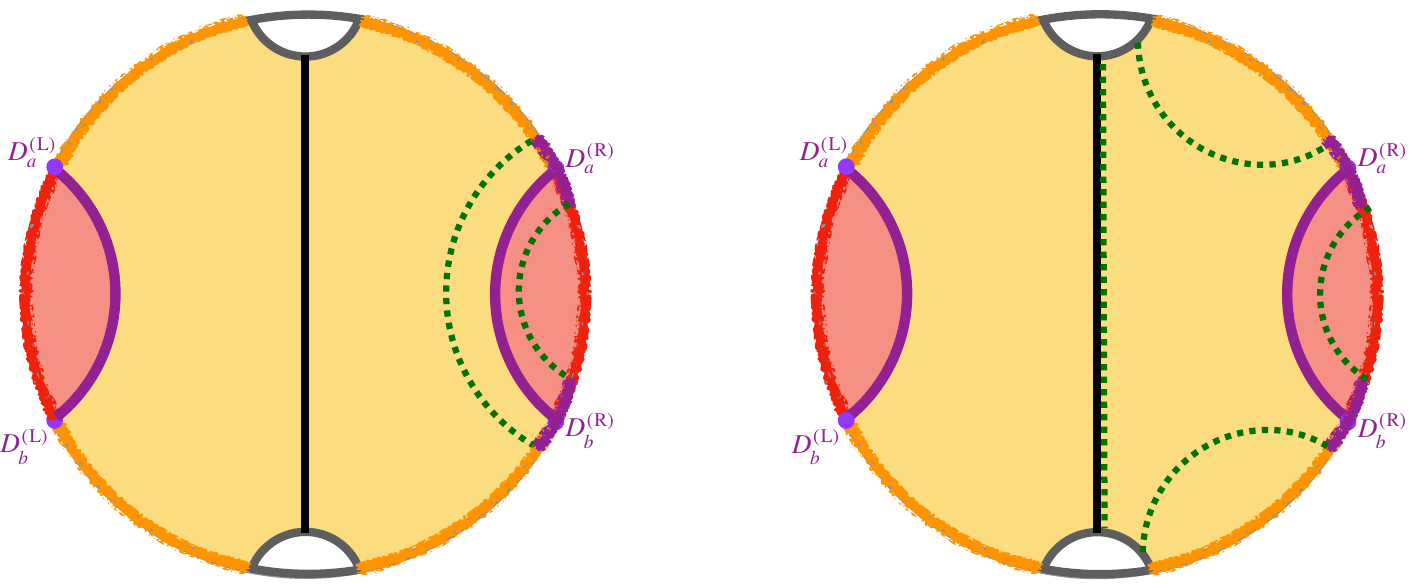}
	\caption{Relevant extremal surfaces in the warm phase (Green lines) for $S_{\mt{EE}}  (  D^{(\mt{R})}_{a,b} )$.} \label{fig:RTWarm01}
\end{figure}

\begin{figure}[t]
	\centering
	\includegraphics[width=4in]{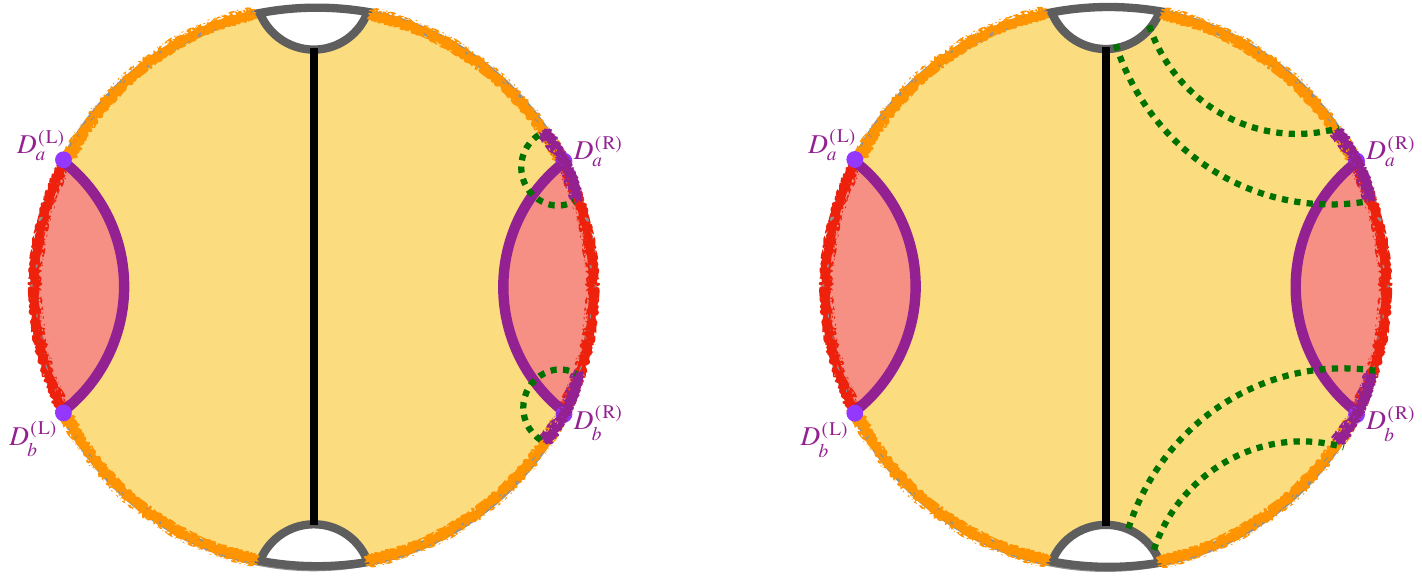}
	\caption{Irrelevant extremal surfaces in the warm phase (denoted by green curves) for $S_{\mt{EE}}  (  D^{(\mt{R})}_{a,b} )$.} \label{fig:RTWarm02}
\end{figure}

First of all, let us consider the entropy of the two defect regions $D^{(\mt{R})}_{a,b}$  on the right boundary. It is obvious that there are two families of extremal surfaces. The first two classes of geodesics stay away from the brane as shown in two plots of figure \ref{fig:RTWarm01}. In contrast, in the second two classes, the  geodesics cross the brane as depicted in figure \ref{fig:RTWarm02}.

Beginning with those shown in the left plot of figure \ref{fig:RTWarm01}, one can easily obtain the corresponding geodesic length because each bulk portion $\mS_i$ is still a locally AdS$_3$ spacetime. As a result, adding the two lengths yields the candidate result of the entanglement entropy 
\begin{equation}
S_{\mt{EE}}  (  D^{(\mt{R})}_{a,b} ) \overset{?}{=}  \f{c}{3} \log \left[ \f{\beta}{\pi \epsilon} \; \sinh \f{\pi ( L'_{\mS_1}- L_1 )}{\beta}\right] + 
 \f{c}{3} \log \left[ \f{L_{\mathcal{S}_{2}}}{\pi \epsilon}  \sin \f{\pi L_{2}}{L_{\mathcal{S}_{2}}}\right] \,, \label{eq:geoRab}
\end{equation}
where the first term corresponds to a geodesic in the BTZ black hole and the second, in global AdS$_3$. The two boundary perimeters in the warm phase are defined as 
\begin{equation}
L'_{\mathcal{S}_{1}} \equiv \Delta \phi_1^{\rm Hor} \ell = L_1 + 2 \Delta \phi_1 \ell  \,,\quad  L_{\mathcal{S}_{2}} \equiv \frac{2\pi\ell}{\s{-m_{2}}}\,. \label{eq:raddef}
\end{equation} 
Similar to the previous calculations, we can evaluate the above formula by taking the series expansion with $m_2 \to -\infty$, which gives rise to 
	\begin{equation}
		\begin{split}
				S_{\mt{EE}}  (  D^{(\mt{R})}_{a,b} ) &\overset{?}{=}    \frac{c}{3} \log \(   \frac{\beta}{ \pi \epsilon}  \sinh \(  \frac{\pi L_2}{\beta}  \frac{k_1}{\pi -k_2}  \)  \) + \frac{c}{3} \log \(  \frac{\sin (\pi - k_2)}{\pi -k_2} \frac{L_2}{\epsilon} \)   + \mathcal{O}\( \frac{L_2^2}{L_1^2} \)    \\
				&\simeq   \frac{c}{3} \log \(  \frac{k_1}{\pi- k_2} \frac{L_2}{\epsilon}  \) + \frac{c}{3} \log \(  \frac{\sin (\pi - k_2)}{\pi -k_2} \frac{L_2}{\epsilon} \)  
				+\mathcal{O}\( \frac{L_2^2}{\beta^2} \) + \mathcal{O}\( \frac{L_2^2}{L_1^2}\)   \,. \\
		\end{split}\label{hanger3}
	\end{equation}
At the leading order, this result for the entanglement entropy $S_{\mt{EE}} ( D^{(\mt{R})}_{a,b})$ in the warm phase is equal to the result \eqref{eq:ColdSDab} in the cold phase. Nevertheless, it should be noted that the subleading terms differ because the corrections in the warm phase involve additional terms which are not suppressed by powers of $L_2/L_1$.  

Now we turn to the geodesics shown in the right plot of figure \ref{fig:RTWarm01}. In this case, the geodesic in $\mS_2$ remains the same as above, while the geodesic in $\mS_1$ bulk now surrounds the black hole. The corresponding geodesic lengths yield 
\begin{equation}
	S_{\mt{EE}}  (  D^{(\mt{R})}_{a,b} ) \overset{?}{=}   \f{c}{3} \log \left( \f{\beta}{\pi \epsilon} \; \sinh \f{\pi L_1 }{\beta}\right)  + S_{\rm horizon}+ 
 \f{c}{3} \log \left[ \f{L_{\mathcal{S}_{2}}}{\pi \epsilon}  \sin \f{\pi L_{2}}{L_{\mathcal{S}_{2}}}\right]\,,
\end{equation}
where $S_{\rm horizon}$ denotes the Bekenstein-Hawking entropy in the warm phase, \ie 
\begin{equation}\label{hanger2}
	\begin{split}
		S_{\rm{horizon}}  =  \frac{  r_h\Delta \phi^{\rm{Hor}}_1 }{4\GN} =  \frac{c }{3} \pi T_{\mt{BH}}  \(  L_1 + 2\Delta \phi_1 \ell  \)
		\simeq \frac{c }{3} \pi T_{\mt{BH}}   \(    L_1  +  \frac{k_1 L_2}{\pi -k_2} \)\,.\\
	\end{split}
\end{equation}
where we have used the leading expansion for $\Delta \phi_1$ from eq.~\eqref{eq:defk2} to obtain the final approximation. It is obvious that we always have the result in eq.~\reef{eq:geoRab} is always smaller than that in eq.~\reef{hanger2} thanks to the inequality $ \log ( \sinh X) < X$ where here we take $X= {2\pi \Delta \phi_1 \ell}/{\beta}$.

On the other hand, we also have the possible geodesics crossing the brane, as shown in figure \ref{fig:RTWarm02}. Considering the simplest example on the left plot of the figure. The minimal geodesic requires the minimization over the crossing point on the brane. As we argued before, the extremal geodesic length would be universally given by eq.~\eqref{eq:lengths} in the limit $\DLD \to 0 $, which is not sensitive to the details of the bulk spacetime. Explicitly, these geodesics can be approximated by  
\begin{equation}
	2S_{\mt{EE}} ( D^{(\mt{R})}_{a}) \simeq  	\frac{2c}{3}    \log \(  \frac{\DLD}{\epsilon} \) + 4 \log g + \mathcal{O} (1)      \,, 
\end{equation}
while those in the right plot yield
\begin{equation}
\frac{c}{3} \log \left( \f{\beta}{\pi \epsilon} \; \sinh \f{\pi L_1 }{\beta}\right) + \frac{c}{3} \log \left( \f{\beta}{\pi \epsilon} \; \sinh \f{\pi L_1 }{\beta}\right) + 4 \log g + \mathcal{O} (1)      \,. 
\end{equation}
However, these candidates for the entanglement entropy do not compete with the previous configurations which do not cross the brane because of the large $\log g$ contributions (in the regime of our interest). 

As a summary, we conclude that the holographic entanglement entropy $S_{\mt{EE}}  (  D^{(\mt{R})}_{a,b} )$ in the warm phase is approximated by 
\begin{equation}\label{eq:SEEDabWarm}
	S_{\mt{EE}}  (  D^{(\mt{R})}_{a,b} ) \simeq \frac{c}{3} \log \(   \frac{\beta}{ \pi \epsilon}  \sinh \(  \frac{\pi L_2}{\beta}  \frac{k_1}{\pi -k_2}  \)  \) + \frac{c}{3} \log \(  \frac{\sin (\pi - k_2)}{\pi -k_2} \frac{L_2}{\epsilon} \)   \,, 
\end{equation}
from eq.~\reef{hanger3}. It is evident that this expression for $S_{\mt{EE}}  ( D^{(\mt{R})}_{a,b} )$ increases with temperature. Conversely, $2S_{\mt{EE}} ( D^{(\mt{R})}_{a})$ remains constant. Thus, it might be inferred that a critical temperature exists above which the latter becomes the dominant. Roughly, this temperature would be given by
\begin{equation}
	 L_2\,T_{\mt{DCFT}}  \simeq \frac{\pi -k_2}{k_1} \log \( \sqrt{\frac{1+T_o\ell}{1-T_o\ell} } \) \,.
\end{equation}
However, from eq.~\reef{eq:WarmCondition}, we recall that the warm phase only exists up to 
\begin{equation}
 L_2\,T_{\mt{DCFT}}  \le C_{\mt{WH}}	 \simeq \frac{(\pi - k_2)^2}{4 \pi}  { \log^{-1}\!\( \sqrt{  \frac{1+T_o \ell}{1-T_o\ell} }\)} \,.
\end{equation}
Therefore, this potential critical temperature is significantly higher than the temperature for the transition from the warm to hot phase, and the dominant holographic entanglement entropy in the warm phase is always determined by eq.~\eqref{eq:SEEDabWarm}. Hence using eq.~\eqref{eq:SEEDabWarm} for the entanglement entropy $S_{\mt{EE}}  (  D^{(\mt{R})}_{a,b} )$ and eq.~\eqref{eq:lengths} for $S_{\mt{EE}}  (  D^{(\mt{R})}_{a})$, we obtain the mutual information in the warm phase as 
\begin{equation}\label{eq:warmIab}
	I_{\mt{R}_a:\mt{R}_b}^{(\mt{W})} \simeq 4 \log g -  \frac{2c}{3}\log \(  \frac{L_2}{\DLD} \,  \frac{\sqrt{k_1 \sin (\pi -k_2)}}{\pi -k_2} \) \,,
\end{equation}
where we have only kept the leading contributions. 

\subsubsection*{Holographic entanglement entropy $S_{\mt{EE}}  (  D^{(\mt{L,R})}_{a,b} )$}
\begin{figure}[t]
	\centering
	\includegraphics[width=4.5in]{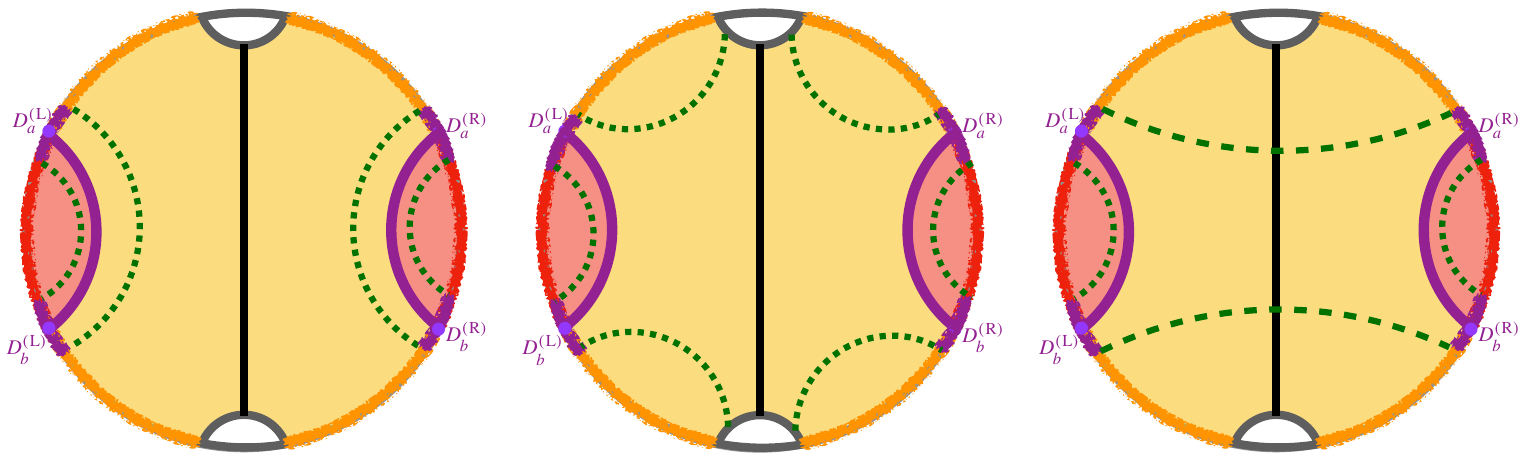}
	\caption{Relevant extremal surfaces in the warm phase (denoted by green curves) for $S_{\mt{EE}}  (  D^{(\mt{L,R})}_{a,b} )$.} \label{fig:RTWarm03}
\end{figure}
\begin{figure}[t]
	\centering
	\includegraphics[width=6in]{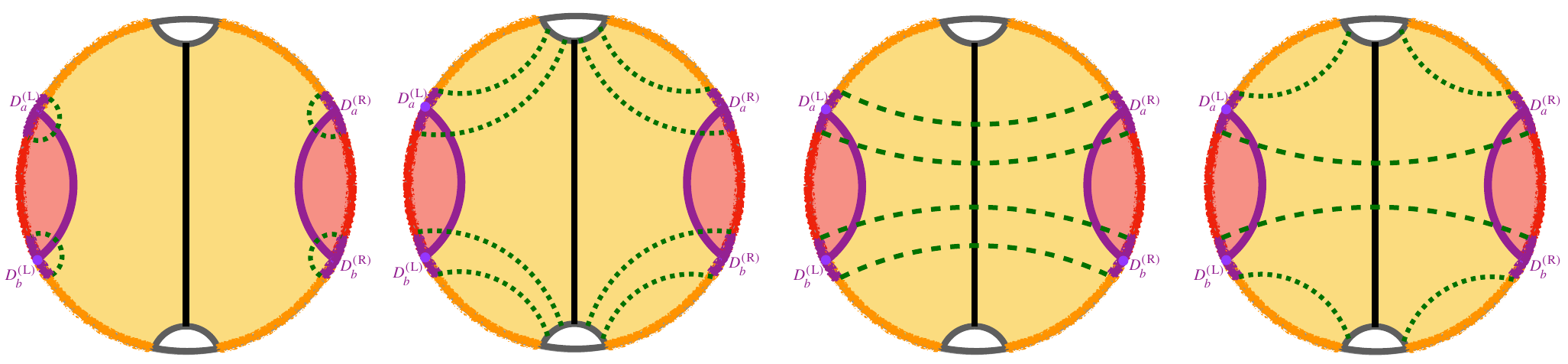}
	\caption{Irrelevant extremal surfaces in the warm phase (denoted by green curves) for $S_{\mt{EE}}  (  D^{(\mt{L,R})}_{a,b} )$.} \label{fig:RTWarm04}
\end{figure}

Let us move now to the holographic calculation of $S_{\mt{EE}}  (  D^{(\mt{L,R})}_{a,b} )$ in the warm phase where we include all four defects on two sides. Similar to the previous analysis in the symmetric case, we will assume the left and right boundary times are $t_{\mt L}, t_{\mt R}$, respectively. Because of the time translation invariance of the TFD state, the final result can only depend on the time difference $(t_{\mt R} - t_{\mt L})$. 

Since we are interested in the regime where the boundary entropy $2\log g$ is large, we split all extremal surfaces into two families according to whether they cross the brane or not, as shown in figures \ref{fig:RTWarm03} and \ref{fig:RTWarm04}. From the previous results, we know the ones crossing the brane acquire contributions proportional to $\log g$ and will not be the minimal extremal surface. So we examine the relevant extremal surfaces depicted in figure \ref{fig:RTWarm03}. Most of the geodesics appearing there have already appeared in the previous calculations of $S_{\mt{EE}}  (  D^{(\mt{R})}_{a,b} )$. The new type is the geodesic that crosses the BTZ black hole horizon and connects the left and right boundaries. Its geodesic length is also known and has been discussed for eq.~\eqref{eq:SEELRhot} for the symmetric case. 

Collecting all previous results, the holographic entanglement entropy $S_{\mt{EE}}  (  D^{(\mt{L,R})}_{a,b} )$ for all four defects is given by the minimization as follows: 
\begin{equation}\label{eq:SEELRwarm}
S_{\mt{EE}}  (  D^{(\mt{L,R})}_{a,b} ) \simeq \frac{2c}{3} \log \(  \frac{\sin (\pi - k_2)}{\pi -k_2} \frac{L_2}{\epsilon} \)  + \min 
	\begin{cases}
		\frac{2c}{3} \log \(   \frac{\beta}{ \pi \epsilon}  \sinh \(  \frac{\pi L_2}{\beta}  \frac{k_1}{\pi -k_2}  \)  \)   \,,  \\ 
		\,\\
	  \f{2c}{3} \log \left( \f{\beta}{\pi \epsilon} \; \sinh \f{\pi L_1 }{\beta}\right)    \,, \\
		\,\\
		\frac{2c}{3}\log \(  \frac{\beta}{ \pi \epsilon}    \cosh \left(  \frac{\pi}{\beta}  (t_{\mt{R}} -t_{\mt{L}} ) \right)   \)    \,. \\
	\end{cases}
\end{equation}
The time growth in the third expression is related to the growth of the wormhole connecting the two sides of the boundary. These three results from top to bottom above correspond to the three candidates for the geodesics from left to right illustrated in figure \ref{fig:RTWarm03}.
The first term which is common to all three expressions arises from the geodesics in the $\mS_2$ region which is common to all three configurations in the figure. As a result, the geodesics in the $\mS_1$ region determine the minimal entropy and the minimization in eq.~\eqref{eq:SEELRwarm} is determined by 
\begin{equation}
	\min  \, \left\{   \sinh \(  \frac{\pi L_2}{\beta}  \frac{k_1}{\pi -k_2} \)\,,  \quad  \sinh \(  \f{\pi L_1 }{\beta}  \) \,, \quad \cosh \left(  \frac{\pi}{\beta}  (t_{\mt{R}} -t_{\mt{L}} ) \right)   \right\}\,. \label{hanger5}
\end{equation}

First of all, it is easy to show that $\sinh \(  \pi L_1 /\beta  \)$ cannot be the minimal quantity in the regime of interest. Noting that the warm phase is preferred when $	T_{\mt{DCFT}} L_1 \ge C_{\mt{CW}}$ and $T_{\mt{DCFT}} L_2 \le C_{\mt{WH}}$, we  always have 
\begin{equation}
	 L_1 \ge \frac{C_{\mt{CW}}}{C_{\mt{WH}}} L_2 \simeq \frac{4\pi L_2}{(\pi-k_2)^2} \, \log \( \sqrt{\frac{1+T_o\ell}{1-T_o\ell} }\) \,. 
\end{equation}
in the warm phase. Due to the appearance of the boundary entropy in this inequality, the second term in eq.~\reef{hanger5} will always be large compared to the first. More explicitly, the second condition for the warm phase implies 
\begin{equation}
	\sinh \(  \frac{\pi L_2}{\beta}  \frac{k_1}{\pi -k_2} \) \le \sinh \(  \frac{k_1 \pi C_{\mt{WH}} }{\pi -k_2}  \)  \simeq  \frac{(\pi - k_2)k_1}{4}\,  { \log^{-1}\! \( \sqrt{  \frac{1+T_o \ell}{1-T_o\ell} }\)}\ll 1\,.
\end{equation} 
While the last inequality above is shown by taking the limit $T_o\ell \to 1$, we can also numerically show that $	\sinh \(  \frac{\pi L_2}{\beta}  \frac{k_1}{\pi -k_2} \) <1 $ for any $T_o\ell \in (\frac{1}{2},1]$. These results also ensure that the first term is also always less than $\cosh \left(  \frac{\pi}{\beta}  (t_{\mt{R}} -t_{\mt{L}} )\)$, the third term in eq.~\reef{hanger5}.

Following the above analysis, we can find that the minimization in eq.~\eqref{eq:SEELRwarm} always selects the first expression and hence the holographic entanglement entropy $S_{\mt{EE}}  (  D^{(\mt{L,R})}_{a,b} )$ reads 
\begin{equation}\label{eq:SEELRwarm2}
	\begin{split}
			S_{\mt{EE}}  (  D^{(\mt{L,R})}_{a,b} ) &= 	S_{\mt{EE}}  (  D^{(\mt{L})}_{a,b} ) + 	S_{\mt{EE}}  (  D^{(\mt{R})}_{a,b} )  \\
			&\simeq
		\frac{2c}{3} \log \(   \frac{\beta}{ \pi \epsilon}  \sinh \(  \frac{\pi L_2}{\beta}  \frac{k_1}{\pi -k_2}  \)  \) + \frac{2c}{3} \log \(  \frac{\sin (\pi - k_2)}{\pi -k_2} \frac{L_2}{\epsilon} \)   \,.   \\ 
	\end{split}
\end{equation}
This tells us that  the mutual information  $	I_{\mt{L:R}} $ in warm phase is always vanishing, \ie 
\begin{equation}
		I_{\mt{L:R}}^{(\mt{W})} = 0 \,. 
\end{equation}
 In the warm phase, two defects on the same boundary are connected by the brane. 
 The brane appears as a thread-like structure that geometrically manifests the entanglement structure among the defects. As a result, any defects situated on two distinct boundaries are not entangled in the warm phase, which leads to the vanishing of mutual information (to leading order in the large $c$ expansion).

\subsubsection{Mutual information in the hot phase}
With the increase in temperature, the bulk spacetime finally locates at the hot phase with a black hole sitting in the middle, as shown in the global picture \ref{fig:global}. More precisely, the hot phase is dominant when 
\begin{equation}
T_{\mt{DCFT}} L_2 \ge C_{\mt{WH}}\,, \qquad T_{\mt{DCFT}} (L_1+L_2) \ge C_{\mt{CH}} \,, 
\end{equation}
where the former applies for $L1/L2 >1$ and the latter, for $L1/L2\lesssim 1$. The
 transition parameters $C_{\mt{WH}}, C_{\mt{CH}}$ are approximately given by eqs.~\eqref{eq:WHtrans} and \eqref{eq:CWtrans}, respectively. The advantage of the hot phase is that the brane profile can be analytically derived as eq.~\eqref{eq:BTZbrane}
which is the same as that in the symmetric setup. This simplification is traced back to the fact that the temperatures of bulk black hole $T_{\mt{BH}}$ in $\mS_1, \mS_2$ are the same. The asymmetric feature of two boundary sizes $L_1, L_2$ is decoded in the area of the black hole horizon in $\mS_1, \mS_2$. From the boundary-bulk dictionary derived in eq.~\eqref{eq:hotmap}, the horizon size of each portion is precisely shown as 
\begin{equation}
	\Delta \phi_i^{\rm Hor}  \ell  = 	L_i + \frac{1}{\pi T_{\mt{BH}}}\, \arctanh \( T_o\ell \)\,,
\end{equation}  
which would be a significant contribution in the limit $T_o \ell \to 0$. For example, the Bekenstein-Hawking entropy $S_{\rm horizon}$ associated with the black hole in the hot phase is given by
 \begin{equation}\label{eq:BHhot02}
	\begin{split}
		S_{\rm{horizon}}  &\equiv  \frac{  r_h\Delta \phi^{\rm{Hor}}_1 +r_h \Delta \phi^{\rm{Hor}}_2}{4\GN}  
		= \frac{c }{3} \pi T_{\mt{BH}}  \(    L_1 +   L_2   \) + 4 \log g   \,.\\
	\end{split}
\end{equation}
In the following analysis, we will find that this fact can help us to simplify the discussion on the relevant extremal surfaces.

\subsubsection*{Holographic entanglement entropy $S_{\mt{EE}}  (  D^{(\mt{R})}_{a,b} )$}

\begin{figure}[t]
	\centering
	\includegraphics[width=6in]{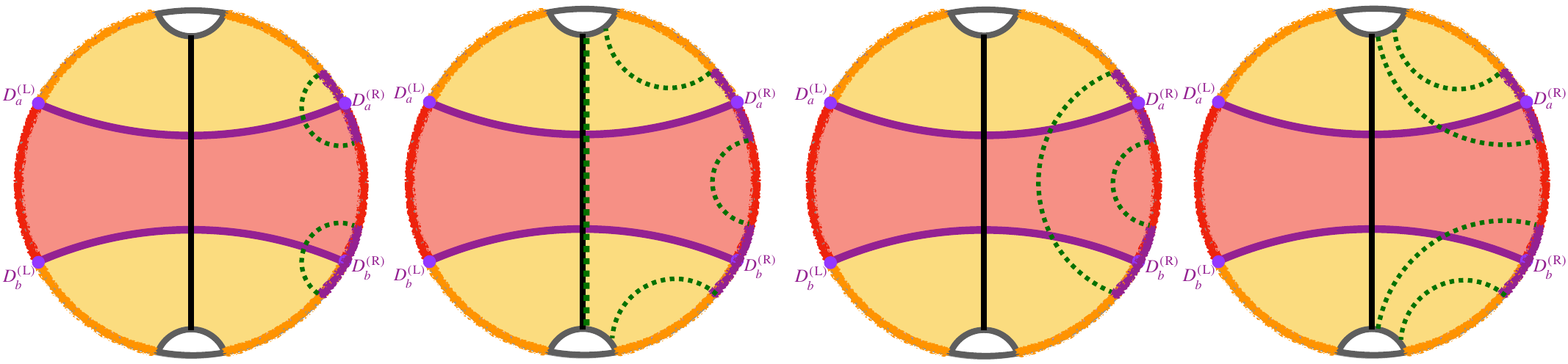}
	\caption{Candidates of extremal surfaces for $S_{\mt{EE}}  (  D^{(\mt{R})}_{a,b} )$ in the hot phase.} \label{fig:RTHot01}
\end{figure}

As in the other phases, for a single defect, we have $S_{\mt{EE}}  (  D^{(\mt{R})}_{a} )=S^\mt{(reg)}_{\rm{def}}$ where the latter was defined in eq.~\reef{hanger7}. Hence we begin here with the extremal surfaces associated with $S_{\mt{EE}}  (  D^{(\mt{R})}_{a,b} )$ for two defects on the same asymptotic boundary. In contrast to the cases in either the cold phase or warm phase, the extremal surfaces here always cross the brane, as depicted in figure \ref{fig:RTHot01}. The two configurations on the far left of the figure were examined in the discussion of the symmetric case -- see eq.~\eqref{eq:SEEsymDab}. Obviously, the length of the geodesics in the first configuration is smaller than that of the second, \ie 
\begin{equation}
		\frac{c}{3} \log \(  \frac{\beta}{ \pi \epsilon}   \sinh \(  \frac{\pi L_1 }{\beta}  \)  \) +	\frac{c}{3} \log \(  \frac{\beta}{ \pi \epsilon}   \sinh \(  \frac{\pi L_2 }{\beta}  \)  \) +S_{\rm{horizon}} > 2\,S^\mt{(reg)}_{\rm{def}}  =2 S_{\mt{EE}}  (  D^{(\mt{R})}_{a} ) \,,  
\end{equation}
where we are considering small defect intervals with $\DLD \ll \ell$.

In contrast to the symmetric case, there are two other candidates of minimal configuration which were ruled out by symmetry argument before. One can also derive analytical results for these, the third and fourth candidates in figure \ref{fig:RTHot01}, by using the explicit brane profile \eqref{eq:BTZbrane} and performing the minimization along the joint point on the brane. However, there is a simpler way to show that the last two candidates will not be the minimal extremal surface in the present case either. The key point is that the geodesic connecting the two branes in either $\mS_1$ or $\mS_2$ is bounded by the size of the horizon. For example, let us consider the geodesic in $\mS_2$ which connects two points $(r_2, \phi_2)$ and $(r_2, -\phi_2)$ on the brane (\ie the third plot in figure \ref{fig:RTHot01}). Using the generic formula for geodesic length in eq.~\eqref{eq:BTZd1}, one finds 
\begin{equation}
	D_{\mt{BTZ}} = 1 +  \frac{2 T_o^2 \ell^2 }{1- T_o^2 \ell^2} \frac{\sinh^2 \(  \frac{r_h \phi_2}{\ell} \)  }{ \sinh^2  \(  \frac{r_h (\phi_2-\phi_{\rm bdy})}{\ell} \)   }  \,, 
\end{equation}
where we have substituted $r_2 (\phi_2)$ by using the brane profile in eq.~\eqref{eq:BTZbrane}. It is then straightforward to show that $\partial_{\phi_2} D_{\mt{BTZ}}  <  0 $. In order words, it means that the minimal geodesic connecting the two branes is located at the intersection between the brane and horizon, \ie the minimal geodesic is the horizon itself. On the other hand, we also know that the area of the minimal geodesic starting from the brane to the defect living in $\mS_1$ is given by $\frac{1}{2} S^\mt{(reg)}_{\rm{def}}$. As a summary, we can conclude that the geodesic length of the third candidate is bounded from below by 
\begin{equation}
	\begin{split}
				&\quad \frac{c}{3} \log \(  \frac{\beta}{ \pi \epsilon}   \sinh \(  \frac{\pi L_2 }{\beta}  \)  \) + S^\mt{(reg)}_{\rm{def}} + \frac{ r_h \Delta \phi^{\rm{Hor}}_2}{4\GN}  \\
				&= \frac{c}{3} \log \(  \frac{\beta}{ \pi \epsilon}   \sinh \(  \frac{\pi L_2 }{\beta}  \)  \)  +  \frac{c}{3} \frac{ \pi L_2 }{\beta} + \frac{c}{3} \log \( \frac{\DLD}{\epsilon} \)  + 4 \log g  \\
				&> 2 S_{\mt{EE}}  (  D^{(\mt{R})}_{a} ) \,, 
	\end{split}
\end{equation}
where the last inequality simply holds since we consider the limit $\DLD \ll \ell$. The contribution of the fourth candidate in figure \ref{fig:RTHot01} is similarly bounded from below by   
\begin{equation}
		\quad \frac{c}{3} \log \(  \frac{\beta}{ \pi \epsilon}   \sinh \(  \frac{\pi L_1 }{\beta}  \)  \) + S^\mt{(reg)}_{\rm{def}} + \frac{ r_h \Delta \phi^{\rm{Hor}}_1}{4\GN}  		> 2 S_{\mt{EE}}  (  D^{(\mt{R})}_{a} ) \,.
\end{equation}
In conclusion, the holographic entanglement entropy $S_{\mt{EE}}  (  D^{(\mt{R})}_{a,b} )$ in the hot phase is fixed as 
\begin{equation}
	S_{\mt{EE}}  (  D^{(\mt{R})}_{a,b} )  = 2 \,S^\mt{(reg)}_{\rm{def}} = 2\,S_{\mt{EE}}  (  D^{(\mt{R})}_{a} )\,,
\end{equation}
which implies the vanishing of mutual information, \ie $I_{\mt{R}_a:\mt{R}_b}^{(\mt{H})}=0$.

\subsubsection*{Holographic entanglement entropy $S_{\mt{EE}}( D^{(\mt{L, R})}_{a,b})$}
\begin{figure}[t]
	\centering
	\includegraphics[width=6in]{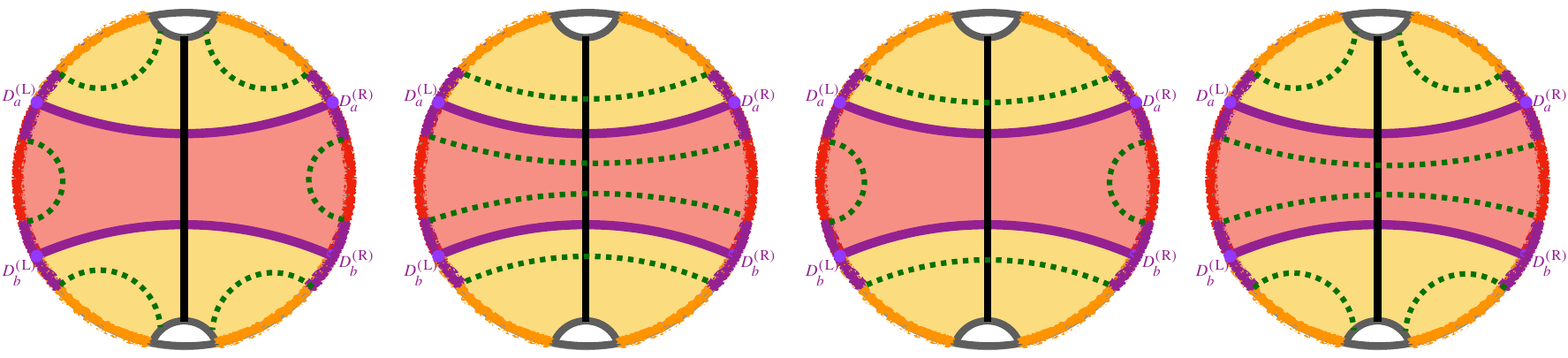}
	\caption{Relevant extremal surfaces for $S_{\mt{EE}}( D^{(\mt{L, R})}_{a,b})$ in the hot phase.} \label{fig:RTHot02}
\end{figure}

\begin{figure}[t]
	\centering
	\includegraphics[width=4.5in]{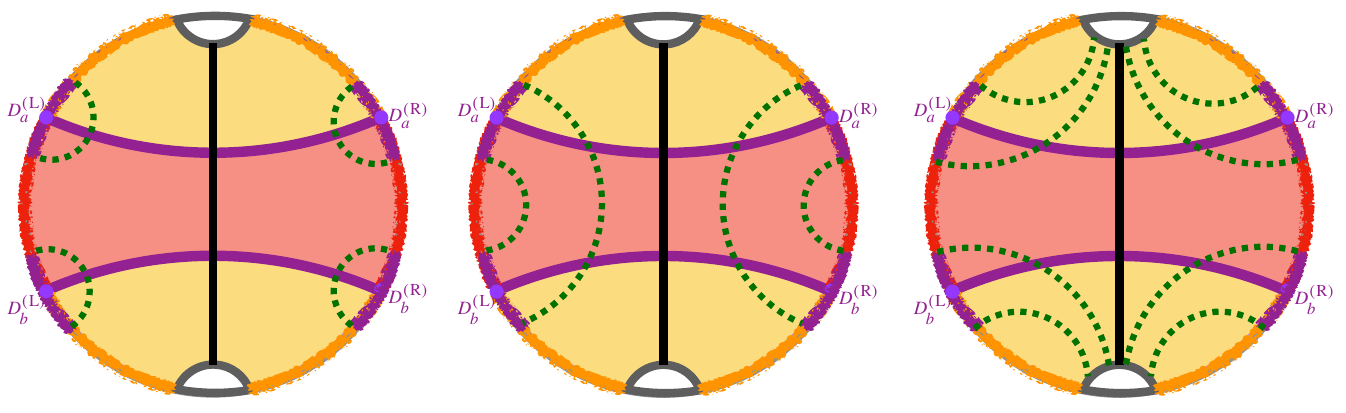}
	\caption{Irrelevant extremal surfaces for $S_{\mt{EE}}( D^{(\mt{L, R})}_{a,b})$ in the hot phase.} \label{fig:RTHot03}
\end{figure}

There are two classes of extremal surfaces for $S_{\mt{EE}}( D^{(\mt{L, R})}_{a,b})$ as depicted in figures \ref{fig:RTHot02} and \ref{fig:RTHot03}.  One observes that the geodesics in the second figure are exactly two copies of those considered for  $S_{\mt{EE}}( D^{(\mt{R})}_{a,b})$ above. It is straightforward to argue that none of these can provide the minimal length because they cross the brane and we are interested in the regime with a large defect entropy. Hence, we focus on the relevant geodesics in figure \ref{fig:RTHot02}. The  holographic entanglement entropy $S_{\mt{EE}}( D^{(\mt{L, R})}_{a,b})$ then comes by picking the minimum area amongst the four relevant extremal surfaces, \ie 
\begin{equation}\label{eq:SEEDLRhot}
	S_{\mt{EE}}^{(\mt{H})}   ( D^{(\mt{R,L})}_{a,b} )= \min 
	\begin{cases}
		S_{1}^{(\mt{H})} = \frac{2c}{3} \log \(  \frac{\beta}{ \pi \epsilon}   \sinh \(  \frac{\pi  L_1 }{\beta}\)  \) +\frac{2c}{3} \log \(  \frac{\beta}{ \pi \epsilon}   \sinh \(  \frac{\pi  L_2 }{\beta}\)  \)  \,,  \\ 
		\,\\
	S_{2}^{(\mt{H})}=	\frac{4c}{3}\log \(  \frac{\beta}{ \pi \epsilon}    \cosh \left(  \frac{\pi}{\beta}  (t_{\mt{R}} -t_{\mt{L}} ) \right)   \)  \,,\\
				\,\\
	S_{3}^{(\mt{H})}=	\frac{2c}{3}\log \(  \frac{\beta}{ \pi \epsilon}    \cosh \left(  \frac{\pi}{\beta}  (t_{\mt{R}} -t_{\mt{L}} ) \right)   \)   + 	\frac{2c}{3} \log \(  \frac{\beta}{ \pi \epsilon}   \sinh \(  \frac{\pi  L_2 }{\beta}\)  \)   \,,  \\ 
				\,\\
	S_{4}^{(\mt{H})}=	\frac{2c}{3}\log \(  \frac{\beta}{ \pi \epsilon}    \cosh \left(  \frac{\pi}{\beta}  (t_{\mt{R}} -t_{\mt{L}} ) \right)   \)   + 	\frac{2c}{3} \log \(  \frac{\beta}{ \pi \epsilon}   \sinh \(  \frac{\pi  L_1 }{\beta}\)  \)   \,,  \\ 
	\end{cases}
\end{equation}
where the four expressions above are denoted by $\{S_{1}^{(\mt{H})},S_{2}^{(\mt{H})},S_{3}^{(\mt{H})},S_{4}^{(\mt{H})}\}$ from top to bottom correspond to the four configurations from left to right shown in figure \ref{fig:RTHot02}. Further, note that we have ignored the corrections involving $\DLD$ by assuming this regulator scale is much smaller than any of the macroscopic scales, \eg $\beta$, $L_1$ and $L^2$. 

First of all, let us note that the entanglement entropy at late times always reduces to the constant, \ie 
\begin{equation}
	S_{\mt{EE}}^{(\mt{H})}   ( D^{(\mt{R,L})}_{a,b} )\big|_{|t_{\mt{R}} -t_{\mt{L}} |  \ge t_{\rm crt1} }  =S_{1}^{(\mt{H})}=	\frac{2c}{3} \log \(  \frac{\beta}{ \pi \epsilon}   \sinh \(  \frac{\pi  L_1 }{\beta}\)  \) +\frac{2c}{3} \log \(  \frac{\beta}{ \pi \epsilon}   \sinh \(  \frac{\pi  L_2 }{\beta}\)  \)  \,,  \\ 
	\,\\ 
\end{equation}
which is equivalent to the sum of the entanglement entropy of two thermal CFTs on regions with sizes $L_1, L_2$. The critical time  $t_{\rm crt1} $ after which this becomes the minimal result in eq.~\reef{eq:SEEDLRhot} is determined by the larger boundary size $L_1$ as
\begin{equation}
t_{\rm crt1} = \frac{\beta}{\pi} \arccosh \(    \sinh \(  \frac{\pi  L_1 }{\beta}\)       \) \simeq L_1   - \frac{2\beta}{\pi}\,e^{ -2\pi  L_1/ \beta  }\,,
\end{equation}
where the last approximation is taken in the large temperature, \ie $\beta/L_1 \to 0$.  Beyond this critical time scale, the mutual information between the left and right defects becomes 
\begin{equation}
	I_{\mt{L:R}}^{(\mt{H})}  \big|_{|t_{\mt{R}} -t_{\mt{L}} |  \ge t_{\rm crt1} }  = 8 \log g  -   \frac{2c}{3} \log \(  \frac{\beta^2}{ \pi^2 \DLD^2}   \sinh \(  \frac{\pi  L_1 }{\beta}\)  \sinh \(  \frac{\pi  L_2 }{\beta}\)   \) \,.
\end{equation}

In the following, let us focus on the situations before this critical time $t_{\rm crt1}$, \eg $t_{\mt{R}} \sim t_{\mt{L}}$. Noting that we have assumed $L_1 \ge L_2$ in this paper, the minimization associated with $S_{\mt{EE}}^{(\mt{H})}   ( D^{(\mt{R,L})}_{a,b} )$ is thus  determined by 
\begin{equation}
	\min \left\{     \cosh \left(  \frac{\pi}{\beta}  (t_{\mt{R}} -t_{\mt{L}} ) \right)    \,, \quad   \sinh \(  \frac{\pi  L_2 }{\beta}\)  \right\} \,. 
\end{equation}
This determines another transition time $t_{\rm crt2}$, \ie 
\begin{equation}
	t_{\rm crt2} = \frac{\beta}{\pi} \arccosh \(    \sinh \(  \frac{\pi  L_2 }{\beta}\)       \)  \,,
\end{equation}
which is related to the smaller boundary size $L_2$. This transition only happens when $  \sinh \(  \frac{\pi  L_2 }{\beta}\) \ge 1$, \eg at high temperatures where $L_2/\beta>1$. The time evolution of the holographic entanglement entropy $S_{\mt{EE}}( D^{(\mt{L, R})}_{a,b})$ is thus summarized as 
\begin{equation}\label{eq:SEEDLRhottime}
	S_{\mt{EE}}^{(\mt{H})}   ( D^{(\mt{R,L})}_{a,b} )=
	\begin{cases}
	S_{2}^{(\mt{H})} \,, \quad  |t_{\mt{R}} - t_{\mt{L}}|  \le t_{\rm crt2}  \,,  \\ 
	S_{3}^{(\mt{H})} \,, \quad   t_{\rm crt2}\le |t_{\mt{R}} - t_{\mt{L}}|  \le t_{\rm crt1}  \,,  \\ 
	S_{1}^{(\mt{H})} \,, \quad    |t_{\mt{R}} - t_{\mt{L}}|  \ge t_{\rm crt1}  \,.   \\ 
	\end{cases}
\end{equation}
Correspondingly, the time evolution of the mutual information between the left and right defects $I_{\mt{L:R}}^{(\mt{H})}$ is recast as 
\begin{equation}\label{eq:ILRhottime}
	I_{\mt{L:R}}^{(\mt{H})} =
	\begin{cases}
			8 \log g  -   \frac{4c}{3} \log \(  \frac\beta{ \pi \DLD}\, \cosh \left(  \frac{\pi}{\beta}  (t_{\mt{R}} -t_{\mt{L}} ) \right)    \)    \,, \qquad\qquad  |t_{\mt{R}} - t_{\mt{L}}|  \le t_{\rm crt2}  \,,  \\ 
			\,\\
			8 \log g  -   \frac{2c}{3} \log \(  \frac{\beta^2}{ \pi^2 \DLD^2} \,\sinh \(  \frac{\pi  L_2 }{\beta} \)  \cosh \left(  \frac{\pi}{\beta}  (t_{\mt{R}} -t_{\mt{L}} ) \right)    \)  \,, \quad   t_{\rm crt2}\le |t_{\mt{R}} - t_{\mt{L}}|  \le t_{\rm crt1}  \,,  \\
		\,\\ 
	8 \log g  -   \frac{2c}{3} \log \(  \frac{\beta^2}{ \pi^2 \DLD^2} \,  \sinh \(  \frac{\pi  L_1 }{\beta}\)  \sinh \(  \frac{\pi  L_2 }{\beta}\)   \)  \,, \qquad    |t_{\mt{R}} - t_{\mt{L}}|  \ge t_{\rm crt1}  \,.   \\ 
	\end{cases}
\end{equation}

\subsubsection{Mutual information in three phases}\label{sec:summary}
\begin{figure}[t]
	\centering
	\includegraphics[width=6in]{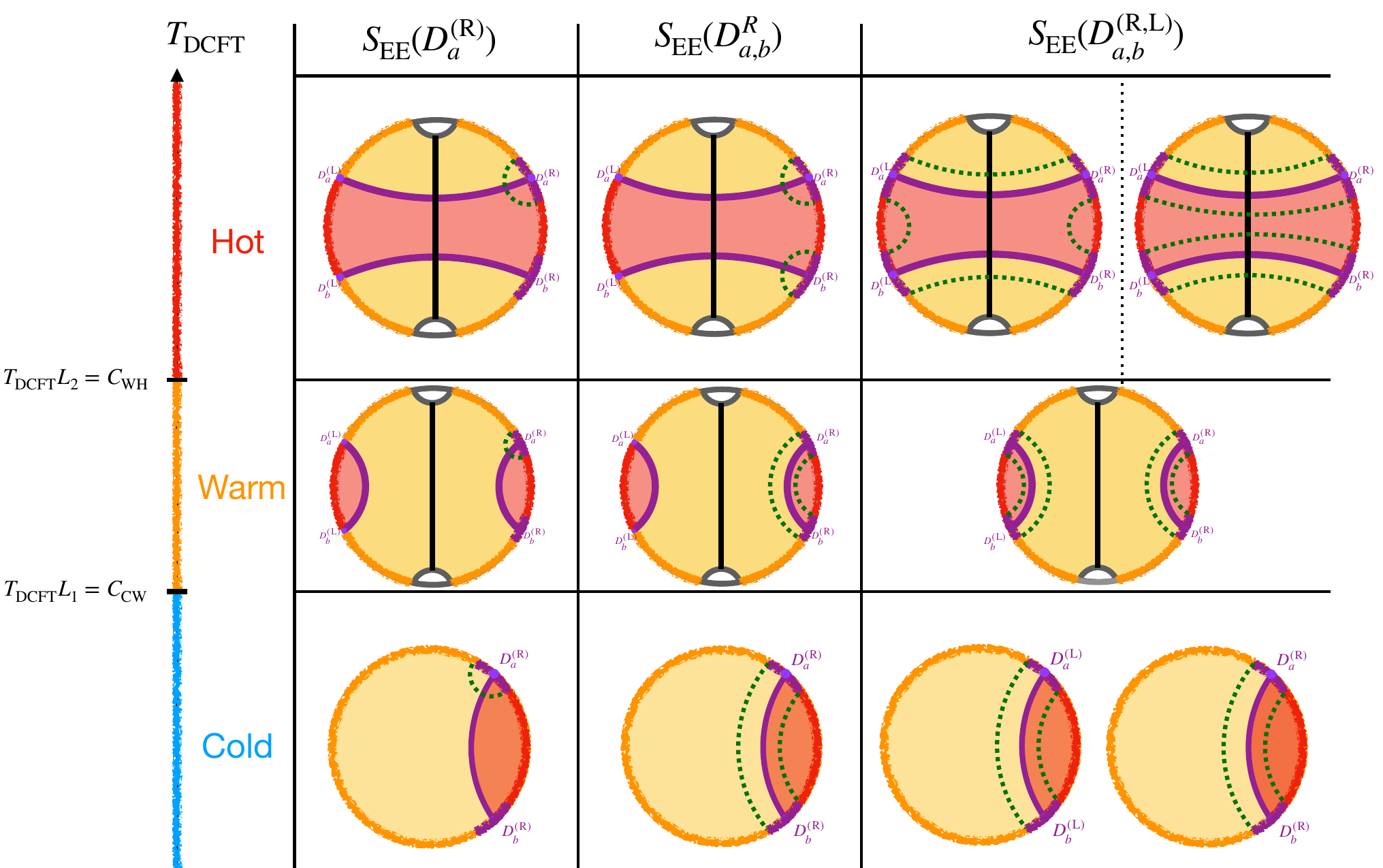}
	\caption{A summary of all relevant extremal surfaces associated with various defects in different phases. These surfaces apply for any time in the cold and warm phases, but only for early times in the hot phase. At late times in the hot phase, the relevant RT surface consists of four disconnected  caps homologous to each of the regulator intervals, \eg as appear in $S_{\mt{EE}}( D^{(\mt{R})}_{a,b} )$ for this phase.} \label{fig:Table}
\end{figure}

Instead of the time evolution, we are more interested in the temperature dependence. Let us consider the case where the boundary sizes $L_1, L_2$ are fixed and the temperature is increased. As shown by the phase diagram in figure \ref{fig:PhaseDiagram}, the bulk dual spacetime passes through two phases transitions from the cold to warm phase and then from the warm to the hot phase. In order to see all three phases, the ratio $L_1/L_2$ cannot be too small. Whereas there are only the cold and hot phases for smaller values of $L_1/L_2$, which is similar to what was shown in the symmetric set-up. The critical ratio is determined by the value of the triple point of the phase diagram \ref{fig:PhaseDiagram}. More explicitly, the approximate value for a generic tension was given in eq.~\eqref{eq:L1L2triple}. We note that it reduces to  
\begin{equation}
	 	\( \frac{L_1}{L_2} \)\Bigg|_{\rm triple} = \frac{C_{\mt{CW}}}{C_{\mt{WH}}}  \simeq \frac{4\pi}{(\pi -k_2)^2}   \log \( \sqrt{  \frac{1+T_o \ell}{1-T_o\ell} } \) - \frac{k_1}{\pi-k_2}  + \mathcal{O}\( \frac{c}{\log g}\)\,,  
\end{equation}
in the large tension limit. Since we are ignoring the time evolution here, we choose $t_{\mt{R}} =t_{\mt{L}}$ to simplify the following analysis. The mutual information $I_{\mt{L:R}}$ between the left and right sides simply vanishes in the cold and warm phases. In the hot phase, the holographic entanglement entropy is further simplified as 
 \begin{equation}\label{eq:SEEDLRhot02}
	S_{\mt{EE}}^{(\mt{H})}   ( D^{(\mt{R,L})}_{a,b} )= \min 
	\begin{cases}
		S_{2}^{(\mt{H})}=	\frac{4c}{3}\log \(  \frac{\beta}{ \pi \epsilon}   \)  \,,\\
		S_{3}^{(\mt{H})}=	\frac{2c}{3}\log \(  \frac{\beta}{ \pi \epsilon}  \)   + 	\frac{2c}{3} \log \(  \frac{\beta}{ \pi \epsilon}   \sinh \(  \frac{\pi  L_2 }{\beta}\)  \)   \,,  \\ 
	\end{cases}
\end{equation}
where either $S_{1}^{(\mt{H})}$ or $S_{4}^{(\mt{H})}$ could be dominant because of $\sinh (\pi L_1/\beta) \ge  \sinh (\pi C_{\mt{CW}}) >1$ at the high temperature phase and the transition between  $S_{2}^{(\mt{H})}$ and $S_{3}^{(\mt{H})}$ is given by 
\begin{equation}
	T_{\mt{DCFT}} L_2 =  \frac{\arcsinh(1)}{\pi} \simeq\frac{0.8814}{\pi}\equiv C_{\mt{H}}  \,.
\end{equation}
As a result, the mutual information $I_{\mt{L:R}}$ with increasing the boundary temperature is given by
\begin{equation}\label{eq:ILRphases}
	I_{\mt{L:R}} \(  t_{\mt{R}} =t_{\mt{L}} \)=
	\begin{cases}
		0   \,, \qquad \qquad  	T_{\mt{DCFT}} L_2   \le C_{\mt{WH}}  \,,  \\ 
		\,\\
		8 \log g  -   \frac{4c}{3} \log \(  \frac{\beta}{ \pi \DLD}\,\sqrt{\sinh \(  \frac{\pi  L_2 }{\beta} \) }    \) \,, \quad  	C_{\mt{WH}}  \le T_{\mt{DCFT}} L_2   \le C_{\mt{H}}  \,,  \\
		\,\\ 
		8 \log g  -   \frac{4c}{3} \log \(  \frac{\beta}{ \pi \DLD}  \) \,, \qquad  T_{\mt{DCFT}} L_2   \ge C_{\mt{H}}  \,.   \\ 
	\end{cases}
\end{equation}
As we increase the temperature, the correlations between two defects on the same asymptotic boundary (say $ D^{(\mt{R})}_{a}$ and $ D^{(\mt{R})}_{b}$) decreases. This can be explicitly seen in the mutual information between them: 
\begin{equation}\label{eq:Iab}
	I_{\mt{R}_a:\mt{R}_b}  \approx 	
	\begin{cases}
		\frac{2c}{3}\log \( \sqrt{\frac{1+T_o\ell}{1-T_o\ell} } \) +  \frac{2c}{3}\log \(  \frac{\DLD}{L_2}   \frac{\pi -k_2}{\sqrt{k_1 \sin (\pi -k_2)}} \) + \mathcal{O}( \(  \frac{L_2}{L_1} \)^2 ) \,,   \quad   	T_{\mt{DCFT}} L_2   \le C_{\mt{WH}}  \,,  \\ 
		\,\\ 
		0 \,, \quad  T_{\mt{DCFT}} L_2   \ge C_{\mt{WH}}  \,.   \\ 
	\end{cases}
\end{equation}
Here we have used the approximate expressions derived in eqs.~\eqref{eq:coldIab}, and \eqref{eq:warmIab} for the cold and warm phases. As a summary, we list all relevant extremal surfaces associated with different defects in figure \ref{fig:Table}.

Finally, it is interesting to ask where is the information about the braneworlds encoded in the boundary theory. In the cold and warm phases, one can see from the extremal surfaces and hence the corresponding entanglement wedges in figure \ref{fig:Table} that the branes are encoded in the two defects on either asymptotic boundary. This is expected of course since the branes connect the two defects on the same boundary. In the hot phase, the brane geometry changes so that it connects defects on different boundaries and the corresponding entanglement wedges now indicate that the braneworlds are encoded in those same defects, \eg $ D^{(\mt{R})}_{a}$ and $ D^{(\mt{L})}_{a}$. At late times in the hot phase, the relevant RT surfaces for any combination of two defects consist of two disconnected caps homologous to each of the regulator intervals, \eg see $S_{\mt{EE}}( D^{(\mt{R})}_{a,b} )$ in figure \ref{fig:Table}. At this stage, the bulk of the braneworlds is encoded in the thermal baths formed by the boundary CFT between the defects -- see \cite{Grimaldi:2022suv}.

Whereas the brane profile only has three phases (cold, warm and hot) depending on the dimensionless parameter $T_{{\rm DCFT}}L_{2}$, it is notable that the entanglement wedge of the four defects $D^{(\mt{L,R})}_{a, b}$ has four relevant phases as shown in figure \ref{fig:Table}. This comes from the fact that the transition temperatures of the brane configurations and the RT surfaces are different. One distinguishing feature of the entanglement wedge transition is that there is an intermediate phase in which all defects are connected in the bulk through the entanglement wedge. That is, the four defects combined also encode a large portion of the bulk between the two branes. This phase is realized in the regime $C_{\mt{WH}}  \le T_{\mt{DCFT}} L_2   \le C_{\mt{H}}$ in the hot phase. It is only in this intermediate phase,  that the entanglement wedge of the four defects is connected. If one further increases the temperature the entanglement wedge splits into two disjoint components, each of which contains one of the branes.


\section{Relation to the Hayden-Preskill Protocol}\label{sec:HP}

\begin{figure}[t]
	\begin{subfigure}{0.3\linewidth}
		\includegraphics[width=2.5in]{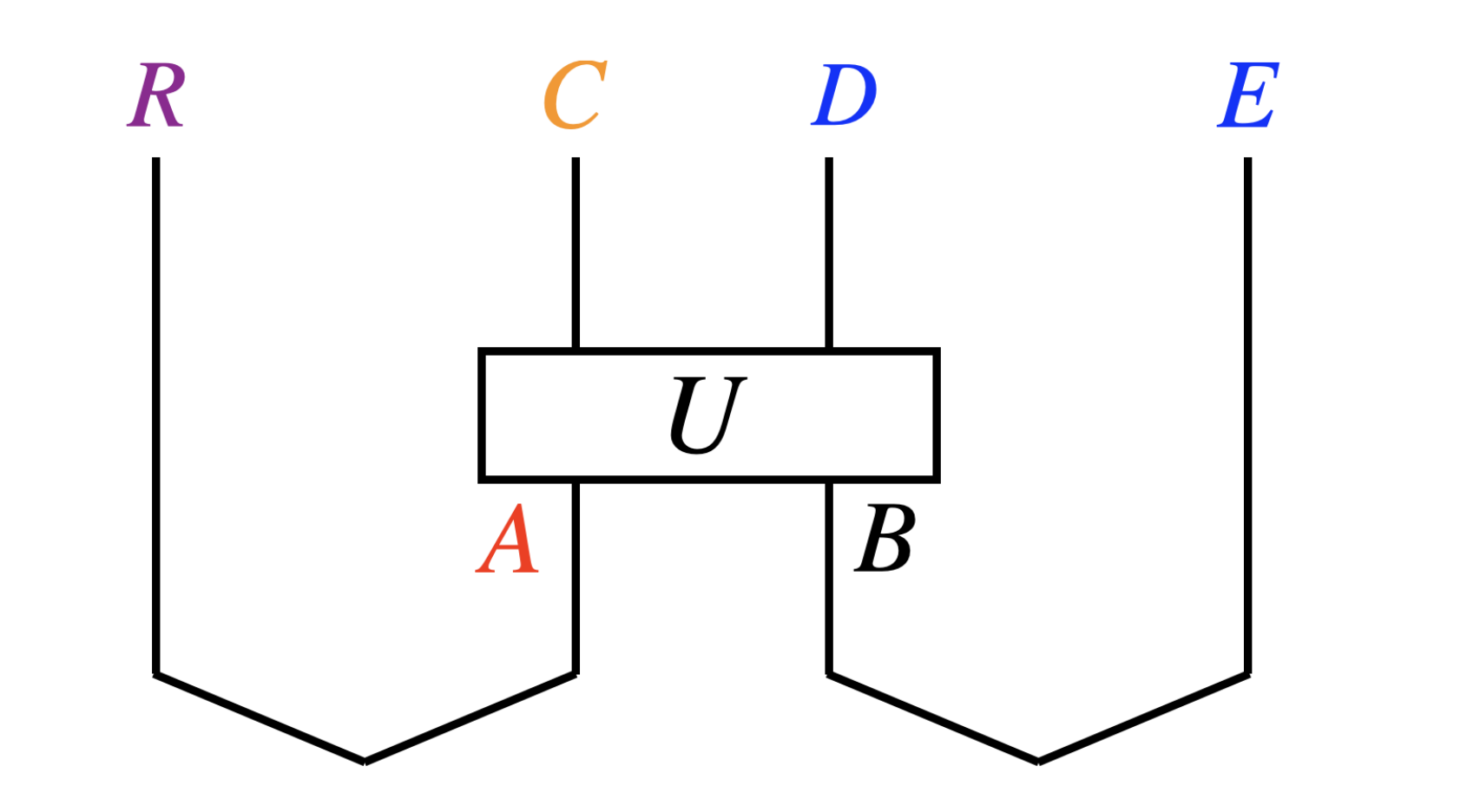}
	\end{subfigure}
	\qquad \qquad 
	\begin{minipage}{0.6\linewidth}
		\begin{subfigure}{0.525\linewidth}
			\includegraphics[width=3in]{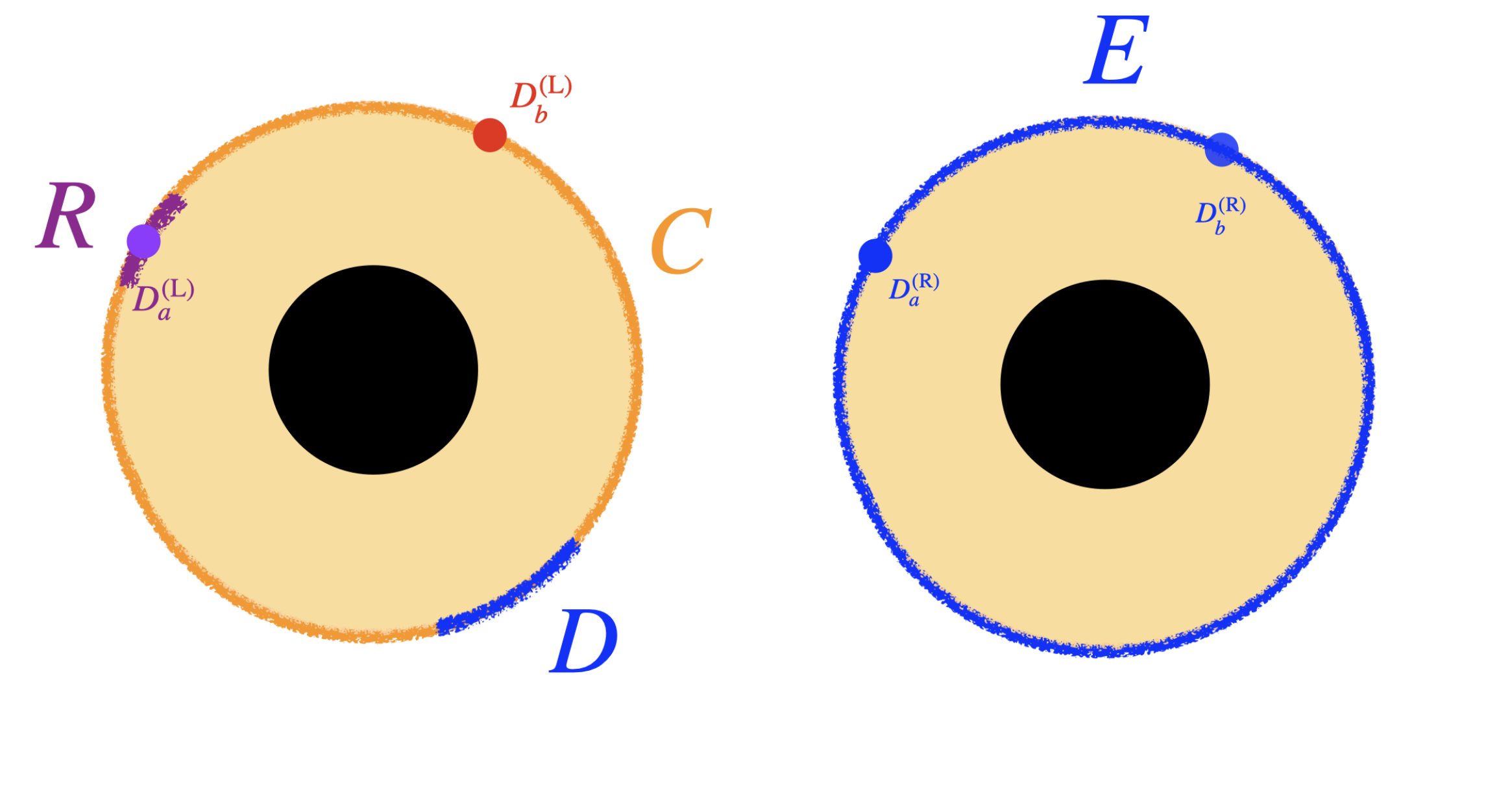}
			\smallskip
		\end{subfigure}\hfill		
		\medskip
		\begin{subfigure}{0.6\linewidth}
			\includegraphics[width=3in]{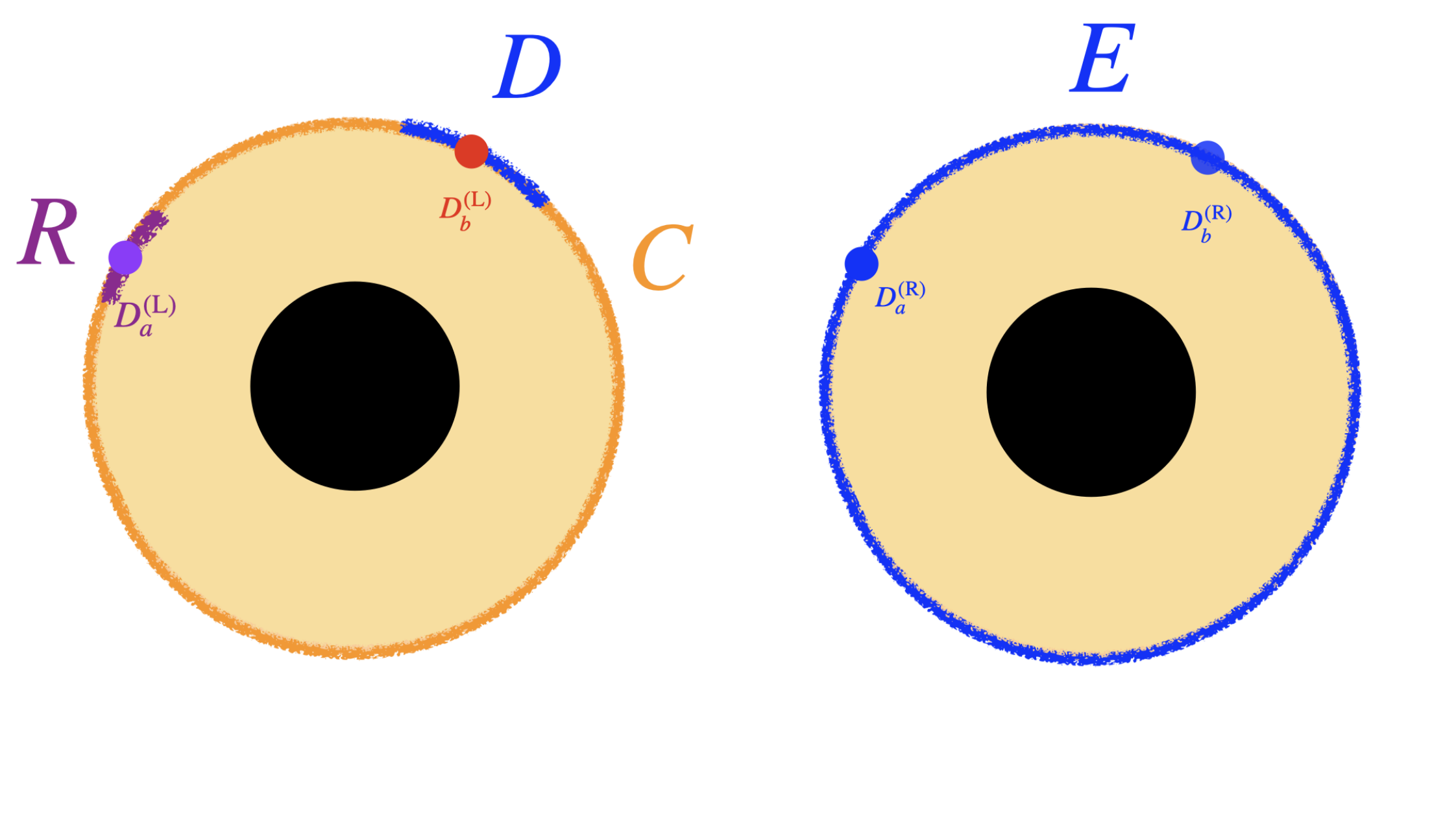}
		\end{subfigure}\hfill		
	\end{minipage}
	\caption{{\bf Left}: Original Hayden-Preskill setup. Initial state consists of the EPR state on $RA$ (The reference and the diary) and $BE$ (the original black hole with early Hawking radiation).  The decoupling theorem \eqref{eq:Decoupling} states that  after the scrambling time, information of the diary is encoded in Hawking radiation $DE$. {\bf Right-Top} A DCFT realization of the Hayden-Preskill setup.  $D^{(\mt{L})}_{a}$ is identified with the reference system $R$ in the HP setup.  In this first case the  region corresponding to the remaining black hole $C$ contains the other defect  $D^{(\mt{L})}_{b} \in C$. The dual bulk configuration can either be the warm phase or the hot phase. {\bf Right-Bottom}  Another DCFT realization. In this case the other defect is in the subsystem corresponding to the late radiation $ D^{(\mt{L})}_{b} \in D$. }
	\label{fig:HP}
\end{figure}

The boundary CFT setup can be considered a holographic realization of the Hayden-Preskill (HP) thought experiment \cite{Hayden:2007cs}. This experiment suggests that after the Page time, information thrown into an evaporating black hole (referred to as the ``diary") immediately comes out through the emitting Hawking radiation. The setup includes the Hilbert space of the diary ($A$), the black hole ($B$), and the early Hawking radiation ($E$) emitted.  Since the black hole is old, it is (almost) maximally entangled with $E$.
Also it is useful to introduce a ``reference" system $R$ which is maximally entangled with the diary $A$. After the black hole $B$ absorbs the diary $A$,  it emits late Hawking quanta $D$ and becomes $C$ (called the remaining black hole). Total Hilbert space on the final time slice  consists of these subsystems, $\mathcal{H}_{\rm tot} =\mH_{\mt{R}} \otimes \mH_{\mt{C}} \otimes \mH_{\mt{D}} \otimes \mH_{\mt{E}}$. Denoting $U_{AB}$ the unitary matrix for this evaporation  dynamics, the resulting state is 
\begin{equation} \label{eq:HP state}	
|\mathrm{HP} \ra = (I_{R} \otimes U_{AB}  \otimes I_{E} ) |\mathrm{EPR} \ra_{RA} \otimes   |\mathrm{EPR} \ra_{BE} \,, 
\end{equation}
where $|\mathrm{EPR} \ra$ denotes the maximally entangled state on the indicated bipartite system, see the left panel of figure \ref{fig:HP}. In \cite{Hayden:2007cs}, it was shown that   if  the time evolution  $U_{AB}$ is approximated by a random unitary,  following decoupling theorem holds,
\begin{equation}\label{eq:Decoupling}	
\int dU || \rho_{RC} -\rho_{R} \otimes \rho_{C} ||_{1}^2 \leq 
\left( \f{d_{A}}{d_{D}}    \right)^{2} \,,
\end{equation}
where $d_{A} = {\rm dim} \( \mH_{A}\), d_{D} ={\rm dim} \(  \mH_{D} \)$. This inequality shows that if we collect sufficiently many Hawking quanta just emitted from the black hole $d_{D} > d_{A}$, then  information of the diary is already in late plus  early Hawking radiation $DE$, since there is no correlation between the remaining black hole $C$ and the reference system $R$ which is maximally entangled with the diary initially. The decoupling between $R$ and $C$, $\rho_{RC} =\rho_{R} \otimes \rho_{C}$  is equivalent to vanishing of the mutual information $I_{R:C}=0$.

Each subsystem has a natural counterpart in our two entangled defect CFT setup.  First, we identify one of the defects in the left boundary with the reference system $R=D^{(\mt{L})}_{a}$. The rest of the left boundary corresponds to the original black hole $B$ which has absorbed the other defect for the diary, $A=D^{(\mt{L})}_{b}$. Since the black hole evaporates, we split the region for $B$ into two subsystems, one for the remaining black hole $C$ and the rest combined with the right boundary is the Hawking radiation system $DE$. We also assume that the size of the Hilbert space for $D$ is small compared to that of the remaining black hole, $d_D \ll d_C$. There are two possibilities for the choice of $DE$, depending on whether it contains $D^{(\mt{L})}_{b}$ or not, see the right panels of figure \ref{fig:HP}.  The degrees of freedom in the right panel correspond to early Hawking radiation. Two defects $D^{(\mt{R})}_{a}, D^{(\mt{R})}_{b} $ in the right boundary will form Bell pairs with the defects on the black hole side, which is the key to decoupling.
\footnote{One can try several other identifications between subsystems in the DCFT setup and the HP. The first possibility is that D and part of E (say $E_{L}$) are on the left boundary, and the rest of E is on the right.  The undesirable feature of this identification is that $C$ and $E_{L}$ can communicate directly.  The second possibility is that part of D is on the left and the rest of D and all of E are on the right. However, in order to keep the entropy of E constant, as in the HP, the size of the region for E must decrease, which is an undesirable property.  However, we want to emphasize that the detailed match with the HP setups is not necessary for archiving the information recovery in C on the left boundary from the degrees of freedom on the right.} 
The purpose of this section is to confirm the decoupling between the two subregions by  computing the following mutual information 
between $R$ and $C$, \ie 
\begin{equation}\label{eq:MIRC}
	I_{R:C}=S_{R}+ S_{C} -S_{RC} \,,
\end{equation}
in both the warm and hot phases. In our setup, the mutual information $I_{RC}$ diverges because two subsystems $R$ and $C$ are adjacent, so the ultraviolet modes are localized near the common boundary of $R$ and $C$. However, these UV modes have nothing to do with the defects, for example the same divergence appears in the mutual information for the thermal field double state without defects. Since we are interested in the correlation between the defect in $R$ and the sub-region $C$ in the current setup, which corresponds to the correlation between the diary and the black hole in the HP, we extract the contribution of the defect to the mutual information by subtracting that of the TFD state,
\begin{equation}\label{eq:MIRCdef}
	\Delta I_{R:C} = I_{R:C} - I^{(0)}_{R:C} \,. 
\end{equation}
Finally, we note that in our setup the Yoshida-Kitaev decoder of Hawking radiation has a simple realization in the DCFT setup.

\subsection{On the relation to the original HP setup}
While the identification between the two setups is natural, a significant difference arises: the HP protocol involves a time-dependent process, whereas our system, consisting of two branes, is static. Nevertheless, we argue that despite this difference, these two systems share a key property that allows us to realize HP decoding in our system: the swap of the initial entanglement shared between the reference system $R$ and the black hole to the entanglement between $R$ and the degrees of freedom.

In the original Hayden-Preskill setup, the initial entanglement exists between the reference $R$ and the diary $A$, and between the initial black hole $B$ and the early Hawking radiation $E$, as shown in the figure \ref{fig:HP}. Shortly after Alice throws her diary $A$ into the initial black hole $B$, $R$ becomes entangled with $B$. Subsequently, due to chaotic dynamics, this entanglement between $R$ and the black hole eventually disappears, as indicated by the inequality \eqref{eq:Decoupling}. Instead, the reference system becomes entangled with the late and early Hawking radiation $DE$. This entanglement swapping occurs because of several factors:
\begin{enumerate}[noitemsep]
	\item  The chaotic dynamics rapidly evolve the initial state to a typical state \eqref{eq:HP state}. The timescale for this evolution is the scrambling time. 
	\item In a typical state, a subsystem $R$ tends to become maximally entangled with the largest subsystem which could be either the remaining black hole $C$ or the radiation $DE$. Since after the Page time $|C| < |DE|$, $R$ must become entangled with the radiation.
	\item Once the entanglement between $R$ and $DE$ is established, the correlation between $R$ and $C$ must vanish due to the monogamy property.
\end{enumerate}
The phenomenon of entanglement swapping manifests itself naturally during the transition from the warm to the hot phase in our doubly holographic setup. For example, in the warm phase, two defects located on the same boundary are connected via the brane in the bulk. This configuration, dictated by the mapping between two different setups, corresponds to the fact that there is a strong correlation between $R$ and $A$ before the scrambling dynamics in the HP setup. This strong correlation is explicitly manifested in the expression for the mutual information $I_{\mathrm{L}_a:\mathrm{L}_b}$ \eqref{eq:Iab} between two defects on the same side. However, as the entanglement between the left and right systems increases, the transition to the hot phase occurs, leading to a remarkable entanglement swapping. Specifically, the correlation is now observed between defects located on different boundaries due to the change in topology of the brane profile. Remarkably, this pattern mirrors the behavior observed in the HP protocol.

The DCFT setup also differs slightly from the original HP for the evaporating black hole in that the entanglement between E on the left boundary and others on the right boundary keeps increasing in the entanglement temperature.  Since information recovery of a black hole is expected to be a general phenomenon that always happens as long as the black hole is highly entangled with the external system, both the HP setup and the DCFT setup are the examples of this situation.  Therefore, one can use the DCFT setup to study aspects of information recovery of the actual evaporating black hole \footnote{Note that the increase in entanglement between C and E in the DCFT setup does not mean that the DCFT state corresponds to the state of the evaporating black hole before the page time. This can be understood by noting how the EPR state between the black hole and the radiation is constructed in the HP setup. Namely, since after the Page time $|\text{BH}|$ (the dimension of the black hole Hilbert space) is smaller than that of the Hawking radiation, by choosing the sub-Hilbert space of HR which is isomorphic to $\mH_{\rm BH}$, one can form an EPR state between them, which is the input state for the HP protocol. Similarly, in the DCFT setup, the right CFT Hilbert space is the sub-Hilbert space of the early radiation that is isomorphic to the BH, \ie the left CFT Hilbert space.  We then form the TFD state between them. In particular, the density matrix $\rho_{\beta}$ on the right CFT obtained by tracing the left one is not the maximally mixed state on the radiation Hilbert space, so the information about the black hole is indeed encoded in the radiation.}.

\subsection{The mutual information $I_{R:C}$}

\subsubsection{$I_{R:C}$ in the warm phase}
\begin{figure}[t]
	\centering
	\includegraphics[width=1.8in]{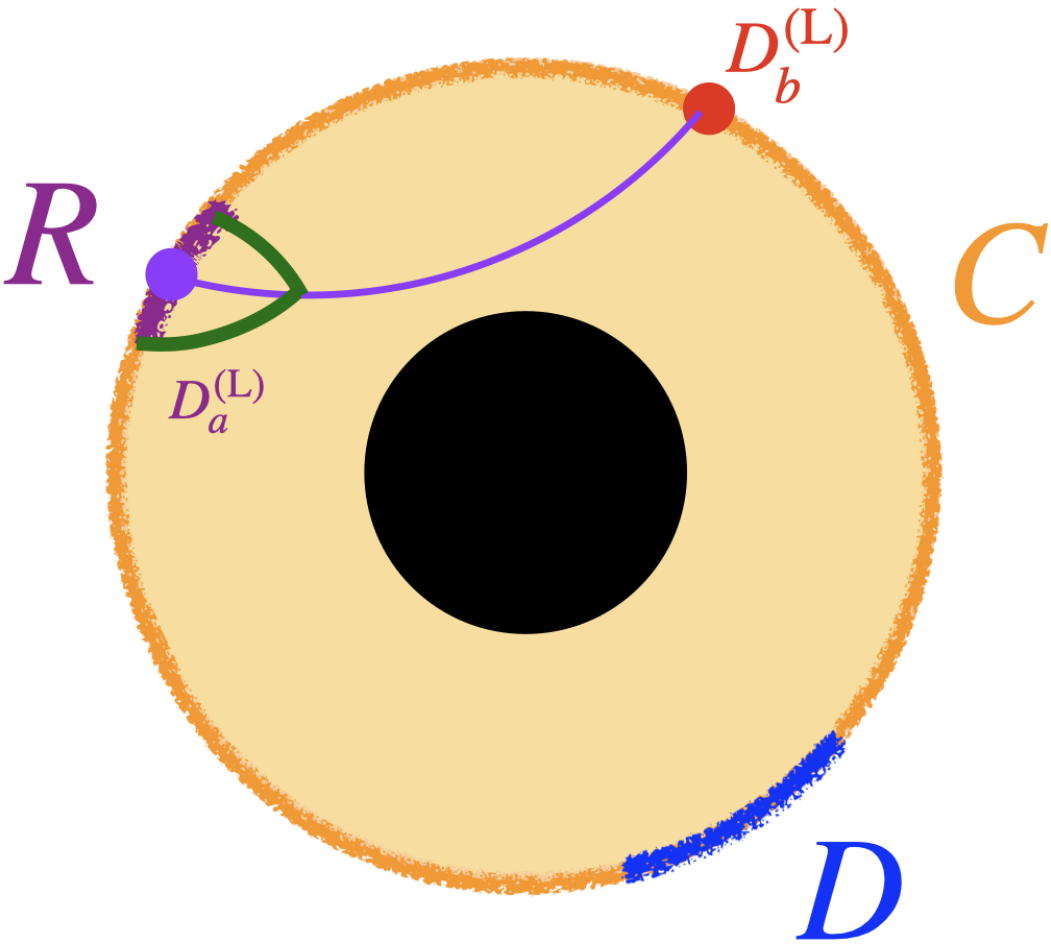}
	\quad
	\includegraphics[width=1.8in]{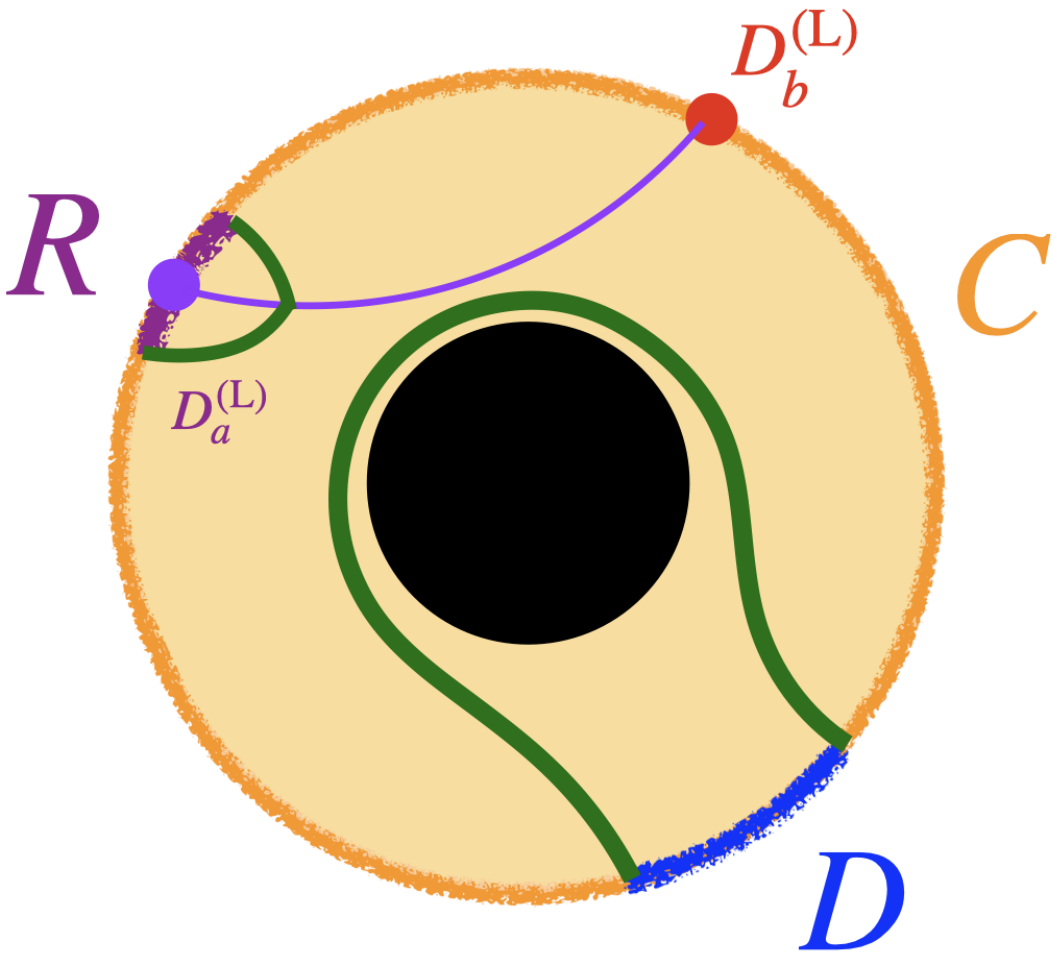}
	\quad
	\includegraphics[width=1.8in]{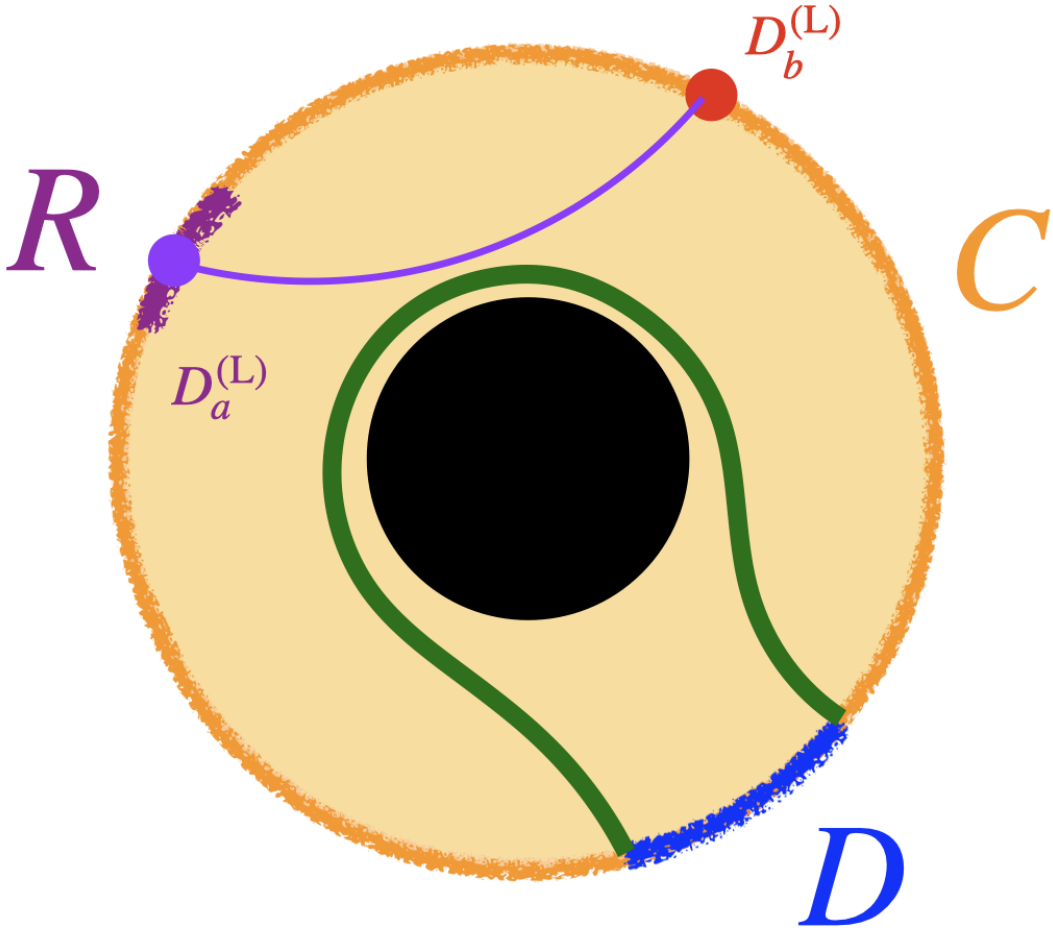}
	\caption{Extremal surfaces  in the warm phase when $D^{(\mt{L})}_{b} \in C$. The purple line connecting $R$ and $A$ is the brane in the bulk. Extremal surfaces are drawn with  thick green lines.  {\bf Left}: The extremal surface for $S_{R}$. {\bf Middle} The extremal surface for $S_{C}$. {\bf Right} The extremal surface for $S_{RC}$. }
	\label{fig:SHPw1}
\end{figure}

Let us proceed to calculate the mutual information between $R$ and $C$ during the warm phase. Given the two possible choices for $D$, depending on whether $D^{(\mt{L})}_{b} \in D$ or $D^{(\mt{L})}_{b} \in C$, we divide our analysis into two distinct cases.

\vskip2ex
\noindent{\bf Case I : $D^{(\mt{L})}_{b} \in C$:} 
We need to evaluate each component of the mutual information \eqref{eq:MIRC}.
First,  $S_{R}$ is nothing but the defect entropy, \ie $S_{R}=S_{{\rm EE}}(D^{(\mt{L})}_{a}) =  S_{\rm defect}$ that is defined in eq.~\eqref{eq:defectentropy}. The relevant extremal surfaces for $S_{C},S_{RC}$ are depicted in figure \ref{fig:SHPw1}. The entropy contributions are determined as follows:
\begin{equation}
S_{C} = S_{{\rm EE}}(D^{(\mt{L})}_{a}) +S_{{\rm EE},\beta}(\; \ell \Delta \phi_1^{\rm Hor} -|D| ) , \quad  S_{RC} = S_{{\rm EE},\beta}(\; \ell \Delta \phi_1^{\rm Hor} -|D| )  \,,
\end{equation}
where $S_{{\rm EE},\beta}(x) $ denotes a thermal entropy:
\begin{equation}
S_{{\rm EE},\beta}(x) =\f{c}{3}\log \left[\f{\beta}{\pi \epsilon} \sinh \left(\f{\pi}{\beta} x\right)\right] \,.
\end{equation}
Here, $|D|$ is the length of the region $D$, and $\Delta \phi_1^{\rm Hor}$ denotes the size of the black hole in $\mathcal{S}_{1}$ as defined in eq.~\eqref{eq:warmL1}. Upon combining the aforementioned expressions, we conclude that the mutual information $I_{R:C}$ during the warm phase assumes a non-zero value:
\begin{equation}
	I_{R:C}  =2 S_{\rm def}^{(\rm{reg})} \approx 
	\frac{2c}{3}  \(   \log \(  \frac{\DLD}{\epsilon} \)+  \log \(   \sqrt{   \frac{ 1+T_o\ell}{1-T_o\ell}  }\)    + \mathcal{O} (1)   \) \,, 
	\label{eq:warmIRC1}
\end{equation}
and is intricately linked to the dimension of the Hilbert space of the diary $\mH_{A}$. This mutual information is diverging because two subregions $R$ and $C$ are adjacent.   To extract the contribution of the defect we subtract the same mutual information of the state without defects $I^{(0)}_{RC}$. Considering the standard BTZ black hole as the holographic dual of CFT without any defects, the background mutual information $I_{R:C}^{(0)} $ is given by 
\begin{equation}
		\begin{split}
			I_{R:C}^{(0)} & \approx 2 S_{R}  + \mathcal{O} (\DLD) \approx \frac{2c}{3} \log \f{\Delta L_{D} }{  \epsilon}  \,.
		\end{split}
\end{equation}
The regularized mutual information $\Delta I_{R:C}$ \eqref{eq:MIRCdef} is thus reduced to 
\begin{equation}
	\Delta I_{R:C} \approx  	\frac{2c}{3}   \log \(   \sqrt{   \frac{ 1+T_o\ell}{1-T_o\ell}  }  \) = 2 S_{\rm def}  \,, 
\end{equation}
which implies that degrees of freedom of the defect in $R$ is highly correlated with $C$. In other words, we can conclude that the decoupling is not achieved yet in the warm phase.

\vskip2ex
\begin{figure}[t]
	\centering
	\includegraphics[width=1.8in]{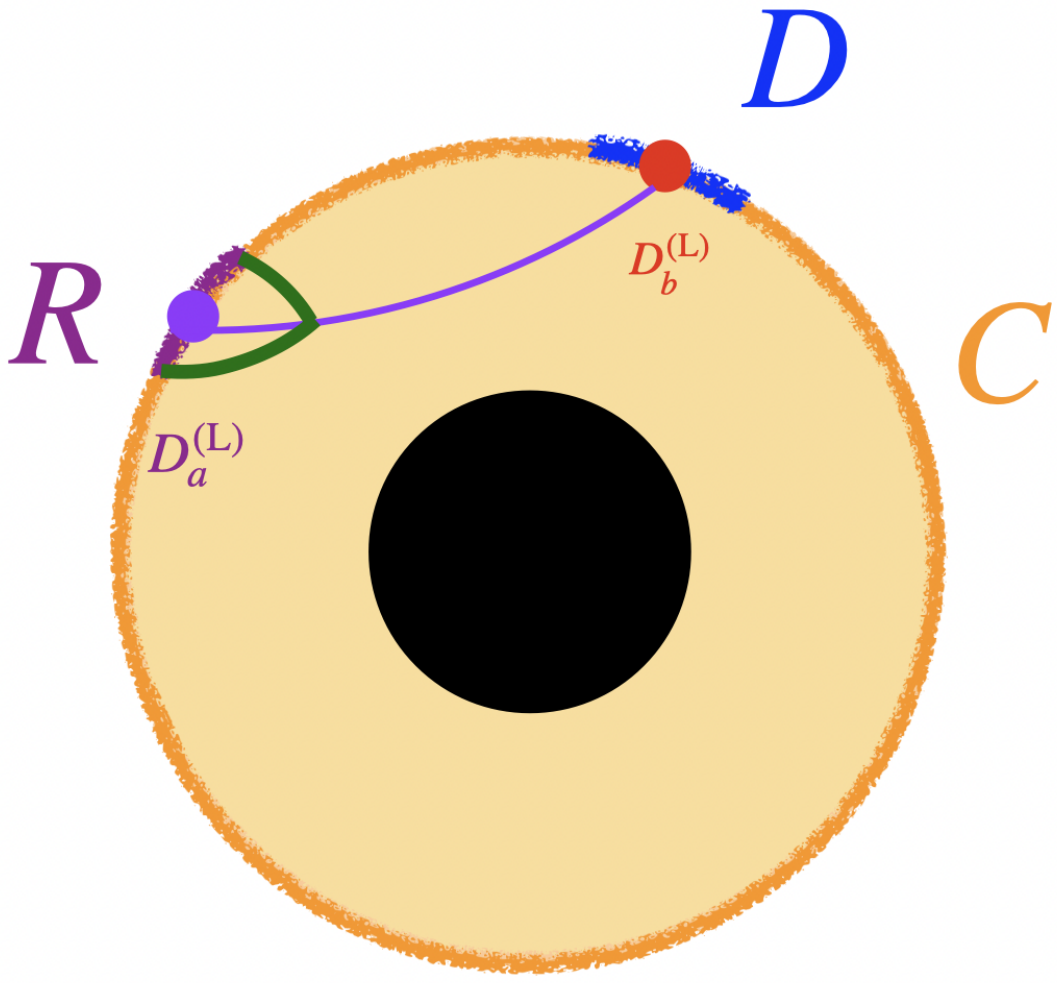}
	\quad
	\includegraphics[width=1.8in]{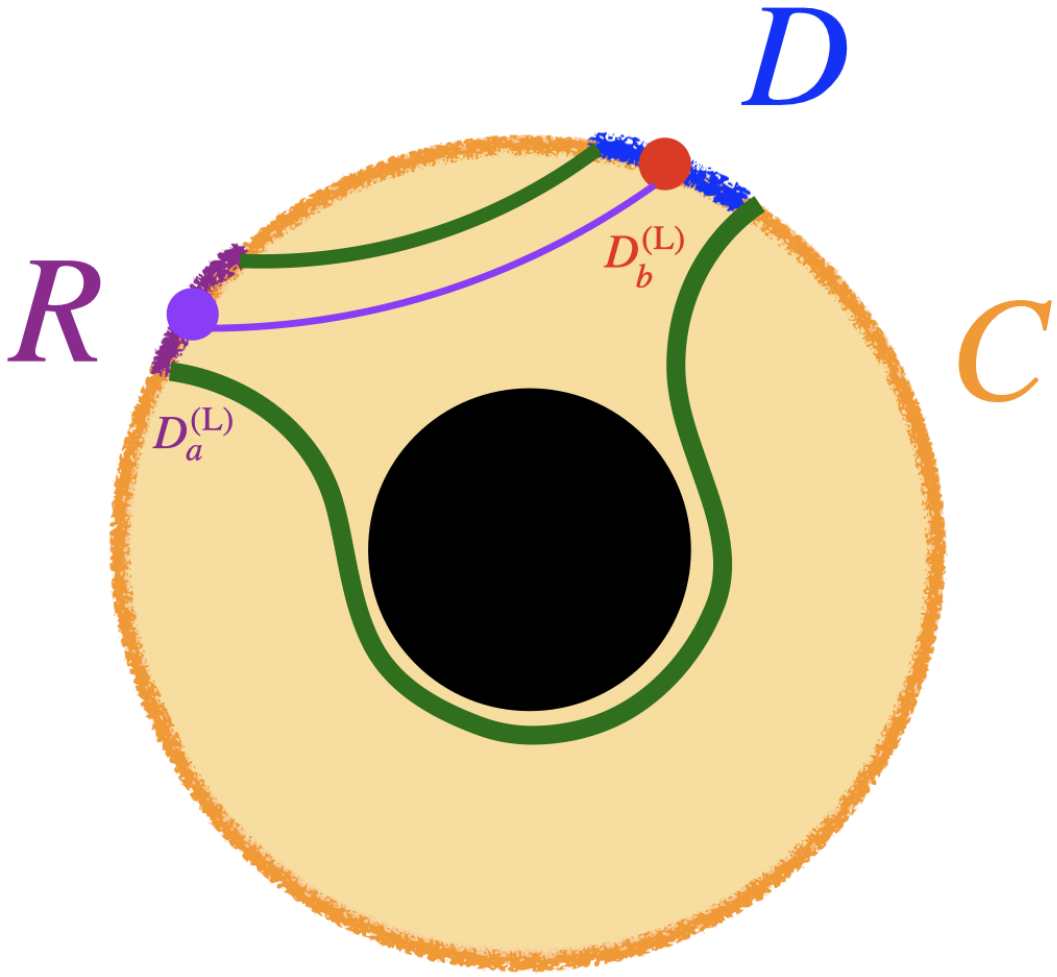}
	\quad
	\includegraphics[width=1.8in]{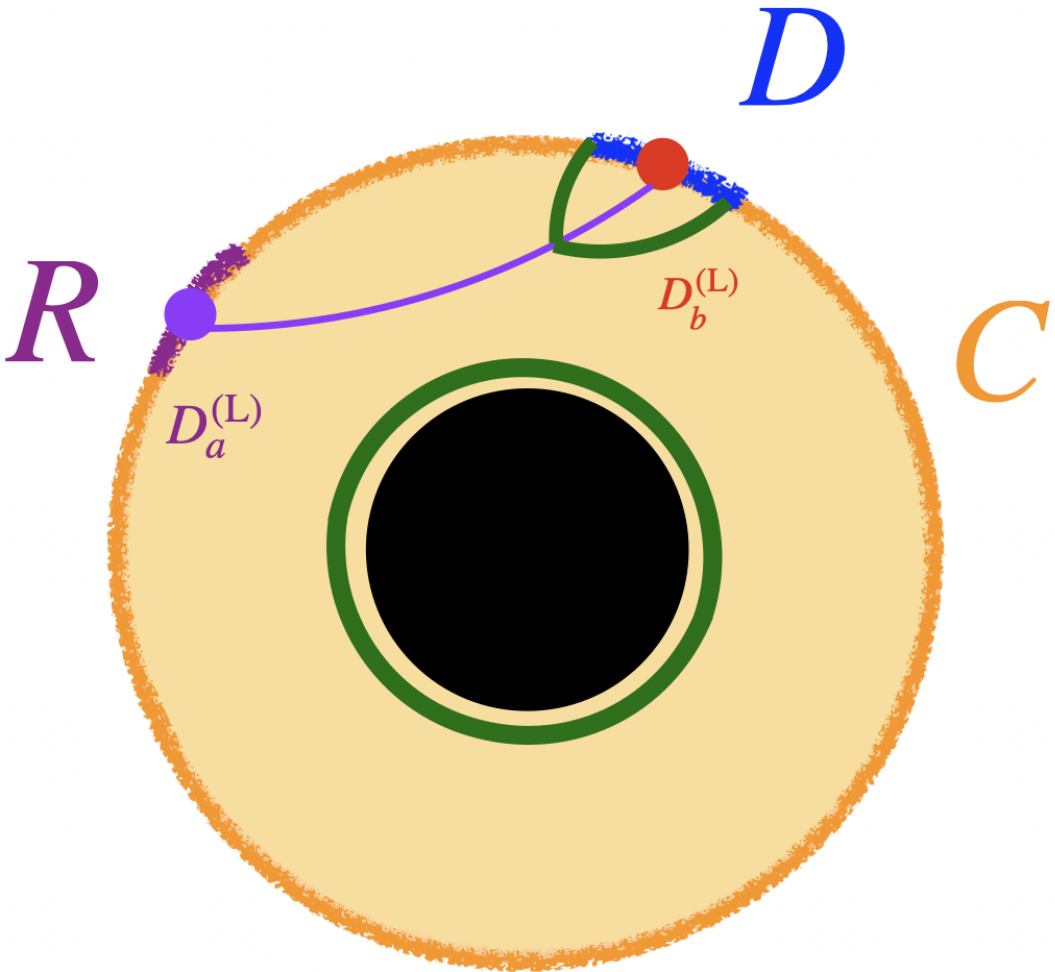}
	\caption{Extremal surfaces  in the warm phase when $D^{(\mt{L})}_{b} \in D$. {\bf Left}: The extremal surface for $S_{R}$. {\bf Middle} The extremal surface for $S_{C}$. {\bf Right} The extremal surface for $S_{RC}$. 
	}
	\label{fig:SHPw2}
\end{figure}
\noindent{\bf Case II: $D^{(\mt{L})}_{b} \in D$:}
The relevant extremal surfaces are illustrated in figure \ref{fig:SHPw2}, where the entropies are expressed as:

\begin{equation}\label{eq:SC}
S_{C} =  \f{c}{3} \log \left[ \f{L_{\mathcal{S}_{2}}}{\pi \epsilon}  \sin \f{\pi L_{2}}{L_{\mathcal{S}_{2}}}\right] +S_{\mt{EE},\beta} (L_{1}) \,, \quad S_{RC} = S_{{\rm EE}}(D^{(\mt{L})}_{b})  + S_{\rm horizon} \,, 
\end{equation}
where the Bekenstein-Hawking entropy in the warm phase $S_{\rm horizon}$ has been derived in eq.~\eqref{hanger2}.
For simplicity, we can choose $|R| =|D| =\Delta L_{D}$ so that the defect contribution in $S_{RC}$ is identical to $S_{R}$. The mutual information thus reduces to 
\begin{equation}
	\begin{split}
		I_{R:C} &=  \f{c}{3} \log \left[ \f{L_{\mathcal{S}_{2}}}{\pi \epsilon}  \sin \f{\pi L_{2}}{L_{\mathcal{S}_{2}}}\right] +S_{\mt{EE},\beta} (L_{1})  - S_{\rm horizon}   \,. 
	\end{split}
\end{equation}
It is straightforward to show that the corresponding contributions from the BTZ background is given by 
\begin{equation}
	{I}_{R:C}^{(0)}  \approx  	\frac{c}{3}  \log \(  \frac{\beta}{ \pi \epsilon}   \sinh \(  \frac{\pi L_1}{\beta}  \) \)  + 	\frac{c}{3}  \log \(  \frac{\beta}{ \pi \epsilon}   \sinh \(  \frac{\pi L_2}{\beta}   \) \)  - \frac{c}{3} \frac{\pi}{\beta} \( L_1 +L_2 \) \,.
\end{equation}
Utilizing previous approximations shown in eqs.~\eqref{eq:coldL1L2} and \eqref{hanger2}, we obtain the approximate contribution of the defect at the first two orders as follows
\begin{equation}\label{eq:miwarm2}
\Delta{I}_{R:C} \approx   \frac{c}{3} \log \(  \frac{\sin (\pi - k_2)}{\pi -k_2}  \)  + \frac{c}{3}\frac{\pi (\pi-k_1-k_2 )}{\pi - k_2} \frac{L_2}{\beta}	  +  \mathcal{O}((L_2)^2)  \,.
\end{equation}
From (\ref{rocket3}) we see that in contrast to the earlier result \eqref{eq:warmIRC1}, where $D^{(\mt{L})}_{a}\in C$, the expression \eqref{eq:miwarm2} remains remarkably small $\Delta{I}_{R:C}  \ll \log g $ in the large tension limit, meaning that most of the degrees of freedom on the defect $R$ are not correlated with $C$. This is due to the entanglement between $D^{(\mt{L})}_{a}$ and $D^{(\mt{L})}_{b}$, which leads to a strong correlation between $R$ and $D$ rather than $C$. However, it's important to note that the decoupling is not fully achieved in the warm phase, as evidenced by the non-vanishing mutual information $\Delta I_{R:C}$. This is because the mutual information between two defects $D^{(\mt{L})}_{a}$ and $D^{(\mt{L})}_{b}$ is large but not maximal, \ie $I_{\mt{L}_a: \mt{L}_b}< 4 \log g$ as manifested in eq.~\eqref{eq:warmIab}, so there is still room for $R=D^{(\mt{L})}_{a}$ and $C$ to correlate. Another way to argue that information recovery is not possible in this case is to check the entanglement wedges of the subsystems. From the right panel of figure \ref{fig:SHPw2} we see that the brane ending on the diary is almost contained in the entanglement wedge of $RC$.

\subsubsection {The mutual information in the  hot phase}
On the other hand, the mutual information in hot phase  is vanishing so the Hayden-Preskill decoupling is actually happening.  

\begin{figure}[t]
	\centering
	\includegraphics[width=1.8in]{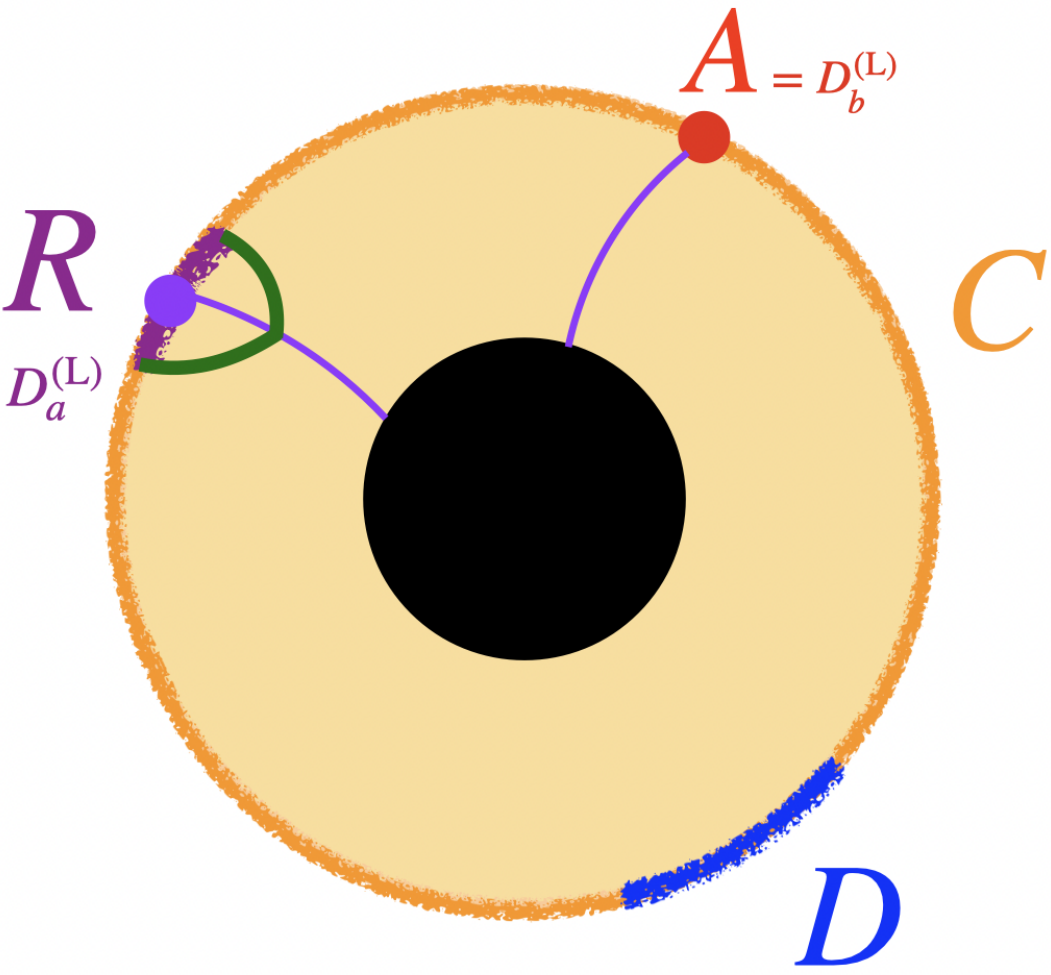}
	\quad
	\includegraphics[width=1.8in]{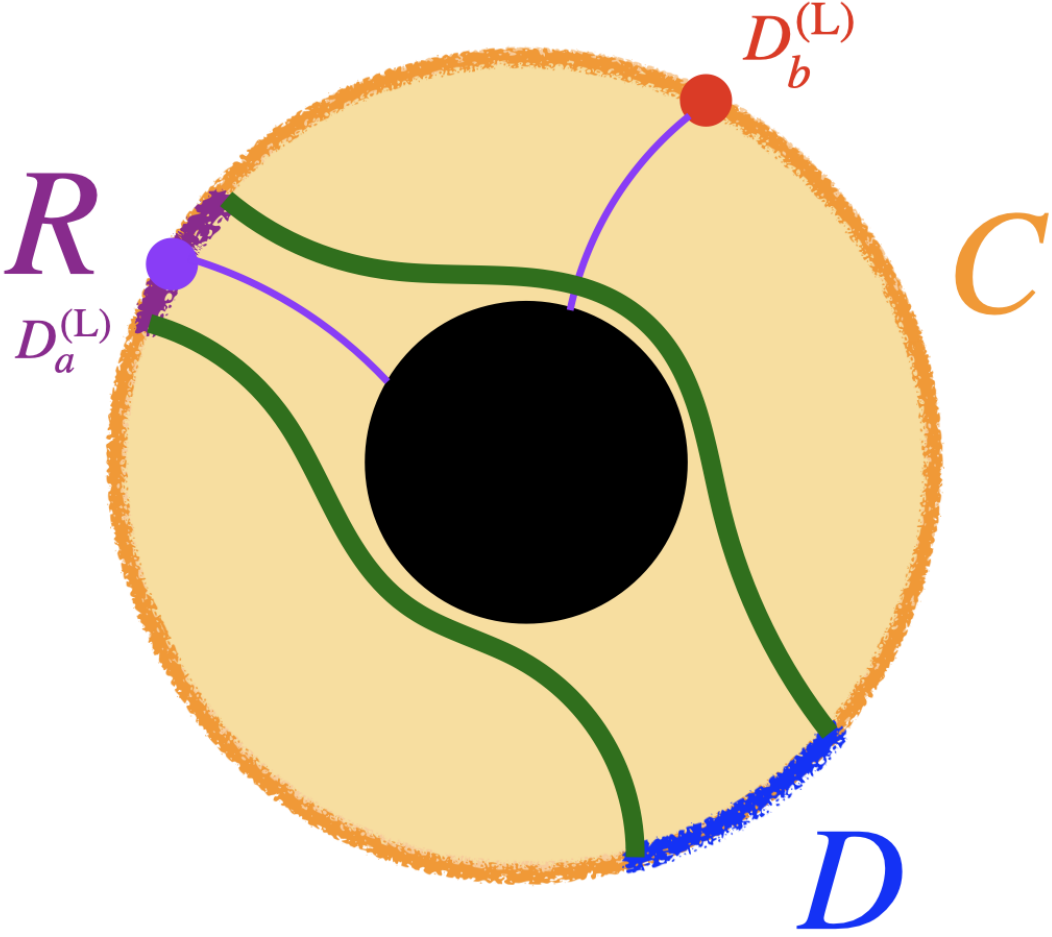}
	\quad
	\includegraphics[width=1.8in]{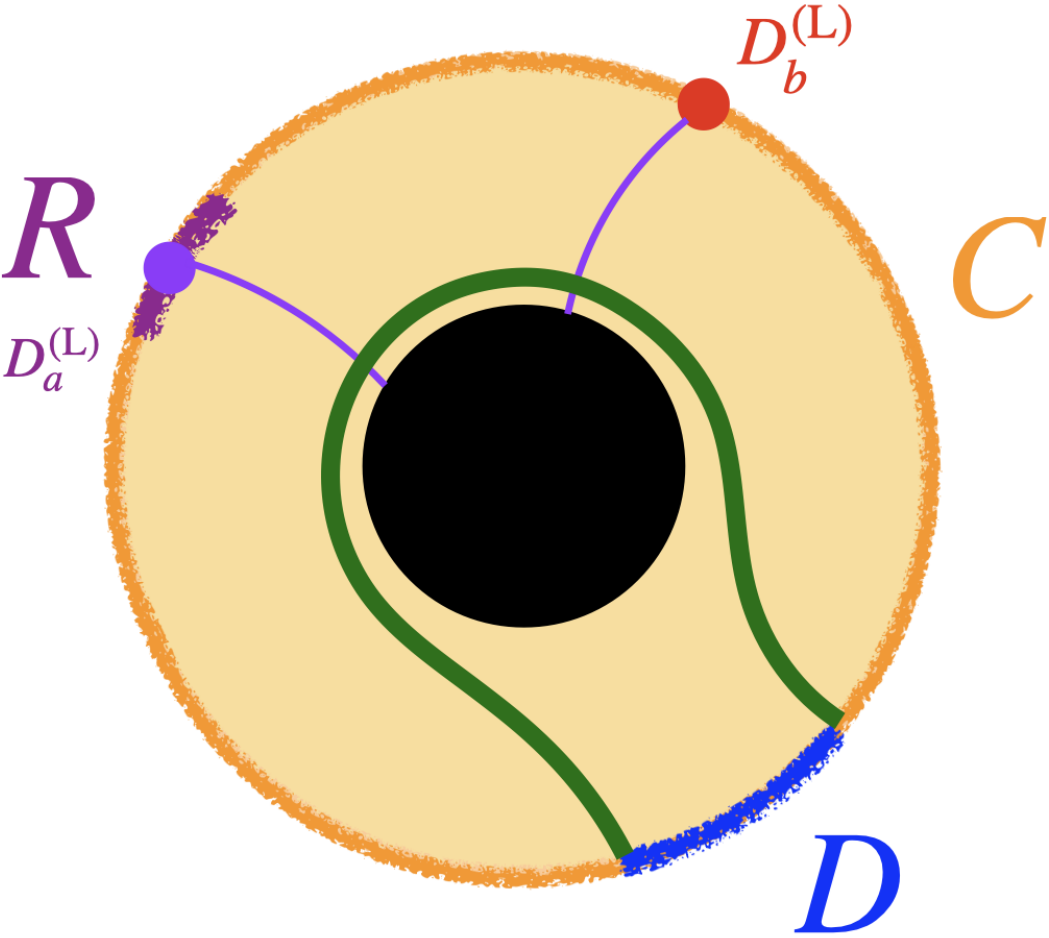}
	\caption{Extremal surfaces  in the hot phase when $D^{(\mt{L})}_{b} \in C$. {\bf Left}: The extremal surface for $S_{R}$. {\bf Middle} The extremal surface for $S_{C}$. {\bf Right} The extremal surface for $S_{RC}$.
}
\label{fig:SHPh1}
\end{figure}

\begin{figure}[t]
	\centering
	\includegraphics[width=3in]{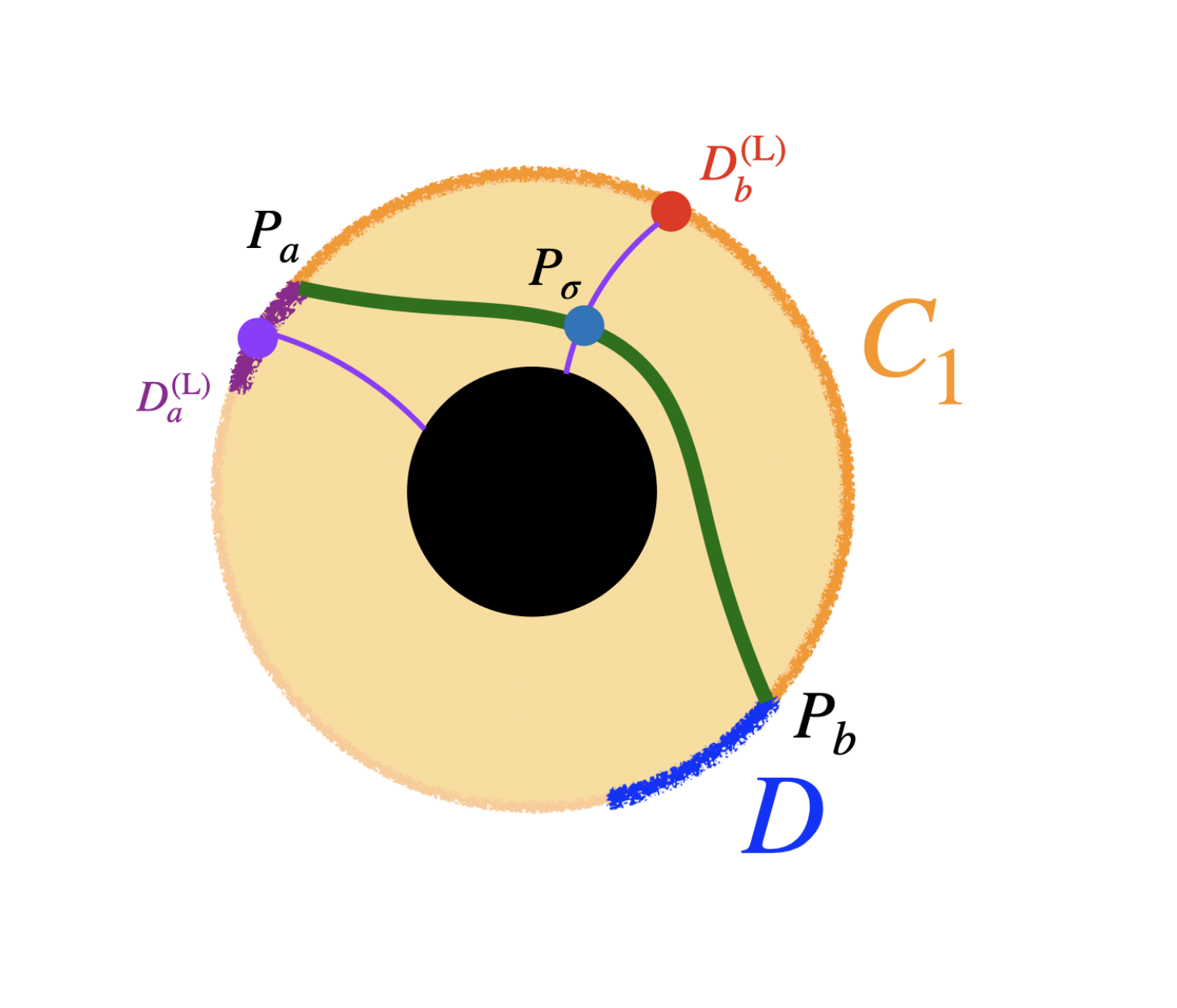}
	\caption{Geodesic  in $\mathcal{S}_{1}$ region  appearing in the calculation of $S_{C}$ in the hot phase, when  $D^{(\mt{L})}_{b} \in C$.The geodesic ends on the brane at $(\phi,r) = (\phi_{1} (\sigma), r_{1} (\sigma))$.  We have a similar contribution from  $\mathcal{S}_{2}$ side. }\label{fig:geo}
\end{figure}

\vskip2ex
\noindent{\bf  Case I : $D^{(\mt{L})}_{b} \in C$:} The extremal surfaces are depicted in figure \ref{fig:SHPh1}.
The curve for  $S_{C}$   shown in Fig  \ref{fig:SHPh1}  intersects with  the branes once.  To compute the geodesic length intersecting with a brane,  we first fix the location on the brane  and compute the geodesic length on each side, add them up and extremize the  total length. In our setting C consists of two disjoint pieces $C_1, C_2$, and let $C1$ be the piece containing the defect $D^{(L)}_{b}$.   Let  $P_{a} : (\phi,r) =( \phi_{\infty}+\phi_{a}, r_{\infty})$ and $P_{b}: (\phi,r) =( \phi_{\infty}+\phi_{b},   r_{\infty})$ be the end points of $C_{1}$ on the boundary cut off surface $r=r_{\infty}$.   Also we pick up a point  $P_{\sigma} (\phi, r)=(\phi_{\infty} +\Delta \phi_{1}(\sigma) , r_{1} (\sigma))$  on the piece of the  brane which ends at the defect $D^{(L)}_{b}$.  $\Delta \phi_{1}(\sigma)$ is 
\be
\Delta \phi_{1}(\sigma) = \int^{\infty}_{\sigma} \phi'_{1} (s) ds,
\ee
and in the hot phase $\Delta \phi_{1}(\sigma) = \Delta \phi_{2}(\sigma) $,  $r_{1} (\sigma)=r_{2}(\sigma)$.
The length of the geodesic connecting $P_{a}$ and $P_{\sigma}$ is given by
\be
D_{a}(\sigma)= \ell \log \left[ \f{r_{\infty}}{R}\right] + \ell \log \left[-\f{\s{r_{1}^{2} (\sigma)-R^{2}}}{R} +\f{r_{1}(\sigma)}{R} \cosh \f{R}{\ell} \left(\Delta \phi_{1}(\sigma)-\phi_{a}\right)  \right],
\ee
see figure \ref{fig:geo}.  The length $D_{b}(\sigma) $ of the geodesic connecting $P_{\sigma}$ and $P_{b}$ is computed similarly.

The location of the intersection $\sigma =\sigma_{*}$ is determined by extremizing  $D_{{\rm tot}}(\sigma)= D_{a}(\sigma) +D_{b}(\sigma)$.
This task gets simplified in the high temperature limit $\;T(L_{1}+L_{2}) \gg 1\;$ where the
the horizon becomes large and the entire curve is getting close to the horizon. This  in particular means $\sigma_{*} \rightarrow 0$. By adding the contribution of other disjoint  geodesic that does not cross the brane,    in this limit  the value of the entropy is given by
\be
S_{C} = \f{c}{3} \log \f{\beta}{\pi \epsilon} +\f{c\pi T_{\mt{BH}} }{3}(L_{1}+L_{2}-|D|)+\f{c}{3} \log \left(\s{\f{1+T_{o}\ell}{1-T_{o}\ell}}\right).
\ee
Other entanglement entropies $S_{R}$, $S_{RC}$  can be computed in a similar manner. The interested mutual information $I_{R:C}$ thus reads 
\begin{equation}\label{eq:mihot}	
I_{R:C}=\frac{c}{3} \log \left(\frac{\DLD \beta}{\pi \; \epsilon^{2}} \right)\,.
\end{equation}
Therefore  the contribution of the defect to the mutual information is vanishing, \ie $\Delta I_{R:C} =0$, and we conclude that the decoupling is achived in the hot phase.

\begin{figure}[ht]
	\centering
	\includegraphics[width=1.8in]{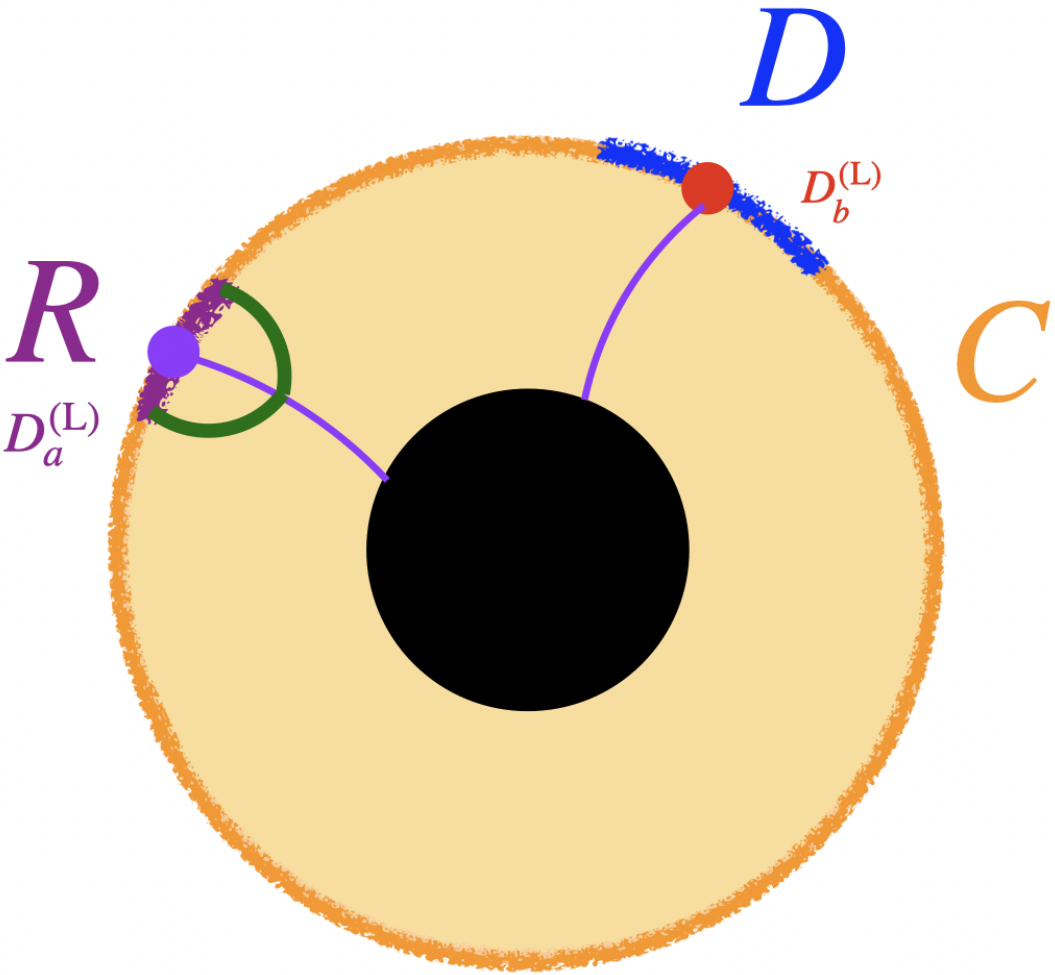}
	\quad
	\includegraphics[width=1.8in]{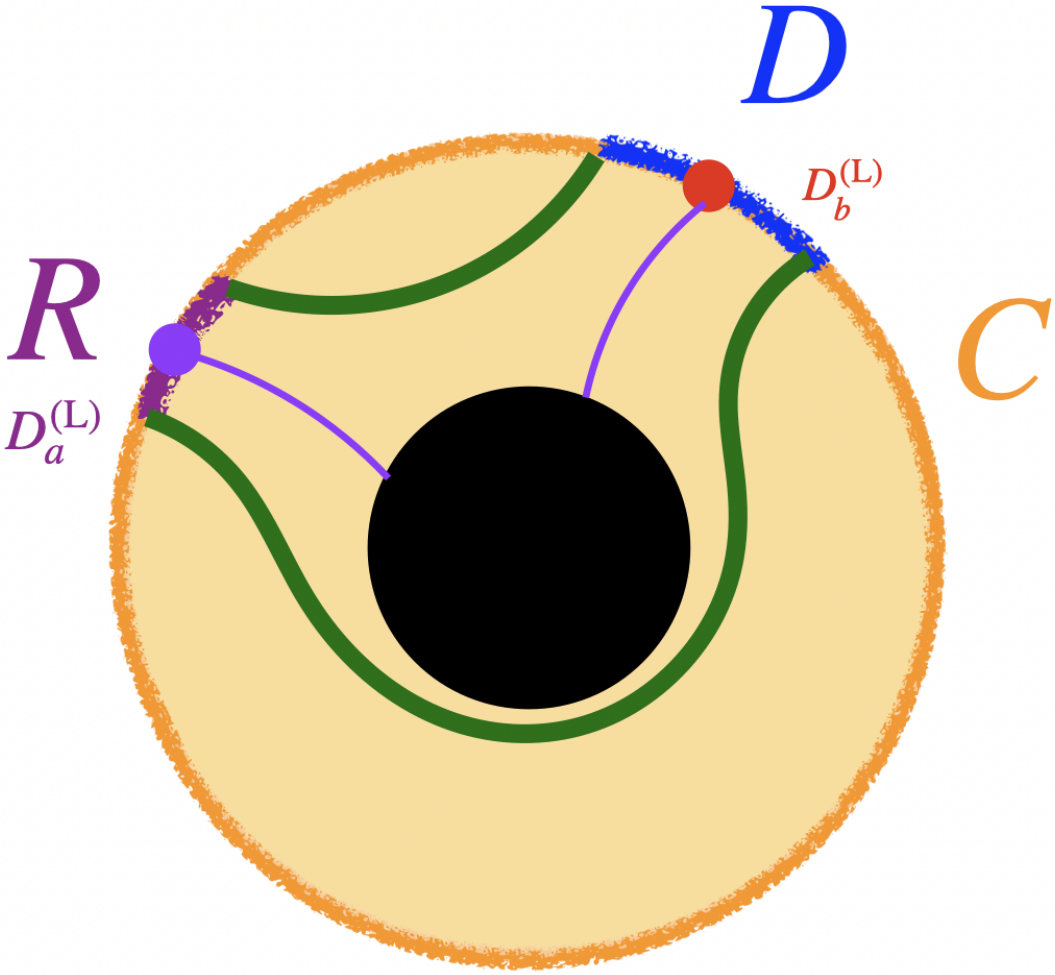}
	\quad
	\includegraphics[width=1.8in]{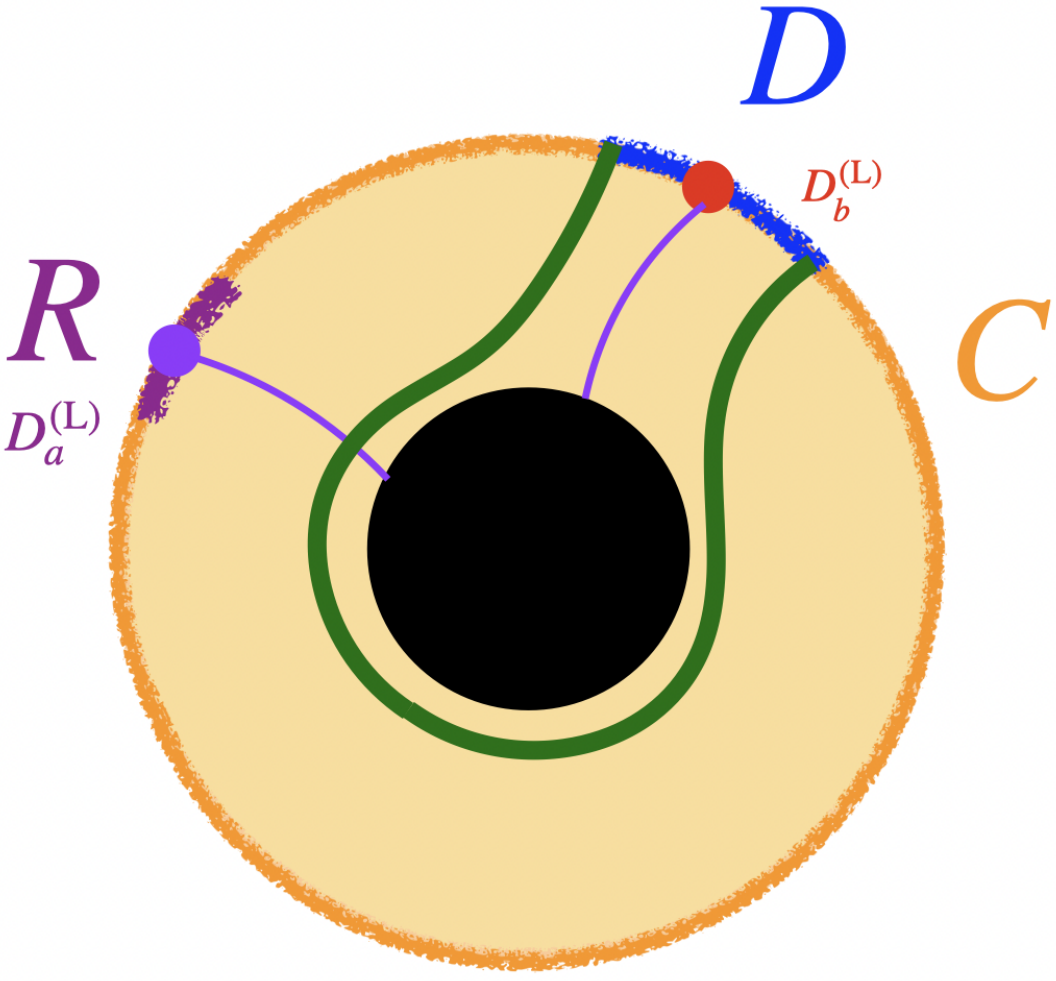}
	\caption{Extremal surfaces  in the warm phase when $D^{(\mt{L})}_{b} \in D$. {\bf Left}: The extremal surface for $S_{R}$. {\bf Middle} The extremal surface for $S_{C}$. {\bf Right} The extremal surface for $S_{RC}$.
}
\label{fig:SHPh2}
\end{figure}

\vskip2ex
\noindent{\bf  Case II : $D^{(\mt{L})}_{b} \in D$:}  The extremal surfaces  are  this case shown in figure \ref{fig:SHPh2}. The extremal surfaces are depicted in figure \ref{fig:SHPh2}. We have
\be
S_{R}=  S_{{\rm EE}}(D^{(\mt{L})}_{a} ) ,\quad  S_{C} = \f{2c}{3} \log \f{\beta}{\pi \epsilon} +\f{c\pi T_{\mt{BH}}}{3}(L_{1}+L_{2}-|D|)
\ee
and 
\be
S_{RC} = \f{c}{3} \log \f{\beta}{\pi \epsilon} +\f{c\pi T_{\mt{BH}}}{3}(L_{1}+L_{2}-|D|)   +\f{c}{3} \log \left(\s{\f{1+T_{o}\ell}{1-T_{o}\ell}}\right) \,.
 \ee
Therefore the mutual information is again given by eq.~\eqref{eq:mihot}, therefore $\Delta I_{R:C}$ is again vanishing.


\subsection{Comments on the relation to the Yoshida-Kitaev decorder }

In \cite{Yoshida:2017non}, Yoshida and Kitaev proposed a decoding protocol of information of the diary $T$ from Hawking radiation $DE$.  The ingredients are in addition to the original system $RCDB$,  a copy of the diary $A’$ and the reference system $R’$. The state on $R’T’$  is again the EPR state.  In this setup, one simulates the dynamics of the black hole on early Hawking radiation and the copy of the diary to get late radiation $D’$ and the remaining black hole $C’$ records the outcome $BT’ \rightarrow C’ D’$.
The next step is the postselection of the total state $ |\mathrm{YK} \ra$ to the EPR state on $D'D$. If this succeeds, the state on $RR'$ is also the EPR state with the probability closed to one, meaning the decoding is successful.  The resulting state is given by 
\begin{equation}
	|\mathrm{YK} \ra = P_{DD'} (I_{R} \otimes U_{AB}  \otimes U^{*}_{A'E} \otimes I_{R'} )  |\mathrm{EPR} \ra_{RA} |\mathrm{HP} \ra_{RCDE} \otimes   |\mathrm{EPR} \ra_{R'A'},
\end{equation}
where $P_{DD'}= |\mathrm{EPR} \ra_{DD'} \la  \mathrm{EPR} |$.
This map is actually equivalent to the Petz recovery map in sufficiently chaotic systems \cite{Nakayama:2023kgr, Yoshidaunp}.

\begin{figure}[t]
	\centering
	\includegraphics[width=2.8in]{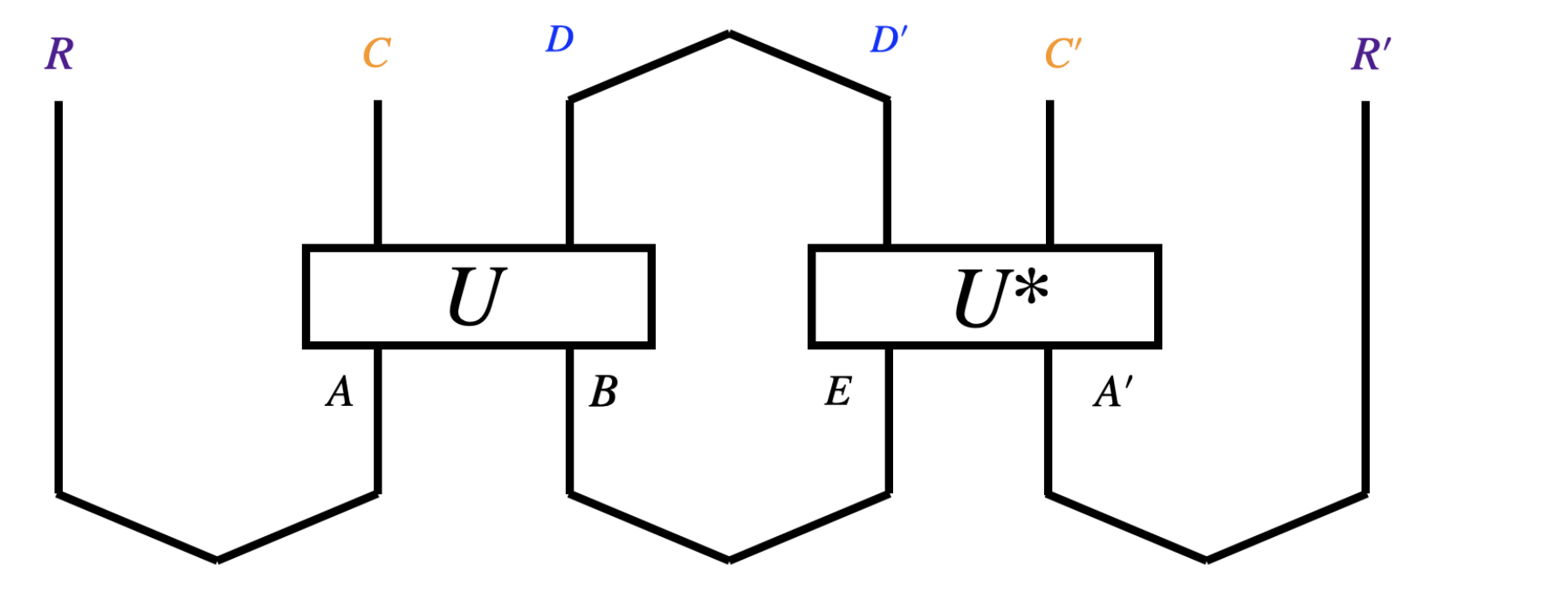}
	\includegraphics[width=3.1in]{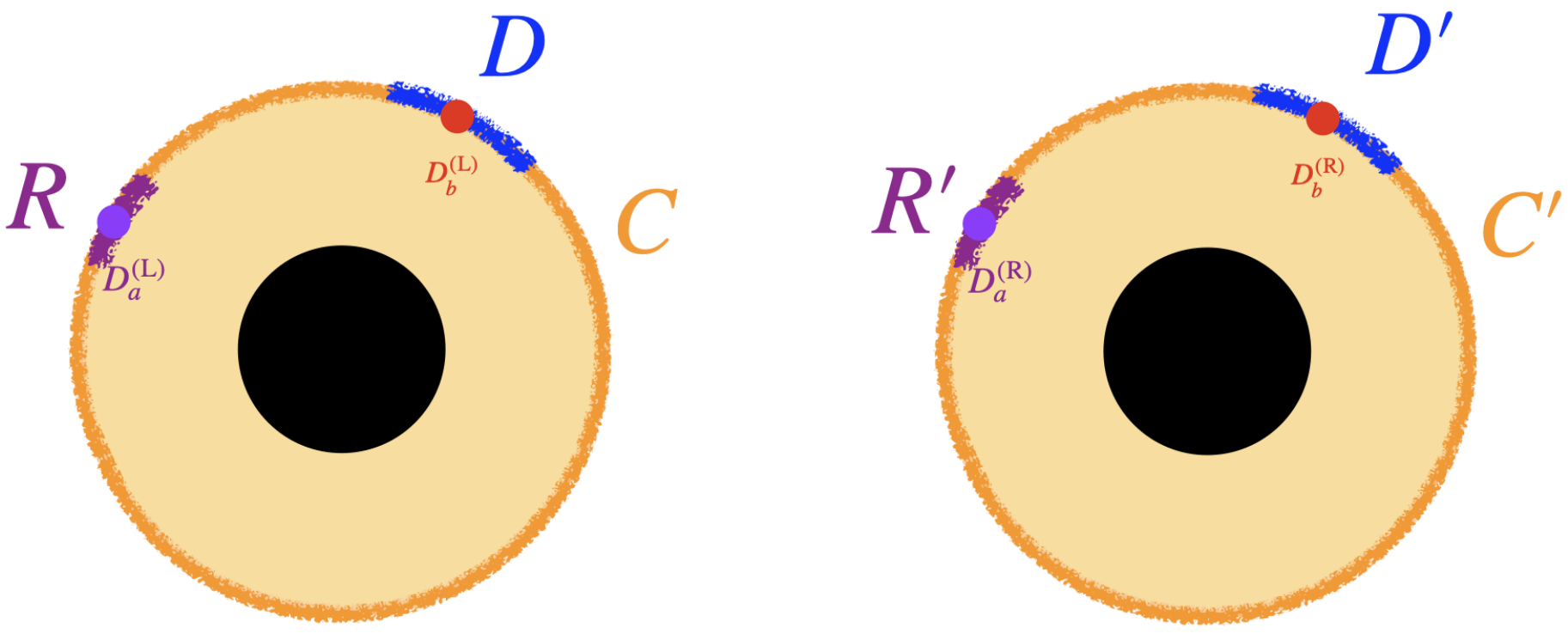}
	\caption{ {\bf Left}: Yoshida-Kitaev recovery map for reconstructing the state of the diary from the Hawking radiation in the Hayden Preskill setup.  The protocol consists of the simulation of the black hole dynamics in $EA'$, and the postselection onto $|\mathrm{EPR} \ra_{DD'}$.    {\bf Right}: Corresponding DCFT setup.  In particular the postselection in the YK protocol corresponds to separating two defects on both boundaries $L$ and $R$.
	}
	\label{fig:YK}
\end{figure}

We  have shown that depending on the choices of the parameter $(\tau, \gamma)$ and the subregion for the late Hawking radiation $D$, the DCFT state can have three distinct entanglement structures: 
\begin{itemize}
	\item $|W \ra_{C}$ : dual to the warm phase when $D^{(\mt{L})}_{b} \in C$. In this case the defect contribution of the mutual information $\Delta I_{R:C}=4 \log g$ is maximal, so the decoupling is not achieved at all. 
	\item $|W \ra_{D}$ : dual to the warm phase when $D^{(\mt{L})}_{b} \in D$. In this case $I_{R:C}$ is small  \ie,  $ I_{R:C} \ll 4 \log g$  and it remains finite but non-vanishing.  In this sense, the decoupling is almost archived in this state.
	\item $|H\ra_{C}$,  $|H\ra_{D}$: The DCFT state dual to the hot phase when  $D^{(\mt{L})}_{b} \in C$ and $D^{(\mt{L})}_{b} \in D$ respectively . In this case, the defect contribution to $I_{R:C}$ is vanishing, and the decoupling is complete in these cases.
\end{itemize}	
When the initial state is $|W \ra_{C}$, as we increase the entanglement between two sides $L$ans $R$ it becomes $|H \ra_{C}$  dual to the hot phase, therefore the decoupling is archived. We identified this with the Hayden Preskill decoupling for evaporating black holes. When the initial state is $|W \ra_{D}$,  an observer collecting all Hawking quanta $DE$ can make it $|H \ra_{D}$ by a LOCC, namely separating  $R$ and $D$ as well as  $R'$ and $D'$. In the hot phase, one of the two branes connects $R =D^{(\mt{L})}_{a} $ and $R'  =D^{(\mt{L})}_{b} $  and the brane hosts a black hole dual to the maximally entangled state on the defect degrees of freedom $|{\rm EPR} \ra_{RR'}$. The same is true for $D =D^{(\mt{L})}_{a} $  and $D' =D^{(\mt{L})}_{b} $. Therefore, $|H \ra_{D} = |{\rm EPR} \ra_{RR'} \otimes |{\rm EPR} \ra_{DD'}\otimes | \psi \ra_{CC'} $ .This is quite similar to the state $|\mathrm{YK} \ra$ obtained by applying the recovery protocol to $| {\rm HP} \ra$. In particular postselection to the maximally entangled state on  $|\mathrm{EPR} \ra_{DD'}$ has a natural counterpart in the DCFT setup, namely separating defects  $ D^{(\mt{L})}_{a}  \leftrightarrow D^{(\mt{L})}_{b} $ and $ D^{(\mt{R})}_{a} \leftrightarrow D^{(\mt{R})}_{b} $ because both of these operations result in  $  |{\rm EPR} \ra_{DD'}$.  This induces quantum teleportation from $R$ to $R'$, implying that the information of the diary appears in early Hawking radiation (right boundary).  The phase transition of the brane configuration in the bulk is a holographic realization of the Yoshida-Kitaev decoding protocol.



\section{Discussion} \label{sec:discuss}
%

This paper investigates a system involving two entangled and gravitational universes using double holography. Our study examines the phase transitions of brane configurations, particularly the topology transitions of brane geometry. This type of transition provides a manifestation of the ER=EPR paradigm. A major focus of our exploration is on the quantum information structure of the entangled universes, specifically the mutual information between defects in the boundary theories. The doubly holographic setup is further utilized to discuss Hayden-Preskill protocols and interpret the Yoshida-Kitaev decoding protocol. 

Section \ref{sec:bulk} provides a comprehensive review of the holographic construction of the boundary system, which corresponds to the TFD state of two-dimensional DCFTs. The section outlines the three bulk phases that the system experiences as temperature increases. The complete phase diagram is presented in figure \ref{fig:PhaseDiagram}. Analytical insights are provided for two specific cases where one of the lengths of the regions between the defects is significantly larger than the other, and when the two lengths are equal, in sections \ref{rocket2} and \ref{sec:sym}, respectively. The subsequent section delves into examining the quantum entanglement between various subsystems within the doubly holographic setup, quantified through mutual information. In particular, we explore how mutual information between defects varies throughout various phase transitions. In the cold and warm phases, we found that the mutual information between defects on the left and right sides vanishes, while it is non-trivial in the hot phase, through a careful analysis of possible candidates for RT surfaces.  All relevant RT surfaces are shown in figure \ref{fig:Table} with the final analytical results summarized in section \ref{sec:summary}.  Section 4 of the paper demonstrates that the doubly holographic setup can emulate the Hayden-Preskill protocol, with successful decoupling evidenced by the vanishing mutual information between the radiation and the remaining black hole. Furthermore, the realization of the Yoshida-Kitaev decoding protocol is discussed by implementing the postselection procedure through the adjustment of distances between two defects in the DCFT setup. This section underscores the potential of the DCFT setup to simulate and study information recovery protocols in the context of black hole evaporation. The calculations in the main part of the paper focus on the bulk perspective. Instead, we explore the brane perspective in Appendix \ref{sec:brane}. In particular, we investigate the phase transition on the braneworld. By defining the holographic stress tensor on the brane, we present the modified relation between the trace of the stress tensor and the Ricci curvature, which includes additional terms arising from the $T\bar{T}$ term. Notably, we explicitly show the equivalence of the three descriptions by evaluating the renormalized action from the brane perspective. Further technical details appear in appendices \ref{sec:corner} and \ref{sec:notes}. In this last section of the paper, we would like to list some other intriguing points as well as potential future directions related to this doubly holographic model.

\subsection*{Double holography of gravitational collapse}

In Appendix \ref{sec:brane}, we investigate the phase transition from the brane perspective. As the temperature increases, the braneworld geometry undergoes a transition from an asymptotically AdS$_2$ spacetime, supported in the cold and warm phases, to a pure AdS$_2$ spacetime in the hot phase, as depicted in figure \ref{fig:branesphases}. Notably, in the cold and warm phases, the brane supports a 'star' or a localized cluster of CFT excitations that deform the AdS$_2$ geometry. In contrast, in the hot phase, the AdS$_2$ braneworld corresponds exactly to the two-sided external black hole, as is evident from the induced metric in eq.~\eqref{eq:AdS2Rindler}. The temperature increase of the thermal bath naturally corresponds to a process of thermalization. From the perspective of braneworld gravity, this thermalization process could be understood as gravitational collapse.

To elaborate, the braneworld supports nontrivial matter contributions in the cold and warm phases (as discussed in Appendix \ref{sec:stresstensor}). As the temperature increases, the CFT star or the accumulation of CFT excitations on the branes grows in size and density at the center of the braneworld. Beyond the critical temperature, this effective star collapses into an AdS$_2$ black hole on the brane. Although our setup represents a static configuration, it serves as a doubly holographic framework for studying gravitational collapse and black hole formation.

Assuming a dynamic process with slowly increasing temperature, the gravitational collapse and black hole formation on the brane can be understood as a dual of the boundary thermalization process, \ie an adiabatic increase of the boundary temperature. With the warm phase, a black hole is formed in the bulk $\mS_1$ region. Then there are two effects as the temperature increases: the horizon radius grows, \ie the black hole becomes larger, and the brane bends closer to the black hole.\footnote{We note that throughout this process the brane and the black hole remain relatively far apart, in the large tension limit. In this regime and in the warm phase, $\sigma_+\propto 1/x$ from eq.~\reef{tab1}, and $T_\mt{DCFT}\propto 1/(L_2 \log(1/x))$ from the discussion around eq.~\reef{eq:ColdHot2}.} Finally, with the transition to the hot phase, the brane collapses, falling into the bulk black hole and creating a low-dimensional black hole on the brane universe. 

One of the most interesting lessons or observations is that the bulk perspective in this double holography provides a bird's eye view for the black hole formation. From a higher-dimensional viewpoint, the gravitational collapse on the braneworld can be interpreted as that the bulk black hole is attracting the braneworld and pulling it closer. The formation of a black hole is a result of the collapse of the braneworld into the higher-dimensional black hole. The low-dimensional black hole on the brane universe is nothing but the intersection of the higher-dimensional black hole with the lower-dimensional braneworld, \eg  refer to the rightmost plot in figure \ref{fig:simp1}. The bulk picture also geometrically illustrates the concept of ER=EPR. When the black hole is formed, the braneworld serves as a wormhole and establishes a connection with its entangled counterpart through the interior of the higher-dimensional black hole in the bulk spacetime. Moreover, this holographic picture for black hole formation can be applied to any type of spacetime, such as asymptotically flat or de Sitter spacetime in any dimension. While technically challenging, it would be very interesting to explore black hole formation in flat or de Sitter braneworlds. 

The braneworlds experience the three distinct phases appearing in the phase diagram shown in figure \ref{fig:PhaseDiagram} if we fix the interval lengths and raise the temperature of the system. Above, we would like to think of the points of the phase diagram as denoting distinct points in an adiabatic process of increasing boundary temperature leading to gravitational collapse on the brane. An alternative way to emulate the process of black hole formation would be to fix the temperature and the total boundary length (\ie $L_1+L_2$). Then starting at large $L_1/L_2$, the system would be in the warm phase where the brane supports a star, and decreasing $L_1/L_2$ would eventually bring us to the hot phase where a black hole appears on the brane. In this scenario, the defects begin relatively close (\ie $L_2\ll L_1$) and they have moved farther apart as the ratio of lengths is decreased. In the bulk, the branes (or in particular, the turning points on the branes) move closer towards the black hole sitting in the middle of the bulk $\mS_1$ region. Of course, this effect is similar to what happens with an increasing temperature, as the size of the black hole increases. In any event, this scenario of moving the defects may give an easier approach to studying gravitational collapse on the braneworld. A starting point for these studies may be previous investigations of the two-dimensional worldsheet in the bulk dual to accelerating Wilson lines in the boundary theory, \eg see \cite{Jensen:2013ora,Sonner:2013mba,Hubeny:2014kma}.

\subsection*{Multi-partite boundary}
\begin{figure}[t]
	\centering
	\includegraphics[width=5.5in]{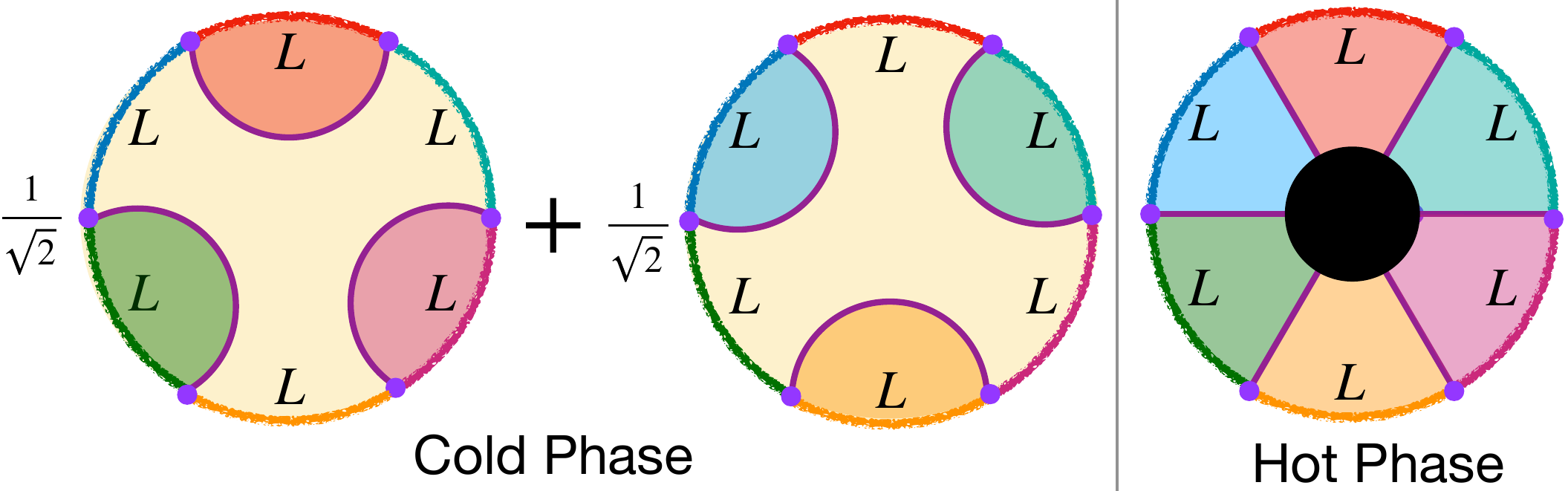}
	\caption{The holographic dual spacetimes of the defect field theory with multi-partite boundary. The left and right is the preferred phase at very low and very high temperature, respectively. Note that the cold phase is a superposition state of two classical spacetimes. A similar warm phase with a black hole at the center may also dominate.}\label{fig:BulkN}
\end{figure}
A straightforward extension of our setup is to consider a multi-partite boundary configuration. In such a scenario, one can expect the emergence of different phases characterized by distinct types of brane profiles. The simplest case arises in the most symmetric configuration. For instance, we can consider a defect conformal field theory with a multi-partite boundary configuration, where each segment of the boundary is of equal length, denoted by $L_i = L_{\text{bdy}}/N$. Figure \ref{fig:BulkN} illustrates an example of this configuration with $N=6$. Due to the strict symmetry constraints, one might naively expect the presence of only cold and hot phases. This guess stems from the expectation that the holographic dual spacetime for a symmetric multi-partite boundary should reflect the same $Z_N$ symmetry, essentially consisting of $N$ replicas of the cold/hot phase explored in subsection \ref{sec:sym} for a symmetric boundary with two defects. However, this simplistic expectation does not always hold. We can find that the warm phase should also dominate for a sufficiently large $N$.

Even more intriguingly, the bulk spacetime may present the replica symmetry breaking. To illustrate this feature explicitly, we display the holographic dual spacetime for very low and very high temperatures in the case of $N=6$ in figure \ref{fig:BulkN}. It's evident that the two candidates of the classical bulk spacetime corresponding to the cold phases do not exhibit replica symmetry $Z_N$ akin to that of the multi-partite defect conformal field theory. However, we propose that instead of considering the holographic bulk dual as a single spacetime, it should be regarded as a superposition of two distinct classical spacetimes, as illustrated in figure \ref{fig:BulkN}. This superposition state allows for the retention of replica symmetry of the boundary theory, as expected for a holographic dual. As the boundary temperature increases, we anticipate that the superposition state for the warm phase, featuring a bulk black hole at the center, could also emerge as the dominant saddle point.

\subsection*{Dynamical branes} 
\begin{figure}[t]
	\centering
	\includegraphics[width=3.5in]{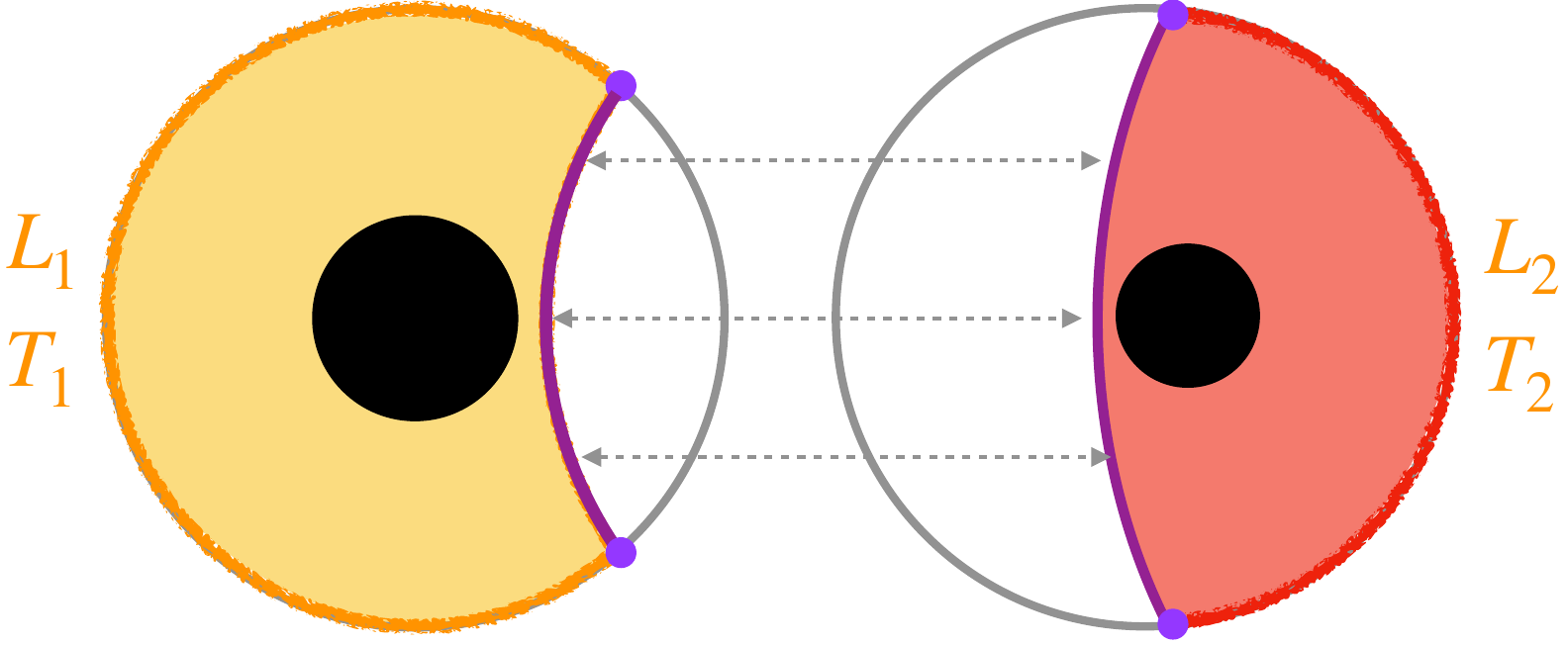}
	\caption{The possible preparation of an initial state involving a doubly holographic spacetime with two black holes may be achieved by gluing two BTZ bulk spacetimes with different masses along a two-dimensional brane.
	}\label{fig:twoBHs}
\end{figure}

The present study is confined to the static configurations, and the corresponding dual quantum state is described by the TFD state defined in eq.\eqref{eq:TFDstate}. Consequently, our examination of brane profiles is restricted to static scenarios. This limitation implies that, in the hot phase, the brane always falls into the BTZ horizon, as illustrated in figure \ref{fig:HotPhase}. Essentially, this setup does not allow for the appearance of a bulk spacetime with two black holes. This restriction arises from the fact that, the two BTZ black holes before gluing possess the same temperature, \ie $m_1=m_2$, which leads to a turning point always occurring at the BTZ horizon located at $r=r_h$ (see eq.\eqref{eq:zerosigma}). This restriction is related to the equilibrium state of the boundary theory, where the temperature associated with the field theory in the $L_1$ and $L_2$ portions is identical.

To overcome this limitation and explore a more general scenario, one can consider two CFTs with different temperatures and join them at the defects. Similar setups have been explored, see \eg \cite{Ugajin:2013xxa, Erdmenger:2017gdk}. It is reasonable to anticipate that the holographic dual spacetime could be constructed by gluing two BTZs at different temperatures, as depicted in figure \ref{fig:twoBHs}. Obviously, this would not be the equilibrium state. Instead, one can view this as the preparation for an initial state. With the thermal evolution, we can anticipate that the two bulk black holes with distinct temperatures should be attracted to each other. Eventually, these two bulk black holes would converge onto the braneworld in the middle and merge into a larger black hole, thereby transitioning to the hot phase.


While it would be intriguing to investigate more generic dynamic branes in the AdS$_3$ bulk spacetime, such as a spinning string in the rotating BTZ black hole background. There has been some previous investigation of this problem in the literature, \eg see \cite{Gubser:2002tv,Kim:2014bga,Kim:2015bba,Bachas:2021tnp,Maxfield:2022rry,Papadopoulos:2023kyd}. The main challenge lies in solving the junction conditions for generic time-dependent branes. Analytical solutions for generic brane profiles become more elusive in dynamic scenarios. However, symmetric setups can be analytically solved since the junction conditions reduce to the Neumann boundary condition. To wit, 
\begin{equation}\label{eq:simplejunction}
	K_{ab}^{(1)} = - T_o h_{ab} = K_{ab}^{(2)} \,. 
\end{equation}
Specifically, for instance, the solutions of junction conditions in global AdS$_3$ simplifies to
the hypersurfaces parametrized as (see \eg \cite{Kawamoto:2023wzj,Akal:2021foz,Akal:2020wfl}):
\begin{equation}\label{eq:CMCglobal}
	(A+E) \cos T + D \sin T +   \frac{B \ell}{\sqrt{R^2+ \ell^2 }}+   \frac{R}{\sqrt{R^2+ \ell^2}} \(  C \sin \Phi + (E-A) \cos \Phi   \)  =0\,,
\end{equation}
where constants $(A, B, C, D, E)$ represent real parameters of the brane profiles. The sign depends on which side of the brane is the choice of our bulk region. It is straightforward to check the extrinsic curvature of the hypersurfaces parametrized by eq.~\eqref{eq:CMCglobal} is the following
\begin{equation}
	K_{ab} = \frac{  \pm B  \, h_{ab} }{\sqrt{|B^2 + C^2 -D^2 -4AE  |}}\frac{1}{\ell} \,, \, \text{with}\quad T_o \ell =   \frac{ \pm B}{\sqrt{|B^2 + C^2 -D^2 -4AE |}} \,. 
\end{equation}

To begin investigating generic time-dependent branes, one can explore solutions for specific backgrounds. For example, considering the case of the hypersurfaces parametrized by eq.\eqref{eq:CMCglobal}, one can observe a rich structure of brane profiles. Moreover, these branes can be studied in any AdS$_3$ spacetime since they are locally the same. For instance, the hypersurfaces given by eq.\eqref{eq:CMCglobal} in the BTZ metric \eqref{eq:BTZ} can be expressed as:
\begin{equation}\label{eq:CMCBTZ}
	A e^{- \frac{2 r_h \phi}{\ell}}  r + B e^{- \frac{ r_h \phi}{\ell}}  r_h + E r + e^{- \frac{ r_h \phi}{\ell}}  \sqrt{r^2 -r_h^2}  \(  C  \cosh \(  \frac{r_h t}{\ell^2} \)  - D  \sinh\(  \frac{r_h t}{\ell^2} \)    \) =0\,,
\end{equation}
where we only show the results related to the non-rotating BTZ black hole for simplicity. 

Despite the analytical challenges associated with time-dependent branes, exploring these solutions could offer valuable insights into gravitational collapse and black hole formation, especially in the context of double holography. Additionally, the generic dynamic brane solutions could serve as a starting point for investigating more complex configurations and dynamical scenarios within the framework of double holography.

\subsection*{Higher-dimensional Spacetime}

The analytical analysis presented so far is limited to the AdS$_3$ bulk spacetime and a two-dimensional braneworld where gravity is mostly topological (see the effective gravitational action of the braneworld in Appendix \ref{sec:brane}). The advantage of this limitation is that we have knowledge of all solutions to the bulk Einstein equations, \ie AdS$_3$ spacetime, and can explore all possible phases. It is natural to explore situations involving higher dimensional spacetime, although this requires more effort to solve the junction conditions and find all possible brane profiles.  Further, the key technical challenge here is that the branes will backreact on the bulk geometry in a nontrivial way, \ie beyond three dimensions, the bulk geometry is no longer locally just AdS$_{d+1}$. This question has been studied extensively in the context of black holes in Randall-Sundrum braneworlds  \cite{Emparan:1999wa,Emparan:1999fd,Emparan:2000fn,Kanti:2004nr,Majumdar:2005ba,Fitzpatrick:2006cd,Creek:2006je,Gregory:2008rf,Figueras:2011gd} -- see also a discussion for the present context in \cite{Grimaldi:2022suv}. 

However, one interesting observation is that the phase transitions shown in figure \ref{fig:simp1} appear to be universal across all dimensions, even though the brane profiles may become more complicated.  One fact about this universality is that the black hole on the codimension-one braneworld exists only at the intersection between the braneworld and the bulk black hole, at least for static black holes. This conclusion is easily proven, as shown below. 

Consider a higher dimensional static bulk spacetime with a metric given by 
\begin{equation}
	ds^2 = -F(r) dt^2 +\frac{dr^2}{G(r)} + r^2 d \Omega^2_{d-1} \,,
\end{equation}
where we can assume that there is a black hole at $r=r_h$ with $F(r_h)=0=G(r_h)$. The braneworld as a hypersurface can be parameterized as 
\begin{equation}
	\Phi = \Phi (r, \phi^i) \,,  
\end{equation} 
where we have assumed a static configuration, \ie no time dependence, and have chosen the Gaussian normal coordinate for the spatial directions. It is obvious to write the induced metric on the brane as 
\begin{equation}
	\begin{split}
		ds^2 \big|_{\mB} &= -F(r) dt^2 +\frac{dr^2}{G(r)} + r^2 \( d \Phi^2 +\gamma_{ij} d\phi^i d\phi^j \) \\
		&= -F(r) dt^2 + \( \frac{1}{G(r)} + r^2 \Phi'(r)^2 \)dr^2 + r^2 \gamma_{ij} d\phi^i d\phi^j \,. 
	\end{split}
\end{equation}
It is nothing but a d-dimensional spacetime whose metric can be recast as 
\begin{equation}
	ds^2 \big|_{\mB} = -f(r) dt^2 +\frac{dr^2}{g(r)} + r^2 \gamma_{ij} d\phi^i d\phi^j \,. 
\end{equation}
with 
\begin{equation}
	f(r)=F(r)\,, \qquad g(r) = \frac{G(r)}{1+ G(r) r^2 \Phi'(r)^2} \,.
\end{equation}
Obviously, there is an event horizon at $r=r_h$ with $f(r_h)=0=g(r_h)$. Furthermore, we can see that the brane black hole has the same temperature as the bulk black hole, since 
\begin{equation}
	T_{\mt{B}} = \frac{\sqrt{f'(r) g'(r)}}{4\pi} \bigg|_{r=r_h} = \frac{\sqrt{F'(r) G'(r)}}{4\pi} \bigg|_{r=r_h} \equiv T_{\mt{BH}} \,.
\end{equation}
Here we have assumed that the function $G(r)\Phi'(r)^2$ vanishes at $r=r_h$, \ie $\Phi'(r)$ is a regular function at $r=r_h$. Otherwise, the divergence $\Phi'(r_h) \to \infty$ indicates the critical case where the brane (without a black hole) is tangent to the bulk horizon. We can see that the dimension or even the sign of cosmological constant does not play any role here. More specific calculations related to the AdS$_3$ case, which is the main focus of this paper, are also shown in the Appendix \ref{sec:brane}.

Put the above argument the other way around. The fact is that the Killing vector $K^\mu$ of the static bulk spacetime is also the Killing vector of the braneworld $k^\alpha = h^\alpha_\mu K^\mu$ after projection onto the brane.  Noticing that the surface gravity $\kappa$ is defined by $K^\nu \nabla_\nu K^\mu = \kappa K^\mu$, we can show that the event horizon on the brane has the same surface gravity due to 
\begin{equation}
	k^\beta D_\beta k^\alpha = \kappa k^\alpha \,. 
\end{equation}

In summary, what we have illustrated above is that the event horizon of the braneworld is (must be) the result of the intersection of the braneworld and the higher dimensional event horizon in the bulk: 
\begin{equation}
	\text{\bf black hole} =	\text{\bf d-dim universe} \cap 	\text{Black Hole in a (d+1)-Dim Bulk}
\end{equation}
This is the holographic interpretation of black hole formation on a d-dimensional braneworld. Considering the generally dynamic braneworld, the interesting picture is that the star on a braneworld collapses to a black hole because the brane universe collides with the black hole in the higher-dimensional bulk.

\subsection*{de Sitter or flat branes}
In Appendix \ref{sec:brane}, we delve into the braneworld geometries in different phases. A noteworthy feature is that the braneworld is consistently (asymptotically) AdS spacetime, as exemplified by eq.~\eqref{eq:braneRicci}, where the intrinsic curvature of the brane geometry is always negative. This characteristic provides a foundation for the double holography built upon two AdS/CFT correspondences. However, it is important to acknowledge that this is contingent on the specific choice of the brane tension satisfying $|T_o \ell| < 1$. The inclusion of more diverse braneworld types becomes possible by allowing the brane tension to extend beyond this range. See \eg \cite{Akal:2020wfl,Akal:2021dqt,Kawamoto:2023nki} for some explorations of other types of branes in AdS$_3$.

Taking the symmetric setup in $(d+1)$-dimensional AdS spacetime as an example, where $K_{ab} \propto h_{ab}$, the Gauss equation of the timelike brane establishes a connection between its intrinsic curvature and the tension, namely 
\begin{equation}
	R[h_{ij}] =  \frac{d}{d-1} (T_o)^2 - \frac{d(d-1)}{\ell^2} \,. 
\end{equation}
The nature of the braneworld—whether it is (asymptotically) AdS, flat, or de Sitter—depends on the regime of the brane tension. Specifically, the braneworld tends to be (asymptotically) AdS, flat, or de Sitter when $|T_o \ell| < d-1$, $|T_o \ell| = d-1$, or $|T_o \ell| > d-1$, respectively. This insight underscores the flexibility in extending the doubly holographic models. By considering $(d+1)$-dimensional AdS spacetime containing a $d$-dimensional asymptotically de Sitter or flat braneworld, we open avenues to explore de Sitter holography or flat holography on the braneworld.
Of course, these setups from the boundary perspective differ from the boundary CFT or defect CFT. One simple evidence is that the dS brane may not intersect with the conformal boundary and the corresponding ``boundary entropy" defined in eq.~\eqref{eq:defg} could even be complex \cite{Akal:2020wfl,Akal:2021dqt}. 
Intriguingly, there is a transition from the AdS/dS braneworld to the flat brane by finely adjusting the brane's tension to the critical case. However, achieving consistent results between the two flat limits of AdS brane and dS brane remains a puzzle.




\begin{acknowledgments}
	We are happy to thank Dongsheng Ge, Jonathan Harper, Yan Liu, Akihiro Miyata, Takato Mori, Tadashi Takayanagi and Zhencheng Wang for useful discussions. We would also like to thank the organizers and participants of the {\it Qubits on the Horizon 2} meeting (April 4-8, 2022) for providing the stimulating environment where this collaboration began. Research at Perimeter Institute is supported in part by the Government of Canada through the Department of Innovation, Science and Economic Development Canada and by the Province of Ontario through the Ministry of Colleges and Universities. RCM is supported in part by a Discovery Grant from the Natural Sciences and Engineering Research Council of Canada, and by funding from the BMO Financial Group.  RCM and SMR were supported by the Simons Foundation through the ``It from Qubit'' collaboration. SMR and TU are also supported by MEXT-JSPS Grant-in-Aid for Transformative Research Areas (A) ``Extreme Universe'', No. 21H05187. SMR is also supported by by JSPS KAKENHI Research Activity Start-up Grant Number JP22K20370.
\end{acknowledgments}

\appendix

\section{Aspects of Brane Perspective}\label{sec:brane}
%
\begin{figure}[h]
	\centering
	\includegraphics[width=3in]{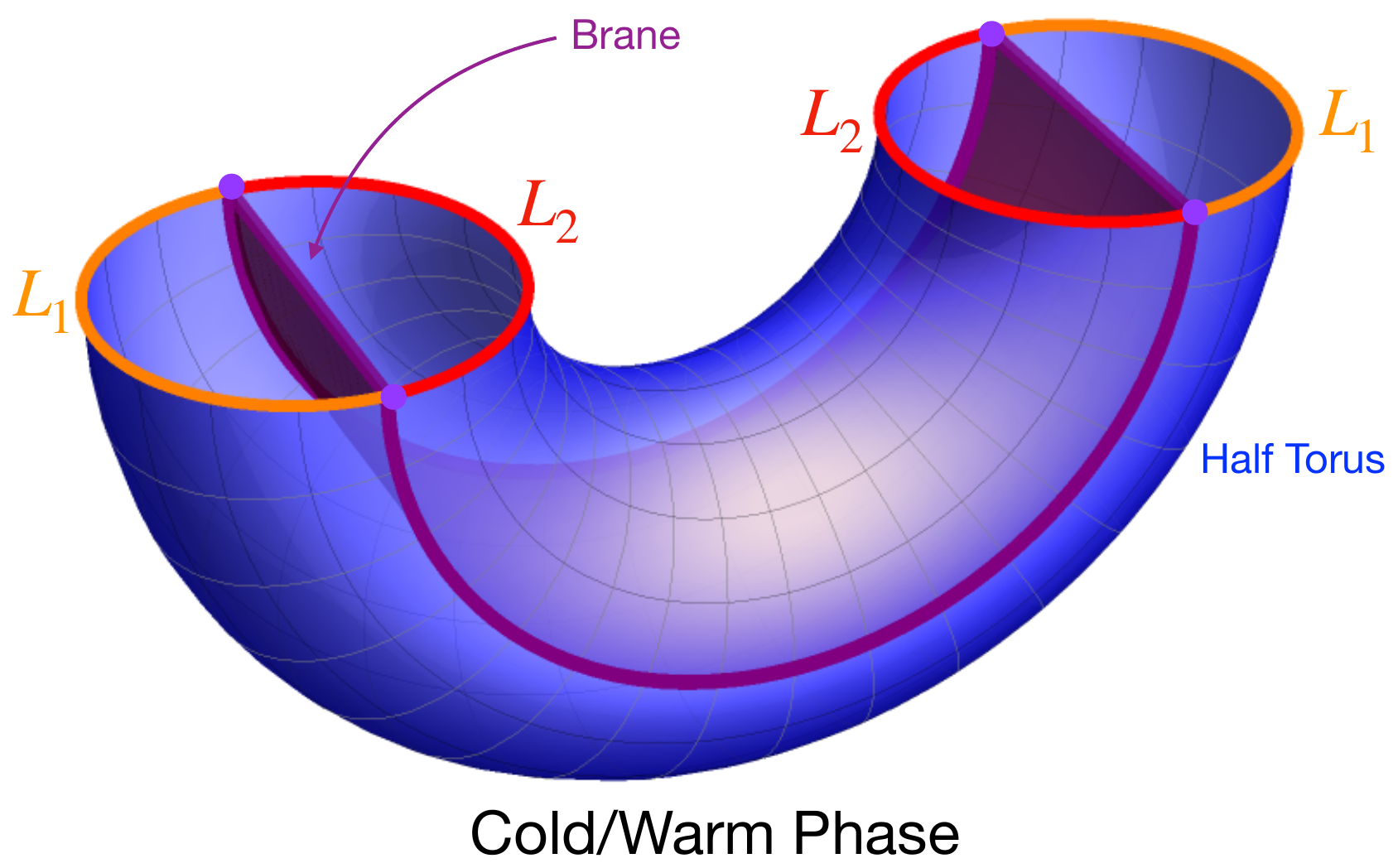}
	\includegraphics[width=2.7in]{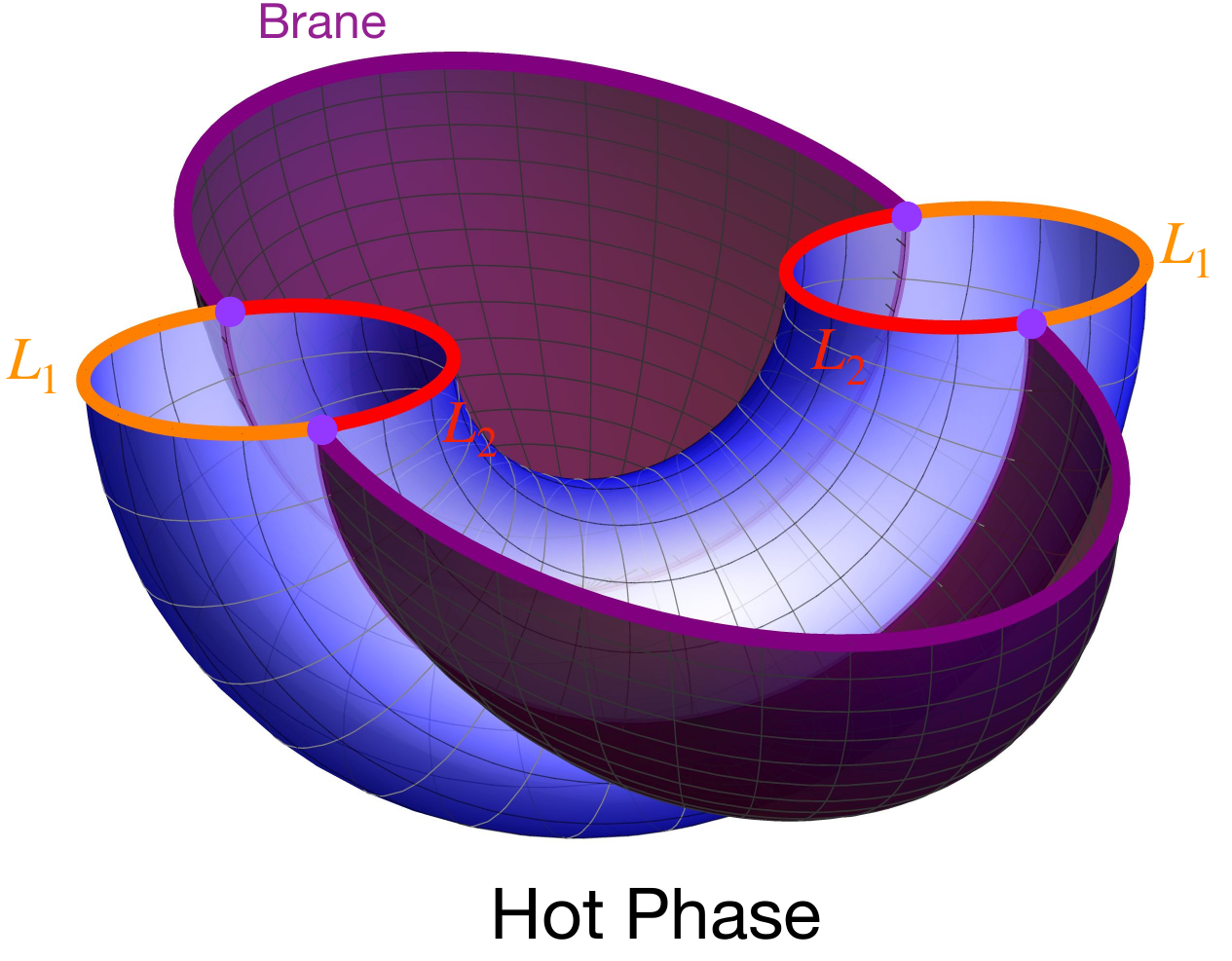}
	\caption{Sketch illustrating the Euclidean path integral preparation of states in cold, warm, and hot phases from the brane perspective.}\label{fig:prep1}
\end{figure}

In this section, we study the three distinct phases from the brane perspective, where CFTs on the left and right boundaries are coupled to gravitational theories residing on two distinct two-dimensional branes, denoted as $\mB_i$. The Euclidean path integral for preparing the states from the brane perspective is described by figure \ref{fig:prep1}. In particular, this figure illustrates the topology change of the brane between these phases. For example, if we double these geometries to evaluate the Euclidean action, the brane in the cold and warm phases has the topology of a single tube,
while in the hot phase, the brane forms two separate disks. In particular, the following analysis serves to illustrate that the braneworld is characterized by an asymptotically AdS$_2$ geometry during both the cold and warm phases. In these phases, each braneworld connects the two defects on one of the asymptotic AdS$_3$ boundaries, such as $D_{a}^{(\mt{L})}$ and $D_{b}^{(\mt{L})}$. As discussed below in section \ref{direction9}, the deformation of the AdS$_2$ geometry is produced by a `star' formed by CFT excitations that accumulate at the center of the brane. As the temperature increases, this star grows in density and shrinks in size. . As the temperature increases, this star grows in density and shrinks in size. However, when the bulk spacetime transitions to the hot phase, the braneworld collapses to form an eternal AdS$_2$ black hole spacetime. This black hole geometry supports a wormhole bridging two defects on different boundaries, for example, spanning from $D_{a}^{(\mt{L})}$ to $D_{a}^{(\mt{R})}$, as shown in figure \ref{fig:prep1}. 

\subsection{Phase transitions on the braneworld}
Beginning with the bulk AdS$_3$ spacetimes \eqref{eq:AdS3} with two mass parameters $m_1, m_2$,  the induced metric on the brane jointing two bulk regions $\mS_1, \mS_2$ is derived as
\begin{equation}\label{eq:ugly}
	\begin{split}
			ds^2 \big|_{\mB} 
		&=-\sigma dt^2 +  \frac{ 4 T_o^2 \ell^4  \, d\sigma^2}{  16  (T_o \ell)^2 (1-(T_o \ell)^2  )\sigma^2   + 8 (m_1 + m_2)(T_o \ell)^2 \sigma -(m_1 -m_2)^2 }   \,,\\
		&= -\sigma dt^2  + \frac{\ellB^2 d\sigma^2}{ 4 (\sigma - \sigma_+)(\sigma-\sigma_-)}\,,
	\end{split}
\end{equation}
where the brane coordinate $\sigma$ is related to the AdS radial direction through $\sigma = \(  \frac{r_i^2}{\ell^2}  - m_i  \) $ according to our gauge choice \eqref{eq:gauge}. Additionally, we have defined the brane length scale $\ellB$, \viz 
\begin{equation}\label{eq:Lbrane}
	\ellB = \frac{\ell}{\sqrt{1- (T_o \ell)^2}} \,.
\end{equation}
Indeed, the induced metric \reef{eq:ugly}  is singular at $\sigma=\sigma_\pm$, in terms of the coordinates $(t, \sigma)$, where $\sigma_\pm$ are defined by eq.~\eqref{eq:sigmapm}, namely
\begin{equation}
	\{ \sigma_+, \sigma_-\} = \frac{\ellB^2}{4\ell^2}  \(  -(m_1+m_2) \pm  \sqrt{ \(\frac{m_1-m_2}{T_o \ell} \)^2+ 4m_1 m_2   }\)\,.
\end{equation}
Since $\sigma_+\ge 0$ and $\sigma_- \le 0$ when $T_o \ell \le 1$, the midpoint on the brane at $\sigma=\sigma_+$ corresponds to the turning point. However, it is important to note that this is only a coordinate singularity. That is, the metric described in terms of either $(t,r)$ or $(t,\sigma)$ coordinate patches only covers half of the brane from the turning point out to the asymptotic boundary.

\begin{figure}[t]
	\centering
	\includegraphics[width=3.5in]{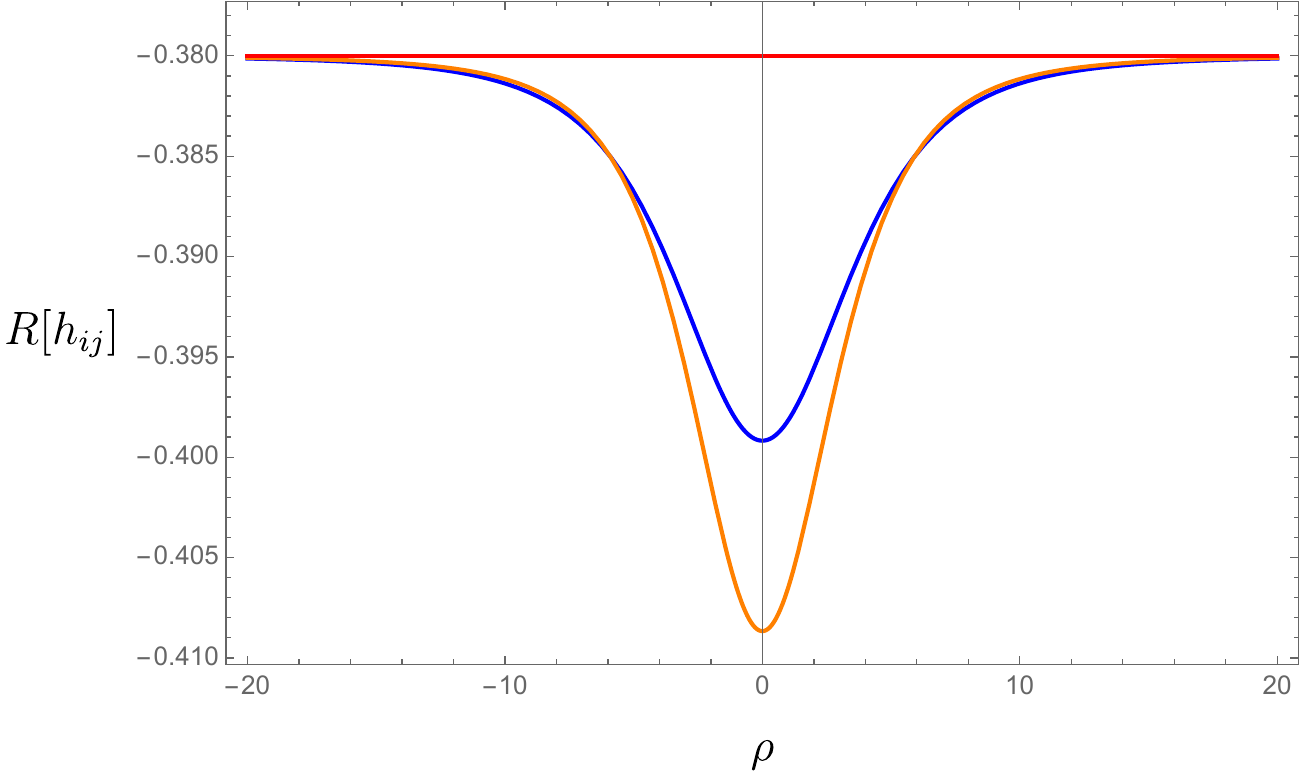}
	\caption{Ricci scalar of the induced geometry on the brane with fixing $T_o \ell=\frac{9}{10},  L_1/L_2=30$. The blue, orange and red curves are examples in the cold, warm and hot phases, respectively. The red curve is simply a constant with $R=-2/\ellB^2$, which matches our finding in eq.~\reef{eq:braneglobal} that the braneworld geometry is precisely that of AdS$_2$.}\label{fig:RicciScalar}
\end{figure}

It is straightforward to ameliorate this situation by introducing a new radial coordinate $\rho$ on the brane, defined as
\begin{equation}
	\rho^2 =\frac{r_i^2}{\ell^2}  - m_i - \sigma_+= \sigma - \sigma_+ \,. 
\end{equation}
In terms of this radial coordinate $\rho$, $\rho>0$ corresponds to one half of the brane profile, while $\rho<0$ represents the other half. The turning point of the brane profile in the bulk corresponds to the origin located at $\rho=0$. Importantly, this new coordinate $\rho$ results in a smooth brane geometry, \viz 
\begin{equation}
		ds^2 \big|_{\mB}  =  - ( \rho^2 + \sigma_+) dt^2 + \frac{\ellB^2}{\rho^2 + \sigma_+ -\sigma_-} d\rho^2  \,. 
		\label{grouper3}
\end{equation}
This expression reveals that the induced geometry asymptotically approaches AdS$_2$ with a curvature scale of $\ellB$. Furthermore, the metric remains well-behaved at the turning point: there are no coordinate singularities such as those seen in the $r$ or $\sigma$ coordinates. With this new radial coordinate $\rho$, the Ricci scalar of the brane geometry can be expressed as 
\begin{equation}\label{eq:braneRicci}
	\begin{split}
		R[h_{ij}] &=   - \frac{2}{\ellB^2}-  \frac{1}{8}  \(  \frac{m_1 -m_2}{ T_o ( r_i^2 - m_i  \ell^2 ) }   \)^2  = - \frac{2}{\ellB^2}  -  \frac{(m_1-m_2)^2}{ 8T_o^2 \ell^4 (\rho^2 +\sigma_+)^2} \, \\
	\end{split}
\end{equation}
which asymptotically approaches the constant $ - \frac{2}{\ellB^2} $ as one moves from the center to the boundary of the brane. That is, the braneworld geometry approaches that of AdS$_2$ in the asymptotic regions. We note that the Ricci scalar has a position-dependent profile of width $\Delta\rho\sim\sqrt{\sigma_+}$. This deformation is created by excitations in the braneworld CFT, as we discuss in section \ref{direction9}. We illustrate this behaviour with some numerical plots in figure \ref{fig:RicciScalar}.

\begin{figure}[t]
	\centering
	\includegraphics[width=3.5in]{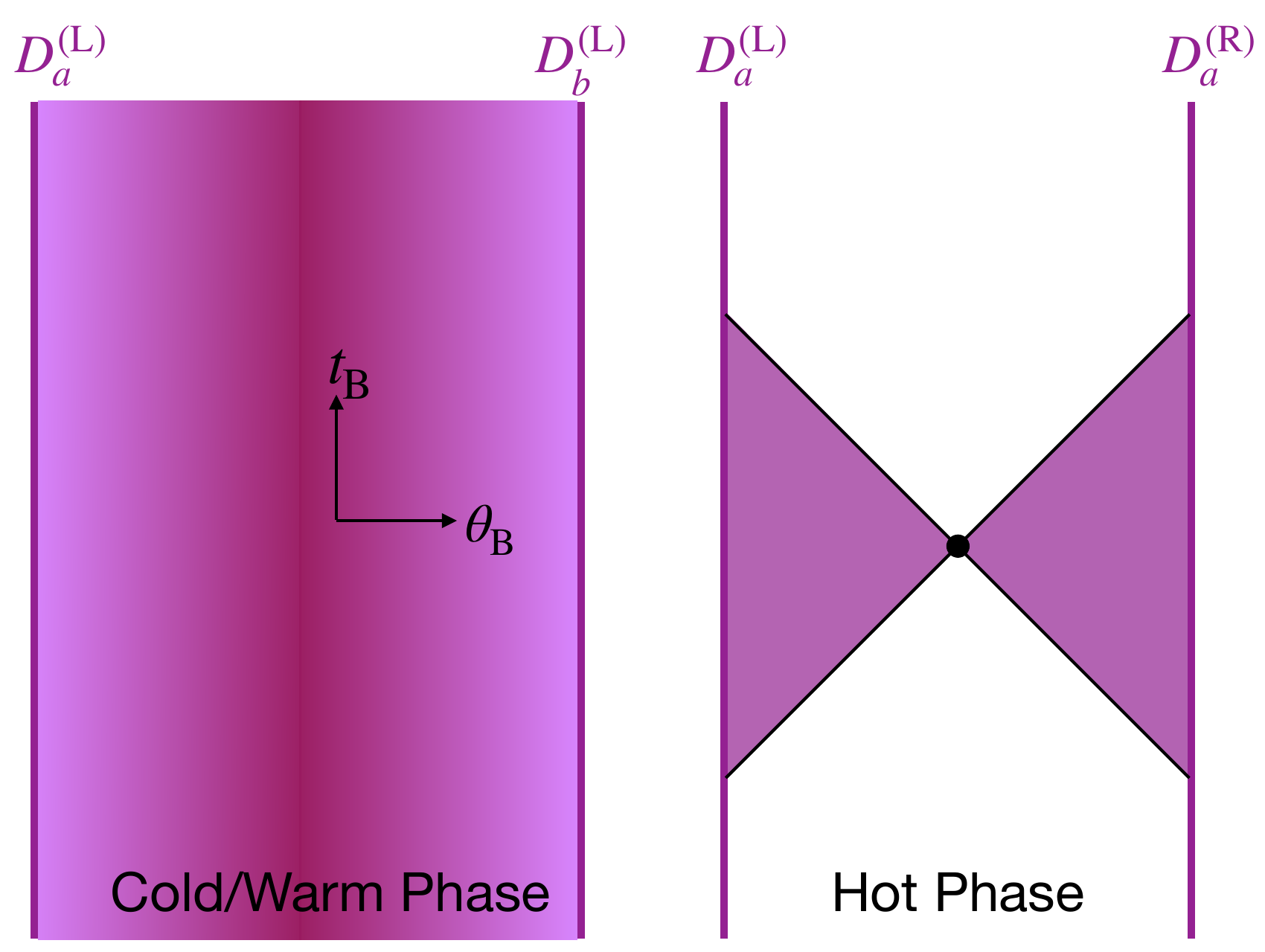}
	\caption{Phases transitions on the braneworld.}\label{fig:branesphases}
\end{figure}

Let us first continue with the analysis of the brane geometry in the cold and warm phases, where $\sigma_+ \ne 0$. The conformal metric of the corresponding brane geometry can be derived as follows:
\begin{equation}\label{eq:braneconformal}
	 ds^2 \big|_{\mB}  = \frac{ \ellB^2 (\rho^2(\theta_{\mt{B}}) + \sigma_+)}{ \sigma_+ -\sigma_-}  \(  - dt_{\mt B}^2 + d\theta_{\mt{B}} ^2 \)\,,
\end{equation}
where the conformal coordinates are given by \footnote{$ \mathbf{F}(z|m)$ is referred to as the incomplete elliptic integral of the first kind and defined by 
\begin{equation}
	 \mathbf{F}(z|m) : = \int^z_0  \frac{dt}{\sqrt{1-m \sin^2 t}}\,, \qquad 0<m<1\,.
\end{equation}
The complete elliptic integral of the first kind is then denoted by $ \mathbf{K}(z)$ with $ \mathbf{K}(m)= \mathbf{F}\(\frac{\pi}{2}|m\)$ and $\mathbf{K}(0)= \frac{\pi}{2}$. 
}
\begin{equation}\label{eq:conformalcoordinate}
	t_{\mt B} = \frac{\sqrt{\sigma_+-\sigma_-}}{\ellB}  t \,,  \quad \theta_{\mt{B}}=- i \, \mathbf{F}\( i \arcsinh\( \frac{\rho}{\sqrt{\sigma_+}} \) \bigg| \frac{\sigma_+}{\sigma_+ - \sigma_-} \)\,.
\end{equation}
The brane geometry is readily recognizable as an asymptotically AdS$_2$ spacetime, characterized by a radius $\ellB$. Taking the infinity $\rho$ limit, it becomes evident that the conformal boundary is located at
\begin{equation}
	 \lim\limits_{\rho \to \infty} \theta_{\mt{B}} =  \mathbf{K} 
	 \(\frac{\sigma_-}{\sigma_- - \sigma_+} \) \ge \frac{\pi}{2}\,.
	\end{equation}
with the inequality being saturated when $\sigma_-=0$. This observation implies that the Penrose diagram depicting the brane geometry in the cold/warm phase forms a wider strip compared to that of AdS$_2$, as visually depicted in figure \ref{fig:branesphases}.

Of particular note is the brane geometry associated with the symmetric boundary configuration where $L_1=L_2$ (or $m_1=m_2$).\footnote{Recall that in the symmetric configuration, we only have the cold phase (\ie the warm phase does not exist) for which $m_1,m_2<0$.} In this scenario, the geometry \reef{eq:braneconformal} simplifies to global AdS$_2$ whose Ricci scalar \reef{eq:braneRicci} reduces to 
\begin{equation}\label{eq:RicciScalar}
	R[h_{ij}]  =-  \frac{2}{\ellB^2}\,, \qquad \text{when} \qquad  m_1=m=m_2\,. 
\end{equation}
The brane coordinates $(t_{\mt{B}}, \theta_{\mt{B}})$ in this case correspond to the global AdS$_2$ coordinates in the conformal gauge. To elaborate further, the global coordinates are given by
\begin{equation}
	t_{\mt B} = \frac{\sqrt{-m}}{\ell}  t \,,  \quad \theta_{\mt{B}}= \arctan \(  \frac{\rho}{\sqrt{\sigma_+}}  \) \,, 
\end{equation}
which are obtained by substituting $\sigma_+= -\frac{m}{1- T_o^2 \ell^2}$ and $\sigma_-=0$ into the coordinate transformations presented in eq.~\eqref{eq:conformalcoordinate}. Consequently, the induced metric on the brane can be simplified to
\begin{equation}\label{eq:braneglobal}
	ds^2 \big|_{\mB}  =  \frac{\ellB^2}{\cos^2 \theta_{\mt{B}}}  \(  - dt_{\mt B}^2 + d\theta_{\mt{B}} ^2 \)\,,
\end{equation}
which clearly indicates that the brane geometry coincides with global AdS$_2$. Notably, the conformal boundary of this geometry is positioned at $\theta_{\mt{B}} = \pm \frac{\pi}{2}$, \ie the locations of the two defects on the conformal boundary of AdS$_3$.

Another way to verify this conclusion is by employing the explicit brane profile in the symmetric configuration, as given by eq.~\eqref{eq:globalbrane}. Substituting this brane profile into the bulk AdS$_3$ metric, we can derive the induced metric in terms of the $(t,r)$ coordinates as follows:
\begin{equation}\label{eq:branetr}
	ds^2 \big|_{\mB}= - \(   \frac{r^2 }{\ell^2}   - m  \)dt^2 +  \frac{  dr^2}{ (\frac{r^2}{\ell^2} - m ) - \frac{T_o^2 }{r^2}(r^2-m \ell^2 )^2 }\,, 
\end{equation} 
whose curvature is a negative constant given by eq.~\eqref{eq:RicciScalar}. To understand how the original $(t,r)$ coordinates cover the same region as the global coordinates $(t_{\mt{B}}, \theta_{\mt{B}})$, we can explicitly express the coordinate transformations between the two systems as 
\begin{equation}
	\frac{r}{ \sqrt{-m}\ell}  = \sqrt{  \frac{1}{(1- (T_o \ell)^2) \cos^2 \theta_{\mt{B}}}  -1 }    \ge \sqrt{ \frac{(T_o \ell)^2}{ 1- (T_o \ell)^2}}\,,
\end{equation}
or equivalently,
\begin{equation}
	\cos \theta_{\mt{B}} = \frac{1}{ 1- (T_o \ell)^2 } \frac{1}{\sqrt{ 1- \frac{r^2}{m \ell^2} }}  \in [0,1]\,.
\end{equation}
Here, it is important to note that the minimal radius is located at the midpoint of the brane. By comparing these transformations, it becomes evident that the induced metric \eqref{eq:branetr} is equivalent to eq.~\eqref{eq:braneglobal}. This demonstrates the consistency of the results and supports the conclusion that the brane geometry related to symmetric boundary corresponds to global AdS$_2$ with the conformal boundary at $\theta_{\mt{B}} = \pm \frac{\pi}{2}$.

In summary, we find that the brane geometry during the cold and warm phases corresponds to an asymptotically thermal AdS$_2$. Notably, the Euclidean time on the brane is compactified. Furthermore, in the case of symmetric configuration with $L_1=L_2$, the brane geometry simplifies to global AdS$_2$.

On the other hand, in the high-temperature regime, the bulk spacetime is dominated by the hot phase. A key characteristic of this phase is that the brane extends from the defects on the conformal boundary to the black hole horizon, as illustrated in figure \ref{fig:HotPhase}. This feature is explicitly reflected by the position of the middle point on the brane, \ie $\sigma=\sigma_+=0$, corresponding to $r=r_h$ due to $m_1=m_2=(2\pi T_{\mt{DCFT}}\ell)^2$. Considering the universal formula \eqref{eq:braneRicci}, we can immediately find that the brane geometry in the hot phase is always AdS$_2$. 
Furthermore, it is natural to expect that the brane also contains a two-dimensional black hole,  stemming from the dimensional reduction of the black hole interior of a three-dimensional BTZ black hole. Consequently, the intersection of the brane and BTZ black hole serves as the horizon of the two-dimensional brane black hole.

For a more detailed description, the brane profile during the hot phase, as given by eq.~\eqref{eq:BTZbrane} 
allows us to derive the induced metric as follows:
\begin{equation}\label{eq:branehot}
	ds^2 \big|_{\mB}= -\frac{r^2 -r_h^2}{\ell^2} dt^2 +  \frac{ \ell^2 dr^2}{ (r^2 -r_h^2) - \frac{T_o^2 \ell^2}{r^2}(r^2 -r_h^2)^2 }\,. 
\end{equation}
Here, the location of the black hole horizon is identified as $r=r_h$, \ie the horizon of the BTZ black hole. It is evident that the curvature of the two-dimensional brane during the hot phase reduces to the constant defined in eq.~\eqref{eq:RicciScalar}. The periodicity of the Euclidean time on the brane remains unchanged and matches that of the BTZ geometry. This is owing to the fact the holographic bulk spacetime in the hot phase corresponds to an equilibrium state. 

The geometry of the brane, as defined by the metric \eqref{eq:branehot}, accommodates a static two-dimensional black hole. The temperature of this black hole reads\footnote{Indeed, the conclusion that black holes on both the brane and in the bulk share the same temperature is quite general and robust. This outcome holds true for any static brane profile that intersects the horizon of the bulk black hole, \eg see \cite{Chen:2020uac,Chen:2020hmv}. This consistency in temperatures between the two perspectives is naturally anticipated due to the nature of static configurations.}
\begin{equation}
	T_{\mt{B}} = \frac{\sqrt{-\partial_r h_{tt}\,\partial_r h^{rr}}}{4\pi} \bigg|_{r=r_h} = \frac{ r_h}{2\pi \ell^2 }   = T_{\mt{BH}} = T_{\mt{DCFT}}\,, 
\end{equation}
which is determined by the periodicity of the Euclidean $t$ coordinate.

To further elucidate, the comparison with the induced metric during the cold phase prompts us to reformulate eq.~\eqref{eq:branehot} in the AdS-Rindler form, \ie 
\begin{equation}\label{eq:AdS2Rindler}
	ds^2 \big|_{\mB}= -  \rho^2 dt^2 + \frac{\ellB^2}{\rho^2  -\sigma_-} d\rho^2 = \ellB^2 \(\frac{\sqrt{m}}{\sinh(\sqrt{m} x_{\mt{B}} \ ) }\)^2 \(  -dt_{\mt{B}}^2 + dx_{\mt{B}}^2   \) \,.
\end{equation}
The dimensionless coordinates $(t_{\mt{B}}, x_{\mt{B}})$ are defined by 
\begin{equation}
	t_{\mt{B}} = \frac{t}{\ell} \,, \quad  x_{\mt{B}} = \frac{  \text{arccoth} \(  \sqrt{1- \frac{\rho^2}{\sigma_-}}  \)  }{\sqrt{m}} = \frac{\ell}{r_h} \arccoth \(   \frac{  \sqrt{ r_h^2 \ellB^2  + (r^2 -r_h^2 )\ell^2 } }{r_h \ellB} \)\,.
\end{equation}
It is important to note that this coordinate system only covers one-sided black hole exteriors, as depicted in figure \ref{fig:branesphases}. In the global coordinates, the two-sided AdS$_2$ black hole connects the two defects on the left and right boundaries. For instance, the constant time slice $t=0$ provides a visualization from the brane perspective, as illustrated by the path integral in figure \ref{fig:prep1}.

In conclusion, the braneworld in the hot phase is characterized by an eternal AdS$_2$ black hole, with a temperature identical to that of the BTZ black hole in the bulk or DCFT on the boundary. Looking from the brane perspective, the transition between the cold/warm phase and the hot phase corresponds to the Hawking-Page transition, where the geometry transforms from an (asymptotically) thermal AdS$_2$ state to an AdS$_2$ black hole configuration. 

We finally remark that the transition on the brane is not the standard Hawking-Page transition for Einstein gravity in AdS$_{d+1}$ spacetime. For the $(d+1)-$dimensional AdS-Schwarzschild black hole, the black hole temperature is given by $T_{\mt{BH}} = \frac{d r_h^2 + (d-2) \ell^2}{4\pi \ell^2 r_h}$ and the corresponding critical temperature associated with the Hawking-Page transition is shown as
\begin{equation}
	T_{\mt{HP}} = \frac{d-1}{2\pi \ell} \,, \qquad \text{with} \qquad    r_h =\ell\,.
\end{equation}
It is clear that this analysis will not apply to our two-dimensional braneworlds (with $d=1$).
The difference is due to the fact that the gravitational action on the brane is not given by the standard Einstein-Hilbert action. To determine the critical temperature, we can evaluate and compare the free energy for the entire two-dimensional system, \ie bath plus gravitational brane universe, for both phases. Of course, we have already derived the same results using the bulk perspective, \ie with the three-dimensional gravity coupled to a brane. In the following subsection \ref{sec:braneaction}, we instead compute the renormalized action from the brane perspective and show that it leads to consistent results from the bulk perspective.

\subsection{Black hole threshold on the braneworld}
The AdS$_2$ black hole on the braneworld could only exist when the corresponding bulk dual spacetime assumes the hot phase. As previously established, the phase transition between the hot and warm phase is determined by the size of the small interval as shown in eq.~\reef{eq:WHtrans}, \ie
\begin{equation}
	T_{\mt{DCFT}} L_2 \simeq  \frac{(\pi -k_2 )\( \sqrt{(\pi-k_2)(\pi -k_1-k_2) + 4 \arctanh^2 (T_o\ell)}  - 2\arctanh(T_o\ell)  \) }{ (\pi -k_1 - k_2)\pi}  \,. 
\end{equation}
Beyond this threshold, the hot phase predominates. Conversely, the hot phase cannot emerge if the size of the smaller interval, namely $T_{\mt{DCFT}} L_2$, falls below the critical size defined above. In those cases, the braneworld cannot give rise to a black hole. As we approach the fusion limit with $L_2 \to 0$, it becomes evident that the Casimir energy associated with interval $L_2$ dominates, and is characterized by 
\begin{equation}
	E_{\rm Casimir} \simeq - \frac{c }{6} \frac{(\pi - \sqrt{-m_2}\, \Delta\phi_2)^2}{\pi L_2}  \simeq      -\frac{c }{6} \frac{(\pi - k_2)^2}{\pi L_2}  \,. 
\end{equation}
The resultant attractive Casimir force acts as an impediment, restraining the boundary intervals from attaining a state of thermal equilibrium with AdS$_2$ black hole on the braneworld.

On the contrary, we remark that the braneworld would only support the black hole phase when the smaller interval $L_2$ is relatively large. More explicitly, a loose upper bound is given by the transition between the hot and cold phase at $L_1 =L_2$, \ie 
\begin{equation}
	T_{\mt{DCFT}} L_2 \ge  T_{\mt{CH}} L_2 =  \frac{1}{2\pi}\( \sqrt{ \pi^2 +4 \, \arctanh^2 \( T_o\ell \)   }  - 2\, \arctanh \( T_o\ell \)  \) \,,
\end{equation}
where the critical temperature $T_{\mt{CH}} $ was derived in eq.~\eqref{eq:TCH}. Beyond this dimensionless size $T_{\mt{DCFT}} L_2$, we always have a black hole on the two-dimensional braneworld.

\subsection{Brane stress tensors}\label{sec:stresstensor}
The standard procedure for evaluating the stress tensor in holographic CFTs involves considering a CFT coupled to a fixed background metric, \eg
\begin{equation}\label{num1}
	\int D\phi\ e^{i I(\phi, g^{(0)}_{ab}) }= e^{i I_{\rm gen}(g^{(0)}_{ab})} \,.
\end{equation}
This equation represents the path integral of the CFT degrees of freedom, resulting in a generating function $I_{\rm gen}(g^{(0)}_{ab})$ for the CFT stress tensor correlators, as denoted by:
\begin{equation}\label{num2}
	\langle T^{ab}\rangle = \frac{2}{\sqrt{-g^{(0)}}}\, \frac{\delta I_{\rm gen}}{\delta
		g^{(0)}_{ab}}\,.
\end{equation}
Nevertheless, the generating function is notably non-local with respect to the background metric $g^{(0)}_{ab}$. Moreover, the result typically exhibits ultraviolet (UV) divergences, with the stress tensor correlation function displaying behaviour like $\langle T^{ab}\rangle \sim 1/\delta^d$, where $\delta$ represents a short-distance cutoff. To remedy this issue, appropriate counterterms, which are local functionals of the metric, are subtracted, \ie
\begin{equation}\label{num3}
	\int D\phi\ e^{i \[I(\phi, g^{(0)}_{ab})- I_{\rm ct}(g^{(0)}_{ab})\]}= e^{i \[I_{\rm gen}(g^{(0)}_{ab})- I_{\rm ct}(g^{(0)}_{ab})\]} = 
	e^{i I'_{\rm gen}(g^{(0)}_{ab})} \,.	
\end{equation}
The expectation value of the`renormalized' stress tensor is properly given by 
\begin{equation}\label{num4}
	\langle T^{ab}\rangle = \frac{2}{\sqrt{-g^{(0)}}}\,\lim_{\delta\to0} \frac{\delta I'_{\rm gen}}{\delta
		g^{(0)}_{ab}}= \frac{2}{\sqrt{-g^{(0)}}}\,\lim_{\delta\to0} \frac{\delta \ }{\delta
		g^{(0)}_{ab}}\[I_{\rm gen}- I_{\rm ct}\]\,.
\end{equation}
In the context of a holographic CFT, the path integral over the boundary CFT degrees of freedom can be replaced with the path integral over the bulk degrees of freedom, particularly the bulk metric. As indicated by the aforementioned notation, this path integral maintains the non-normalizable mode of the metric, \ie the boundary metric $g^{(0)}_{ab}$, fixed. Furthermore, following the standard assumption, it is assumed that the latter path integral is evaluated in a saddle point approximation. 

We will now discuss the doubly holographic models as described in \cite{Chen:2020uac,Chen:2020hmv} which involve integrating out the bulk theory on both sides of the brane to produce two boundary CFTs and a dynamical theory of gravity on the brane. In terms of the boundary theory, the corresponding path integral can be expressed as 
\begin{equation}\label{raz1}
	\begin{split}
			\int D\tg\,\int D\phi_\mt{L}\,D\phi_\mt{R} e^{i \[ I_{\mB}(\tg_{ab})+\sum_i I(\phi_i, \tg_{ab}) \]} &= e^{i \[ I_{\mB}(\tg_{ab})+\sum_i I_{{\rm{gen}},i}(\tg_{ab})\]}
	\\
		&=\int D\tg\, e^{i \[ I_{\mB}(\tg_{ab})+\sum_i I_{ct,i}(\tg_{ab})+\sum_i \(I_{{\rm gen},i}(\tg_{ab})-I_{{\rm{ct}},i}(\tg_{ab})\)\]}
		\\
		&=\int D\tg\,e^{i \[ I_{\rm grav}(\tg_{ab})+\sum_i I'_{{\rm gen},i}(\tg_{ab})\]} \,,
	\end{split}
\end{equation}
Here, $I_{\mB}$ denotes the intrinsic action on the brane, which typically takes the form of the worldvolume term:\footnote{However, we may also add various DGP terms, as described in \cite{Chen:2020uac,Chen:2020hmv}.}
\begin{equation}\label{bran1}
	I_{\mB}=- \frac{T_o}{4 \pi \GN }\,\int d^dx\,\sqrt{-\tg}\,. 
\end{equation}
We note that the expressions are written in terms of the induced metric on the brane $\tg_{ab}$, rather than the non-normalizable mode $g^{(0)}_{ab}$ which appears above in the standard holographic discussion. Moreover, we assume that the brane tension is tuned to be near the critical value, which yields a small UV cutoff $\delta$ and puts us in a regime where the gravitational theory on the brane is well approximated by Einstein gravity. Of course, in contrast to the above discussion, the cutoff $\delta$ remains finite throughout.

As previously discussed, the `divergences' that arise from integrating out the boundary CFTs or the bulk metrics on either side of the brane are what lead to the gravitational action on the brane. Specifically, the gravitational brane action can be expressed as
\begin{equation}\label{grav1}
	I_{\rm grav}=I_{\mB} + \sum_i I_{{\rm{ct}},i}\,,
\end{equation}
where we have separated the contributions from the two CFTs. However, it should be noted that these UV divergences are all local functionals of the brane metric $\tg_{ab}$ and therefore will be identical for both theories.\footnote{While this conclusion is implied by the fact that the bulk theory, including the cosmological constant, is the same on both sides of the brane, if we were to consider cases where the bulk theories on either side of the brane were different, the nature of the two boundary CFTs would also change. In such a scenario, while the form of the divergences would remain the same, the precise coefficients of the individual terms may not be identical.} We have introduced the counterterm action as an accounting device to rearrange the contributions in an interpretable way. In particular, apart from the gravitational action \reef{grav1}, we have identified the generating functions
\begin{equation}\label{CFT1}
	I'_{{\rm{gen}},i}=I_{{\rm{gen}},i}-I_{{\rm{ct}},i}\,.
\end{equation}
which are identical to the expressions derived in the standard holographic approach. 
We can further interpret 
\begin{equation}\label{CFT2}
	\langle T^{ab}_{(i)} \rangle= \frac{2}{\sqrt{-\tg}}\, \frac{\delta I'_{{\rm{gen}},i}}{\delta
		\tg_{ab}}
\end{equation}
as the expectation value of the CFT with a finite cutoff. These expectation values serve as sources in the gravitational equations of motion on the brane.

Let us turn our attention to the situation outlined in the main text, specifically, the case with $d=2$ and a brane metric $h_{ab}$. In this scenario, the counterterm action comprises a single term, as expressed in the following equation:
\begin{equation}\label{count1}
	I_{{\rm{ct}},i}=\frac{1}{8\pi\GN\,\ell}\int d^2x \,\sqrt{-h} =
	\frac{1}{16\pi G_\mt{eff}}\int d^2x\,\sqrt{-h} \,\frac2{\ell^2}  \,,
\end{equation}
where $\GN$ and $G_\mt{eff}=\GN /\ell$ represent the Newton's constants for the bulk and brane theories \cite{Chen:2020uac,Chen:2020hmv,Grimaldi:2022suv}.\footnote{The Einstein term on the brane is topological and hence the definition of $G_\mt{eff}$ is a convention, \eg see \cite{Grimaldi:2022suv}.} The term mentioned combines with the tension term to produce the effective cosmological constant term
\begin{equation}\label{const2}
	I_{\rm grav}= \frac{1}{4\pi\GN}\int d^2x\,\sqrt{-\tg} \(\frac{1}{\ell}-T_o\) = 	\frac{1}{16\pi G_\mt{eff}}\int d^2x\,\sqrt{-h}\,\frac{2}{\ell_\mt{eff}^2}\,, 
\end{equation}
leading to the gravitational equation of motion \viz 
\begin{equation}\label{eom1}
	\begin{split}
		0&=\frac1{8\pi G_\mt{eff}}\,\frac{1}{\ell_\mt{eff}^2} \,h_{ab}+\langle T_{ab}^{(1)}   \rangle+
		\langle T_{ab}^{(2)}\rangle  \\
		&=\frac{1}{4\pi \GN}\(\frac{1}{\ell}-T_o\)h_{ab}+ \langle T_{ab}^{(1)}   \rangle+
		\langle T_{ab}^{(2)}\rangle\,.
	\end{split}
\end{equation}
The contribution of the stress tensors here corresponds to the contribution coming from finite terms, \ie the $R\log R$ term and higher powers of the curvature \cite{Chen:2020uac,Chen:2020hmv,Grimaldi:2022suv}.

On the other hand, the total action on the brane takes a simple form: 
\begin{equation}\label{simp2}
	I_{\rm brane}=\frac{1}{8\pi\GN} \int d^2x\,\sqrt{-h}\(K^{(1)}+K^{(2)}-2T_o\)\,,
\end{equation}
and the corresponding boundary equation of motion corresponds to the Israel junction condition
\begin{equation}\label{eom2}
	0=\frac{1}{8\pi\GN} \(K_{ab}^{(1)} + K_{ab}^{(2)} -  \( K^{(1)}+ K^{(2)}- 2 T_o \) h_{ab} \) \,.
\end{equation}
Comparing eqs.~\reef{eom1} and \reef{eom2}, we can arrive at the following identification:
\begin{equation}\label{stress9}
	\langle T_{ab}^{(i)}  \rangle=-\frac{1}{8\pi\GN}\(K^{(i)}_{ab}- K^{(i)} h_{ab}+\frac{1}{\ell}h_{ab}\)\,.
\end{equation}
as elaborated in \cite{Kawamoto:2023wzj}. If we take the $\delta\to0$ limit, the brane stress tensor \eqref{stress9} reduces to the quasi-local stress tensor \cite{Balasubramanian:1999re} \eqref{stress8} appearing in the standard holographic setup, \viz 
\begin{equation}\label{stress8}
	\langle T^{ab}\rangle= \frac{2}{\sqrt{-h}}\frac{\delta I_{\rm gen}'}{\delta \tg_{ab}} =-\frac{1}{8\pi\GN}\(K^{ab}-K\,\tg^{ab}\) + \frac{2}{\sqrt{-h}}\frac{\delta I_{\rm ct}}{\delta h_{ab}} \,.
\end{equation}

\subsection{Brane stress tensors in different phases} \label{direction9}
Focusing on the doubly holographic models considered in the main text, we explicitly calculate the corresponding brane stress tensor \eqref{stress9} in different phases in the following. With using the explicit profiles of the codimension-one brane, we can also obtain the extrinsic curvature of the brane, \eg 
\beqa
K_{ab}^{(i)}= \begin{pmatrix}
	K_{tt} & 0\\ 
	0 & K_{{r_i}{r_i}}
\end{pmatrix} &=& 
 \begin{pmatrix}
	  \( T_o + \frac{(-)^i(m_2-m_1)}{4 T_o \(  r_i^2  - m_i \ell^2 \) } \) h_{tt} & 0\\ 
	0 &   \( T_o - \frac{(-)^i(m_2-m_1)}{4 T_o \(  r_i^2  - m_i \ell^2 \) } \) h_{{r_i}{r_i}}
\end{pmatrix}\,,
\eeqa
in $(t,r_i)$ coordinates. Similarly, we have
\begin{equation}
	K_{ab}^{(i)}= \begin{pmatrix}
		K_{tt} & 0\\ 
		0 & K_{\rho\rho}
	\end{pmatrix} =  
 \begin{pmatrix}
	 \( T_o + \frac{(-)^i(m_2-m_1)}{4 T_o \ell^2 (\rho^2 +\sigma_+)} \) h_{tt} & 0\\ 
	0 & \( T_o - \frac{(-)^i(m_2-m_1)}{4 T_o \ell^2 (\rho^2 +\sigma_+) } \) h_{\rho\rho}
\end{pmatrix}  
\end{equation}
in $(t,\rho)$ coordinates. It is easy to see that the brane is a CMC slice \cite{Kawamoto:2023wzj} with 
\begin{equation}
	K^{(i)} \equiv K^{(i)}_{ab}h^{ab} = 2 T_o = \text{constant}\,.
\end{equation}
However, as we show below, the holographic stress tensor on the brane would only be proportional to the induced metric, \ie $K^{(i)}_{ab} \propto h_{ab}$ when we have $m_1=m_2$. However, we further note that $\Delta K_{ab}=K^{(1)}_{ab}+K^{(2)}_{ab}=2 T_o\,h_{ab}$, which we comment on below.

It is straightforward to evaluate the brane stress tensor and one obtains
\begin{equation}\label{eq:defineTab}
	\begin{split}
			T_{ab}^{(i)} = 
	\frac{1}{8 \pi \GN} \ 	\begin{pmatrix}
			\( T_o- \frac{1}{\ell} - \frac{(-)^i(m_2-m_1)}{4 T_o (\rho^2 +\sigma_+)} \) h_{tt} & 0\\ 
			0 &   \( T_o - \frac{1}{\ell}+\frac{(-)^i(m_2-m_1)}{4 T_o (\rho^2 +\sigma_+) } \) h_{\rho\rho}
		\end{pmatrix}\,,
	\end{split}
\end{equation} 	
whose trace is given by 
\begin{equation}\label{eq:traceT}
	\(T^{(i)}\)^a{}_a =  - \frac{1}{8 \pi \GN}  \(  \frac{2}{\ell}  - K \) =  \frac{1}{4 \pi \GN } \(   T_o- \frac{1}{\ell} \)\,.
\end{equation}
In the hot phase or a symmetric configuration, the stress tensor  reduces to  simply
\begin{equation}
	T_{ab}^{(i)} = -\frac{1-T_o \ell}{ 8 \pi \GN \ell} h_{ab}\,, \qquad \text{when} \qquad m_1=m_2\,,
\end{equation}
which becomes small in the large tension limit (\ie with $T_o \ell\sim1$).

In contrast to the standard CFT$_2$ result, the trace of the brane stress tensor in eq.~\eqref{eq:traceT} is not simply proportional to the Ricci curvature of the brane geometry derived in eq.~\eqref{eq:braneRicci}. As shown in \cite{Kawamoto:2023wzj}, the relation between $T^a{}_a$ and $R[h]$ is modified by the $T\bar{T}$ term as follows\footnote{Our definition of the brane stress tensor \eqref{eq:defineTab} corresponds to choosing a vanishing potential of the Liouville field. More explicitly, here, we can apply eq.(2.22) in \cite{Kawamoto:2023wzj} with taking $T^{(a)}= \frac{1}{\ell}$ and $\mu^{(a)}=0$. } 
\begin{equation}\label{eq:TTbar}
	\begin{split}
		 \(T^{(i)}\)^a{}_a&=  \frac{\ell}{ 16 \pi \GN }\(    R[h] + (8 \pi \GN)^2 \(   T^{(i) ab}\,T^{(i)}_{\,ab} -  \( T^{(i) a}{}_a \)^2     \)   \)\,,\\
		&=\frac{c}{ 24 \pi }\(    R[h] + \(\f{12\pi\ell}{c}\)^2 \(   T^{(i) ab}\,T^{(i)}_{\,ab} - \( T^{(i) a}{}_a \)^2    \)   \)\,. 
	\end{split}
\end{equation}
We note that the second term, \ie the $T\bar{T}$ term is suppressed in the larger tension limit due to the fact that it is quadratic (with including two traces).

More importantly, the above constraint equation \eqref{eq:TTbar} can be understood as the Einstein equation of the brane geometry. Of course, it is nothing but the Hamiltonian constraint (Gauss equation) of the timelike brane in AdS$_3$ bulk spacetime, namely 
\begin{equation}
	 R[h] = (K^{(i)})^2 -K^{(i) ab} K^{(i)}_{\,ab}  - \frac{2}{\ell^2} \,. 
\end{equation}
Using the explicit solutions of $T_{ij}$ and $R[h]$ derived before, one can easily verify the equality shown in eq.~\eqref{eq:TTbar}.

Examining the stress tensors in eq.~\reef{eq:defineTab}, we note the position dependent contributions with the profile $\frac{(m_2-m_1)}{4 T_o (\rho^2 +\sigma_+)}$. These excitations of the braneworld CFTs constitute the `star' to which we referred in earlier discussions and through the gravitational equation of motion \reef{eq:TTbar} deform the brane geometry. This profile has a peak value of ${(m_2-m_1)}/{(4 T_o \sigma_+)}$ at $\rho=0$, and a width of $\Delta\rho\sim\sqrt{\sigma_+}$. Hence the parameter $\sigma_+$, which determines the turning point of the brane in the bulk, also determines the key features of the CFT star in the cold and warm phases. We note from eq.~\reef{tab1} that this parameter is large in the large tension limit. Fixing the lengths $L_1$ and $L_2$ in the cold phase, this profile remains unchanged as the temperature varies in the cold phase. However, in the warm phase, it is straightforward to see that $\sigma_+$ decreases with increasing temperature in the warm phase. Hence one finds that the profile changes with a growing peak and a shrinking width as the temperature is increased in this phase. Hence, it is natural to think that this behaviour is a precursor to gravitational collapse, as discussed in section \ref{sec:discuss}.

Before concluding this section, we comment on a remarkable property of the gravitational equation of motion \reef{eq:TTbar}. First, we emphasize that the braneworld supports two copies of the boundary CFT, each with a UV cutoff and which weakly interact with each other, \eg see \cite{Chen:2020uac,Chen:2020hmv}. Of course, these are dual to the bulk geometries on either side of the brane. In eq.~\reef{eq:defineTab}, we evaluate the stress tensor associated with each CFT separately. An outstanding feature is that the position-dependent profile comes with an overall factor of $(-)^i$ where $i=1,2$ indicates whether the CFT is dual to the bulk region $\mS_1$ or $\mS_2$. This is quite remarkable as since the CFT star discussed above generally corresponds to a positive energy density in CFT$_1$ and a precisely matching negative energy density in CFT$_2$.\footnote{The signs correspond to the choice of the interval lengths $L_1>L_2$.} As a result, the total stress tensor on the brane is
\beq
T_{ab}^{(\rm tot)} = T_{ab}^{(1)}+ T_{ab}^{(2)} =
-\frac{1-T_o\ell}{4 \pi \GN \ell} \ h_{ab}\,.		
\label{direction8}
\eeq
That is, the total stress tensor is simply proportional to the induced metric on the brane. 

The feature of the gravitational equation of motion \reef{eq:TTbar} which we wish to highlight is that the geometry is determined by $T_{ab}^{(1)}$ or $T_{ab}^{(2)}$ separately, and not the total stress tensor \reef{direction8}. Further, even though these individual stress tensors differ (\ie $T_{ab}^{(1)}\ne T_{ab}^{(2)}$), they consistently yield the same result for the Ricci scalar. In fact, the present analysis only provides an explicit example of what was found as a general feature in \cite{Kawamoto:2023wzj}. For example, eq.~\reef{direction8} is a result that will hold in general. Of course, this feature of our two-dimensional gravity theory on the brane contrasts with the intuition that in higher dimensions, Einstein gravity couples to the total stress tensor, \ie to all forms of energy and momentum simultaneously.

\subsection{Renormalized action from the braneworld}\label{sec:braneaction}
The phase diagram is determined by evaluating the Euclidean action of each phase, which is obtained from the three-dimensional gravitational action given by
\begin{equation}\label{eq:3Daction02}
	I_{\rm tot} =I_{\rm{bulk}}+I_{\rm{brane}} +I_{\rm{boundary}}\,.
\end{equation}
The total action is derived from the bulk perspective in the main text. In accordance with the principle of double holography, we expect to obtain consistent results from both the brane perspective and the boundary perspective. In this appendix, we consider the simplest case, namely the symmetric setup with equal lengths $L_1=L_2$, from the brane perspective. Of course, we would reproduce the total action derived in the bulk perspective, namely 
	\begin{equation}
	\begin{split}
		I_{\mt{E}}^{\rm cold} &= -\frac{c\pi}{6} \frac{1}{T_{\mt{BH}} (L_1 +L_2)}  \,,\\
		I_{\mt{E}}^{\rm hot} &=- \frac{c }{6} \pi T_{\mt{BH}}  \(    L_1 +   L_2   \) - \frac{2c }{3}  \log \(   \sqrt{   \frac{ 1+T_o\ell}{1-T_o\ell}  }\)   \,,\\
	\end{split}
\end{equation}
for the cold and hot phases. 

The total (Euclidean) action on the brane perspective, denoted as $\mathcal{I}_{\mt E}$, can be decomposed into three distinct components: 
\begin{equation} \mathcal{I}_{\mt E} = \mathcal{I}_{\mt E}^{\mt{bdy CFT}} + \mathcal{I}_{\mt{E}}^{\rm brane \,gravity}+ \mathcal{I}_{\mt{E}}^{\rm brane\, CFT} \,, 
\end{equation} 
where each component corresponds to the contributions from the two-dimensional conformal field theory residing on the conformal boundary, the effective gravity on the brane, and the CFTs existing on the braneworld, respectively.

Let us consider the thermal conformal field theory residing on a circular domain with circumference $L_{\rm bdy}$ ($=L_1+L_2$ in the present context). The partition function of this thermal CFT can be obtained in either the low-temperature or high-temperature limit. The universal expressions for the thermal partition function are as follows:
 \begin{equation} 
 	- \ln Z_{\rm thermal}^{\mt{CFT}} =
 	 \begin{cases}  - \frac{c \pi}{6 } \frac{1}{T_{\mt{CFT}} L_{\rm bdy}} \,, \qquad  T_\mt{CFT} \to 0 \,, \\ \,\\  - \frac{c\pi}{6}  L_{\rm bdy} T_{\mt{CFT}} \,, \qquad T_\mt{CFT} \to \infty \,.
 	\end{cases} 
\end{equation}
The two expressions are related to each other by the modular transformation $	T_\mt{CFT} \leftrightarrow 1/(L_{\rm bdy}^2 T_\mt{CFT})$. 

The leading behaviour (at $\mathcal{O}(c)$ order) of the thermal partition function for CFT with large $c$ is universal, assuming the spectrum of light states is sparse, which is typically true for holographic theories. In the symmetric case, the DCFT reduces to the boundary conformal field theory on each side of the defect. Therefore, the partition function of the boundary CFT living on two boundary circles in the brane perspective can be identified as follows: 
	\begin{equation}
	\mathcal{I}_{\mt E}^{\mt{bdy CFT}} =  
	\begin{cases}
		- \frac{c \pi}{6} \frac{1}{T_{\mt{DCFT}} (L_1 +L_2)}  +\mathcal{O}(c^0)  \,, \qquad  \text{Cold phase} \,,  \\ 
		\,\\
		- \frac{c \pi}{6}  (L_1 +L_2)T_{\mt{DCFT}} +\mathcal{O}(c^0)  \,, \quad    \text{Hot phase} \,. \\ 
	\end{cases}
\end{equation}
On the contrary, the two-dimensional brane gravity theory is coupled to the boundary CFT and can be segregated into the gravitational and CFT sectors. To calculate the total induced action on the brane, one needs to conduct the radial integral of the bulk action. In situations where symmetry prevails, the total {\it unrenormalized} action of the brane is derived as
\begin{equation}
	\mathcal{I}_{\mt{E}}^{\mt{AdS}_2} \equiv \mathcal{I}_{\mt{E}}^{\rm brane \,gravity}+	\mathcal{I}_{\mt{E}}^{\rm brane\, CFT}  =    - \frac{\ell}{16 \pi \GN } \int_{\brane }   \log \( { \frac{1+ T_o \ell}{1-T_o\ell} } \)  \( - \frac{2}{\ellB^2}   \)  \sqrt{h} \, d^2y  \,, 
\end{equation}
where $h$ and $\ell_B$ represent the induced metric and AdS radius of the brane $\mathcal{B}$. Note that we have combined the contributions from two separate branes here. Intuitively, this expression looks like the standard Einstein-Hilbert term with $R[h]=  - \frac{2}{\ellB^2} $ and an effective Newton constant 
\begin{equation}
	\frac{1}{G_{\rm eff}} \sim 	\frac{\ell}{\GN}  \log \( { \frac{1+ T_o \ell}{1-T_o\ell} } \) \,.
\end{equation}
However, the corresponding equation of motion from Einstein-Hilbert action with an effective Newton constant would not be consistent after this simple identification. Since the only intrinsic geometric quantity for a two-dimensional braneworld is its Ricci scalar, we can generally rewrite the effective gravitational action on the braneworld as a $f(R)$ gravity. The total action $	\mathcal{I}^{\mt{AdS}_2}$ could be interpreted as the second-derivative scalar-tensor theory by introducing an extra scalar field $\Phi$. In particular, the special case with $d=2$ is nothing but the two-dimensional dilaton gravity without a kinematic term \footnote{The kinematic term can emerge after performing the Weyl transformation $h_{ij} \to \Omega^{2} h_{ij}$.}. The corresponding action of the dilaton gravity (or classically equivalent scalar-tensor theory) is given by 
\begin{equation}\label{jet2}
	\begin{gathered}
		I_{{\rm brane}}^{s t}=\frac{1}{16 \pi  G_{\rm brane} } \int d^{1+1} y \sqrt{-h}\left( f(\Phi)+f^{\prime}(\Phi)({R}[h_{ij}]-\Phi)\right) \\
		\quad+\frac{1}{8 \pi G_{\rm brane}  } \oint dy  \sqrt{-\gamma} K  f^{\prime}(\Phi) .
	\end{gathered}
\end{equation}
with the Newton constant on the brane chosen as
\begin{equation}
	\frac{1}{  G_{\rm brane} } = \frac{\ell}{\GN }
\end{equation}
and 
\begin{equation}\label{eq:fRgravity}
 f(\Phi) = \Phi \log \( \frac{\sqrt{4 + 2 \Phi \ell^2} + 2}{\sqrt{4 + 2 \Phi \ell^2} -  2} \) + \frac{1}{T_o\ell} \( \Phi + \frac{2}{\ellB^2} \) + \sum_{n>1} c_n  \( \Phi + \frac{2}{\ellB^2} \)^n  \,.
\end{equation}
The important fact here is that the coefficient in the linear term of $f(R)$ function has to be fixed as 
\begin{equation}
	c_1 = \frac{1}{T_o \ell} \,, 
\end{equation}
which needs to reproduce the consistent answer for the intrinsic curvature of the brane $R[h_{ij}]$. On the contrary, higher-order terms cannot be fixed since they do not change either the on-shell action or the solution of field equations. 

For example, we can derive the equation of motion with respect to the scalar field $\Phi$ as
\begin{equation}
	\Phi = R[h_{ij}]
\end{equation}
Varying the metric yields the field equations, \ie 
\begin{equation}
	0=f(R) - R f'(R)  =  \frac{R}{\sqrt{1+ \frac{\ell^2 R}{2}}} + \frac{2}{\ell^2} \(  \frac{1}{T_o\ell} - T_o\ell  \) \,, 
\end{equation}
which is an algebraic equation for $R$ . This algebraic equation has a negative solution 
\begin{equation}
	R =- \frac{2}{\ell^2} + 2 T_o^2=-  \frac{2}{\ellB^2} \,,
\end{equation}
as we expected for the braneworld. The coefficient associated with the boundary term is derived as 
\begin{equation}
	\begin{split}
		f'(\Phi) &= \log \( \frac{\sqrt{4 + 2 \Phi \ell^2} + 2}{\sqrt{4 + 2 \Phi \ell^2} -  2} \)  + \frac{1}{T_o\ell} - \sqrt{\frac{2}{2+ \Phi \ell^2}} = \log \( { \frac{1+T_o\ell}{1-  T_o \ell} }  \) \,, 
	\end{split}
\end{equation}
where we imposed the on-shell condition, \ie $\Phi=R =- 2 \frac{1- (T_o\ell)^2}{\ell^2} $.  Finally, let us also note that the effective $f(R)$ gravity on the brane with 
\footnote{To make an explicit connection to the higher curvature gravity shown in \cite{Chen:2020uac,Grimaldi:2022suv}, we note that the on-shell action (in Lorentzian signature) on the braneworld with the large tension limit ($T_o\ell \to 1$) reduces to
	\begin{equation}\label{sock99}
		\begin{split}
			\mathcal{I}^{\mt{AdS}_2}  &= \frac{\ell}{16\pi \GN }  \int_{\brane }  R \log \( \frac{\sqrt{4 + 2 R \ell^2} + 2}{\sqrt{4 + 2 R \ell^2} -  2} \) \sqrt{-h} d^2 y  \\
			&\simeq  \frac{\ell}{16\pi \GN }  \int_{\brane }   \(  \frac{2}{\ellB^2} + R   - R \log \(- \frac{R\ell^2}{8} \)   +  \frac{R^2 \ell^2}{4}  + \mathcal{O}(R^3) \)\sqrt{-h} d^2 y \,,
		\end{split}
	\end{equation}
	where we perform the series expansion with $R\ell^2\ll 1$ to obtain the second line and the first two terms $ \frac{2}{\ellB^2} + R$ are added by hand because they vanish for on-shell brane solutions. However, the choice of the first two terms is not unique, which would depend on higher-order terms. In the previous work \cite{Chen:2020uac,Grimaldi:2022suv}, the cosmological constant is chosen as  $\frac{2}{\ell_{\rm eff}^2} = \frac{4}{\ell^2} \( 1-  T_o\ell \)$,  resulting in 
		\begin{equation}\label{sock99}
			\mathcal{I}^{\mt{AdS}_2} \simeq  \frac{1}{16\pi G_{\rm brane} }  \int_{\brane }   \(  \frac{2}{\ell^2_{\rm eff}} + R   - R \log \(- \frac{R\ell^2}{8} \)   +  \frac{R^2 \ell^2}{8}  + \mathcal{O}(R^3) \)\sqrt{-h} d^2 y  \,. 
	\end{equation}
At this perturbative level (\eg $\mathcal{O}(R^2)$), these results are equivalent due to 
	\begin{equation}
	0=	\frac{2}{\ellB^2} + R   = \frac{2}{\ell^2_{\rm eff}} + R  - \frac{\ell^2}{2\ell^4_{\rm eff}} \approx \frac{2}{\ell^2_{\rm eff}} + R  - \frac{R^2\ell^2}{8} +\mathcal{O}(R^3) \,. 
\end{equation}
in the large tension limit. 
}
\begin{equation}
	 f(R) = R \, \log \( \frac{\sqrt{4 + 2 R \ell^2} + 2}{\sqrt{4 + 2 R \ell^2} -  2} \) + \frac{1}{T_o\ell} \( R+ \frac{2}{\ellB^2} \) + \sum_{n>1} c_n  \( \ R + \frac{2}{\ellB^2} \)^n  \,,
\end{equation}
or the equivalent scalar-tensor theory \eqref{jet2} indicates that the Wald entropy, \ie the area term appearing in the island formula should be given by 
\begin{equation}
``\text{Area term}"= \frac{f'(R)}{4 G_{\rm brane}} \bigg|_{R= - \frac{2}{\ellB^2}} = \frac{\ell}{ 4 \GN} \log \( { \frac{1+T_o\ell}{1-  T_o \ell} }  \) = \frac{c }{6}  \log \(   {   \frac{ 1+T_o\ell}{1-T_o\ell}  }\)  \equiv  S_{\rm def}\,,
\end{equation}
which is the defect entropy $S_{\rm def}$ \reef{eq:defg} of a single defect.

On the other hand, the induced action on the brane, denoted by $ \mathcal{I}_{\mt{E}}^{\mt{AdS}_2} $, is evidently divergent since the brane geometry is simply AdS$_2$. To obtain the renormalized action, a regularization scheme must be introduced for the braneworld theory. The regulator occurs naturally due to the bulk cut-off surface positioned at $r = \frac{\ell^2}{\delta}$. Similarly, the two-dimensional braneworld incorporates a cut-off surface located at $r = \frac{\ell^2}{\delta}$ on the brane $\mB$. Consequently, it is essential to consider the boundary term in eq.~\reef{jet2}, defined as
 \begin{equation}\label{jet3}
	\begin{split}
		\mathcal{I}_{\mt E}^{\mt{cut-off}} &=  -  \frac{\ell}{8 \pi \GN}  \log \( { \frac{1+T_o\ell}{1-  T_o \ell} }  \) \int_{\rm cut-off}  K  \sqrt{\gamma}  \, d\tau  \,.  \\
	\end{split}
\end{equation}

In the cold phase, we have demonstrated that the brane geometry is exclusively global AdS$_2$. The induced metric can be represented as 
\begin{equation} 
	ds^2 \big|_{\mB}= - \(   \frac{r^2 }{\ell^2}   - m  \)dt^2 +  \frac{  dr^2}{ (\frac{r^2}{\ell^2} - m ) - \frac{T_o^2 }{r^2}(r^2-m \ell^2 )^2 }\,, 
\end{equation} 
in which the minimal radius equals $r_{\rm min} = \sqrt{-m}\ell \sqrt{ \frac{(T_o \ell)^2}{ 1- (T_o \ell)^2}}$. It is simple to perform the radial integral and derive
\begin{equation}
	\mathcal{I}_{\mt{E}}^{\mt{AdS}_2}  =   \frac{\Delta \tau \ell}{4 \pi \GN}   \log \( \sqrt{ \frac{1+T_o\ell}{1-  T_o \ell} }  \)  \sqrt{ \frac{1}{\delta^2}   + T_o^2 \(  m- \frac{\ell^2}{\delta^2} \) }  \,. 
\end{equation}
with utilizing the maximal radius on the brane to be $r = \frac{\ell^2}{\delta}$.\footnote{In fact, the choice of the cutoff surface is not really important for the following calculations.} Of course, the Euclidean time circle is identified as the inverse temperature $\beta = \Delta \tau$. On the contrary, it can be observed that the contribution from the bulk is canceled out by the boundary term on the cut-off surface. By substituting the extrinsic curvature at the cutoff radius, which is represented as 
 \begin{equation}
 	K \sqrt{\gamma} =  \sqrt{  (\frac{r^2}{\ell^2} - m ) - \frac{T_o^2 }{r^2}(r^2-m \ell^2 )^2  } \, \partial_r \sqrt{\gamma} =  \sqrt{  \frac{r^2}{\ell^4} + T_o^2   \(m - \frac{r^2}{\ell^2}\)  } \,, 
 \end{equation}
we find 
\begin{equation}
	\mathcal{I}_{\mt E}^{\mt{cut-off}}  = -  \frac{\beta  \ell}{8 \pi \GN }  \log \( \sqrt{ \frac{1+T_o\ell}{1- T_o \ell} }  \) \sqrt{ \frac{1}{\delta^2}   + T_o^2 \(  m- \frac{\ell^2}{\delta^2} \) }  \times 2 = - 	\mathcal{I}_{\mt{E}}^{\mt{AdS}_2}   \,, 
\end{equation}
which gives rise to the vanishing renormalized action on the brane. We also emphasize that it is not necessary for this approach to add any additional counterterms on the brane since there are no extra divergences after we account for the boundary term $\mathcal{I}_\mt{E}^{\mt{cut-off}} $ on the cut-off surface. 

Once the critical temperature is surpassed, the bulk spacetime transitions to the hot phase in which the brane geometry is exemplified by an AdS$_2$ eternal black hole. The corresponding induced metric is derived as follows 
 \begin{equation}
 	ds^2 \big|_{\mB}= -\frac{r^2 -r_h^2}{\ell^2} dt^2 + \frac{ \ell^2 dr^2}{ (r^2 -r_h^2) - \frac{T_o^2 \ell^2}{r^2}(r^2 -r_h^2)^2 }\,. 
 \end{equation}
 By integrating the radius from the horizon at $r=r_h$ to the cut-off surface, we can reduce the Euclidean action of the braneworld to: 
\begin{equation}
	\mathcal{I}_{\mt{E}}^{\mt{AdS}_2}  =   \frac{\Delta \tau }{8 \pi \GN \ell }  \log \( \sqrt{ \frac{1+T_o\ell}{1- T_o \ell}  } \)  \(   \sqrt{  r_{\rm max}^2 - T_o^2\ell^2(r_{\rm max}^2 -r_h^2)   }  -r_h  \)  \times 2  \,. 
\end{equation}
Using the extrinsic curvature at a constant radius in the AdS$_2$ black hole background, specifically, 
\begin{equation}
	K   \sqrt{\gamma}=  \frac{\sqrt{ r^2 - T_o^2 \ell^2  \(r^2 - r_h^2\)  } }{\ell^2}  \,, 
\end{equation}
we immediately obtain the contribution of the boundary term: 
\begin{equation}
	\mathcal{I}_{\mt E}^{\mt{cut-off}}  = -  \frac{\beta }{4 \pi \GN  \ell }  \log \( \sqrt{ \frac{1+T_o\ell}{1- T_o \ell} }  \) \sqrt{ r_{\rm max}^2 - T_o^2 \ell^2  \(r^2_{\rm max} - r_h^2\)  }     \,. 
\end{equation}
Combining the two individual contributions yields the {\it renormalized gravitational action} on the braneworld in the hot phase, \ie 
\begin{equation}
	\mathcal{I}_{\mt{E}}^{\mt{AdS}_2} + 	\mathcal{I}_{\mt E}^{\mt{cut-off}} =  - \frac{\beta r_h }{4 \pi \GN \ell }  \log \( \sqrt{ \frac{1+T_o\ell}{1- T_o \ell}  } \)  \,. 
\end{equation}

As a summary, we conclude that the renormalized action from the brane perspective is expressed as
	\begin{equation}
	\mathcal{I}_{\mt E}^{\mt{bdy CFT}}  + 	\mathcal{I}_{\mt{E}}^{\mt{AdS}_2} + 	\mathcal{I}_{\mt E}^{\mt{cut-off}}  =
	\begin{cases}
		- \frac{c \pi}{6}\, \frac{1}{T_{\mt{DCFT}} (L_1 +L_2)} \,, \qquad\qquad  \text{cold phase} \,,  \\ 
		\,\\
		- \frac{c \pi}{6} \,{T_{\mt{DCFT}} (L_1 +L_2)}  	- \frac{c}{6}  \log \(   {   \frac{ 1+T_o\ell}{1-T_o\ell}  }\) \,, \quad    \text{hot phase} \,.
	\end{cases}
\end{equation}
At the leading order $\mathcal{O}(c)$, the summation of all contributions from the brane perspective agrees with the analogous Euclidean actions in eqs.~\reef{eq:coldaction} and \reef{eq:hotaction}, which were derived with three-dimensional gravitational integrals from the bulk perspective. Hence, the transition temperature found here for the symmetric configuration will agree with that found in eq.~\reef{eq:ColdHot} with $ T_{\text{DCFT}} (L_1 + L_2)=C_{\mt{CH}} $.


\section{Around the corner of the defect} \label{sec:corner}
%


In section \ref{sec:bulk}, we decomposed the gravitational action into various constituent components: the bulk term, the boundary term, and the brane term. However, we also noted that there were two additional boundary terms associated with the branes. The first is called the joint term or Hayward term \cite{Hayward:1993my,Brill:1994mb}, which may arise at the intersection between the chosen cut-off surface and the brane. The second is a boundary counterterm that is added as part of the brane action. 

\subsection{Boundary terms on the brane}
\begin{figure}[!]
	\centering
	\includegraphics[width=2.2in]{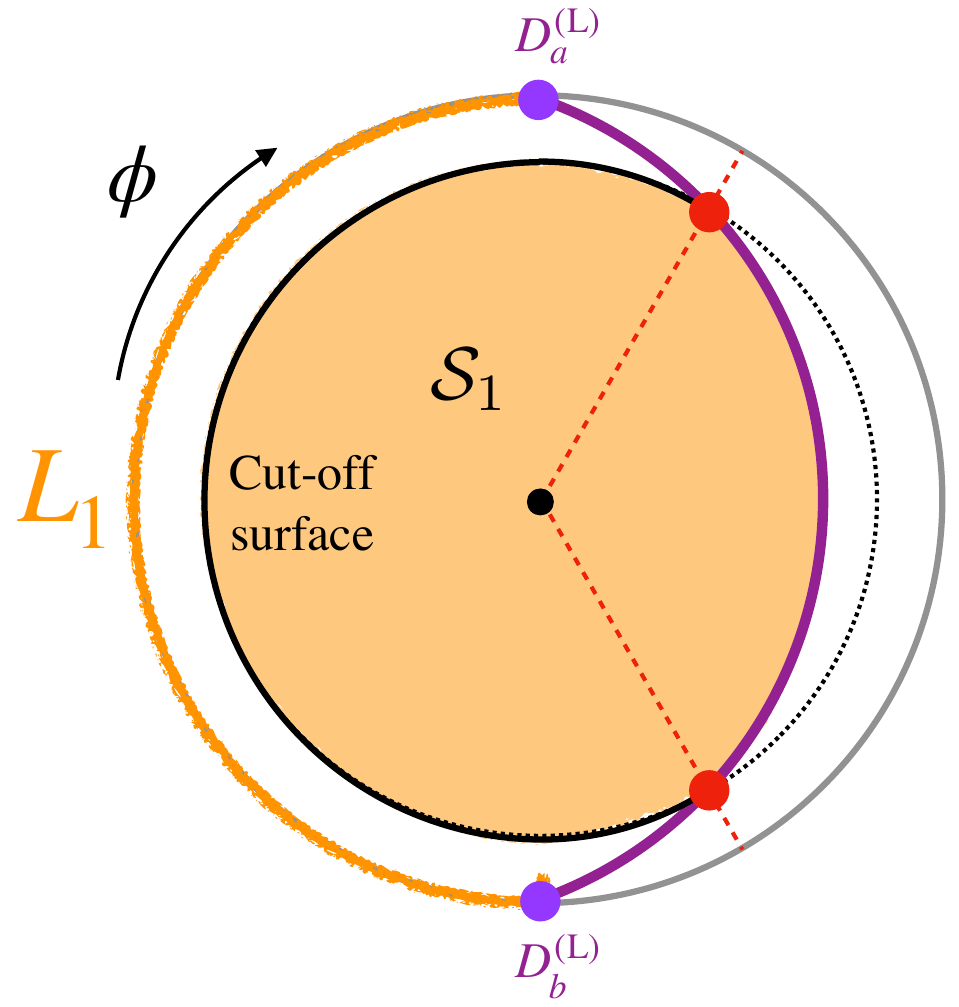}
	\quad
	\includegraphics[width=3.5in]{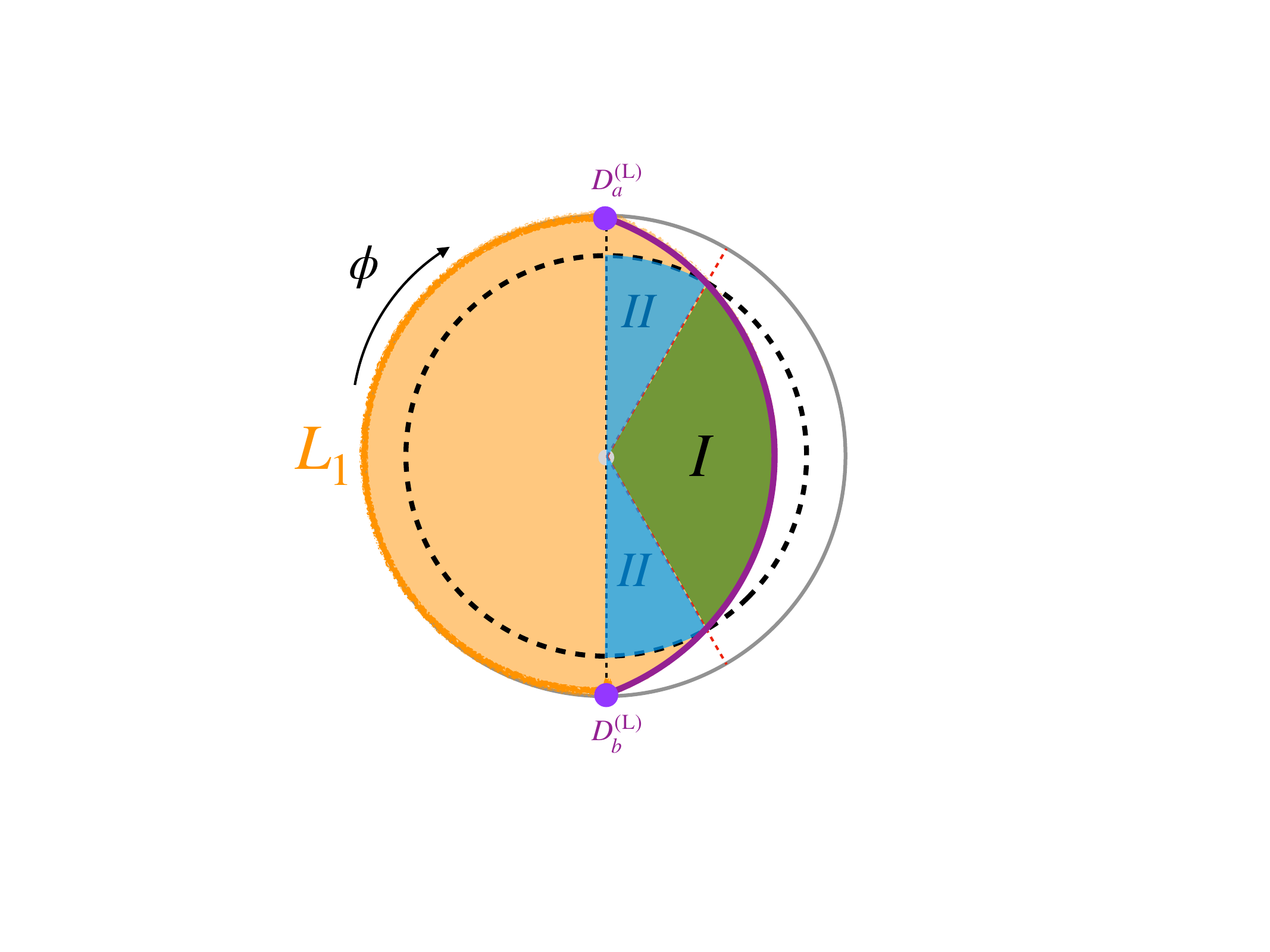}
	\caption{Left: the bulk spacetime $\mS_1$ with a naive cut-off surface located at $r=\frac{\ell}{\epsilon}$. Right: the spacetime geometry $\mS_1 \cup \mS_2$ near the defect. The black curve represents the cut-off surface.}\label{fig:corner}
\end{figure}

To furnish a concrete example, let us consider the symmetric scenario (with $m_1=m_2$) for clarity. Without loss of generality, we can parameterize the bulk regions $\mathcal{S}_i$ employing Euclidean global coordinates, as follows:
\begin{equation}
	ds^2 =   \(1 + \frac{R^2}{\ell^2}\) d\tau^2 + \frac{dR^2 }{1+\frac{R^2}{\ell^2}} + d\Phi^2 \,.
\end{equation}
Here, the AdS$_2$ brane is situated at
\begin{equation}
	| R \cos \Phi | = \frac{T_o \ell^2}{\sqrt{1-T_o^2 \ell^2}}  \,. 
\end{equation}
A commonly adopted albeit simple choice for the cut-off surface is expressed as
\begin{equation}
	R_{\rm cut} = \frac{\ell^2}{\epsilon} \,. 
\end{equation}
as depicted in figure \ref{fig:corner}. Given the non-smooth nature of this cut-off surface around the brane, a kink arises on the brane, depicted by the red dot in figure \ref{fig:corner}. Consequently, it becomes imperative to incorporate the (timelike) joint term, often referred to as the Hayward term \cite{Hayward:1993my,Brill:1994mb}, at the intersection between the cut-off surface and the brane. One can think of this contribution as arising from $\delta$-function contribution to the extrinsic curvature, which appears in the boundary action \reef{Iboundary}. For each side $\mathcal{S}_i$, this term takes the following form: 
\begin{equation}
	I_{\rm joint} (\mS_i)  =  -   \frac{1}{8 \pi \GN} \int  \theta  \sqrt{\gamma} d\tau  \,,
\end{equation}
where $\gamma$ denotes the induced metric on the corner, with $\sqrt{\gamma}= \sqrt{1+ R_{\rm cut}^2/\ell^2}$, and $\theta $ represents the angle between the two timelike hypersurfaces, as illustrated in figure \ref{fig:corner}. By using the brane profile, it is straightforward to derive the angle $\theta$, \ie\footnote{The gravitational action would not be additive due to obstacles arising from timelike joints.}
\begin{equation}
	\theta \equiv \arccos \( n_{\rm cut} \cdot n_{\mB} \) = \arccos \(  T_o\ell \sqrt{1+ \epsilon^2/\ell^2}  \) \,. 
\end{equation}
Evidently, the joint term $ I_{\rm joint} $ is divergent (because of the measure factor $\sqrt{\gamma}$), necessitating the introduction of a counterterm at the same joint. The simplest choice for this counterterm is given by 
\begin{equation}
	I_{\rm ct}^{\rm joint} (\mS_i) =  +  \frac{\theta_o}{8 \pi \GN} \int    \sqrt{\gamma}  \, d\tau \,,
\end{equation}
with $\theta_o = \arccos (T_o \ell)$. The sum of these two joint terms yields 
\begin{equation}
	I_{\rm joint} + I_{\rm ct}^{\rm joint}  \approx + \frac{T_o}{\sqrt{1-T_o^2 \ell^2}}   \cdot  \frac{\beta  \epsilon }{16 \pi \GN} + \mathcal{O}(\epsilon^3) \to 0  \,. 
\end{equation}
As a summary, we observe that there is not necessary to concern ourselves with the Hayward term at the boundary of the brane since it is canceled by the corresponding counterterm on the joint upon taking the $\epsilon \to 0$ limit.

\subsection{Smooth cut-off surface}
\begin{figure}[!]
	\centering
	\includegraphics[width=3.5in]{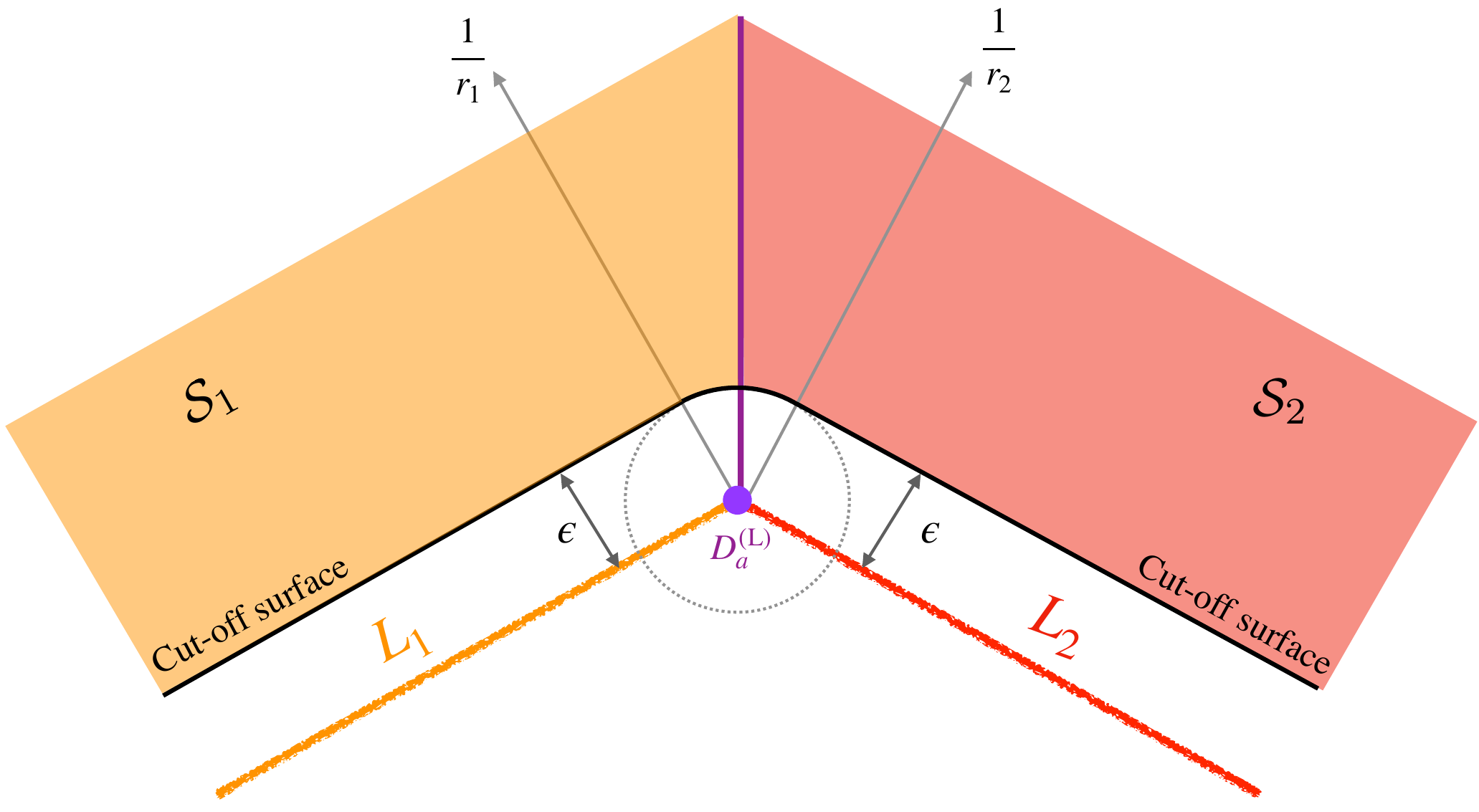}
	\caption{A smooth cut-off surface with regularizing the small part (at the order of $\mathcal{O}(\epsilon)$) near the defects.}\label{fig:cutoff}
\end{figure}

The previous example, which focused on the symmetric case, does not demonstrate the cancellation of these boundary terms on the brane in general. It is important to note that a simple regulator like $r_i=\ell^2/\epsilon$ does not result in a smooth or continuous cutoff surface. Using eq.~\reef{eq:gauge}, it is straightforward to see that in a nonsymmetric configuration, these cutoff surfaces in the bulk regions $\mS_1$ and $\mS_2$ do not intersect the brane as the same radius, \ie at different values of the brane coordinate $\sigma$.

To ensure that the cut-off surfaces from $\mS_1$ to $\mS_2$ are connected, we must align the cut-off surfaces at $\mS_1$ and $\mS_2$ according to
\begin{equation}\label{eq:cutoff12}
	\frac{\ell}{\epsilon_1} = \frac{r_{1} (\sigma)}{\ell} \bigg|_{\rm cut} = \sqrt{\sigma+ m_1}  \,, \qquad  \frac{\ell}{\epsilon_2} = \frac{r_{2} (\sigma)}{\ell}  \bigg|_{\rm cut} = \sqrt{\sigma+ m_2} \,.
\end{equation}
However, we would like there to be a single universal cutoff scale in the defect conformal field theory. Denoting the latter as $\epsilon$, we can achieve a smooth cut-off surface (as shown in figure \ref{fig:cutoff}) by considering two slightly different cut-off surfaces as follows 
\begin{equation}
	\frac{1}{\epsilon_i} = \frac{1}{\epsilon} + A_i + B_i \epsilon +\mathcal{O} (\epsilon) \,.
\end{equation}
The specifics of these parameters depend on the chosen smooth cut-off surface but do not affect any physical quantities (after renormalization) as considered in the main text. Moreover, a notable advantage of the smooth cut-off surface is the guarantee of cancellation between the joint term and its corresponding counterterm at the intersection between the brane and the cut-off surface. This holds true due to the following equality
\begin{equation}
		I_{\rm ct}(\mS_1 \cup \mS_2) =   -  \frac{\pi}{8 \pi \GN} \int    \sqrt{\gamma}  \, d\tau =	- I^{\rm joint}_{\rm ct} (\mS_1 \cup \mS_2)   \,,
\end{equation}
for any smooth or connected cut-off surface. As a result, we can effectively disregard the contributions from these boundary terms on the brane.

Lastly, it is worth noting that the choice of the smooth cut-off surface could influence the calculations of the renormalized action at the order of $\mathcal{O}(\epsilon)$. A simple choice to achieve a completely smooth cutoff surface would involve using an extremal hypersurface in the vicinity of the brane for regularization, as illustrated in figure \ref{fig:cutoff} for the symmetric case. Since its trace of the extrinsic curvature vanishes, Gibbons-Hawking-York term on that portion of the cutoff surface vanishes.



\section{Thermodynamics in the Large Tension Limit} \label{sec:notes}
%

In this appendix, we examine the thermodynamics of our holographic system of defect CFTs, however, we focus on the large tension limit, \ie $T_o\ell\simeq 1$. From the brane perspective, this limit ensures that contributions coming from the higher curvature terms in the effective gravitational action \reef{sock99} on the branes are suppressed, \eg see \cite{Grimaldi:2022suv}. From the bulk perspective,
the brane moves far for the center of the AdS geometry in this limit, as is evident from the growth of the turning point $\sigma_+$ in eq.~\reef{eq:sigmapm} -- see also eq.~\reef{tab1} below. From the boundary perspective, this limit corresponds to a large number of degrees of freedom on the defect. That is, the boundary entropy is very large, 
\begin{equation}\label{orca}
	\log g=\frac{c}{12}\,\log\(\frac{1+T_o\ell}{1-T_o\ell}\)\,.
\end{equation}
Further, with the ratio of $\log g/c\gg1$, one finds 
that information leaks slowly from the defect to the bath in this regime \cite{Sully:2020pza}.

 Introducing the  parameter $x=1-T_o\ell\ll1$, we  can re-express various results in our previous analysis using this parameter. For example, the angle between the boundary defect and the midpoint of the brane in eq.~\reef{eq:deltaphi} becomes
\begin{equation}\label{dddphi}
	\Delta \phi_i  =  \frac{\left[ 4 (1-x)^2 \, \mathbf{\Pi} \(- \frac{m_i}{\sigma_+}, \frac{\sigma_-}{\sigma_+}\) +\frac{m_{i}-m_{j}}{m_{i}} \left( \mathbf{K}\( \frac{\sigma_-}{\sigma_+}\) - \mathbf{\Pi} \(-\frac{m_i}{\sigma_+}, \frac{\sigma_-}{\sigma_+}\)  \right)\right]}{4(1-x)\sqrt{2-x}\,\sqrt{\vphantom{2}x\,\sigma_+}  } \,.
\end{equation}
Similarly, $\sigma_\pm$ defined in eq.~\reef{eq:sigmapm} become
\begin{equation} \label{eq:sigma}
	\sigma_\pm = \frac{1}{4(2-x)x}  \(  -(m_1+m_2) \pm  \sqrt{ \(\frac{m_1-m_2}{1-x} \)^2+ 4m_1 m_2   }\)\,. 
\end{equation}

In preparing for a more detailed analysis of the various phases, let us examine the cold/warm phase where $m_1+m_2 <0$. Considering $x\ll1$, eq.~\eqref{eq:sigma} leads to the following approximations:
\begin{equation} \label{tab1}
\begin{split}
	\sigma_+&\simeq\frac{|m_1+m_2|}{4\,x}+\frac{m_1^2+m_2^2}{4\,|m_1+m_2|}+ \mathcal{O}(x)
	\,,\\
	\sigma_-&\simeq\frac{(m_1-m_2)^2}{8\,|m_1+m_2|}+x\,\frac{(m_1-m_2)^2}{(m_1+m_2)^2}\,(3m_1+m_2)(m_1+3m_2)+ \mathcal{O}(x^2)  \,.
\end{split}
\end{equation}
Note that $\sigma_+\propto1/x\gg1$, indicating the brane's turning point occurs at very large $\sigma$. Furthermore, in this regime, $\sigma_-/\sigma_+,\ m_i/\sigma_+\propto x\ll1$. Now turning to eq.~\reef{dddphi}, we may utilize the generalized power series expansion of the complete elliptic integral of the third kind, \ie 
\begin{equation}
\mathbf{\Pi}(n, m)=  \frac{\pi}{2} \sum^{\infty}_{k=0} \sum_{j=0}^{k}   \frac{(2k!)(2j!)m^j n^{k-j}}{4^k 4^j (k!)^2(j!)^2}  \,, \qquad  |n|<1,\ |m|<1 \,. 
\end{equation}
In the limit as $n, m \to 0$, this series expansion becomes
\begin{equation}
\mathbf{\Pi} (n,m) \simeq \frac{\pi}{2} \(  1 + \frac{m}{4} + \frac{9}{64}m^2 + \frac{n}{2}+ \frac{3}{8}n^2 + \frac{3}{16} mn       \) + \mathcal{O}(n^3, m^3, m^2n , n^2 m) \,. 
\end{equation}
Applying this series expansion to eq.~\reef{dddphi} then yields
\begin{equation} \label{royal1}
	\begin{split}
		\Delta \phi_1 \simeq \frac{\pi}{\sqrt{2}}   \( \frac{1}{\sqrt{|m_1+m_2|}}  + \frac{(m_1- m_2) (m_1 + 7 m_2)}{8|m_1+m_2|^{5/2}}  x   \)  + \mathcal{O}(x^2) \,, \\ 
		\Delta \phi_2 \simeq \frac{\pi}{\sqrt{2}}   \( \frac{1}{\sqrt{|m_1+m_2|}}  + \frac{(m_2- m_1) (7m_1 +  m_2)}{8|m_1+m_2|^{5/2}}  x   \)  + \mathcal{O}(x^2) \,. \\ 
	\end{split}
\end{equation}
for the cold/warm phase with $m_1 +m_2 <0$.

Let us note here that in the warm phase, the choice with $m_1>0,m_2<0$ but $m_1 +m_2 >0$ is unphysical. In this parameter regime and in the large tension limit with $x\ll1$, 
the expansions of $\sigma_\pm$ in eq.~\reef{eq:sigma} interchange their character so that
$\sigma_-\sim{\mathcal O}(1/x)$ and $\sigma_+\sim{\mathcal O}(1)$. As a result,
we would find that $L_2 <0$ since 
	\begin{equation}\label{rocket6}
		\Delta \phi_2 \simeq  - \frac{\log x}{\sqrt{2}\sqrt{m_1 +m_2}}  \to +\infty \,,
	\end{equation}
using $\mathbf{K}(- \frac{1}{x})   \sim \frac{\sqrt{x}}{2}\log(\frac{16}{x}) \sim \mathbf{\Pi} (n, - \frac{1}{x}) $ in the limit $x \to 0$.

On the other hand, let us consider $m_1 >0, m_2>0$ which leads to different approximations:
\begin{equation}
	\begin{split}
		\sigma_+&\simeq+\frac{(m_1-m_2)^2}{8(m_1+m_2)}+x\,\frac{(m_1-m_2)^2}{(m_1+m_2)^2}\,(3m_1+m_2)(m_1+3m_2)+  \mathcal{O}(x^2) \,, \\
		\sigma_-&\simeq	-\frac{m_1+m_2}{4\,x}-\frac{m_1^2+m_2^2}{4(m_1+m_2)}  + \mathcal{O}(x^2)\,,  \\ 
	\end{split}
\end{equation}
indicating the brane's turning point is located at the middle of bulk spacetime. Consequently, the expression \reef{dddphi} for $\Delta \phi_i $ with elliptic integrals yields
\begin{equation}
	\Delta \phi_i \simeq  \frac{  \log \(  \frac{32}{x}  \)  +2 \log \( \frac{m_1+m_2}{|m_1-m_2|}  \) }{\sqrt{2(m_1+m_2)}} - \frac{1}{\sqrt{m_i}} \arcsinh \(   \frac{  \sqrt{8m_i (m_1+m_2)} }{|m_1-m_2|} \) + \mathcal{O} (x^2) \,. 
\end{equation}
Of course, we are only interested in the case with $m_1=m_2>0$ for the hot phase. Then taking the limit $m_2 \to m_1$ above yields
\begin{equation}
\Delta \phi_i \simeq   \frac{\log \( \frac{2}{x} \)}{2\sqrt{m_i}}  + \mathcal{O}(x^2) \,,
\end{equation}
This same result can also be derived by taking the large tension limit of the precise expression in eq.~\eqref{eq:deltaphihot}.

\vskip2ex
\noindent{\bf Cold phase ($m_i<0$):} 
The renormalized Euclidean action was given in eq.~\eqref{eq:ColdAction}, as follows:
\begin{equation}
	I^\mt{cold}_\mt{E}=\frac{c}{24\pi T_{\mt{DCFT}}}\(L_1\,m_1
	+L_2\,m_2\)\,\frac{1}{\ell^2} \,.
\end{equation}
The parameters $m_i$ (or rather $m_i/\ell^2$) may be determined in terms of boundary data using eq.~\reef{eq:boundarybulk}, \ie
\begin{equation}\label{LI}
	 L_i  = 2\ell\(\frac{\pi }{\sqrt{-m_i}} - \Delta \phi_i \)  \,. 
\end{equation}
We are considering the $\Delta\phi_i$ given by eq.~\eqref{royal1} in the limit $x\ll1$. To leading order, we then find
\begin{equation}\label{royal2}
	L_1 \simeq 2 \pi \ell  \(  \frac{1}{\sqrt{|m_1|}}  - \frac{1}{\sqrt{2\,|m_1+m_2|}}  \) \,, \quad 	L_2\simeq 2 \pi \ell  \(  \frac{1}{\sqrt{|m_2|}}  - \frac{1}{\sqrt{2\,|m_1+m_2|}}  \)  \,. 
\end{equation}
The solutions $m_i (L_1, L_2)$ are analytical but complex functions and hence we refrain from explicitly presenting the general expressions here. Utilizing standard thermodynamic relations, we derive the energy and entropy as 
\begin{equation}\label{parrot1}
	\begin{split}
		E^{\rm cold}&=-T_\mt{DCFT}^2\frac{\partial I^\mt{cold}_\mt{E}}{\partial T_\mt{DCFT}}=
		\frac{c}{24\pi }\(L_1\,m_1
		+L_2\,m_2\)\,\frac{1}{\ell^2}\,,\\
		S^{\rm cold}&=-\frac{\partial (T_\mt{DCFT}\, I^\mt{cold}_\mt{E})}{\partial T_\mt{DCFT}}=0\,.
	\end{split}
\end{equation}
As expected, the entropy vanishes to leading order in a large $c$ expansion, while the energy adopts a Casmir-like form. That is,  $E^{\rm cold}$ is  purely a function of the boundary geometry, \ie, $L_1$ and $L_2$, but independent of the temperature. Further, the energy is negative since $m_i<0$.

Two simple cases, which we might examine further, are: $m_1=m_2=m$ and $|m_1|\ll |m_2|$. In the first case, we see from eq.~\reef{royal2} that $L_1=L_2=\pi\ell/\sqrt{|m|}$, which agrees with eq.~\reef{royal3} in section \ref{sec:sym} for the symmetric boundary. Further, as expected then, the action reduces to the same expression found in eq.~\reef{eq:coldaction} and the Casimir form of the energy becomes evident with
\beq
E^{\rm cold}=-\frac{\pi\,c}6\,\frac{1}{L_1+L_2}\,.
\eeq
We note that this is the precise result that does not rely on the small $x$ limit. 

In the regime $|m_1|\ll |m_2|$, we find from eq.~\reef{royal2} that to leading order
\beq\label{rocket1}
L_1\simeq \frac{2\pi\ell}{\sqrt{|m_1|}}\,,\qquad
L_2\simeq \frac{2\pi\ell}{\sqrt{|m_2|}} \(1-\frac{1}{\sqrt{2}}\)\,.
\eeq
Hence $L_2\ll L_1$ and we have arrived at the fusion limit considered in section \ref{rocket2}. Using the above expressions \reef{rocket1} to simplify the energy in eq.~\reef{parrot1}, we arrive at
\beq
E^{\rm cold}\simeq
 -\frac{\pi \, c}{6}\[\frac{1}{L_2}\(1- \frac1{\sqrt{2}}\)^2 +\frac1{L_1}+\mathcal{O}\(\frac{L_2}{L_1^2}\)\]\,.
 \label{rocket11}
\eeq
Comparing this expression with $E^{\rm cold}$ in eq.~\reef{rocket5}, we see that the two energies agree when we substitute $k_2\simeq\frac{\pi}{\sqrt{2}}+\mathcal{O}(x)$ from eq.~\reef{rocket3}. 

Hence with these two simple limits, we confirm the present analysis for the large tension limit by comparing to our previous results in sections \ref{rocket2} and \ref{sec:sym}. However, in principle, we have access to the thermodynamic behaviour for general configurations with arbitrary values of $L_1$ and $L_2$. For example, one could easily study small deviations from the symmetric configuration, \ie $m_1-m2=\delta m\ll1$, in this regime. One could also evaluate corrections in a small $x$ expansion.

\vskip2ex
\noindent{\bf Warm phase ($m_1>0$, $m_2<0$):} The renormalized Euclidean action is now represented by eq.~\eqref{eq:WarmAction} as:
\begin{equation}\label{eq:WarmActionC}
I^\mt{warm}_\mt{E}=\frac{c}{24\pi T_{\mt{DCFT}}}\((L_1-2\ell\,\Delta\phi_1^\mt{Hor})\,m_1 +L_2\,m_2\)\,\frac{1}{\ell^2}	\,.
\end{equation}
 We need to determine $m_i$ and $\Delta\phi_1^\text{Hor}$ in terms of the boundary data. In this scenario, a BTZ black hole exists in region $\mathcal{S}_1$, and hence $m_1$ is determined by 
\begin{equation}\label{swallow1}
	\frac{m_1}{\ell^2}=(2\pi\,T_\mt{DCFT})^2\,,
\end{equation}
as in eq.~\reef{eq:m1T}. Meanwhile, $m_2$ is determined using eq.~\eqref{LI}
\begin{equation}\label{LI2}
	L_2  = 2\ell\(\frac{\pi }{\sqrt{-m_2}} - \Delta \phi_2 \)  \,,
\end{equation}
as in the cold phase.
Lastly, $\Delta\phi_1^\text{Hor}$ is determined by eq.~\eqref{eq:warmL1}, resulting in
\begin{equation}\label{swallow2}
	\ell\,\Delta\phi_1^\mt{Hor}=L_1+2\ell\,\Delta\phi_1 \,.
\end{equation}

Using standard thermodynamic relations, one finds the energy and entropy are given by
\beqa
E^{\rm warm}=-T_\mt{DCFT}^2\frac{\partial I^\mt{warm}_\mt{E}}{\partial T_\mt{DCFT}}&=&
\frac{c}{6}\(\pi\,T_\mt{DCFT}^2\,(L_1+4\ell\,\Delta\phi_1) +\frac{L_2}{4\pi\,}\,\frac{m_2}{\ell^2}\)
\label{parrot2}\\
&&\quad +\frac{c}{6}\(4\pi\ell\,T_\mt{DCFT}^3\,\frac{\partial\Delta\phi_1}{\partial T_\mt{DCFT}} -\frac{L_2\,T_\mt{DCFT}}{4\pi\,\ell^2}\,\frac{\partial m_2}{\partial T_\mt{DCFT}} \)
\nonumber\\
S^{\rm warm}=-\frac{\partial (T_\mt{DCFT}\, I^\mt{warm}_\mt{E})}{\partial T_\mt{DCFT}}&=&\frac{\pi\,c}{3}\,T_\mt{DCFT}\,(L_1+4\ell\,\Delta\phi_1) \,.
\nonumber\\
&&\quad +\frac{c}{6}\(4\pi\ell\,T_\mt{DCFT}^2\,\frac{\partial\Delta\phi_1}{\partial T_\mt{DCFT}} -\frac{L_2}{4\pi\,\ell^2}\,\frac{\partial m_2}{\partial T_\mt{DCFT}} \)
\nonumber
\eeqa
Hence we see that region $\mS_1$ contributes a thermal energy and entropy (proportional to $L_1$) that is characteristic of the deconfined phase of the CFT. The two contributions with $m_2$ and $\Delta\phi_1$ are more complicated because generally these parameters now depend on the temperature through the appearance of $m_1$ in eqs.~\reef{dddphi} and \reef{eq:sigma}.  

One can make these observations more explicit but recalling some aspects of the warm phase from section \ref{sec:Phasediagram}. From the discussion there and the phase diagram in figure \ref{fig:PhaseDiagram}, we expect $L_2\ll L_1$ to be in this phase. That is, we are in the  the fusion limit described in section \ref{rocket2}. Further to be below the phase transition between the hot and warm phases, we have $T_\mt{DCFT}L_2<C_\mt{WH}$ but in the large tension limit, eq.~\reef{rocket7} shows that $C_\mt{WH}\sim c/\log g\ll1$ and hence we have $T_\mt{DCFT}L_2\ll1$. However, to be above the cold-warm transition, we must have $T_\mt{DCFT}L_1\gtrsim1$ -- see discussion around eq.~\reef{eq:Ccw}. That is, we must satisfy three constraints to be in the warm phase
\beq
L_1\gg L_2\,,\qquad T_\mt{DCFT}L_2\ll1\,,\qquad T_\mt{DCFT}L_1\gtrsim1\,.
\label{rocket8}
\eeq
Now expanding the expressions in eq.~\reef{parrot2} to leading order for small $T_\mt{DCFT}L_2$, we find
\beqa
E^{\rm warm}&=&
\frac{\pi\,c}{6}\(T_\mt{DCFT}^2\,L_1-\frac{1}{L_2}\,\(1-\frac{1}{\sqrt{2}}\)^2+\frac{T_\mt{DCFT}^2\,L_2}{\sqrt{2}-1}+{\mathcal O}(T_\mt{DCFT}^4\,L_2^3)\)\,,
\label{rocket9}\\
S^{\rm warm}&=&\frac{\pi\,c}{3}\(T_\mt{DCFT}\,L_1+\frac{T_\mt{DCFT}\,L_2}{\sqrt{2}-1}+{\mathcal O}(T_\mt{DCFT}^3\,L_2^3)\)\,,
\label{rocket9}\\
\nonumber
\eeqa
The first term in these expressions corresponds to the thermal energy and entropy of the deconfined phase of the CFT in the $\mS_1$ region of length $L_1$ -- see eq.~\reef{pigeon3} below. The second term in $E^{\rm warm}$ corresponds to a Casimir energy for the $\mS_2$ region of length $L_2$. We note that this term precisely matches the $1/L_2$ contribution found in eq.~\reef{rocket11} for the fusion limit in the cold phase.
The terms proportional to $L_2$ (and higher powers) are corrections indicating that there is an interaction between the thermal phase in the $\mS_1$ region and the Casimir phase in the $\mS_2$ region. In fact, the linear terms have the form of thermal contributions for a region of length $L_2$, however, we note that the coefficient is larger than that expected for the thermal CFT because of the factor $(\sqrt{2}-1)^{-1}\simeq 2.41$. Examining the entropy, one finds that it is precisely the horizon entropy of the black hole in the bulk $\mS_1$ region, as expected -- see eq.~\reef{hanger2} and also eq.~\reef{rocket2} for the large tension values of $k_1$ and $k_2$. A final observation is that in contrast to the hot phase below, there is no defect contribution to the entropy here.
\vskip2ex
\noindent{\bf Hot phase ($m_i>0$):} In this case, we have
\beq
\frac{m_1}{\ell^2}=\frac{m_2}{\ell^2}=(2\pi\,T_\mt{DCFT})^2\,,
\label{swallow3}
\eeq
from eq.~\reef{parrot77}. With this and other simplifications, the action reduces to
\beq
I_{\mt{E}}^{\rm hot} =- \frac{\pi\,c  }{6} T_{\mt{DCFT}}  \(   L_1 +   L_2  \) -   4 \log g   
\eeq
as shown in eq.~\reef{eq:HotAction}. Again using standard thermodynamic relations, one finds the energy and entropy are given by
\beqa\label{pigeon3}
E^{\rm hot}&=&-T_\mt{DCFT}^2\frac{\partial I^\mt{hot}_\mt{E}}{\partial T_\mt{DCFT}}=
\frac{\pi\,c}{6}\,T_\mt{DCFT}^2\,\(L_1+L_2\)\,,
\label{parrot3}\\
S^{\rm hot}&=&-\frac{\partial (T_\mt{DCFT}\, I^\mt{hot}_\mt{E})}{\partial T_\mt{DCFT}}=\frac{\pi\,c}{3}\,T_\mt{DCFT}\,\(L_1+L_2\)+4\log g \,.
\nonumber
\eeqa
Hence we see that both regions contribute a thermal energy and entropy (proportional to $L_1$ and $L_2$) that is characteristic of the deconfined phase of the CFT. The two defects do not contribute to the energy but each contribute a constant defect entropy $S_{\rm def}=2\log g$ -- recall eq.~\reef{eq:defg}.


\bibliographystyle{jhep}
\bibliography{bibliography_notes}

\providecommand{\href}[2]{#2}\begingroup\raggedright\begin{thebibliography}{10}

\bibitem{Ryu:2006bv}
S.~Ryu and T.~Takayanagi, \emph{{Holographic derivation of entanglement entropy
  from AdS/CFT}},
  \href{https://doi.org/10.1103/PhysRevLett.96.181602}{\emph{Phys. Rev. Lett.}
  {\bfseries 96} (2006) 181602}
  [\href{https://arxiv.org/abs/hep-th/0603001}{{\ttfamily hep-th/0603001}}].

\bibitem{Ryu:2006ef}
S.~Ryu and T.~Takayanagi, \emph{{Aspects of Holographic Entanglement Entropy}},
  \href{https://doi.org/10.1088/1126-6708/2006/08/045}{\emph{JHEP} {\bfseries
  08} (2006) 045} [\href{https://arxiv.org/abs/hep-th/0605073}{{\ttfamily
  hep-th/0605073}}].

\bibitem{Hubeny:2007xt}
V.~E. Hubeny, M.~Rangamani and T.~Takayanagi, \emph{{A Covariant holographic
  entanglement entropy proposal}},
  \href{https://doi.org/10.1088/1126-6708/2007/07/062}{\emph{JHEP} {\bfseries
  07} (2007) 062} [\href{https://arxiv.org/abs/0705.0016}{{\ttfamily
  0705.0016}}].

\bibitem{Almheiri:2019hni}
A.~Almheiri, R.~Mahajan, J.~Maldacena and Y.~Zhao, \emph{{The Page curve of
  Hawking radiation from semiclassical geometry}},
  \href{https://doi.org/10.1007/JHEP03(2020)149}{\emph{JHEP} {\bfseries 03}
  (2020) 149} [\href{https://arxiv.org/abs/1908.10996}{{\ttfamily
  1908.10996}}].

\bibitem{Engelhardt:2014gca}
N.~Engelhardt and A.~C. Wall, \emph{{Quantum Extremal Surfaces: Holographic
  Entanglement Entropy beyond the Classical Regime}},
  \href{https://doi.org/10.1007/JHEP01(2015)073}{\emph{JHEP} {\bfseries 01}
  (2015) 073} [\href{https://arxiv.org/abs/1408.3203}{{\ttfamily 1408.3203}}].

\bibitem{Faulkner:2013ana}
T.~Faulkner, A.~Lewkowycz and J.~Maldacena, \emph{{Quantum corrections to
  holographic entanglement entropy}},
  \href{https://doi.org/10.1007/JHEP11(2013)074}{\emph{JHEP} {\bfseries 11}
  (2013) 074} [\href{https://arxiv.org/abs/1307.2892}{{\ttfamily 1307.2892}}].

\bibitem{Almheiri:2019psf}
A.~Almheiri, N.~Engelhardt, D.~Marolf and H.~Maxfield, \emph{{The entropy of
  bulk quantum fields and the entanglement wedge of an evaporating black
  hole}}, \href{https://doi.org/10.1007/JHEP12(2019)063}{\emph{JHEP} {\bfseries
  12} (2019) 063} [\href{https://arxiv.org/abs/1905.08762}{{\ttfamily
  1905.08762}}].

\bibitem{Penington:2019npb}
G.~Penington, \emph{{Entanglement Wedge Reconstruction and the Information
  Paradox}},  \href{https://arxiv.org/abs/1905.08255}{{\ttfamily 1905.08255}}.

\bibitem{Penington:2019kki}
G.~Penington, S.~H. Shenker, D.~Stanford and Z.~Yang, \emph{{Replica wormholes
  and the black hole interior}},
  \href{https://arxiv.org/abs/1911.11977}{{\ttfamily 1911.11977}}.

\bibitem{Almheiri:2019qdq}
A.~Almheiri, T.~Hartman, J.~Maldacena, E.~Shaghoulian and A.~Tajdini,
  \emph{{Replica Wormholes and the Entropy of Hawking Radiation}},
  \href{https://doi.org/10.1007/JHEP05(2020)013}{\emph{JHEP} {\bfseries 05}
  (2020) 013} [\href{https://arxiv.org/abs/1911.12333}{{\ttfamily
  1911.12333}}].

\bibitem{Balasubramanian:2021wgd}
V.~Balasubramanian, A.~Kar and T.~Ugajin, \emph{{Entanglement between two
  gravitating universes}},
  \href{https://doi.org/10.1088/1361-6382/ac3c8b}{\emph{Class. Quant. Grav.}
  {\bfseries 39} (2022) 174001}
  [\href{https://arxiv.org/abs/2104.13383}{{\ttfamily 2104.13383}}].

\bibitem{Miyata:2021qsm}
A.~Miyata and T.~Ugajin, \emph{{Entanglement between two evaporating black
  holes}}, \href{https://doi.org/10.1007/JHEP09(2022)009}{\emph{JHEP}
  {\bfseries 09} (2022) 009}
  [\href{https://arxiv.org/abs/2111.11688}{{\ttfamily 2111.11688}}].

\bibitem{Anderson:2020vwi}
L.~Anderson, O.~Parrikar and R.~M. Soni, \emph{{Islands with gravitating baths:
  towards ER = EPR}},
  \href{https://doi.org/10.1007/JHEP10(2021)226}{\emph{JHEP} {\bfseries 21}
  (2020) 226} [\href{https://arxiv.org/abs/2103.14746}{{\ttfamily
  2103.14746}}].

\bibitem{Geng:2020fxl}
H.~Geng, A.~Karch, C.~Perez-Pardavila, S.~Raju, L.~Randall, M.~Riojas et~al.,
  \emph{{Information Transfer with a Gravitating Bath}},
  \href{https://arxiv.org/abs/2012.04671}{{\ttfamily 2012.04671}}.

\bibitem{Geng:2021iyq}
H.~Geng, S.~L\"ust, R.~K. Mishra and D.~Wakeham, \emph{{Holographic BCFTs and
  Communicating Black Holes}},
  \href{https://arxiv.org/abs/2104.07039}{{\ttfamily 2104.07039}}.

\bibitem{Balasubramanian:2021xcm}
V.~Balasubramanian, B.~Craps, M.~Khramtsov and E.~Shaghoulian,
  \emph{{Submerging islands through thermalization}},
  \href{https://doi.org/10.1007/JHEP10(2021)048}{\emph{JHEP} {\bfseries 10}
  (2021) 048} [\href{https://arxiv.org/abs/2107.14746}{{\ttfamily
  2107.14746}}].

\bibitem{Liu:2022pan}
Y.~Liu, Z.-Y. Xian, C.~Peng and Y.~Ling, \emph{{Black holes entangled by
  radiation}}, \href{https://doi.org/10.1007/JHEP11(2022)043}{\emph{JHEP}
  {\bfseries 09} (2022) 179}
  [\href{https://arxiv.org/abs/2205.14596}{{\ttfamily 2205.14596}}].

\bibitem{Balasubramanian:2023xyd}
V.~Balasubramanian, Y.~Nomura and T.~Ugajin, \emph{{de Sitter space is
  sometimes not empty}},  \href{https://arxiv.org/abs/2308.09748}{{\ttfamily
  2308.09748}}.

\bibitem{Balasubramanian:2020coy}
V.~Balasubramanian, A.~Kar and T.~Ugajin, \emph{{Entanglement between two
  disjoint universes}},
  \href{https://doi.org/10.1007/JHEP02(2021)136}{\emph{JHEP} {\bfseries 02}
  (2021) 136} [\href{https://arxiv.org/abs/2008.05274}{{\ttfamily
  2008.05274}}].

\bibitem{Balasubramanian:2020xqf}
V.~Balasubramanian, A.~Kar and T.~Ugajin, \emph{{Islands in de Sitter space}},
  \href{https://doi.org/10.1007/JHEP02(2021)072}{\emph{JHEP} {\bfseries 02}
  (2021) 072} [\href{https://arxiv.org/abs/2008.05275}{{\ttfamily
  2008.05275}}].

\bibitem{Maldacena:2013xja}
J.~Maldacena and L.~Susskind, \emph{{Cool horizons for entangled black holes}},
  \href{https://doi.org/10.1002/prop.201300020}{\emph{Fortsch. Phys.}
  {\bfseries 61} (2013) 781} [\href{https://arxiv.org/abs/1306.0533}{{\ttfamily
  1306.0533}}].

\bibitem{Chen:2020uac}
H.~Z. Chen, R.~C. Myers, D.~Neuenfeld, I.~A. Reyes and J.~Sandor,
  \emph{{Quantum Extremal Islands Made Easy, Part I: Entanglement on the
  Brane}}, \href{https://doi.org/10.1007/JHEP10(2020)166}{\emph{JHEP}
  {\bfseries 10} (2020) 166}
  [\href{https://arxiv.org/abs/2006.04851}{{\ttfamily 2006.04851}}].

\bibitem{Chen:2020hmv}
H.~Z. Chen, R.~C. Myers, D.~Neuenfeld, I.~A. Reyes and J.~Sandor,
  \emph{{Quantum Extremal Islands Made Easy, Part II: Black Holes on the
  Brane}}, \href{https://doi.org/10.1007/JHEP12(2020)025}{\emph{JHEP}
  {\bfseries 12} (2020) 025}
  [\href{https://arxiv.org/abs/2010.00018}{{\ttfamily 2010.00018}}].

\bibitem{Hernandez:2020nem}
J.~Hernandez, R.~C. Myers and S.-M. Ruan, \emph{{Quantum extremal islands made
  easy. Part III. Complexity on the brane}},
  \href{https://doi.org/10.1007/JHEP02(2021)173}{\emph{JHEP} {\bfseries 02}
  (2021) 173} [\href{https://arxiv.org/abs/2010.16398}{{\ttfamily
  2010.16398}}].

\bibitem{Grimaldi:2022suv}
G.~Grimaldi, J.~Hernandez and R.~C. Myers, \emph{{Quantum extremal islands made
  easy. Part IV. Massive black holes on the brane}},
  \href{https://doi.org/10.1007/JHEP03(2022)136}{\emph{JHEP} {\bfseries 03}
  (2022) 136} [\href{https://arxiv.org/abs/2202.00679}{{\ttfamily
  2202.00679}}].

\bibitem{Chen:2019uhq}
H.~Z. Chen, Z.~Fisher, J.~Hernandez, R.~C. Myers and S.-M. Ruan,
  \emph{{Information Flow in Black Hole Evaporation}},
  \href{https://doi.org/10.1007/JHEP03(2020)152}{\emph{JHEP} {\bfseries 03}
  (2020) 152} [\href{https://arxiv.org/abs/1911.03402}{{\ttfamily
  1911.03402}}].

\bibitem{Rozali:2019day}
M.~Rozali, J.~Sully, M.~Van~Raamsdonk, C.~Waddell and D.~Wakeham,
  \emph{{Information radiation in BCFT models of black holes}},
  \href{https://arxiv.org/abs/1910.12836}{{\ttfamily 1910.12836}}.

\bibitem{Balasubramanian:2020hfs}
V.~Balasubramanian, A.~Kar, O.~Parrikar, G.~Sárosi and T.~Ugajin,
  \emph{{Geometric secret sharing in a model of Hawking radiation}},
  \href{https://arxiv.org/abs/2003.05448}{{\ttfamily 2003.05448}}.

\bibitem{Chen:2020jvn}
H.~Z. Chen, Z.~Fisher, J.~Hernandez, R.~C. Myers and S.-M. Ruan,
  \emph{{Evaporating Black Holes Coupled to a Thermal Bath}},
  \href{https://doi.org/10.1007/JHEP01(2021)065}{\emph{JHEP} {\bfseries 01}
  (2021) 065} [\href{https://arxiv.org/abs/2007.11658}{{\ttfamily
  2007.11658}}].

\bibitem{Caceres:2020jcn}
E.~Caceres, A.~Kundu, A.~K. Patra and S.~Shashi, \emph{{Warped Information and
  Entanglement Islands in AdS/WCFT}},
  \href{https://arxiv.org/abs/2012.05425}{{\ttfamily 2012.05425}}.

\bibitem{Krishnan:2020fer}
C.~Krishnan, \emph{{Critical Islands}},
  \href{https://arxiv.org/abs/2007.06551}{{\ttfamily 2007.06551}}.

\bibitem{Sully:2020pza}
J.~Sully, M.~Van~Raamsdonk and D.~Wakeham, \emph{{BCFT entanglement entropy at
  large central charge and the black hole interior}},
  \href{https://arxiv.org/abs/2004.13088}{{\ttfamily 2004.13088}}.

\bibitem{Omiya:2021olc}
H.~Omiya and Z.~Wei, \emph{{Causal structures and nonlocality in double
  holography}}, \href{https://doi.org/10.1007/JHEP07(2022)128}{\emph{JHEP}
  {\bfseries 07} (2022) 128}
  [\href{https://arxiv.org/abs/2107.01219}{{\ttfamily 2107.01219}}].

\bibitem{Neuenfeld:2021bsb}
D.~Neuenfeld, \emph{{Homology conditions for RT surfaces in double
  holography}}, \href{https://doi.org/10.1088/1361-6382/ac51e7}{\emph{Class.
  Quant. Grav.} {\bfseries 39} (2022) 075009}
  [\href{https://arxiv.org/abs/2105.01130}{{\ttfamily 2105.01130}}].

\bibitem{Neuenfeld:2021wbl}
D.~Neuenfeld, \emph{{The Dictionary for Double Holography and Graviton Masses
  in d Dimensions}},  \href{https://arxiv.org/abs/2104.02801}{{\ttfamily
  2104.02801}}.

\bibitem{Chu:2021gdb}
J.~Chu, F.~Deng and Y.~Zhou, \emph{{Page curve from defect extremal surface and
  island in higher dimensions}},
  \href{https://doi.org/10.1007/JHEP10(2021)149}{\emph{JHEP} {\bfseries 10}
  (2021) 149} [\href{https://arxiv.org/abs/2105.09106}{{\ttfamily
  2105.09106}}].

\bibitem{Li:2021dmf}
T.~Li, M.-K. Yuan and Y.~Zhou, \emph{{Defect extremal surface for reflected
  entropy}}, \href{https://doi.org/10.1007/JHEP01(2022)018}{\emph{JHEP}
  {\bfseries 01} (2022) 018}
  [\href{https://arxiv.org/abs/2108.08544}{{\ttfamily 2108.08544}}].

\bibitem{Wang:2021xih}
Z.~Wang, Z.~Xu, S.~Zhou and Y.~Zhou, \emph{{Partial reduction and cosmology at
  defect brane}}, \href{https://doi.org/10.1007/JHEP05(2022)049}{\emph{JHEP}
  {\bfseries 05} (2022) 049}
  [\href{https://arxiv.org/abs/2112.13782}{{\ttfamily 2112.13782}}].

\bibitem{Geng:2021wcq}
H.~Geng, Y.~Nomura and H.-Y. Sun, \emph{{An Information Paradox and Its
  Resolution in de Sitter Holography}},
  \href{https://arxiv.org/abs/2103.07477}{{\ttfamily 2103.07477}}.

\bibitem{Ling:2021vxe}
Y.~Ling, P.~Liu, Y.~Liu, C.~Niu, Z.-Y. Xian and C.-Y. Zhang, \emph{{Reflected
  entropy in double holography}},
  \href{https://doi.org/10.1007/JHEP02(2022)037}{\emph{JHEP} {\bfseries 02}
  (2022) 037} [\href{https://arxiv.org/abs/2109.09243}{{\ttfamily
  2109.09243}}].

\bibitem{Geng:2022slq}
H.~Geng, A.~Karch, C.~Perez-Pardavila, S.~Raju, L.~Randall, M.~Riojas et~al.,
  \emph{{Jackiw-Teitelboim Gravity from the Karch-Randall Braneworld}},
  \href{https://doi.org/10.1103/PhysRevLett.129.231601}{\emph{Phys. Rev. Lett.}
  {\bfseries 129} (2022) 231601}
  [\href{https://arxiv.org/abs/2206.04695}{{\ttfamily 2206.04695}}].

\bibitem{Suzuki:2022xwv}
K.~Suzuki and T.~Takayanagi, \emph{{BCFT and Islands in two dimensions}},
  \href{https://doi.org/10.1007/JHEP06(2022)095}{\emph{JHEP} {\bfseries 06}
  (2022) 095} [\href{https://arxiv.org/abs/2202.08462}{{\ttfamily
  2202.08462}}].

\bibitem{Karch:2022rvr}
A.~Karch, H.~Sun and C.~F. Uhlemann, \emph{{Double holography in string
  theory}}, \href{https://doi.org/10.1007/JHEP10(2022)012}{\emph{JHEP}
  {\bfseries 10} (2022) 012}
  [\href{https://arxiv.org/abs/2206.11292}{{\ttfamily 2206.11292}}].

\bibitem{Deng:2022yll}
F.~Deng, Y.-S. An and Y.~Zhou, \emph{{JT gravity from partial reduction and
  defect extremal surface}},
  \href{https://doi.org/10.1007/JHEP02(2023)219}{\emph{JHEP} {\bfseries 02}
  (2023) 219} [\href{https://arxiv.org/abs/2206.09609}{{\ttfamily
  2206.09609}}].

\bibitem{Anous:2022wqh}
T.~Anous, M.~Meineri, P.~Pelliconi and J.~Sonner, \emph{{Sailing past the End
  of the World and discovering the Island}},
  \href{https://doi.org/10.21468/SciPostPhys.13.3.075}{\emph{SciPost Phys.}
  {\bfseries 13} (2022) 075}
  [\href{https://arxiv.org/abs/2202.11718}{{\ttfamily 2202.11718}}].

\bibitem{Neuenfeld:2023svs}
D.~Neuenfeld and M.~Srivastava, \emph{{On the causality paradox and the
  Karch-Randall braneworld as an EFT}},
  \href{https://doi.org/10.1007/JHEP10(2023)164}{\emph{JHEP} {\bfseries 10}
  (2023) 164} [\href{https://arxiv.org/abs/2307.10392}{{\ttfamily
  2307.10392}}].

\bibitem{Chang:2023gkt}
J.-C. Chang, S.~He, Y.-X. Liu and L.~Zhao, \emph{{Island formula in Planck
  brane}}, \href{https://doi.org/10.1007/JHEP11(2023)006}{\emph{JHEP}
  {\bfseries 11} (2023) 006}
  [\href{https://arxiv.org/abs/2308.03645}{{\ttfamily 2308.03645}}].

\bibitem{Liu:2023ggg}
Y.~Liu, Q.~Chen, Y.~Ling, C.~Peng, Y.~Tian and Z.-Y. Xian, \emph{{Entanglement
  of defect subregions in double holography}},
  \href{https://arxiv.org/abs/2312.08025}{{\ttfamily 2312.08025}}.

\bibitem{Basak:2023bnc}
J.~K. Basak, D.~Basu, V.~Malvimat, H.~Parihar and G.~Sengupta,
  \emph{{Holographic Reflected Entropy and Islands in Interface CFTs}},
  \href{https://arxiv.org/abs/2312.12512}{{\ttfamily 2312.12512}}.

\bibitem{Jeong:2023lkc}
H.-S. Jeong, K.-Y. Kim and Y.-W. Sun, \emph{{Entanglement entropy analysis of
  dyonic black holes using doubly holographic theory}},
  \href{https://doi.org/10.1103/PhysRevD.108.126016}{\emph{Phys. Rev. D}
  {\bfseries 108} (2023) 126016}
  [\href{https://arxiv.org/abs/2305.18122}{{\ttfamily 2305.18122}}].

\bibitem{Kawamoto:2023wzj}
T.~Kawamoto, S.-M. Ruan and T.~Takayanagia, \emph{{Gluing AdS/CFT}},
  \href{https://doi.org/10.1007/JHEP07(2023)080}{\emph{JHEP} {\bfseries 07}
  (2023) 080} [\href{https://arxiv.org/abs/2303.01247}{{\ttfamily
  2303.01247}}].

\bibitem{Bachas:2021fqo}
C.~Bachas and V.~Papadopoulos, \emph{{Phases of Holographic Interfaces}},
  \href{https://doi.org/10.1007/JHEP04(2021)262}{\emph{JHEP} {\bfseries 04}
  (2021) 262} [\href{https://arxiv.org/abs/2101.12529}{{\ttfamily
  2101.12529}}].

\bibitem{Freedman:2016zud}
M.~Freedman and M.~Headrick, \emph{{Bit threads and holographic entanglement}},
  \href{https://doi.org/10.1007/s00220-016-2796-3}{\emph{Commun. Math. Phys.}
  {\bfseries 352} (2017) 407}
  [\href{https://arxiv.org/abs/1604.00354}{{\ttfamily 1604.00354}}].

\bibitem{Yoshida:2017non}
B.~Yoshida and A.~Kitaev, \emph{{Efficient decoding for the Hayden-Preskill
  protocol}},  \href{https://arxiv.org/abs/1710.03363}{{\ttfamily 1710.03363}}.

\bibitem{Affleck:1991tk}
I.~Affleck and A.~W.~W. Ludwig, \emph{{Universal noninteger 'ground state
  degeneracy' in critical quantum systems}},
  \href{https://doi.org/10.1103/PhysRevLett.67.161}{\emph{Phys. Rev. Lett.}
  {\bfseries 67} (1991) 161}.

\bibitem{Takayanagi:2011zk}
T.~Takayanagi, \emph{{Holographic Dual of BCFT}},
  \href{https://doi.org/10.1103/PhysRevLett.107.101602}{\emph{Phys. Rev. Lett.}
  {\bfseries 107} (2011) 101602}
  [\href{https://arxiv.org/abs/1105.5165}{{\ttfamily 1105.5165}}].

\bibitem{Fujita:2011fp}
M.~Fujita, T.~Takayanagi and E.~Tonni, \emph{{Aspects of AdS/BCFT}},
  \href{https://doi.org/10.1007/JHEP11(2011)043}{\emph{JHEP} {\bfseries 11}
  (2011) 043} [\href{https://arxiv.org/abs/1108.5152}{{\ttfamily 1108.5152}}].

\bibitem{Israel:1966rt}
W.~Israel, \emph{{Singular hypersurfaces and thin shells in general
  relativity}}, \href{https://doi.org/10.1007/BF02710419,
  10.1007/BF02712210}{\emph{Nuovo Cim.} {\bfseries B44S10} (1966) 1}.

\bibitem{Emparan:1999pm}
R.~Emparan, C.~V. Johnson and R.~C. Myers, \emph{{Surface terms as counterterms
  in the AdS / CFT correspondence}},
  \href{https://doi.org/10.1103/PhysRevD.60.104001}{\emph{Phys. Rev. D}
  {\bfseries 60} (1999) 104001}
  [\href{https://arxiv.org/abs/hep-th/9903238}{{\ttfamily hep-th/9903238}}].

\bibitem{Hayward:1993my}
G.~Hayward, \emph{{Gravitational action for space-times with nonsmooth
  boundaries}}, \href{https://doi.org/10.1103/PhysRevD.47.3275}{\emph{Phys.
  Rev. D} {\bfseries 47} (1993) 3275}.

\bibitem{Brill:1994mb}
D.~Brill and G.~Hayward, \emph{{Is the gravitational action additive?}},
  \href{https://doi.org/10.1103/PhysRevD.50.4914}{\emph{Phys. Rev. D}
  {\bfseries 50} (1994) 4914}
  [\href{https://arxiv.org/abs/gr-qc/9403018}{{\ttfamily gr-qc/9403018}}].

\bibitem{Bachas:2007td}
C.~Bachas and I.~Brunner, \emph{{Fusion of conformal interfaces}},
  \href{https://doi.org/10.1088/1126-6708/2008/02/085}{\emph{JHEP} {\bfseries
  02} (2008) 085} [\href{https://arxiv.org/abs/0712.0076}{{\ttfamily
  0712.0076}}].

\bibitem{Hartman:2014oaa}
T.~Hartman, C.~A. Keller and B.~Stoica, \emph{{Universal Spectrum of 2d
  Conformal Field Theory in the Large c Limit}},
  \href{https://doi.org/10.1007/JHEP09(2014)118}{\emph{JHEP} {\bfseries 09}
  (2014) 118} [\href{https://arxiv.org/abs/1405.5137}{{\ttfamily 1405.5137}}].

\bibitem{Cooper:2018cmb}
S.~Cooper, M.~Rozali, B.~Swingle, M.~Van~Raamsdonk, C.~Waddell and D.~Wakeham,
  \emph{{Black Hole Microstate Cosmology}},
  \href{https://doi.org/10.1007/JHEP07(2019)065}{\emph{JHEP} {\bfseries 07}
  (2019) 065} [\href{https://arxiv.org/abs/1810.10601}{{\ttfamily
  1810.10601}}].

\bibitem{Hayden:2007cs}
P.~Hayden and J.~Preskill, \emph{{Black holes as mirrors: Quantum information
  in random subsystems}},
  \href{https://doi.org/10.1088/1126-6708/2007/09/120}{\emph{JHEP} {\bfseries
  09} (2007) 120} [\href{https://arxiv.org/abs/0708.4025}{{\ttfamily
  0708.4025}}].

\bibitem{Nakayama:2023kgr}
Y.~Nakayama, A.~Miyata and T.~Ugajin, \emph{{The Petz (lite) recovery map for
  the scrambling channel}},
  \href{https://doi.org/10.1093/ptep/ptad147}{\emph{PTEP} {\bfseries 2023}
  (2023) 123B04} [\href{https://arxiv.org/abs/2310.18991}{{\ttfamily
  2310.18991}}].

\bibitem{Yoshidaunp}
B.~Yoshida, \emph{{Unpublished note}}, .

\bibitem{Jensen:2013ora}
K.~Jensen and A.~Karch, \emph{{Holographic Dual of an Einstein-Podolsky-Rosen
  Pair has a Wormhole}},
  \href{https://doi.org/10.1103/PhysRevLett.111.211602}{\emph{Phys. Rev. Lett.}
  {\bfseries 111} (2013) 211602}
  [\href{https://arxiv.org/abs/1307.1132}{{\ttfamily 1307.1132}}].

\bibitem{Sonner:2013mba}
J.~Sonner, \emph{{Holographic Schwinger Effect and the Geometry of
  Entanglement}},
  \href{https://doi.org/10.1103/PhysRevLett.111.211603}{\emph{Phys. Rev. Lett.}
  {\bfseries 111} (2013) 211603}
  [\href{https://arxiv.org/abs/1307.6850}{{\ttfamily 1307.6850}}].

\bibitem{Hubeny:2014kma}
V.~E. Hubeny and G.~W. Semenoff, \emph{{String worldsheet for accelerating
  quark}}, \href{https://doi.org/10.1007/JHEP10(2015)071}{\emph{JHEP}
  {\bfseries 10} (2015) 071} [\href{https://arxiv.org/abs/1410.1171}{{\ttfamily
  1410.1171}}].

\bibitem{Ugajin:2013xxa}
T.~Ugajin, \emph{{Two dimensional quantum quenches and holography}},
  \href{https://arxiv.org/abs/1311.2562}{{\ttfamily 1311.2562}}.

\bibitem{Erdmenger:2017gdk}
J.~Erdmenger, D.~Fernandez, M.~Flory, E.~Megias, A.-K. Straub and P.~Witkowski,
  \emph{{Time evolution of entanglement for holographic steady state
  formation}}, \href{https://doi.org/10.1007/JHEP10(2017)034}{\emph{JHEP}
  {\bfseries 10} (2017) 034}
  [\href{https://arxiv.org/abs/1705.04696}{{\ttfamily 1705.04696}}].

\bibitem{Gubser:2002tv}
S.~S. Gubser, I.~R. Klebanov and A.~M. Polyakov, \emph{{A Semiclassical limit
  of the gauge / string correspondence}},
  \href{https://doi.org/10.1016/S0550-3213(02)00373-5}{\emph{Nucl. Phys. B}
  {\bfseries 636} (2002) 99}
  [\href{https://arxiv.org/abs/hep-th/0204051}{{\ttfamily hep-th/0204051}}].

\bibitem{Kim:2014bga}
J.~Kim and M.~Porrati, \emph{{Long string dynamics in pure gravity on
  AdS$_{3}$}}, \href{https://doi.org/10.1134/S1063776115030097}{\emph{J. Exp.
  Theor. Phys.} {\bfseries 120} (2015) 477}
  [\href{https://arxiv.org/abs/1410.3424}{{\ttfamily 1410.3424}}].

\bibitem{Kim:2015bba}
J.~Kim and M.~Porrati, \emph{{More on long string dynamics in gravity on
  AdS$_3$ : Spinning strings and rotating BTZ black holes}},
  \href{https://doi.org/10.1103/PhysRevD.91.124061}{\emph{Phys. Rev. D}
  {\bfseries 91} (2015) 124061}
  [\href{https://arxiv.org/abs/1503.06875}{{\ttfamily 1503.06875}}].

\bibitem{Bachas:2021tnp}
C.~Bachas, Z.~Chen and V.~Papadopoulos, \emph{{Steady states of holographic
  interfaces}}, \href{https://doi.org/10.1007/JHEP11(2021)095}{\emph{JHEP}
  {\bfseries 11} (2021) 095}
  [\href{https://arxiv.org/abs/2107.00965}{{\ttfamily 2107.00965}}].

\bibitem{Maxfield:2022rry}
H.~Maxfield and Z.~Wang, \emph{{Gravitating spinning strings in AdS$_{3}$}},
  \href{https://doi.org/10.1007/JHEP07(2022)075}{\emph{JHEP} {\bfseries 07}
  (2022) 075} [\href{https://arxiv.org/abs/2203.02492}{{\ttfamily
  2203.02492}}].

\bibitem{Papadopoulos:2023kyd}
V.~Papadopoulos, \emph{{Membranes, holography, and quantum information}},
  \href{https://arxiv.org/abs/2310.18521}{{\ttfamily 2310.18521}}.

\bibitem{Akal:2021foz}
I.~Akal, Y.~Kusuki, N.~Shiba, T.~Takayanagi and Z.~Wei, \emph{{Holographic
  moving mirrors}},
  \href{https://doi.org/10.1088/1361-6382/ac2c1b}{\emph{Class. Quant. Grav.}
  {\bfseries 38} (2021) 224001}
  [\href{https://arxiv.org/abs/2106.11179}{{\ttfamily 2106.11179}}].

\bibitem{Akal:2020wfl}
I.~Akal, Y.~Kusuki, T.~Takayanagi and Z.~Wei, \emph{{Codimension two holography
  for wedges}}, \href{https://doi.org/10.1103/PhysRevD.102.126007}{\emph{Phys.
  Rev. D} {\bfseries 102} (2020) 126007}
  [\href{https://arxiv.org/abs/2007.06800}{{\ttfamily 2007.06800}}].

\bibitem{Emparan:1999wa}
R.~Emparan, G.~T. Horowitz and R.~C. Myers, \emph{{Exact description of black
  holes on branes}},
  \href{https://doi.org/10.1088/1126-6708/2000/01/007}{\emph{JHEP} {\bfseries
  01} (2000) 007} [\href{https://arxiv.org/abs/hep-th/9911043}{{\ttfamily
  hep-th/9911043}}].

\bibitem{Emparan:1999fd}
R.~Emparan, G.~T. Horowitz and R.~C. Myers, \emph{{Exact description of black
  holes on branes. 2. Comparison with BTZ black holes and black strings}},
  \href{https://doi.org/10.1088/1126-6708/2000/01/021}{\emph{JHEP} {\bfseries
  01} (2000) 021} [\href{https://arxiv.org/abs/hep-th/9912135}{{\ttfamily
  hep-th/9912135}}].

\bibitem{Emparan:2000fn}
R.~Emparan, R.~Gregory and C.~Santos, \emph{{Black holes on thick branes}},
  \href{https://doi.org/10.1103/PhysRevD.63.104022}{\emph{Phys. Rev. D}
  {\bfseries 63} (2001) 104022}
  [\href{https://arxiv.org/abs/hep-th/0012100}{{\ttfamily hep-th/0012100}}].

\bibitem{Kanti:2004nr}
P.~Kanti, \emph{{Black holes in theories with large extra dimensions: A
  Review}}, \href{https://doi.org/10.1142/S0217751X04018324}{\emph{Int. J. Mod.
  Phys. A} {\bfseries 19} (2004) 4899}
  [\href{https://arxiv.org/abs/hep-ph/0402168}{{\ttfamily hep-ph/0402168}}].

\bibitem{Majumdar:2005ba}
A.~S. Majumdar and N.~Mukherjee, \emph{{Braneworld black holes in cosmology and
  astrophysics}}, \href{https://doi.org/10.1142/S0218271805006948}{\emph{Int.
  J. Mod. Phys. D} {\bfseries 14} (2005) 1095}
  [\href{https://arxiv.org/abs/astro-ph/0503473}{{\ttfamily
  astro-ph/0503473}}].

\bibitem{Fitzpatrick:2006cd}
A.~L. Fitzpatrick, L.~Randall and T.~Wiseman, \emph{{On the existence and
  dynamics of braneworld black holes}},
  \href{https://doi.org/10.1088/1126-6708/2006/11/033}{\emph{JHEP} {\bfseries
  11} (2006) 033} [\href{https://arxiv.org/abs/hep-th/0608208}{{\ttfamily
  hep-th/0608208}}].

\bibitem{Creek:2006je}
S.~Creek, R.~Gregory, P.~Kanti and B.~Mistry, \emph{{Braneworld stars and black
  holes}}, \href{https://doi.org/10.1088/0264-9381/23/23/004}{\emph{Class.
  Quant. Grav.} {\bfseries 23} (2006) 6633}
  [\href{https://arxiv.org/abs/hep-th/0606006}{{\ttfamily hep-th/0606006}}].

\bibitem{Gregory:2008rf}
R.~Gregory, \emph{{Braneworld black holes}},
  \href{https://doi.org/10.1007/978-3-540-88460-6_7}{\emph{Lect. Notes Phys.}
  {\bfseries 769} (2009) 259}
  [\href{https://arxiv.org/abs/0804.2595}{{\ttfamily 0804.2595}}].

\bibitem{Figueras:2011gd}
P.~Figueras and T.~Wiseman, \emph{{Gravity and large black holes in
  Randall-Sundrum II braneworlds}},
  \href{https://doi.org/10.1103/PhysRevLett.107.081101}{\emph{Phys. Rev. Lett.}
  {\bfseries 107} (2011) 081101}
  [\href{https://arxiv.org/abs/1105.2558}{{\ttfamily 1105.2558}}].

\bibitem{Akal:2021dqt}
I.~Akal, T.~Kawamoto, S.-M. Ruan, T.~Takayanagi and Z.~Wei, \emph{{Page curve
  under final state projection}},
  \href{https://doi.org/10.1103/PhysRevD.105.126026}{\emph{Phys. Rev. D}
  {\bfseries 105} (2022) 126026}
  [\href{https://arxiv.org/abs/2112.08433}{{\ttfamily 2112.08433}}].

\bibitem{Kawamoto:2023nki}
T.~Kawamoto, S.-M. Ruan, Y.-k. Suzuki and T.~Takayanagi, \emph{{A half de
  Sitter holography}},
  \href{https://doi.org/10.1007/JHEP10(2023)137}{\emph{JHEP} {\bfseries 10}
  (2023) 137} [\href{https://arxiv.org/abs/2306.07575}{{\ttfamily
  2306.07575}}].

\bibitem{Balasubramanian:1999re}
V.~Balasubramanian and P.~Kraus, \emph{{A Stress tensor for Anti-de Sitter
  gravity}}, \href{https://doi.org/10.1007/s002200050764}{\emph{Commun. Math.
  Phys.} {\bfseries 208} (1999) 413}
  [\href{https://arxiv.org/abs/hep-th/9902121}{{\ttfamily hep-th/9902121}}].

\end{thebibliography}\endgroup

\end{document}